# Decoding Starlight with Big Survey Data, Machine Learning, and Cosmological Simulations

Kirsten N. Blancato

Submitted in partial fulfillment of the
requirements for the degree
of Doctor of Philosophy
under the Executive Committee of the
Graduate School of Arts and Sciences

COLUMBIA UNIVERSITY

2020

# ABSTRACT

# Decoding Starlight with Big Survey Data, Machine Learning, and Cosmological Simulations

Kirsten N. Blancato

Stars, and collections of stars, encode rich signatures of stellar physics and galaxy evolution. With properties influenced by both their environment and intrinsic nature, stars retain information about astrophysical phenomena that are not otherwise directly observable. In the time-domain, the observed brightness variability of a star can be used to investigate physical processes occurring at the stellar surface and in the stellar interior. On a galactic scale, comparatively fixed properties of stars, including chemical abundances and stellar ages, serve as a multi-dimensional record of the origin of the galaxy. In the Milky Way, together with orbital properties, this informs the details of the subsequent evolution of our Galaxy since its formation. Extending beyond the Local Group, the attributes of unresolved stellar populations allow us to study the diversity of galaxies in the Universe.

By examining the properties of stars, and how they vary across a range of spatial and temporal scales, this Dissertation connects the information residing within stars, to global processes in galactic formation and evolution. We develop new approaches to determine stellar properties, including rotation and surface gravity, from the variability that we directly observe. We offer new insight into the chemical enrichment history of the Milky Way, tracing different stellar explosions, that capture billions of years of evolution. We advance knowledge and understanding of how stars and galaxies are linked, by examining differences in the initial stellar mass distributions comprising galaxies, as they form. In building up this knowledge, we highlight current tensions between data and theory. By synthesizing numerical simulations, large observational data sets, and machine learning techniques, this work makes valuable methodological contributions to maximize insights from diverse ensembles of current and future stellar observations.

# Contents

# List of Figures

# List of Tables

# ACKNOWLEDGMENTS

*The young women of today to whom all doors are ajar can scarcely realize the difficulties or experience the enthusiasm of those who more than half a century ago pushed open the doors of opportunity in the various departments of learning.*

– Annie Jump Cannon, 1927, writing on the passing of Sarah Frances Whiting (Cannon 1927a)

While the Ph.D. is often viewed as an individual accomplishment, this Dissertation was only possible with the help of an incredible team in my corner.

First and foremost, thank you to Melissa Ness for taking me on as your first Ph.D. student at Columbia. I've been truly fortunate to have learned from such an incredible scientist and kind person. Thank you for encouraging me to express and be myself, and for providing a space to become comfortable with asking for help and saying "I don't know", which is often when the real learning and growth begins. Thank you as well to Kathryn Johnston, for mentoring me alongside Melissa, and for always providing a trusted perspective on all aspects of my work, graduate school, and beyond. Thank you to Shy Genel and Greg Bryan for your kindness in guiding me through my first research project at Columbia, and in the development of my thesis proposal.

Thank you to further committee members and research collaborators: David Schiminovich, Megan Bedell, Jan Rybizki, Daniel Huber, Yuxi(Lucy) Lu, and Ruth Angus, all of whom I've enjoyed working with and learned a lot from. Thank you to the astronomy department staff, Mildred Kramer-Garcia and Ayoune Payne, as well as the Pupin Hall janitorial staff. The good-mornings and passing chats always brightened my day. And thank you to Moiya McTier, Yong Zheng, Rayna Rampalli, Nicole Melso, Adrian Lucy, Daniel DeFelippis, Mihir Kulkarni, Alex Teachey, Julia Blue Bird, Jorge Cortés, Steven Mohammed, Stephanie Douglas, Adam Wheeler, and Rebecca Oppenheimer, for your friendship, advice, and help at various points throughout this experience.

I would not have pursued a Ph.D. in astronomy if it weren't for Kim McLeod. As a first-generation college student, I arrived at Wellesley and immediately felt out of place and in over my head. Kim's support, guidance, and kindness made a world of a difference and helped make the astronomy department at Wellesley feel like a home. Spending so much time observing with the 24-inch telescope and the beautiful, historical 12-inch refractor are some of my fondest memories of Wellesley, and Whitin Observatory will always hold a special place in my heart. Thank you as well to Anicia Arrendondo, James Battat, Wendy Bauer, Steve Slivan, Carol Gagosian, Kaća Bradonjić, Yue Hu, Robbie Berg, and all of my wonderful Wellesley friends.

Two and a half years ago I stepped outside of my comfort zone and into Mendez Boxing gym for my first boxing lesson with Coach Reese. I've had so much fun being a student of "the sweet science", working through all the lows, highs, and setbacks that come along with it, and learning how to slowly push myself past my limits while being kind to myself in the process. Nothing compares to those moments in the ring when everything connects, and the rest of the world melts away. Seeing Reese bring her vision of Women's World of Boxing to life has been so special. The kindness she meets everyone with that she trains is something I admire deeply and hope to bring to whatever I do. It has been such a privilege to get to know everyone at the gym. Boxing helped me through the final rounds of my Ph.D. and gave me so many new ways to view myself and life.

Thank you to my Mom, my sister Tori, Steve, Dad, and Wanda for your support while I've spent the last 9 years musing over stars and galaxies. Mom, you fought so hard for me to be able to pursue an education and the choices that come along with one. Thanks to you, I have been able to dream bigger than I ever could have imagined. While I value my formal education deeply, I have also realized that the education I've gained listening to the communities and people around me is just as, if not more, valuable, and that what matters most is how I can use my own education to help open doors for others.

xxi

Last, but not least, thank you to my partner and soon to be wife, Rose. I'm so incredibly lucky that you followed me from Wellesley to Columbia. Thank you for being my greatest champion over the past six years in everything that I do, for always encouraging me to finish the Ph.D., and for being a big goofball with me all along the way. My life with you is the greatest gift that astronomy has brought me, and I can't wait to tackle whatever may be next with you by my side.

<div style="text-align: right">June 2020, Manhattan</div>

# Chapter 1

# Introduction

*Since 1882, with increasing skill, astronomers have been able to photograph star light in such a manner that the marvelous wireless message from the distant body may be deciphered... The results will help to unravel some of the mysteries of the great universe, visible to us, in the depths above. They will provide material for investigation of those distant suns of which we know nothing except as revealed by the rays of light, travelling for years with great velocity through space, to be made at last to their magical story on our photographic plates.*
– Annie Jump Cannon, 1915[1]

## 1.1 Historical context and overview

### 1.1.1 The legacy of astronomy at Wellesley College

In the mid-1800s, after the afflictions of the wars in 1776 and 1812 abated, the foundation of what would become modern Western astronomy in the United States was beginning to develop. Enabled by technological advances that facilitated the production and dissemination of printed works, scientific texts of astronomy were being circulated more widely than ever

---
[1] http://academics.wellesley.edu/Astronomy/Annie/understanding.html (Website author: Logan Hennessey)

before (Cottam & Orchiston 2015). In 1855, Hannah Mary Bouvier Peterson authored one of the earliest original astronomical texts in the United States, "Familiar Astronomy: or, An Introduction to the Study of the Heavens" (Peterson 1856), which was used to teach astronomy in schools[2]. During this time, a number of women's colleges were being established to expand access to higher education at a time when most universities were closed to women, and some would remain so for nearly another century. Chartered in 1870, Wellesley College in Massachusetts, one of the women's colleges comprising the "Seven Sisters", opened its doors and began admitting students in 1875.

The founders of Wellesley College, Pauline and Henry Durant, were driven to offer an excellent science education alongside traditional liberal arts studies. Inspired by his friend Edward Pickering, who established the first teaching-focused physics laboratory at Massachusetts Institute of Technology (MIT), Mr. Durant sought to bring a similar education to the women of Wellesley. He identified Sarah Frances Whiting, a then teacher at the Brooklyn Heights Seminary school, to come to Wellesley and establish its physics department. Attending Pickering's lectures at MIT, Whiting learned various experimental techniques and developed the curriculum for the country's first undergraduate experimental physics lab accessible to women (Cannon 1927b; Stahl 2005). As recalled by one of her most notable students, Annie Jump Cannon,

> "Miss Whiting had the pleasure of doing many interesting things, none of which perhaps was more exciting than her early experiments with X-rays. After reading in a Boston morning paper of the existence and discovery of these Röntgen rays, she immediately set up an old Crookes' tube. A picture hook and a coin within a flat purse were placed over a photographic plate in its holder, slide inserted, and, after the holder was wrapped in black cloth, an exposure was made to the rays

---

[2]http://www.gardenstatelegacy.com/files/Ambassadors_to_the_Heavens_Malpas_GSL1.pdf

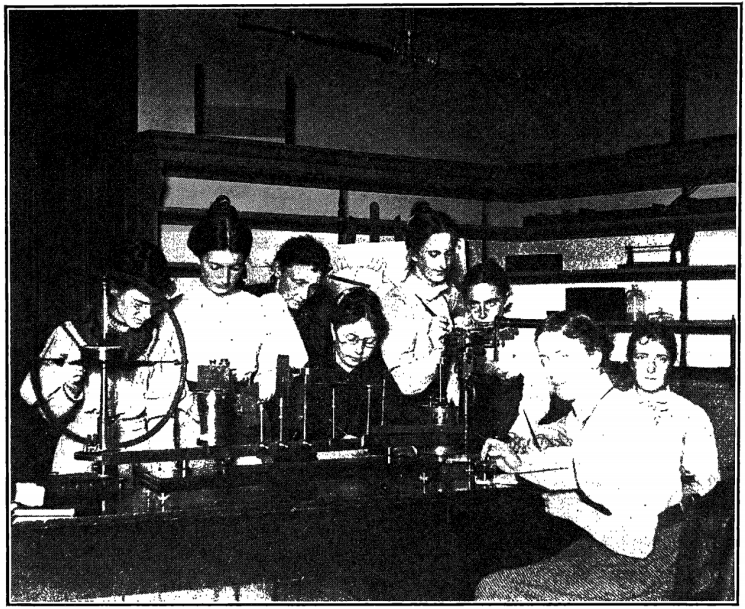

MISS WHITING AND HER CLASS IN ADVANCED PHYSICS, 1895-96.

Figure 1.1: Physics lab at Wellesley College in 1895/1896 with Sarah Francis Whiting and Annie Jump Cannon. Image credit: Image 2 of Cannon (1927a) which is provided by the NASA Astrophysics Data System courtesy of Maria Mitchell Observatory.

emanating from the tube. What uncanny feelings the shadows of that picture hook and coin produced on those who developed that first plate!" (Cannon 1927a)

Whiting's enthusiasm for the experimental study of light soon led to her interest in astronomy. In 1879, she began her curriculum in astronomy with a particular focus on hands-on instruction in instrumentation and the methods of photometry, optics, and spectroscopy. Before Wellesley's Observatory was built, she focused on instructional activities that could be performed during the day, having students sketch spectral features of the sun and other elemental substances, chart the motion of stars using celestial globes, and analyze photographic plates obtained from Mr. Pickering, who by then had moved to the Harvard College Observatory (Stahl 2005). Sarah Frances Whiting published many of her astronomy activities as a series of articles, including "Spectroscopic Work for Classes in Astronomy" and "The Use of Photographs in Teaching Astronomy" (Whiting 1905d,c,a,b). She also wrote of her pedagogical philosophy in "A Pedagogical Suggestion for Teachers of Astronomy" (Whiting 1912) and views on the purpose of education in astronomy in "Daytime and evening exercises in astronomy, for schools and colleges", articulating that,

> "Hope expressed that these exercises may form a slight contribution to the better teaching of the only science dealing with matter which takes thought off this little planet, and gives those larger conceptions of time and space which stretch the mind and furnish proper perspective for other subjects." (Whiting 2018)

In 1898, at an annual "Float night" dinner hosted by Whiting, one of Wellesley's trustees, Mrs. Whitin, learned about the college's wish to build a proper observatory. At this party, Mrs. Whitin, who had a longtime interest in the stars, agreed to finance and help plan the construction of an observatory for Wellesley (Cannon 1927a). Built from white marble, Whitin Observatory opened its doors to students on October 8th, 1900, with Sarah Frances Whiting as the first director. The 20-foot telescope dome housed a 12-inch refractor, originally built by Henry Fitz and later refurbished by Alvan Clark & Sons, which the college purchased from a banker in Brooklyn (Sewall 1997). The telescope remains today, and over the past century, it has provided a view of the celestial heavens for thousands of Wellesley students.

In 1880, Annie Jump Cannon, the foremother of stellar classification, began her career in astronomy at Wellesley College. Cannon, born in Dover Delaware in 1863, learned the constellations from her mother as a child and was encouraged by her to pursue further education (Jardins 2010). Cannon arrived at Wellesley to study physics, working closely under the guidance of Whiting to learn the methods of photometry and spectroscopy. Cannon wrote of Whiting that,

> "For of all branches of physics and astronomy, she was most keen on the spectroscopic development. Even at her Observatory receptions, she always had the spectra of various elements on exhibition. So great was her interest in the subject that she infused into the mind of her pupil who is writing these lines, a desire to continue the investigation of spectra." (Cannon 1927a)

After her education at Wellesley, Cannon moved back to Delaware until she eventually returned to Massachusetts in 1894. Over the next few years, she helped teach astronomy labs at Wellesley and began doing research in Edward Pickering's group at the Harvard College Observatory. She became a part of what is now referred to as the "Harvard Computers", a group of women hired by Mr. Pickering to perform detailed studies of stellar spectra captured on photographic plates. Cannon built on the work of Nettie Farrar, Williamina Fleming, and Antonia Maury, and by-eye she examined the spectra of over 250,000 stars[3]. Studying the detailed absorption patterns of tens of thousands of stars, she developed a classification scheme in which stars are categorized into eight spectral classes, denoted by the letters OBAFGKM, which corresponds to a continuum of hot and young O-type stars to the cool and older M-type stars. By 1915, the prospect of using stellar spectra to learn about various other stellar properties was starting to be realized, with Cannon writing "In very recent years remarkable relations have been found to exist between the class of spectrum and other properties of the stars, such as distances and motions."[4] Cannon's classification scheme is still used by astronomers today, and her work serves as a foundation of modern stellar astrophysics.

## 1.1.2  From Annie Jump Cannon to this Dissertation

Since the founding of Whitin Observatory and the pioneering work of Annie Jump Cannon, the fundamental measurements we make of stars, namely photometric, spectroscopic, and astrometric measurements, has remained unchanged. However, the quantity and quality of these measurements have significantly advanced. This is primarily due to innovations in astronomical instrumentation and has also been facilitated by innovations in computer technology. The level of granularity that we can probe in our stellar observations is dependent on our instrument capabilities, and the physical limitations of our observing circumstances. In the Milky Way and nearby, we can resolve individual stars. Beyond the extent of the Milky Way, we typically cannot resolve large numbers of individual stars. Instead, we make measurements of the integrated stellar light emitted by galaxies, and then use these measurements to infer properties of the overall population of stars. These stellar observations have pushed forward our understanding of stars, and have uncovered new ways in which the properties of stars can be used to probe a variety of astrophysical objects and phenomena.

---

[3] https://www.sdsc.edu/ScienceWomen/cannon.html
[4] http://academics.wellesley.edu/Astronomy/Annie/understanding.html (Website author: Logan Hennessey)

In tandem, with access to large amounts of data, recent developments in our computational power have revolutionized many pursuits both within and outside of science. With these new capabilities come new challenges, and we are being confronted with difficult ethical and legal questions, from fairness and bias in algorithmic decisions to data ownership and privacy. Astrophysics is no exception and is also being transformed by advancements in computer technology, giving us new and more sophisticated tools that we can use to examine large and complex astronomical data sets. There are also challenges to leveraging these tools in astronomy, including training in these new techniques, as well as collaboration both across scientific disciplines and within the astronomy community. One such methodological challenge is the interpretability of machine learning models, as many "off-the-shelf" approaches have not been developed with the goals of science in mind. However, with a combination of effective collaboration and utilization of specialized domain knowledge, machine learning has the potential to reveal new connections between observations and underlying astrophysical theory.

Against this historical and technological background, in this Dissertation, we use a variety of approaches to extract and interpret information encoded in stellar photometry, spectroscopy, and astrometry, with the goal of better understanding the light we receive from stars.

In Section 1.2, we begin our discussion starting with individual stars. We provide some basic physical understanding of stars and their evolution over time, and then detail the three primary ways in which we make measurements of individual stars, photometry, spectroscopy, and astrometry, with a particular focus on highlighting past, current, and future surveys that have and will contribute to our growing assortment of stellar observations.

Building from this knowledge, in Section 1.3 we then discuss some properties of stellar populations and galaxies, each of which contains collections of individual stars. We begin with a basic overview of singular simple stellar populations and then build further to discuss galaxies, whose stellar contents are each composed of a mixture of numerous stellar populations. We also overview some of the surveys that have made measurements of these stellar collections.

Next, in Section 1.4, we present a high-level overview of some of the progress in computer technology that has been occurring over the past few decades. This progress has increased our computational capacity, enabling the continual development and improvement of tools that we can use to make interpretations of astronomical data sets. We discuss two such tools that are relevant to this astronomy and this Dissertation: cosmological simulations and machine learning. We conclude this section with a brief discussion of the recent Milky Way science that these tools are enabling.

Finally, in Section 1.5, we outline the original contributions of this Dissertation. This work is built on a foundation consisting of our physical understanding of stars, the various stellar and galactic observations we have collected, and technological innovation that is advancing the tools we have to make sense of this theory and data.

## 1.2 Stars

A cornerstone modern astronomy research has been to interpret the various properties of stars that we can measure, including their chemical abundances, ages, orbital parameters, as well as the types of stars that make up a galaxy, and understand how these properties can be used to make sense of various astrophysical phenomena. For example, in the context of the Milky Way, this pursuit, termed "Galactic archaeology", aims to use stellar properties as a multi-dimensional fossil record to probe the current structure of the Milky Way, as well as better understand how it has arrived at this current state. To make sense of the wealth of information encoded in stellar properties, we must first understand stars themselves. This includes the physics of their interiors, how they evolve over time, and how collections of stars interact with their broader galactic environment. For more comprehensive reviews of stars and their evolution, see Kippenhahn et al. (2012) and Prialnik (2000).

### 1.2.1 Stellar evolution and the theoretical HR diagram

A star is a self-gravitating, approximately spherical body, that radiates energy produced within itself. This energy is typically generated from the release of gravitational potential as the star contracts, and from nuclear fusion reactions taking place inside the star. As stars release energy over time, their internal structure and elemental composition evolve. The evolutionary path that an individual star follows over its lifetime, as well as its final end-state, is dependent on the star's initial mass. Low-mass stars ($M_* < 8\ M_\odot$) generally end their lives as a white dwarf. High-mass stars ($M_* > 8\ M_\odot$), on the other hand, have the eventual fate of becoming either a neutron star or black hole. The initial mass, as well as the surface chemical composition of a star (for most elements), is set at the time of stellar birth, with stars forming from the fragmentation and gravitational collapse of giant molecular clouds in the interstellar medium (ISM). The earliest stars to form, hypothesized as population III stars, were formed from gas directly produced by the Big Bang, composed of hydrogen and helium, and some trace amounts of lithium, that has not yet been enriched by subsequent generations of star-formation. However, as these first stars exploded and injected their metals into the ISM, the next generations of stars to form inherit the chemical composition of this enriched gas.

The Hertzsprung-Russell (HR) diagram is a fundamental tool for understanding stars and their evolutionary paths. The theoretical HR diagram demonstrates what regions of the temperature-luminosity parameter space that stars of all ages inhabit. Shown along the $x$-axis is typically the effective temperature, $T_{\text{eff}}$, which is an estimate of a star's surface temperature. The stellar classification scheme developed by Annie Jump Cannon corresponds to a sequence in temperature, with O-type stars being the hottest and bluest, reaching temperatures of $T_{\text{eff}} \sim 25{,}000$ K and M-type stars being the coolest and reddest, having temperatures as low as $T_{\text{eff}} \sim 3{,}000$ K. Shown along the $y$-axis is typically luminosity, $L$, which is a measure of how bright a star is, or more precisely the amount of energy a star emits per unit of time. The luminosity of a star is usually expressed relative to the luminosity of the Sun ($L_\odot = 4 \times 10^{26}$ Watts), with stellar luminosity values generally ranging from $10^{-4} L_\odot$ to $10^6 L_\odot$. Distinct structure is exhibited in many HR diagrams, with stars clustering into groups and sequences. The Yerkes stellar classification scheme (also known as the MK classification scheme) classifies stars based on these clusters in the $T_{\text{eff}}$-$L$ plane, including five main types of stars: main sequence stars, subgiant stars, giant stars, supergiant stars, and white dwarf stars.

The main sequence is the most prominent feature of the HR diagram, and where stars spend the majority of their lives. It spans diagonally across the $T_{\rm eff}$-$L$ plane, from stars with $T_{\rm eff} \sim 25{,}000$ K and $L = 10^6 L_\odot$ to stars with $T_{\rm eff} \sim 3{,}000$ K and $L = 10^{-4} L_\odot$. Stars in the main sequence are in hydrostatic and thermal equilibrium and generate energy through the fusion hydrogen (H) to helium (He) in their cores. The location of an individual star along the main sequence is dependent on its initial mass, with high-mass stars being hotter and more luminous than low-mass stars, roughly following the relation $L \propto M_*^4$. Stars on the lower main sequence, with masses generally $M_* > 1.1\ M_\odot$, have radiative cores and convective envelopes, and primarily fuse H to He via the proton-proton chain. Stars on the upper main sequence, with masses generally $M_* < 1.1\ M_\odot$, have convective cores and radiative envelopes, and primarily fuse H to He via the CNO (carbon-nitrogen-oxygen) cycle. The timescales at which main sequence stars burn through their hydrogen is dependent on initial stellar mass as well, with the main sequence lifetime, $t_{\rm MS}$, roughly following the relation $t_{\rm MS} \propto M_*^{-3}$. This is because nuclear fusion is highly dependent on stellar core temperature, making it so that high-mass stars fuse H to He much more rapidly than low-mass stars. The main sequence lifetime of the Sun is $\sim 10$ Gyrs, while $t_{\rm MS} \sim 10^4$ Gyrs for a 0.1 $M_\odot$ star and $t_{\rm MS} \sim 0.01$ Gyrs for a 10 $M_\odot$ star.

For low- and intermediate-mass ($< 8 M_\odot$), the remaining groupings of stars on the theoretical HR diagram roughly correspond to their evolutionary path. Subgiant stars, which are more luminous than main sequence stars at a given temperature, are the precursors to giant stars. As main sequence stars with $M_* < 8\ M_\odot$ burn through the final stores of hydrogen in their cores, their temperature, and thus brightness, increases. After hydrogen is completely exhausted, the cores of these stars collapse, and hydrogen burning commences in their shells. With this increased generation of energy in the shell, these stars begin to expand and cool to lower temperatures, evolving to giant stars. The time stars spend in the subgiant phase is relatively short ($\sim 10^6$ yrs), which is manifested in the HR diagram as a sparsely populated gap between the main sequence and giant branch.

Giant stars, which form what is commonly referred to as the red giant branch on the HR diagram, generally have temperatures of $T_{\text{eff}} < 5{,}000$ K (G-, K-, and M-type stars) and span a range of luminosities from $10^3$ to $10^5$ $L_\odot$. The precise evolutionary path of red giant branch stars is dependent on both the initial mass of the star, as well as its metallicity. However, generally, these stars are now composed primarily of helium core, surrounded by a shell fusing hydrogen through the CNO cycle. To counterbalance the increase in produced energy, these stars expand in size, typically to $R_* = 10$ - $100$ $R_\odot$. As their size increases, they ascend the red giant branch, becoming more and more luminous. At the most luminous end of the red giant branch, the temperature of the stellar core reaches nearly $\sim 10^8$ K. At this point, helium fusion is ignited in what is called the "helium flash", and in the core, helium begins to fuse to carbon via the triple-$\alpha$ process. Proceeding He fusion in the core, these stars then follow along the horizontal branch and then asymptotic giant branch, and their cores are eventually composed of primarily carbon and oxygen. As the C-O cores of post-RGB stars contract, the stellar envelope is separate from the core and ejected outwards as a planetary nebula. The C-O core rapidly increases in temperature at constant luminosity as it continues to contract until the core becomes a degenerate electron gas with a size of $R_* \sim 0.01$ $R_\odot$ and emits much less energy. At this point, the star is now in its final evolutionary state as a white dwarf, with temperatures typically $T_{\text{eff}} > 6000$ K and luminosities $L < 1$ $L_\odot$.

High-mass O- and B-type stars, with $M_* > 8M_\odot$, follow a different evolutionary path than low-mass stars. After their relatively short lifetimes on the upper main sequence, the cores of high-mass stars contract as their hydrogen exhausts, which causes them to cool at near-constant luminosity as they become red supergiants with luminosities $L > 10^5$ $L_\odot$. These stars soon begin fusing H to He in their shells, and once their core temperature reaches $\sim 2\times 10^8$ K, the He in their cores begin to fuse to carbon and oxygen. At these temperatures, which are higher than what low-mass stars are able to achieve, heavier elements begin being produced in concentric "onion-like" shells. Once He is exhausted, C burning begins, generating O, Ne, and Mg, and eventually, the core temperature is high enough for Ne burning, then O burning, and finally Si burning, which produces Ni and Fe. When the central core of the star is composed of inert Fe at temperatures of $> 10^9$ K, stellar nucleosynthesis has reached the "iron peak", the point at which the production of heavier elements would be an endothermic reaction requiring more energy than available. Eventually, when the mass of the core reaches $\sim$1.2-1.4 $M_\odot$, it collapses in on itself, and the processes of photodisintegration and neutronization release energy. This release of energy causes the star to explode as a supernova, and its envelope is ejected into the ISM in a luminous $L > 10^{10}$ $L_\odot$ explosion, leaving behind either a neutron star or black hole. The nucleosynthetic reactions that take place during the supernova explosion, via the $s$-process and $r$-process, is what produces nearly all of the elements in the Universe heavier than Fe.

### 1.2.2 Stellar measurements and data

Our current theoretical understanding of stars has developed alongside and is dependent on our ability to make observations of stars. The basic techniques for making measurements of stars have remained relatively the same since the times of Annie Jump Cannon, with the advancements coming in the way of the quality and quantity of stellar observations. The primary methods for making measurements of stars are photometry, spectroscopy, and astrometry, which we describe in the following sections. Most of our knowledge of stars, as well as the galaxies they live in, has been derived from these observations.

### 1.2.2.1 Photometry

Arguably the most fundamental measurement that can be made of a star is of how bright it is. The light emitted by a star can be measured in several filters, using specialized instrumentation optimized to measure stellar flux in a specific wavelength regime. In the visible and infrared, CCDs are commonly used to take images of the sky and record stellar fluxes. Photometric measurements are reported in the magnitude system, with the apparent (observed) brightness of a star, denoted by $m$, typically computed by counting the total flux emitted by a star within some aperture. Since a star's apparent magnitude depends on its distance from us on Earth, to compare the intrinsic brightnesses of stars we compute an absolute magnitude, denoted by $M$. The absolute magnitude of a star is defined as the magnitude it would have if it were 10 parsecs away from the Earth, computed as $M = m - 5\times\log(d/10)$, where $d$ is the distance to the star in parsecs. To compute the absolute magnitude of a star in a given filter, the distance to the star must be known. One of the primary methods of measuring stellar distances, via astrometric parallaxes, will be discussed in Section 1.2.2.3.

Over the past few decades, there have been dozens of photometric sky surveys, a few of which we briefly describe here. The Two Micron All Sky Survey (2MASS) survey (Skrutskie et al. 2006), operational from 1997 to 2001, surveyed ∼99% of the sky with two telescopes, one in the northern hemisphere located at Mount Hopkins, Arizona and one in the southern hemisphere located at the Cerro Tololo Inter-American Observatory in Chile. To a limiting magnitude of $J = 15.8$, 2MASS surveyed over 500 million point sources, including stars and galaxies, in three infrared bands: the $J$-band centered at 1.23 $\mu$m, the $H$-band centered at 1.66 $\mu$m, and the $K_s$-band centered at 2.16 $\mu$m. The Wide-field Infrared Survey Explorer (WISE) survey (Wright et al. 2010), commencing in 2009, is a space-based mission surveying the sky at even longer wavelengths than 2MASS. To a typical limiting magnitude of $W1 \sim 16.6$, the WISE mission imaged nearly the entire sky in four infrared bands: the $W1$-band centered at 3.4 $\mu$m, the $W2$-band centered at 4.6 $\mu$m, the $W3$-band centered at 12 $\mu$m, and the $W4$-band centered at 22 $\mu$m. The final WISE data product includes over 500 million point sources with a signal-to-noise ratio greater than 5. Operating at shorter wavelengths, the space-based Galaxy Evolution Explorer (GALEX) mission (Martin et al. 2005) was functional from 2003 to 2012. GALEX was the first survey to image the sky in the ultraviolet, with two channels including a far-UV band spanning 0.135-0.175 $\mu$m and a near-UV band spanning 0.175-0.2 $\mu$m. The all-sky survey, with a limiting magnitude of $m_{AB} \sim 20$-21, covers 26,000 deg$^2$ of the sky and includes over 500 million sources (Bianchi et al. 2017), while the targeted deep surveys, with a limiting magnitude of $m_{AB} \sim 26$, covers 80 deg$^2$ of the sky.

Photometric surveys, including those just described, have played a key role in advancing our understanding of stars, as well as a number of other astrophysical objects. The science enabled by stellar photometry is vast. The most basic stellar property that broadband photometry allows us to probe is stellar temperature, which, by comparing the magnitude of a star in a red versus blue filter, provides insight into a star's spectral type. In combination with models of stellar evolution, stellar photometry is a key ingredient in observationally estimating stellar properties such as age, mass, and metallicity. This practice of isochrone fitting uses models of stellar evolution (Dotter et al. 2008) to determine the location of same-aged stars in the HR diagram, which varies with the mass and metallicity of a star. Comprehensive studies of stellar Spectral Energy Distributions (SEDs), which is the measurement of stellar magnitudes across a range of wavelength range from the UV to IR, can be used to probe even more detailed information about stars and improve estimates of bolometric luminosities. Photometric surveys also push forward other types of stellar observations. For example, the photometry and positions of stars as measured by the 2MASS survey serve as the basis of most spectroscopic survey targeting (which will be described further in Section 1.2.2.2). At the forefront of future stellar surveys is the planned SPHEREx mission[5], which is a joint optical, IR, and UV photometric, as well as a spectroscopic, survey that will cover more than 100 million stars.

Going beyond one-time measurements, photometric surveys are also carried out in the time domain, regularly monitoring the light emitted by objects over a period of time. While many astrophysical events take place on timescales much longer than we could observe, there are a number of more transient phenomena that can be observed on timescales from ranging from days to years, including asteroid orbits, exoplanet transits, stellar flares and variable stars, relativistic beaming of AGN, and microlensing events. Time domain surveys targeted specifically at measuring the brightness variability of stars have largely been advanced with exoplanet science, via the transit method, as a primary goal. The Convection, Rotation and planetary Transits (CoRoT) mission[6], operational from 2006 to 2013, was the first space-based survey commissioned for exoplanet science and asteroseismology. The mission continuously observed small patches of the sky in the $V$-band, targeting stars with $V$=11-16, for long runs of 150 days, or short runs of 30 days. The mission resulted in 32 confirmed exoplanets, and another ∼500 exoplanet candidates. In addition to advancing exoplanet science, CoRoT was also a fundamental step forward for asteroseismology. Targeting stars with $V$=6-9, the 150-day baseline observations result in a frequency resolution of ∼0.1 $\mu$Hz, while the 30-day baseline observations result in a ∼0.6 $\mu$Hz resolution. This stellar variability data collected by CoRoT resulted in the first detection of solar-like oscillations for an ensemble of stars other than the Sun (Baglin et al. 2006; Michel et al. 2008), and was also used to probe stellar rotation (Mosser et al. 2009).

The *Kepler* space telescope (Borucki et al. 2008) produced arguably the most prolific exoplanet science survey to date. The *Kepler* mission resulted in the discovery of ≈ 3,000 exoplanet candidates, and also served as a key driver of asteroseismology. Operating from 2009 to 2018, *Kepler* observed a 116 deg$^2$ patch of the sky with a broad bandpass, covering 0.42 to 0.90 $\mu$m. *Kepler* delivered flux measurements for ∼150,000 stars over a continuous ≈ 4-year baseline to a precision of ∼20 parts per million. All of the ≈ 150,000 stars, spanning V = 9-15, were observed with a cadence of 29.4 minutes, and a smaller sub-sample ($< 1\%$) were also observed at a shorter cadence of 1-minute. The *Kepler* data has enabled the empirical connection between time domain variability and various stellar properties, which is essential for characterizing exoplanet systems but has also advanced the field of stellar astrophysics. Based on the *Kepler* time series data, the primary asteroseismology parameters, $\nu_{\max}$ and $\Delta\nu$, have been measured in the power spectra for thousands of stars (e.g. Gilliland et al. 2010; Bedding et al. 2010; Stello et al. 2013; Yu et al. 2018). As will be discussed further in Chapter 4, these two parameters of the stellar power spectrum can be used to derive precise measurements of stellar mass and radius (e.g. Kjeldsen & Bedding 1995; Stello et al. 2009a,b; Huber et al. 2011). The *Kepler* light curves have also been used to infer the surface gravity (log $g$) and temperature ($T_{\rm eff}$) of red giant branch stars. As demonstrated in Ness et al. (2018), with a modified version of The Cannon (Ness et al. 2015), log $g$ can be recovered to $< 0.1$ dex and $T_{\rm eff}$ can be recovered to $< 100$ K. And finally, the *Kepler* observations have also enabled the measurement of stellar rotation periods for many stars, which has been exciting for the prospect of gyrochronology (Barnes 2003). Spanning the largest sample of stars, McQuillan et al. (2014) derive rotation periods for ∼30,000 main sequence stars with a peak identification procedure in the autocorrelation function (ACF) domain.

---

[5]https://www.jpl.nasa.gov/missions/spherex/
[6]https://sci.esa.int/web/corot

Continuing the success of time domain surveys like CoRoT and *Kepler*, over the next decade the number of stars with photometric time series data is projected to increase by a hundredfold. The ongoing Transiting Exoplanet Survey Satellite (TESS) mission (Ricker et al. 2014) will deliver light curves for the order of $10^5$ stars, while the ground-based Large Synoptic Survey Telescope (LSST) (LSST Science Collaboration et al. 2009) is planned to deliver time domain data for $\sim 10^8$ stars. As seen in Figure 1.2, which includes past and future space-based missions, the number of red giant stars with detected solar-like oscillations is expected to be on the order of $10^6$ upon the completion of the TESS survey. In the more distant future, with planned missions like WFIRST and PLATO, this number is expected to increase to $10^7$. For fainter stars, namely dwarfs and subgiants, TESS will bring the number of stars with solar-like oscillations detected to $10^4$ and eventually to $10^5$ with future missions. With this projected increase in stellar time domain data, it's apt to prepare for and assess the prospects of efficiently estimating stellar parameters from this data, spanning different baselines and cadences. This pursuit is the subject of Chapter 4 using the *Kepler* time domain data.

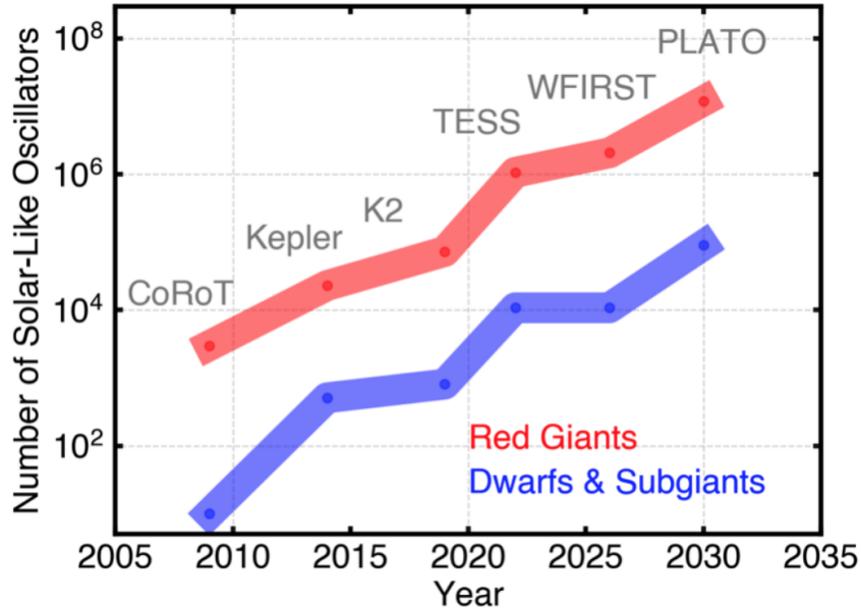

Figure 1.2: The current and projected number of stars with solar-oscillations detected by space-based missions. Image credit: Figure 1 (right) of Huber et al. (2019).

### 1.2.2.2 Spectroscopy

Spectroscopy is arguably the most powerful tool we have for studying stars. The spectrum of a star provides a wealth of information, allowing us to measure the radial velocities, chemical compositions, surface gravities and temperatures of stars. The basic technique of spectroscopy is the dispersion of light into its component wavelengths, producing a "spectrum" of the intensity of radiation across a range of wavelengths. This produces a detailed characterization of the light emitted by an object, differing from photometry in which the total flux is measured within a broad wavelength bin. Stellar spectra are typically absorption spectra, in which the radiation emitted by the hotter stellar core passes through the cooler gas of the envelope, with photons of specific energies being absorbed by the gas. The absorption of photons at certain energy levels are manifested as absorption lines in stellar spectra, with a decreased intensity compared to the continuum at the corresponding wavelength. The photon energies which are absorbed are dependent on the elements present and the temperature of the stellar envelope, making stellar spectra a rich tool for studying the chemical composition of stars.

The Sun's spectrum was the first stellar spectrum to be studied extensively, and in the early 19th century hundreds of absorption lines were recorded in the Solar spectrum. By the end of the century, spectra were measured for thousands of stars, typically using prisms to disperse starlight onto photographic plates. This data served as the foundation of Annie Jump Cannon's work classifying stars according to the absorption patterns exhibited by their spectra. Around this same time, another early astronomer Cecilia Payne-Gaposchkin was the first to discover that stars are primarily composed of hydrogen and helium. She came to this conclusion by studying stellar spectra as a part of her Ph.D. thesis carried out at the Harvard College Observatory. Her findings were revolutionary and overturned conventional wisdom of the time that the Sun and stars were chemically similar to the Earth (Wayman 2002). In the late 20th century CCDs became the preferred detector for astronomical imaging, and the use of diffraction gratings instead of prisms improved the quality and extent of spectroscopic observations. Our theoretical understanding of stellar spectra has advanced alongside these technological innovations. The motions, namely radial velocities, of stars are measured by comparing the observed wavelengths of stellar absorption features to the theoretical rest wavelength of those features. Stellar surface temperatures, gravities, and chemical compositions are estimated by comparing observed spectra to physical models of stellar spectra determined across a range of stellar types.

Modern spectroscopic surveys have been instrumental to our understanding of stars and the Milky Way, as well as our broadening view of the Universe. The most influential survey to date is the Sloan Digital Sky Survey (SDSS), which, since its commencement in 2000, has measured the spectra of over 3 million objects as well as produced deep imaging of nearly one-third of the sky in multiple photometric bands. The most recent SDSS campaign, SDSS-IV (Blanton et al. 2017), includes three surveys: the new Mapping Nearby Galaxies at APO (MaNGA) integral field unit (IFU) survey to obtain spectra as a function of radius across the extent of galaxies, as well as updates to the existing the Baryon Oscillation Spectroscopic Survey (BOSS) survey measuring the distances to quasars to chart the volume of the Universe and the Apache Point Observatory Galactic Evolution Experiment (APOGEE) survey focused on obtaining large numbers of stellar spectra. The APOGEE survey in particular has greatly contributed to our knowledge of stars by delivering chemical abundances, temperatures, surface gravities, and well as stellar age estimates (Ness et al. 2016; Martig et al. 2016; Ho et al. 2017) for thousands of stars in the Milky Way. The APOGEE campaign, using the APOGEE near-IR spectrograph (Wilson et al. 2010) (with a resolution of $R \sim 22,500$) on the 2.5 meter SDSS telescope (Gunn et al. 1998) located at Apache Point Observatory, obtained the spectra for ∼250,000 stars spanning a wavelength range from 1.5 - 1.7 $\mu$m. In this wavelength range, 19 individual chemical abundances are measured, including C/CI, N, O, Na, Mg, Al, Si, P, S, K, Ca, Ti/TiII, V, Cr, Mn, Fe, Co, Ni, and Rb (Abolfathi et al. 2018; García Pérez et al. 2016; Ness et al. 2015). This Dissertation makes use of APOGEE abundance measurements in Chapter 3.

Beyond APOGEE, spectroscopic surveys like RAVE, GALAH, and LAMOST have delivered spectra for many more stars and have facilitated the measurement of additional abundances. The RAdial Velocity Experiment (RAVE) survey (Steinmetz et al. 2006), focused primarily on obtaining stellar radial velocities, obtained stellar spectra with the Australian Astronomical Observatory (AAO) 1.2-m UK Schmidt Telescope in the infrared wavelength range of 0.841 to 0.8794 $\mu$m with a resolution of $R \sim 7,000$. RAVE delivered velocities for nearly 500,000 stars to a precision of 1.5 km s$^{-1}$, and also provided Mg, Al, Si, Ti, Fe, Ni, and $\alpha$ abundances for ∼40,000 of these stars (Kunder et al. 2017; Casey et al. 2017). With the HERMES spectrograph on AAO's 3.9 m Anglo-Australian Telescope, the primary goal of the ongoing GALactic Archaeology with HERMES (GALAH) survey is to measure the spectra of ∼1 million stars (De Silva et al. 2015; Martell et al. 2017). With a resolution of $R \sim 28,000$, GALAH is obtaining stellar spectra in four wavelength regimes: blue from 0.4718 to 0.4903 $\mu$m, green from 0.5649 to 0.5873 $\mu$m, red from 0.6481 to 0.6739 $\mu$m, and infrared from 0.7590 to 0.7890 $\mu$m, which have been optimally selected to derive ∼25 chemical abundances including Li, C, O, Na, Al, K, Mg, Si, Ca, Ti, Sc, V, Cr, Mn, Fe, Co, Ni, Cu, Zn, Y, Zr, Ba, La, Nd, and Eu. The second GALAH data release includes ∼350,000 stars, comprising 35% of its eventual sample size (Buder et al. 2018). Lastly, the Large Sky Area Multi-Object Fibre Spectroscopic Telescope (LAMOST) survey (Newberg et al. 2012; Zhao et al. 2012) is a low-resolution ($R \sim 1,800$) spectroscopic survey providing spectra for ∼5 million stars over the wavelength range of 0.369 to 0.910 $\mu$m. While the LAMOST survey is the largest spectroscopic survey to date, the abundances that can be measured from the low-resolution spectra are generally limited to a bulk $\alpha$ abundance, and a handful of other elements like C, N, and Fe. However, recent work in using data-driven modeling to derive more detailed abundance information from low-resolution spectra based on a reference set of high-resolution spectra appears promising (e.g. Xiang et al. 2019; Wheeler et al. 2020). This will be discussed further in Section 1.4.2.

From the surveys described above, we now have abundance measurements for hundreds of thousands of stars spanning five chemical families: light elements, $\alpha$ elements, Fe peak elements, odd-Z elements, and neutron-capture elements, each with distinct nucleosynthetic origins, which we briefly summarize here[7]. The light elements, including C, N, and O, are ejected into the ISM via type II supernovae and AGB winds, with C and O produced in the hydrostatic He burning of high-mass stars. C is also produced in low mass stars, and N is produced via the CNO cycle. The $\alpha$ elements, including Mg, Ca, Si, Ti, S, and O, are ejected into the ISM via type II supernovae and produced by $\alpha$-particle capture in the C, He, and Ne burning of massive stars. The Fe peak elements, including Mn, Ni, Co, Cr, and Zn, are ejected into the ISM via type Ia supernovae and are produced similarly to Fe. The odd-Z elements, including Na, Al, K, V, P, and Cu, are ejected into the ISM via type II and type Ia supernovae through a variety of mechanisms including C, O, and Ne burning. And finally, the neutron-capture elements, including Sr, Y, Zr, Ba, La, Eu, and Nd, are ejected into the ISM via type II and type Ia supernovae, as well as AGB winds, and are produced from the $s$- and $r$-processes.

This high-dimensional chemical abundance information, spanning many different production mechanisms, is a rich data set for Galactic archaeology pursuits. Future surveys, namely the planned SDSS-V Milky Way Mapper Galactic Genesis survey, will open many additional avenues for studying the Galaxy by delivering spectra for an unprecedented 4-5 million stars with a resolution of $R \sim 22,500$, which, as shown in Figure 1.3. This will allow us to probe a much larger extent of the Milky Way's stars compared to APOGEE. These spectra, in addition to abundances, can also be used to estimate stellar ages. For example, for $\sim$70,000 red giant branch stars covered by the APOGEE survey, Ness et al. (2016) find that spectra are informative to stellar mass (primarily through CN absorption strength) to a precision of $\sim$0.07 dex, which corresponds to age estimates with $\sim$40% errors. Since the chemical composition of a star's envelope is inherited from the ISM at the time of stellar birth and remains effectively constant throughout its lifetime (with the exception of C and N), chemical abundances are a probe of stellar birth conditions. This makes stellar age estimates, in combination with highly-dimensional chemical abundance information, a rich avenue for studying the chemical enrichment history of the Milky Way. However, recent work has suggested that the information content of multi-dimensional abundances is much lower than expected (Ness et al. 2019b). This underlies the significant challenges in connecting detailed abundances variations to an interpretable physical origin (Blancato et al. 2019), which is the topic of Chapter 3.

---

[7]For a more comprehensive discussion of the chemical families, see the Ph.D. thesis of Professor Keith Hawkins here: http://www.as.utexas.edu/∼khawkins/attachments/thesis.pdf

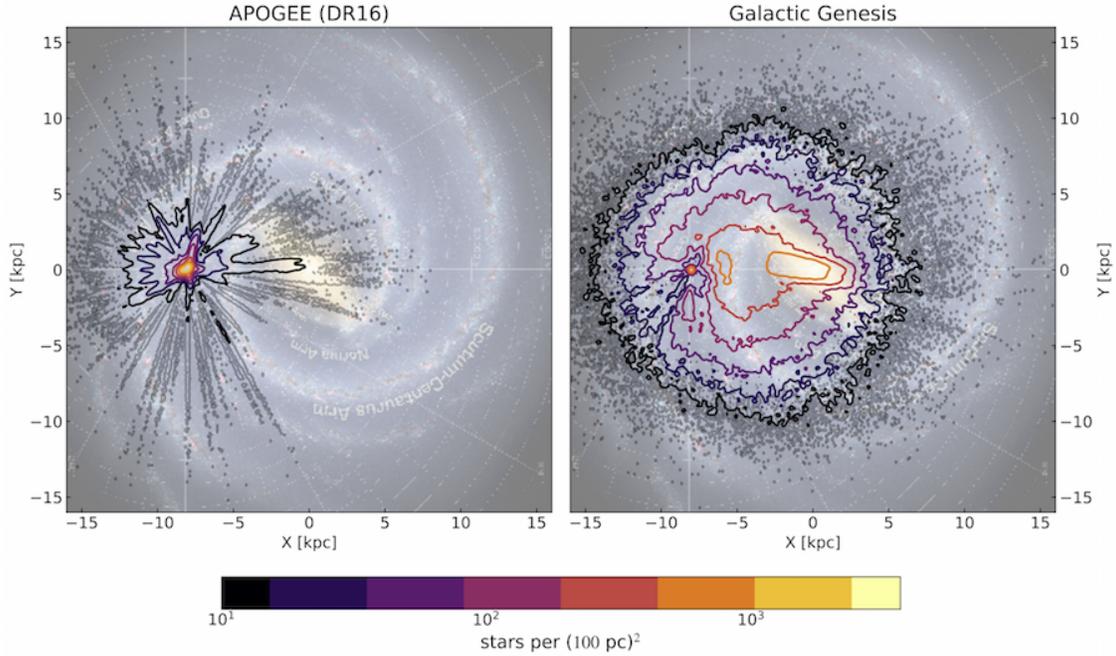

Figure 1.3: The extent of the Milky Way's spectroscopic coverage from the latest APOGEE data release (left) and the planned Milky Way Mapper Galactic Genesis survey (right). Image credit: NASA/JPL-Caltech/R. Hurt (background), J. Bird.

#### 1.2.2.3 Astrometry

Lastly, another fundamental measurement that can be made of a star is its position in the sky. Stellar astrometry, which surveys the positions and motions of stars, aims to map stars throughout the Milky Way. This provides insight into the Galaxy's structure, kinematics, and dynamics. One of the fundamental goals of astrometry is to obtain stellar distances. These can be estimated from measuring the relative shift of a star's position in the sky due to the motion of the Earth around the Sun. For stars increasingly distant from the Earth, this apparent shift termed the parallax angle, becomes increasingly small, requiring increasingly precise measurements of stellar position.

The practice of astrometry predates modern science, but until the 1980s when CCD technology became readily available, positions for only $\sim 10^3$ stars had been recorded. The first space-based astrometric mission, European Space Agency's (ESA) *Hipparcos*[8], which finished its survey in 1993, measured the precise positions of $\sim 10^6$ stars. Monitoring stars to a limiting magnitude of $V = 12.4$ mag, the final *Hipparcos Catalogue* has a parallax accuracy of 0.002 arcseconds, with $\sim$20,000 stars having distances measured to an accuracy of 10% and another $\sim$50,000 stars having their distances measured to an accuracy of 20% (Perryman et al. 1997). In addition to parallaxes and distances, the survey also included photometric measurements of stars in the visible over the mission's 3.5-year duration. The *Hipparcos Catalogue* served as an important step forward in several areas of astronomy, pro-viding high-quality distances and luminosities for the largest sample of stars at the time. The science resulting from the catalog includes: calibrating the cosmic distance ladder (Feast & Catchpole 1997), associating open clusters allowing age dating of these stellar populations (Perryman et al. 1998), using the proper motions of stars in the solar neighborhood to probe Milky Way's rotation and gravitational potential (Dehnen & Binney 1998), the identification and characterization of binary star systems (Söderhjelm 1999), and furthering our understanding of stellar properties and evolution (Pietrinferni et al. 2004).

---

[8]https://sci.esa.int/web/hipparcos

The following, and most recent, advance in stellar astrometry is currently being enabled by ESA's *Gaia* mission[9], which was launched in 2013 and is still operating. The goals of *Gaia* are similar to those of its predecessor, *Hipparcos*. However, it's far more ambitious, with the ultimate goal of surveying over $\sim 10^9$ stars. The main science objective of *Gaia* is to better understand the structure and formation of the Milky Way by building an extensive 3D map of the Galaxy. To this end, the *Gaia* satellite houses a multi-purpose instrument, capable of measuring angular positions, brightnesses, and radial velocities. The publicly released *Gaia* DR2 catalog (Gaia Collaboration et al. 2018b) contains high precision measurements for almost 1.7 billion stars, which includes measurements of parallax, 2D positions, and 2D proper motions (Lindegren et al. 2018), as well as radial velocity (Katz et al. 2019) and $G$-band magnitudes to a limiting magnitude of $G = 21$. The average astrometric precision is $\sim 10^{-6}$ arcseconds, with nearly 50 million stars in DR2 having distances measured with uncertainties less than 10%.

Showcasing the extraordinary extent of the mission, Figure 1.4 is the color-magnitude diagram for 66 million stars constructed from *Gaia* astrometry and photometry (Gaia Collaboration et al. 2018a). The color-magnitude diagram, with the $G$-band absolute magnitude ($M_{\rm G}$) of stars along the $y$-axis and the blue minus red color ($G_{\rm BP}$ - $G_{\rm RP}$) of stars along the $x$-axis, is the observational counterpart to the theoretical HR diagram described in Section 1.2.1. Stars with lower $M_{\rm G}$ values are more luminous, while stars with higher $M_{\rm G}$ values are less bright, and stars with lower $G_{\rm BP}$ - $G_{\rm RP}$ values are bluer, corresponding to higher effective temperatures, while stars with higher $G_{\rm BP}$ - $G_{\rm RP}$ values are redder, corresponding to lower effective temperatures. The main sequence is clearly visible from $M_{\rm G} \sim 0$ and $G_{\rm BP}$ - $G_{\rm RP} = 0$ to $M_{\rm G} \sim 12$ and $G_{\rm BP}$ - $G_{\rm RP} = 4$. The red giant branch is also present, with $M_{\rm G} < 2.5$ and $G_{\rm BP}$ - $G_{\rm RP} > 1$, as well as the population of white dwarfs with $M_{\rm G} > 10$ and $G_{\rm BP}$ - $G_{\rm RP} < 2$.

---

[9]https://sci.esa.int/web/gaia

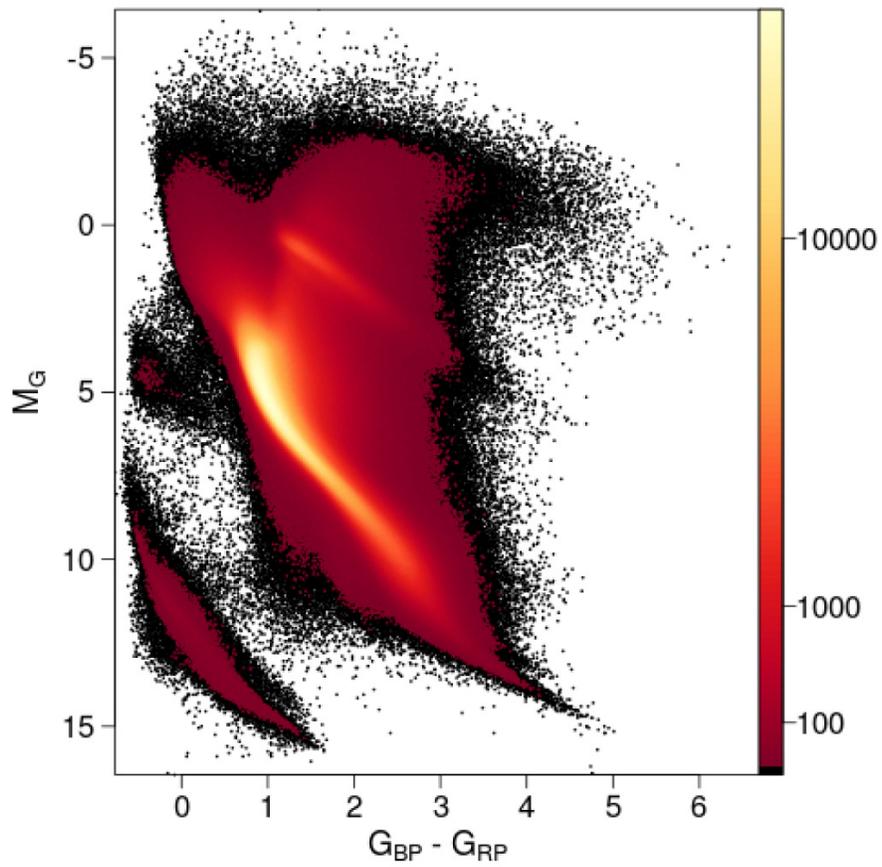

Figure 1.4: The *Gaia* DR2 color-magnitude diagram for ≈66 million stars. Image credit: Figure 1 of Gaia Collaboration et al. (2018a), C. Babusiaux/Gaia, Creative Commons License 4.0

As the mission is still in its initial stages, the full extent of *Gaia*'s science impact is yet to be realized. However, there have been numerous results based on the DR2 catalog, including studies of the kinematics of dwarf galaxies and globular clusters (Gaia Collaboration et al. 2018c), kinematic studies of the Milky Way's disk and stellar halo (Belokurov et al. 2018), combined kinematic, chemical, and age studies of Milky Way stellar populations (Helmi et al. 2018), further refinement of constraints on the Hubble Constant using distances to Cepheids (Riess et al. 2018), and calibration of asteroseismology measurements with *Gaia* parallaxes (Huber et al. 2017). This Dissertation makes use of *Gaia* data in Chapter 3.

## 1.3 Collections of stars

Individual stars do not form, nor exist, in isolation. At a given time, the collapse of cool gas clouds in the ISM results in the creation of many stars, formed from a common natal material. The types of stars comprising, as well as the overall properties of, a coeval population of stars formed in a burst of star-formation is in part governed by the physics of how molecular clouds fragment. This makes the properties of these stellar populations a probe of the physics of star-formation. Going a step further, a collection of many coeval stellar populations is what comprises the stellar content of a galaxy. By studying a galaxy's stellar populations, which formed in different environments and conditions throughout the history of the Universe, we can begin to piece together how galaxies evolved to their present-day state. In this section, we briefly discuss stellar populations, the stellar content of galaxies, and the observations made to probe unresolved populations of stars. This serves as the observational foundation of the work carried out in Chapter 2, as well as the Galactic chemical evolution modeling pursued in Chapter 3.

### 1.3.1 Simple stellar populations

A simple stellar population (SSP) is defined as a chemically homogeneous population of stars that formed in the same star-formation event, from the same natal gas. This same-aged stellar population includes stars spanning a range of stellar masses, from low-mass ($M_* \sim 0.01\ M_\odot$) to high-mass stars ($M_* \sim 100\ M_\odot$). In the theoretical HR diagram detailed in Section 1.2.1, the stars belonging to an SSP fall along an isochrone. This is a curve on the HR diagram tracing the $T_{\text{eff}}$-$L$ values of same-aged stars with a range of masses. Since a star's mass determines how long it takes for hydrogen to be exhausted in its core, and therefore when a star will evolve off the main sequence, isochrones can be used to measure stellar age. For a population of stars assumed to be the same age (e.g. stellar clusters), examining the location in the CMD where the main sequence ends and begins to form the subgiant branch, referred to as "the main sequence turn-off point", tells us what mass of star has most recently exhausted its hydrogen. This implies the stellar age for the entire cluster. This age-dating technique is commonly used to age open clusters as well as globular clusters.[10]

---

[10]However, some of these seemingly simple globular cluster stellar populations have been observed to contain two or more generations of star-formation, as indicated by multiple isochrone tracks (Piotto et al. 2007).

As mentioned, stars belonging to an SSP span a range of initial birth masses. The relative number of stars formed at each stellar mass is typically described by the stellar initial mass function (IMF). The exact form of this initial stellar mass distribution is challenging to observe, making it difficult to characterize the number of stars formed at each stellar mass. Up until the last decade or so, the functional form of the IMF was typically assumed to be universal, in which every star-formation event produced the same fraction of low-mass to high-mass stars. While the evidence for this is mostly supported by measurements of resolved stellar populations in the Milky Way and Local Group (see Offner et al. (2014) for an overview), the physics setting the form of the IMF still remains uncertain. This is due to deficient knowledge of the physical theory governing how molecular clouds collapse and fragment to form protostars, and the subsequent accretion of gas onto these protostars.

As will be discussed in Chapter 2, recent observations of more distant stellar systems stand to challenge the notion of a universal IMF. In the past few years, dozens of studies have obtained evidence for systematic IMF variations. While understanding the origin of IMF variations will certainly provide insight into star-formation physics (e.g. Bastian et al. 2010; Offner et al. 2014), the IMF plays a role in a number of astrophysical domains. This includes the formation and occurrence of compact objects (e.g. Belczynski et al. 2014, 2017), the production of heavy elements essential for planet formation (e.g. Santos et al. 2001; Adibekyan et al. 2012a), as well as the interpretation of measurements made of stellar populations and galaxies (e.g. Bell et al. 2003; McGee et al. 2014; Clauwens et al. 2016). If proven robust, a variable IMF could have significant consequences for our current understanding of how stellar populations evolve. This is because the fraction of low-mass to high-mass stars influences the energy available for supernovae feedback, the amount of metals injected into the ISM, and the mass of baryons trapped in low-mass stars.

#### 1.3.1.1 The evolution of stellar populations

Proceeding the birth of a new stellar population, nuclear fusion, and the stellar evolution processes, as described in Section 1.2.1, commence. Since the stars belonging to a stellar population span a range of masses, at any given time a range of stellar evolutionary states is represented. The mass-dependent nature of stellar evolution, and the varying end-states of high-mass versus low-mass stars, has implications for how the stars belonging to a stellar population chemically enrich the ISM over time. There are three primary channels of stellar nucleosynthesis, through which stars dispel various metals into the ISM: core-collapse supernovae (CC-SN), supernovae type Ia (SN Ia), and winds from asymptotic giant branch (AGB) stars. For a more comprehensive review of stellar nucleosynthesis, see Clayton (1983).

The highest mass stars ($M_* > 8\ M_\odot$) are the first to evolve off the main sequence, pass through the giant phase, and then promptly explode as CC-SN. As discussed in Section 1.2.1, throughout their lifetimes, high-mass stars fuse elements up to Fe within their cores. When high-mass stars explode, these metals get expelled into the ISM, as well as the metals heavier than Fe that are generated during the explosion. The most abundant chemical yields from CC-SN are generally $\alpha$-elements, H, He, C, N, O, Fe, and some odd-$Z$ elements (e.g. Nomoto et al. 2013; Chieffi & Limongi 2004). However, the relative amounts vary depending on both the mass of the dying star, as well as its initial metallicity. Since the lifetimes of massive stars are relatively short, the enrichment from CC-SN is often characterized as occurring "instantaneously" in chemical evolution models. As will be discussed in Chapter 3, there are subtle differences in how elements even belonging to the same chemical family are produced in CC-SN (Hasselquist et al. 2017). Focusing on $\alpha$-elements specifically, some $\alpha$-elements, like O and Mg, are produced hydrostatically, in the outer shells of the star during the hydrostatic burning stage. Other $\alpha$-elements, like Si and Ca, are produced closer to the stellar core in the nucleosynthesis leading up to the CC-SN explosion.

Compared to high-mass stars, low-mass stars ($M_* < 8\ M_\odot$) spend much more time on the main sequence, as they more slowly burn through their hydrogen cores. Dependent on the specific mass, some low-mass stars do evolve off the main sequence, while the very low-mass stars have main sequence lifetimes longer than the age of the Universe. For the low-mass stars that do eventually reach their end-state as white dwarfs, a fraction of these will explode as SN Ia. This occurs under certain circumstances, for example, if a white dwarf is in a binary pair with an evolved giant star and if the stars are close enough for the matter from the extended envelope of the giant star to transfer onto the white dwarf. From this transfer, once the mass of the white dwarf reaches the Chandrasekhar limit (1.4 $M_\odot$), the white dwarf is no longer able to support itself against collapse. This commences a runaway fusion of C and O in the white dwarf, producing an enormous amount of energy, which results in the explosion of the white dwarf as an SN Ia. The most abundant elements expelled into the ISM from these explosions are iron-peak elements, C, O, Fe, Si, and some odd-Z elements, with the relative amounts dependent on the initial metallicity of the star (e.g. Tinsley 1979; Seitenzahl et al. 2013; Thielemann et al. 2003). The timescale of SN Ia enrichment is more difficult to characterize since it depends on the longer timescales of low-mass star evolution and the stellar binarity fraction. Given this, yields from SN Ia are typically parameterized as a power-law delay time distribution (DTD), in which the chemical enrichment of the ISM increases over timescales of up to a few Gyrs as more SN Ia explosions occur (Maoz et al. 2010).

The ISM is also enriched by the winds of AGB stars. For stars with masses from $M_* \sim$ 1 - 8 $M_*$, enough of their lifetime is spent on the AGB for winds from these stars to dispel a significant amount of material out to the ISM. Following H exhaustion, the envelopes of stars expand in size as they evolve to the RGB and then AGB. The mass residing in the outermost part of these stellar envelopes is far away enough from the core to become gravitationally unbound from the star, and this material is transported out into the ISM via stellar winds. The most abundant elements that AGB winds deposit into the ISM are H, He, C, N, Na, Mg, as well as some *s*-process elements, with the relative amounts dependent on the initial mass and metallicity of the star (e.g. Karakas 2010; Ventura et al. 2013).

#### 1.3.1.2 Modeling chemical enrichment

There are two main approaches to modeling the chemical evolution of stellar populations: hydrodynamical simulations and analytical formalisms. Both of these approaches depend on an input set of nucleosynthetic yield tables which prescribe the metals produced through CC-SN (e.g. Nomoto et al. 2013; Chieffi & Limongi 2004), SN Ia (e.g. Seitenzahl et al. 2013; Thielemann et al. 2003), and AGB winds (e.g. Karakas 2010; Ventura et al. 2013), as described above.

Hydrodynamical simulations include empirical prescriptions of chemical enrichment, as well as the gravitational dynamics of gas, stars, and dark matter (e.g. Pillepich et al. 2018; Grand et al. 2017). These simulations are typically computationally intensive, which limits the parameter space describing chemical evolution that can be explored. Analytical formalisms instead describe chemical enrichment with closed-form equations, requiring strong simplifying assumptions to do so (e.g. Spitoni et al. 2017). While this approach is less detailed and deviates more from first principles compared to hydrodynamical simulations, it's often much less computationally expensive. This allows for a larger parameter space to be explored. Many models describing Galactic chemical evolution (GCE) (e.g. Tinsley 1979; Schönrich & Binney 2009; Matteucci 2012) take an analytical approach, combining physically motivated models of star-formation and stellar evolution with galactic ISM physics to predict the time evolution of the ISM's chemical abundance. The most simple GCE models include a single zone of star-formation surrounded by a basin of gas, and this gas becomes enriched over time from the products of stellar nucleosynthesis.

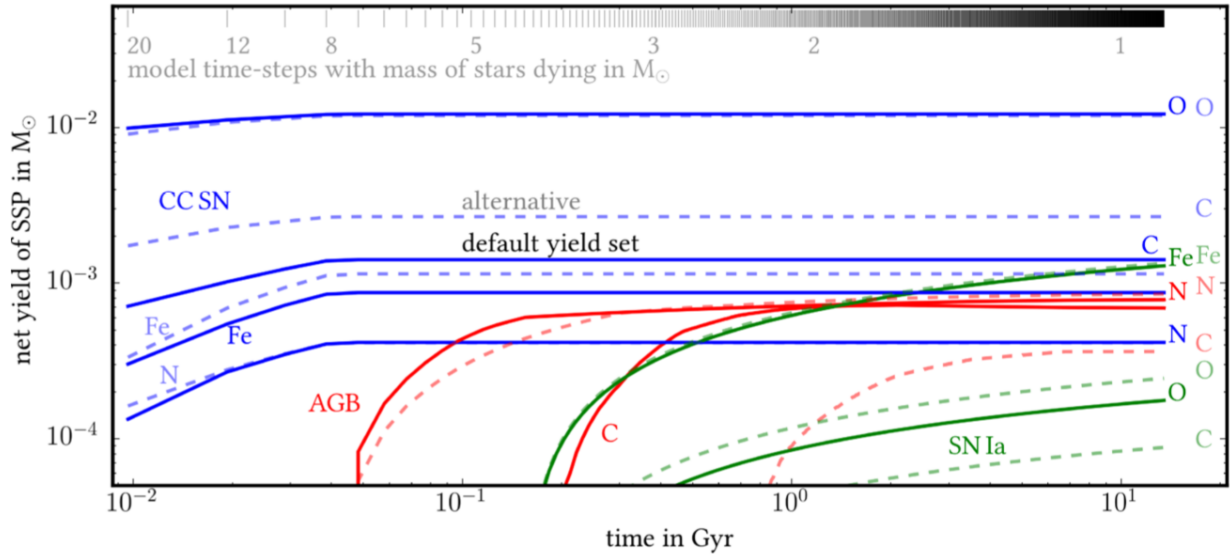

Figure 1.5: The cumulative C, N, O, and Fe yield contributions from CC-SN, SN Ia, and AGB winds from a solar-metallicity and solar-mass simple stellar population, as modeled with the *Chempy* Galactic chemical evolution model. Image credit: Figure 4 of Rybizki et al. (2017).

The *Chempy* GCE model (Rybizki et al. 2017), which is used for the work of Chapter 3, is a recently developed GCE code that permits flexible Bayesian inference of chemical evolution parameters given observed stellar abundances. The parameters that can be fit include the high-mass slope of the IMF, the SN Ia time delay, a normalization constant for the number of exploding SN Ia, the mass of the gas reservoir, the star-formation efficiency, the peak of the star-formation rate, as well as the fraction of stellar yields that outflow to the gas reservoir. Assuming observationally motivated values for the parameters listed above, Figure 1.5 shows the cumulative C, N, O, and Fe net yields for a 1 $M_\odot$ SSP of solar-metallicity, separating the contributions from CC-SN, SN Ia, and AGB stars. As seen in the figure, metals produced via CC-SN are the first to enrich the ISM as massive stars explode. After the time at which stars with $M_* = 8\ M_\odot$ explode, CC-SN no longer contribute to ISM enrichment. Next, as intermediate-mass stars evolve off the main sequence, elements carried by AGB winds begin to dispel into the ISM. These yields rise sharply at first as the more massive stars expel a relatively large amount of mass, and afterward, the yields only marginally increase as lower-mass stars enter the AGB phase. Yields from SN Ia begin to be produced shortly after AGB enrichment commences, as the intermediate-mass stars evolve to white dwarfs and a fraction of these explode as SN Ia. Parameterized by a power-law, the yields from SN Ia increase over time. The GCE parameters listed above can be constrained by determining which parameter values result in chemical evolution tracks that best match observed stellar abundances over a range of stellar ages. However, as demonstrated in Chapter 3, known tensions between theoretical and empirical yields (e.g. Fink et al. 2014; Müller 2016; Philcox et al. 2018) limit our physical interpretation of detailed stellar abundance patterns.

### 1.3.2 Galaxies

Galaxies are collections of gravitationally bound stars, gas, dust, and dark matter. The stellar content of a galaxy is composed of numerous stellar populations, each with the properties as described in the previous section. With galaxy surveys, such as the Hubble Space Telescope (HST) CANDELS survey[11] and the various SDSS redshift surveys[12], it's now estimated that there are on the order of 100 billion galaxies in the Universe, spanning a diverse range of morphologies and evolutionary paths. Yet, up until the early 1900s, we had little conception of galaxies, with our view of the Universe limited to the galaxy in which we reside.

The first observational evidence of a Universe beyond the extent of the Milky Way was obtained by Edwin Hubble using the Hooker telescope on Mount Washington in California.

---

[11] http://arcoiris.ucolick.org/candels/
[12] https://www.sdss.org/surveys/

In his observations, Hubble took note of small patches of dust, which he called "nebulae" (Hubble 1936, 1958). At first, these nebulae were assumed to be a part of the Milky Way. However, the work of Henrietta Swan Leavitt, another member of the Harvard computers, enabled the placement of one of these nebulae far beyond the reaches of our Galaxy. Establishing the period-luminosity relationship of Cepheid variable stars (Leavitt 1908), she showed that Cepheids can be used as "standard candles" in which the distance to one of these stars can be inferred from its observed period and brightness alone. Building on this insight, Hubble was able to show that one of the nebulae he observed resided at a distance far greater than the current size estimate of the Milky Way at the time. This nebula came to be known as the Andromeda galaxy and this realization was a pivotal step forward in realizing that entire galaxies exist beyond the reaches of our own Milky Way.

Since the 1900s, our knowledge of galaxies has advanced significantly. This has been driven by various large galactic observing campaigns, as well as our theoretical and computational ability to model galaxy formation and evolution. Redshift surveys, like SDSS and the 2dF Galaxy Redshift Survey[13] (2dFGRS), have measured the distances to hundreds of thousands of galaxies. Surveys like these map the 3D distribution of galaxies in the local Universe and have revealed large-scale structures and clusters of galaxies. Imaging surveys, including those carried out with HST and SDSS, have demonstrated the large diversity of visual galaxy morphologies. While these observations are critical to our understanding galaxies, piecing together how galaxies form and evolve through time is a more difficult task since the timescales of galaxy evolution are on the order of $10^9$ years.

In the absence of observing the real-time evolution of individual galaxies, we instead characterize populations of galaxies at different times in the Universe's history. The measured redshift of a galaxy, $z$, is a function of distance, where galaxies that are farther from us have larger observed redshifts. Since the speed of light is finite, when we observe galaxies at higher redshifts, we are seeing them not as they appear today, but as they were earlier in time. Given this, by observing galaxies across a range of redshift ranges, we can characterize galaxy properties throughout the history of the Universe. This provides a view of how populations of galaxies evolve, but it's challenging to link this understanding to the evolutionary path that individual galaxies follow. Numerical simulations of galaxies, including those of isolated galaxies, and large-volume fully cosmological simulations like the Illustris[14] and IllustrisTNG[15], are powerful tools that can be used to link observations of galaxy populations to how individual galaxies evolve and the physical theory underlying their evolution.

---

[13] http://www.2dfgrs.net/

In the remainder of this section, we briefly overview some of the basic properties of elliptical and spiral galaxies, both of which we gain some insight into in this Dissertation by examining the properties of their stellar populations.

#### 1.3.2.1 Elliptical galaxies

Elliptical galaxies, also referred to as early-type galaxies (ETGs), were initially thought of as the simplest of the galaxy types due to their smooth, featureless appearance. However, these stellar systems, and how they are posited to have formed, have revealed themselves to be quite complex. The stellar light profiles of ETGs generally follow de Vaucouleurs' law (de Vaucouleurs 1948), where the center of the galaxy exhibits the highest measured flux, which then decreases as a function of galactocentric radius. The sizes of ETGs are typically characterized by their half-light radius ($r_e$), the elliptical radius in which that half the total flux of the galaxy is contained within. The measured sizes of ETGs typically span a fraction of a kpc up to ∼100 kpc.

ETGs exhibit little net rotation, as their stars are characterized by random orbits. The distribution of stellar velocities closely follows a Gaussian distribution, where the standard deviation of this distribution referred to as the velocity dispersion of the galaxy, $\sigma$. The velocity dispersion is a fundamental property, enabling the estimation of the galaxy's mass. The virial theorem, which relates the average kinetic energy of a system to its average potential energy, is used to derive the dynamical (gravitational) mass ($M_{\text{dyn}}$) of ETGs based on measurements of the galaxy's velocity dispersion and half-light radius, with $M_{\text{dyn}} = \beta \sigma^2 r_e / G$. The estimated masses of ETGs can reach up to $10^{13}$ $M_\odot$. The constant $\beta$ is typically assumed to be unity, although detailed comparisons of dynamical mass estimates indicate that this may not always be true (Magorrian & Ballantyne 2001; Koopmans et al. 2009). Based on observations of ETG surface brightnesses, luminosities, $\sigma$, and $r_e$, it has been found that tight empirical relationships exist between the properties. The Faber-Jackson relation describes the power-law relationship between luminosity and $\sigma$ (Faber & Jackson 1976), while the fundamental plane describes the relationship among $r_e$, $\sigma$, and average surface brightness.

---

[14]https://www.illustris-project.org/
[15]https://www.tng-project.org/

ETGs are primarily composed of stars and dark matter, containing little dust and gas. The luminosity output of ETGs is primarily from evolved K-type giant stars, suggesting that many of their stellar populations have evolved off of the main sequence, with high-mass O- and B-type stars no longer contributing to the galaxy's stellar light. For this reason, ETGs are typically referred to as "red and dead", as the giant stars light peaks towards visible red wavelengths.

Lastly, as relevant to the work in Chapter 2, recent observations of ETGs suggest that the stellar populations of these galaxies formed according to a bottom-heavy IMF, i.e. having a higher fraction of low-mass to high-mass stars compared to the inferred IMF in the Milky Way.

### 1.3.2.2 Spiral galaxies

Spiral galaxies, including the Milky Way, are seemingly more complex than ETGs, at least in terms of their visual morphology. Spiral galaxies are composed of a prominent disk, but can also exhibit spiral arms, a central bulge, and a bar. The stellar light profiles of spiral galaxies are generally modeled as two components, a de Vaucouleurs' profile to describe the central bulge and an exponential profile to describe the disk. The stellar light from the disk decreases with radius, with a characteristic scale length $h_r$, which is often used to characterize the sizes of spiral galaxies. These surface brightness profiles are usually corrected for the observed viewing inclination of the disk, as well as averaged over the spiral arms. The typical sizes of spiral galaxies are usually a few times $h_r$, ranging from about 5 to 50 kpc.

Unlike ETGs, the disks of spiral galaxies are characterized ordered stellar rotation. Dynamical masses of spiral galaxies can be measured from these rotation curves, as $M_{\text{dyn}} = v_{\text{rot}}^2 R/G$, where $v_{\text{rot}}^2$ is the stellar circular velocity at radius $R$. The typical masses of spiral galaxies can reach masses of up to $10^{12}$ $M_\odot$. The inner disks of spiral galaxies typically exhibit solid-body rotation, with the orbital velocities of stars increasing with distance from the center of the galaxy. At some radius, differential rotation takes over, and in the outer regions of the disk, stellar velocities remain nearly constant with increasing galactocentric radius. Based on measured stellar rotation curves of spiral galaxies, in the 1970s Vera Rubin was the first astronomer to demonstrate the existence of dark matter. From luminous matter alone (stars, gas, and dust) stellar velocities should decrease with galactocentric radius in the outer disk in accordance with Keplerian motion. The stellar velocities, remaining constant into the outer regions of the disk, she argued necessitated the existence of non-luminous matter (Rubin et al. 1980). As with ETGs, scaling relations have been empirically demonstrated to tightly correlate global properties of spiral galaxies, such as the Tully-Fisher relationship between luminosity and rotational velocity (Tully & Fisher 1977).

The stellar populations of spiral galaxies are quite different than those of ETGs. The disks of spiral galaxies are composed of gas, dust, and primarily young stars, and in the spiral arms hot O- and B-type stars and gas are observed as evidence of active star formation. This recent and ongoing star-formation makes spiral galaxies generally bluer in color than the red and dead ETGs. The central bulges of spiral galaxies have similar stellar populations as ETGs, mostly old and evolved stars. The outer, less dense, regions of these bulges are often referred to as the "halo" which also contains old stars and little gas and dust. Considering the mass distribution of stars, IMF variations have been suggested for spiral galaxies as well, although these observations are less certain given the challenges associated with disentangling stellar mass of spiral galaxies from their gas mass and dust (Li et al. 2017).

### 1.3.2.3 Brief overview of galaxy evolution

Until the advent of large-volume cosmological simulations, there were primarily two emerging scenarios aiming to describe how the different types of galaxies arrived at their observed states. These scenarios are monolithic collapse and hierarchical merging. In the case of monolithic collapse, also referred to as the "top-down" model of galaxy formation, the formation of galaxies is a result of the collapse and fragmentation of the large-scale matter distribution (Eggen et al. 1962). This scenario can be summarized as follows. In the early Universe, fluctuations in the density of dark matter led to the fragmentation of matter on large scales, which would eventually become clusters of galaxies. These large clusters of matter fragmented further, with primordial gas collapsing to form stars, resulting in the creation of galaxies. These galaxies subsequently evolve in relative isolation. Hierarchical merging, also referred to as the "bottom-up" model of galaxy formation, poses a different galaxy evolution scenario (Toomre 1977). In the early Universe, fluctuations in the density of dark matter led to the fragmentation of matter on much smaller scales, leading to the creation of many low mass star clusters. Over time these small stellar systems began to merge together, successively so, until enough stellar mass coalesced and evolved to produce the galaxies we observe today. In particular, massive elliptical galaxies are posited to be a result of the interaction and merging of individual spiral galaxies.

While there still remain many open questions regarding the formation and evolution of galaxies, we have made progress in our understanding due to the increasing quality and quantity of galaxy observations, as well our ability to interpret these observations in the context of cosmological simulations (see Conselice (2014), Naab & Ostriker (2017), and Somerville & Davé (2015) for reviews). It is now generally thought that both the monolithic collapse and hierarchical merging formation mechanisms each play a role. For instance, demonstrating how galaxies may build up their stellar mass over cosmic time, Figure of 10 Oser et al. (2010) shows the stellar mass assembly history for simulated galaxies of different masses. For the more massive galaxies examined, it is suggested that their present-day stellar content is primarily composed of stars that formed at early times, ex-situ of the galaxy itself, and gradually accreted onto the galaxy over its lifetime. On the other hand, for the less massive galaxies in the simulation, it is found that a higher fraction of their stellar content was formed in-situ. This has implications for massive ETGs examined in Chapter 2 of this Dissertation, as well as the Milky Way galaxy, which we focus on in Chapters 3 and 4 of this Dissertation.

### 1.3.3 Measuring unresolved stellar populations

As outlined in Section 1.2.2, there are three primary ways in which we make measurements of individual stars. These are photometry, spectroscopy, and astrometry. However, beyond our own Milky Way and its local neighborhood, making these measurements become more challenging due to the physical limitations of our observing circumstances. Most stars beyond the Local Group cannot be individually resolved with our current telescope and instrument capabilities. Instead, we measure the "integrated light" emitted by collections of stars, which is the summation of the light emitted by each star individually and how this light interacts with gas and dust that it encounters. While the circumstances are slightly different, we use similar methods of photometry, spectroscopy, and astrometry to make measurements of galaxies and other unresolved stellar populations.

### 1.3.3.1 Redshift surveys

First, charting the 3D positions of galaxies throughout the Universe, redshift surveys can be thought of as the extragalactic counterpart of stellar astrometry. Using multi-object spectrographs, redshift surveys measure the precise wavelength positions of prominent galactic spectral features and infer the distances to galaxies by comparing these observed feature positions to their theoretical, rest wavelength positions. In the local Universe, the 2dF Galaxy Redshift Survey (Colless et al. 2001) obtained more than 200,000 galactic redshifts and SDSS obtained nearly 1 million galactic redshifts, each with a mean redshift of $z = 0.1$ and extending out to $z \sim 0.3 - 0.4$. More recent redshift surveys have also probed the structure of the early Universe, with surveys such as the DEEP2 (Deep Extragalactic Evolutionary Probe) Galaxy Redshift Survey (Newman et al. 2013) obtaining galactic spectra for $\sim$50,000 galaxies at $z \sim 1$, spanning $z = 0$ to $z = 1.4$.

These surveys, and others, revealed that galaxies aren't uniformly distributed in space, but are arranged to form long filaments, separated by voids, on scales much larger than the size of the Milky Way. This has placed strong constraints on the distribution of matter in the Universe, as well as on cosmological constants including the Hubble parameter.

### 1.3.3.2  Photometry

Photometric measurements can also be made of unresolved stellar populations and galaxies. Imaging surveys, such as those carried out with HST, SDSS, Spitzer, and GALEX, have made measurements of the luminosity output stellar populations across several wavelength regimes, as well as the 2D surface brightness profiles of these systems. As with photometric measurements of individual stars, imaging surveys have facilitated aperture photometry of galaxies and stellar populations. The Extended Groth Strip (EGS), which is the most extensively imaged patch of sky, has been imaged in numerous bandpasses from the X-ray (3.1 Å) to the radio (20 cm), in a coordinated effort from facilities including Chandra, GALEX, CFHT, HST, Palomar, Spitzer, and the VLA (Davis et al. 2007). From images like these, the magnitudes of galaxies in multiple bands can be measured and galaxy color-magnitude diagrams, as well as full SEDs, can be constructed. For example, UV-optical galaxy CMDs based on NUV photometry from GALEX, which has been empirically demonstrated to be sensitive to radiation from young, newly formed stars, have revealed galaxies to clearly separate in into two clusters, one with redder colors and the other with bluer colors (Wyder et al. 2007). These clusters correspond to the two primary galaxy types: ETGs with their emitted light dominated by old giant stars that peak in the red, and spiral galaxies with their emitted light dominated by young and massive stars that peak in the blue. However, CMDs like this one have also revealed new populations of galaxies, in particular, what has been termed the "green valley" galaxies (Wyder et al. 2007). This demonstrates that galaxies, and their colors and properties, exist on a continuum, rather than falling neatly into distinct categories.

Beyond CMDs, full SEDs can be constructed for galaxies as well as star clusters. Since stellar populations emit light over the entire electromagnetic spectrum, more detailed infor-mation about unresolved stellar populations can be probed by observing these populations in multiple bands over a large wavelength range. As will be discussed stellar population synthesis (SPS) models predict the SEDs of stellar populations over a range of population parameters. By comparing observed SEDs to theoretical ones, we attempt to constrain the star-formation histories, total stellar mass, the IMF, as well as the distances (via photometric redshifts) to unresolved stellar populations. As an example, the `SLUG` (Stochastically Lighting Up Galaxies) SPS code simulates and fits galaxy and star cluster SEDs in a probabilistic manner that takes into account the stochastic nature of star-formation, which is especially pertinent for estimating the properties of low-mass star clusters (Krumholz et al. 2015).

In addition to measuring the colors and SEDs of stellar populations, photometry can also be used to characterize how a galaxy's stellar light is distributed across its observable face. The primary goal of the largest HST campaign, the Cosmic Assembly Near-infrared Deep Extragalactic Legacy Survey (CANDELS) survey, was to further our understanding of galaxy evolution by probing galactic structure in the high redshift Universe (Koekemoer et al. 2011; Grogin et al. 2011). To this goal, CANDELS imaged ∼250,000 galaxies from $z = 8$ to $z = 1.5$ in five fields (including the EGS), using HST's Wide Field Camera 3 (WFC3) and Advanced Camera for Surveys (ACS) detectors, obtaining deep galaxy images in several IR bands. This imaging revealed that galaxies in the early Universe, just ∼1 Gyr since the Big Bang, appear quite different than galaxies in the local ($z \sim 0$) Universe, with high redshift galaxies are often appearing more irregular than present-day galaxies, exhibiting asymmetries, extended tidal arms, double nuclei, and nearby companions. These images of irregular galaxies, which are thought to capture the various stages of galaxy mergers, provides evidence for the hierarchical nature of galaxy evolution, and by studying populations of galaxies across $z = 8$ to $z = 1.5$, our understanding of how ETG and spiral galaxies formed has been greatly advanced (Bell et al. 2012; Bruce et al. 2012).

### 1.3.3.3 Spectroscopy

As for individual stars, spectroscopy is an immensely useful tool for learning about stellar populations and galaxies. For example, the spectrum of a galaxy contains a wealth of information, not only probing its stellar makeup, but also its gas, dust, and total mass content. The spectrum of a stellar population measures the superimposed light of the numerous individual stars that make up the population. This combined starlight is further modified by the gas and dust content surrounding the stellar population. This means that these integrated spectra encode information about the types of stars that compose stellar populations and galaxies, as well as information about the broader environment in which these stars reside. Consequently, beyond their photometric and morphological differences, much of what we know about ETGs and spiral galaxies have been derived from their spectra. For a more detailed overview of galactic spectra see Kennicutt (1992).

The spectra of ETGs typically exhibit a strong continuum component due to the range of blackbody radiation from the underlying stellar population, with a characteristic break in this continuum at wavelengths shorter than $\sim$4000Å. This is because of the lack of hot O- and B-type stars that peak in this bluer wavelength range, as well as the absorption of this high-frequency radiation by stellar atmospheres. Dependent on the temperature and metal content of the stellar atmospheres, certain wavelengths of radiations are additionally absorbed, resulting in prominent absorption features along the radiation continuum. Stellar radiation is also absorbed by cold ISM gas, resulting in additional absorption lines. Typical absorption features observed in the spectra of ETGs include the G-band at 4304Å, calcium (Ca) H+K lines at 4934Å and 3969Å, Mg at 5175Å, sodium (Na) at 5894Å, and titanium oxide (TiO) at 7150Å. The absence of emission lines in the spectra of typical ETGs is indicative of little recent star-formation activity, implying the old age of their stellar populations, as well as a lack of hot ISM gas permeating the galaxy.

The spectra of spiral galaxies are markedly different than those of ETGs. Spiral galaxy spectra also exhibit a strong continuum component reflecting the range of stellar types belonging to its stellar population, however, these spectra typically do not exhibit the same 4000Å break as ETGs. This is because hot O- and B-type stars are still present in their stellar populations, contributing to the blue continuum of the spectrum. In addition to absorption lines, strong emissions lines are a characteristic feature of spiral galaxy spectra. These emission lines are produced by hot gas that is heated and ionized by massive stars, that recombines and re-radiates energy at specific wavelengths. Typical emissions lines observed in spiral galaxy spectra include the Balmer lines (e.g. H$\alpha$, H$\delta$, H$\gamma$, H$\beta$), OIII at 4363Å, and OII at 3737Å. The presence of these emission lines is indicative of a galaxy with a young stellar population with ongoing star-formation, as well as the presence of hot ISM gas. The further presence of absorption lines as well indicates that an older stellar population is also present. The relative strengths of these emissions lines are often used to estimate the star-formation rate of a galaxy.

In addition to probing the types of stars that belong to a galaxy, providing insight into its overall age and star-formation history, galaxy spectra also contain information about the motion of the galaxy as a whole, as well as the motion of the stars it contains. As mentioned earlier, redshift surveys have measured the spectra for hundreds of thousands of galaxies with the goal of measuring their spectroscopic redshifts. This is done by measuring the precise positions of prominent spectral features (such as those listed above), and comparing the observed wavelengths of these features to their expected rest wavelengths. A galaxy's internal stellar motion is also imprinted in its spectrum. The velocity dispersion, which is the dispersion of stellar velocities around their mean motion, results in the Doppler broadening of spectral features. This means that a galaxy's velocity dispersion can be derived by measuring the width of galactic spectral features, which, as described in Section 1.3.2, can be used to estimate the total mass of a galaxy. Beyond measuring the spectrum of a galaxy at one (typically central) location, integral field unit (IFU) spectroscopy instead measures spectra at numerous points across a galaxy's 2-dimensional observable face. These measurements provide much information, probing how stellar populations vary within a galaxy, as well as producing 2D maps of a galaxy's stellar kinematics that probe its gravitational potential as a function of position. IFU surveys include the SDSS MaNGA survey (Bundy et al. 2015), which is measuring the 2D spectra of ∼10,000 nearby galaxies, and the ATLAS$^{3D}$ project (Cappellari et al. 2011) which obtained IFU spectroscopy of ∼300 local ETGs.

One tool for interpreting spectroscopic measurements here, and the photometric measurements described in Section 1.3.3.2, are SPS models, which aim to link these measurements to the physical parameters governing star-formation. The crux of SPS modeling is to determine what mix of individual stars gives rise to the integrated light we observe from galaxies and other unresolved stellar populations. To do this requires modeling the interaction of several physical phenomena that describe the formation and evolution of stellar populations. This includes the IMF describing the mass distribution of stars, a library of either theoretical or empirical stellar spectra to determine the emitted light and abundances of stars (e.g. the MILES (Sánchez-Blázquez et al. 2006) and ELODIE (Prugniel & Soubiran 2001) spectral libraries), isochrones derived from stellar evolution models (e.g. the Padova (Bertelli et al. 1994) and BaSTI (Pietrinferni et al. 2004) isochrones), the star-formation history of a stellar population (e.g. exponentially declining), models for how dust attenuates stellar light, as well as how these phenomena depend on metallicity. While many successful SPS models have been developed and used to interpret the light we receive from unresolved stellar populations (e.g. the Bruzual and Charlot models (Bruzual & Charlot 2003), the Maraston models (Maraston 2005), PEGASE (Fioc & Rocca-Volmerange 1997), and FSPS (Conroy et al. 2009a)), discrepancies do exist among the stellar population properties predicted from different SPS models, such as stellar mass estimates. These discrepancies are thought to be due to the uncertain physics describing many of the ingredients of SPS models, each of which is still an ongoing pursuit of research.

## 1.4 Rising computational power and the astrophysics it enables

Up until now, we have focused our discussion on stars and collections of stars. This has included a review of our basic knowledge of stars and their evolutionary processes, as well as the various ways we measure the properties of individual stars, stellar populations, and galaxies. Along the way, we touched on some of the analytical, empirical, and numerical tools we can use to make sense of these varied observations.

We highlighted some of the pivotal astronomical surveys that have collected the vast amount of stellar observations that we now have, including large photometric surveys (e.g. 2MASS, WISE, GALEX), time domain surveys (e.g. CoRoT, *Kepler*, TESS), spectroscopic surveys (e.g. APOGEE, RAVE, GALAH, LAMOST), and astrometric surveys (e.g. *Hipparcos*, *Gaia*). These surveys have measured stellar properties for hundreds of thousands of stars and stellar populations, and upcoming surveys such as SphereX, LSST, and the Milky Way Mapper Galactic Genesis survey will deliver observations for many more millions of stars throughout the Milky Way and beyond.

With these current and forthcoming astronomical data sets, the goal is to make sense of this large volume of highly-dimensional data, to be able to articulate interpretable connections to astrophysical models and theory. Facilitating this ambition are more recent developments in our computing capabilities, as well as automated methods for learning from large amounts of data.

Over the past two decades, computing power has grown exponentially. Since 1993, the TOP500[16] list has ranked the most powerful supercomputing clusters in the world according to their performance on a single task, the Linpack benchmark test. This test measures the speed at which a system can solve a set of linear equations, providing a single, easily comparable measure of a system's floating point computational ability. This performance is typically measured in terms of FLOPS (floating-point operations per second), which scales with both the number of nodes and cores, as well as the cycle speed, of a computing cluster. Tracking the increase in computing capability, Figure 1.6 shows the performance of the TOP500 supercomputer clusters over the last 26 years, as indicated by the performance of the #1 and #500 ranked supercomputers, as well as the total computing power of all 500 clusters. As seen in the figure, the combined performance of the TOP500 supercomputers has increased by nearly six orders of magnitudes over the last $\sim$30 years, from one trillion ($10^{12}$) FLOP/s in 1993 to one quintillion ($10^{18}$) FLOPS/s in 2019. The performance of the top-ranked supercomputer has similarly increased by six orders of magnitude, from one billion ($10^9$) FLOP/s in 1993 to one quadrillion ($10^{15}$) FLOPS/s in 2019.

A part of the exponential increase in computing power has been driven by the use and development of GPUs. Graphic processing units (GPUs), originally developed for the high-resolution rendering of video and images, have become the primary hardware for machine learning. Compared to central processing units (CPUs), which typically have a few tens of cores, a single GPU is composed of thousands of cores. This makes GPUs much more highly parallel than CPUs, being able to perform thousands of simultaneous operations at once. This property makes GPUs well suited for training machine learning models, which often rely on hundreds of thousands of independent gradient computations. Additionally, while GPUs typically have a higher latency than CPUs, given the number of cores and their higher throughput, GPUs typically outperform CPUs in terms of memory performance as well. This feature is also well suited for machine learning tasks, which often require large amounts of data for training.

---

[16]https://www.top500.org/

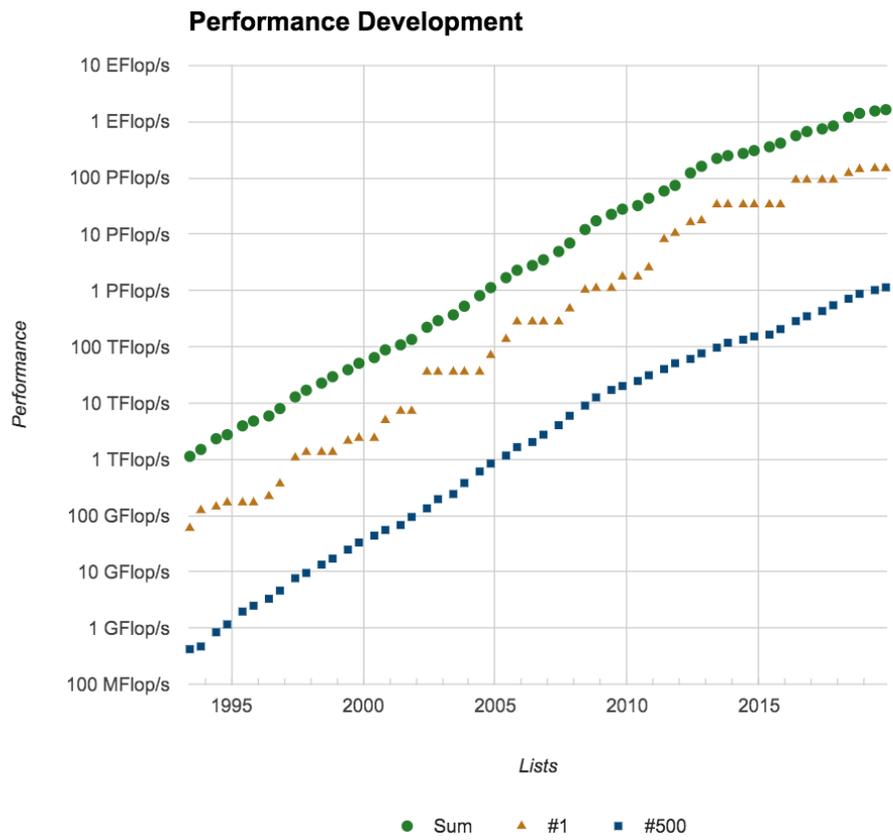

Figure 1.6: The performance of the top 500 supercomputers in the world from 1993 to 2019. Image credit: TOP500, https://www.top500.org/statistics/perf-devel/

The increase in computing power over the past ∼25 years has revolutionized nearly every pursuit of society. In astronomy, this has resulted in increasingly detailed numerical simulations of astrophysical phenomena, as well as the ability to more efficiently process and identify patterns and structure in large amounts of data. In the remainder of this section, we briefly outline both of these pursuits, first discussing cosmological simulations in Section 1.4.1 and then machine learning in Section 1.4.2. Then, in Section 1.4.3, we discuss some of the recent Milky Way science that these advances have enabled.

### 1.4.1 Cosmological simulations

The ability to simulate astrophysical processes is essential to understanding the Universe and our observations of it. Many phenomena, such as the formation and evolution of galaxies, arise from physical processes that are highly complex, often non-linear, and occur over several orders of magnitude in mass, distance, and time (see Somerville & Davé (2015) and Naab & Ostriker (2017) for reviews). This includes stellar evolution, and how stellar populations enrich the ISM over time, as discussed in Sections 1.2 and 1.3, respectively. A primary tool for understanding complex astrophysical phenomena is cosmological simulations, which include numerical prescriptions for how dark matter and baryons interact and evolve from the initial conditions of the early Universe to form the galaxies and structure we observe today. Over the past few decades, cosmological simulations have advanced significantly. This advancement has been driven by the increase in computational power available, which has enabled higher mass and spatial resolutions of these simulations in large cosmological volumes, as well as the implementation of more detailed physical models. By allowing us to make detailed observational predictions, these simulations have provided insight into numerous areas of astrophysics including the large-scale structure and the distribution of dark matter, circumgalactic medium (CGM) dynamics, galaxy evolution and formation, and star-formation in the Universe.

One of the early large-volume cosmological simulations, including prescriptions for dark matter as well as baryonic physics, was the Millennium Run[17], which tracked the evolution of $10^{10}$ particles from $z = 127$ with a mass resolution of $\sim 10^9$ $M_\odot$ (Springel et al. 2005). In a volume of $\sim 600$ Mpc$^3$, $\sim 10^6$ galaxies formed by $z = 0$ with structure resolved at the $\sim 10$ kpc scale. The total number of FLOPs required to carry out the simulation was $\sim 5 \times 10^7$. The Millennium Run was based on the `GADGET` cosmological code (Springel et al. 2001b; Springel 2005), which includes a TreePM algorithm (Xu 1995) to solve the $N$-body problem and smoothed-particle hydrodynamics (SPH) to model fluid flow (Lucy 1977; Gingold & Monaghan 1977).

Facilitated by the increasing computational power, as well as improved numerical techniques, more recent cosmological simulations have pushed the envelope in terms of both the total simulation volume and the mass and spatial resolution. At the forefront of this advancement is the Illustris Project and its successor, IllustrisTNG. The Illustris Project includes a suite of simulations at three mass resolutions, each evolving the Universe from $z \sim 127$ to $z \sim 0$ within a $(106.5 \text{ Mpc})^3$ cube (Vogelsberger et al. 2014a,b; Genel et al. 2014). The highest resolution simulation includes $1.8 \times 10^{10}$ dark matter particles with a mass resolution of $\sim 6.3 \times 10^6$ $M_\odot$, as well as $1.8 \times 10^{10}$ baryonic resolution elements with a mass resolution of $\sim 1.3 \times 10^6$ $M_\odot$, resulting in $\sim 4$ million galaxies by $z = 0$. These simulations compute gravitational forces with TreePM technique (Xu 1995), and carry out hydrodynamic computations with the moving-mesh code `AREPO` code (Springel 2010), which implements a Lagrangian technique to solve the hydrodynamic equations via the finite volume method, where the conservation of energy, mass, and momentum are followed across an adaptive Voronoi tessellation mesh.

---

[17]https://wwwmpa.mpa-garching.mpg.de/galform/virgo/millennium/

In addition to the gravity and hydrodynamic computations, the Illustris simulations include various prescriptions to model galaxy formation and evolution. This includes descriptions of radiative cooling, black hole seeding, star-formation, stellar evolution, and chemical enrichment, as well as galactic winds and AGN feedback, all of which are described in Vogelsberger et al. (2013a). As the focus of this Dissertation is on stars, here we provide a brief description of the Illustris star-formation and stellar evolution models. Stellar particles form according to the Kennicutt-Schmidt relation (Kennicutt 1989) from dense ISM gas over timescales of 2.2 Gyr once densities reach $n \approx 0.13$ cm$^{-3}$. Each stellar particle formed represents an SSP, which returns mass and metals to the surrounding ISM gas cells according to expected stellar lifetimes. Post-main sequence evolution occurs instantaneously, with low-mass stars enriching the ISM via AGB winds and SN Ia, and high-mass stars enriching the ISM via CC-SN, distributed across 9 elements, including H, He, C, N, O, Ne, Mg, Si, and Fe. Through these different channels, the metals ejected from stellar particles over time are computed using yield tables (such as those described in Section 1.3.1.2), which depend on the stellar particle's metallicity, mass, and the assumed IMF.

Continuing the successes of the Illustris Project is the IllustrisTNG Project (herein "TNG") (Pillepich et al. 2018; Weinberger et al. 2017). The TNG Project includes simulations at three different volumes, (302.6 Mpc)$^3$, (110.7 Mpc)$^3$, and (51.7 Mpc)$^3$, named TNG300, TNG100, and TNG50, respectively, at various mass resolutions. The highest resolution TNG300 simulation includes $1.5\times10^{10}$ dark matter as well as $1.5\times10^{10}$ baryonic resolution elements, with mass resolutions of $\sim 6\times10^7$ M$_\odot$ and $\sim 1\times10^7$ M$_\odot$, respectively.

The highest resolution TNG100 simulation includes $6\times10^9$ dark matter and baryonic resolution elements, with mass resolutions of $\sim 7.5\times10^6$ $M_\odot$ and $\sim 1.4\times10^6$ $M_\odot$, respectively. And the highest resolution TNG50 simulation includes $1\times10^{10}$ dark matter and baryonic resolution elements, with mass resolutions of $\sim 4.5\times10^5$ $M_\odot$ and $\sim 8.5\times10^4$ $M_\odot$, respectively. Based on the `AREPO` code, these simulations further include the effects of magnetism, solving the idealized magnetohydrodynamics (MHD) equations over the simulation mesh. TNG's prescription for galaxy formation is built on the physics included in the original Illustris simulations, with updates as described in Vogelsberger et al. (2013b), Pillepich et al. (2018), and Weinberger et al. (2017). Some of the primary changes include an improved galactic winds model, updates to the feedback prescription from black holes, as well as updates yield tables, which result in more realistic galaxies that better match observational constraints.

From large-volume cosmological simulations like Illustris and TNG, we have been able to study the coevolution of dark and baryonic matter starting from the initial conditions of the early Universe, to the formation of the first galaxies, to the present-day properties of galaxies at $z \sim 0$. This includes a diverse range of topics in galaxy evolution including the formation of massive ETGs (Wellons et al. 2015, 2016), the stellar mass assembly of galaxies (Rodriguez-Gomez et al. 2016), the size evolution of galaxies (Genel et al. 2018), the angular momentum of the GCM (DeFelippis et al. 2020), and the distribution of magnesium and europium throughout Milky Way-like galaxies (Naiman et al. 2018). Further, Chapter 2 of this Dissertation uses the Illustris simulation suite to explore putative IMF variations in the context of the hierarchical buildup of massive galaxies.

### 1.4.2 Machine learning in astronomy

Many astrophysical pursuits are increasingly intersecting with data and computational sciences. This is being seen across the study of transient phenomenon (Bellm et al. 2019), exo-planets (Borucki et al. 2003), stars (*Gaia* Collaboration et al. 2016), black holes (Kollmeier et al. 2017), galaxies (York et al. 2000) and cosmology (Abbott et al. 2018). Data and simulations from these realms, and beyond, cannot be reasonably inspected, let alone analyzed and understood, by ocular means (although citizen science routes have been deployed with some success (Fortson et al. 2012)). Machine learning, whereby tasks such as detection, classification, and regression, are automated, affords a promising solution to parse and make sense of large and complex astronomical data sets.

Machine learning is often categorized into supervised and unsupervised frameworks. Unsupervised machine learning describes the extraction of features and characteristics in data, through modeling of their ensemble variance. Unsupervised learning has been used within spectroscopy for dimensionality reduction (e.g. Ting et al. 2012; Price-Jones & Bovy 2018), and for outlier and novel source detection (e.g. Giles & Walkowicz 2018). Supervised learning algorithms typically seek to classify, or label, some data using regression tasks. While classification organizes data into different groups, regression tasks label data with continuous quantities. Popular algorithms for these tasks include random forest and neural networks. Neutral networks have an expansive hierarchy of varieties (e.g. convolutional, recurrent, generative adversarial networks). Similarly, random forest, and other tree-based methods, also allow for a number of possible algorithmic choices.

Employing a supervised machine learning model typically involves the use of training, validation, and test sets of data. These are used to train, optimize, and test the performance of a model for a given task. Generally, models are built with the goal to understand and label new data. In training, the relationship between the data and the labels that describe that data is captured by the model. A validation set is used to optimize the parameters of the model (the so-called hyperparameters). The test set of data is unseen data that are labeled using the model. Test sets are typically used to assess the performance of the model. There are numerous metrics that can be used to quantify the performance of a supervised machine learning model. This includes metrics specific to classification (e.g. accuracy and precision) and regression (e.g. $r^2$ and bias). The choice of evaluation metrics is often specific to the particular uses and goals of the model. For a more complete review of machine learning within astronomy, recent articles by Baron (2019) and Ivezić et al. (2019) provide thorough pedagogical overviews.

In many realms of astronomy, the pace of data collection is outpacing developments in theoretical models (for example, see Chapter 3). As a consequence, the use of machine learning has tended toward "data-driven" approaches. Simple data-driven machine learning methods, that employ linear or polynomial regression models, have led to significant advances within the spectroscopic domain in particular. This includes in labeling stars, and in getting new information out of spectra and photometry (e.g. Wheeler et al. 2020; Eilers et al. 2019; Casagrande et al. 2019). Neural networks are also being increasingly used in many domains data-driven astronomy (for example, see Chapter 4). Although they are often very effective, a frequent problem encountered in using neural networks for data-driven modeling is that of over-fitting. This is often a consequence of the infrastructure they afford exceeding the simple physical parameter space of the data. Another is that many "off-the-shelf" implementations (such as those built using PyTorch and TensorFlow) do not necessarily incorporate errors on input data and labels. This can impact how well models trained on high fidelity data transfer to lower, or variable quality data, at test time.

The inclusion of machine learning within astronomy has been accompanied by efforts to understand which methods perform best in specific domains (Jamal & Bloom 2020). As the field becomes more embedded in automated approaches, astronomy will likely drive the development of the algorithms themselves. There is a particular opportunity to utilize, as well as build, a methodology for the current landscape. The numbers of relatively low-quality data (low signal to noise or short baseline) far exceed the high. Methods that leverage the high fidelity data to learn from the low will be of particularly high utility. This opportunity is a focus of Chapter 4.

### 1.4.3 Opportunities in Milky Way science

As reviewed in the previous sections, rising computational power over the past few decades has facilitated innovation in many realms of astrophysics, particularly through improved simulations of galaxy evolution (including cosmological simulations) and the use of machine learning in astronomy. We now highlight some of the opportunities in Milky Way science and Galactic archaeology (relevant to Chapters 3 and 4) that these tools, and the data sets described in Section 1.2.2, are enabling.

The Milky Way is the galaxy that we can study in the most detail, using the photometry, spectroscopy, astrometry, as well as the time-varying brightness of its individual stars (see Section 1.2.2). Assuming the Milky Way is a typical spiral galaxy (see Section 1.3.2.2), in its study we can learn about processes of galaxy formation in the wider Universe. Most of the stellar mass in the Milky Way is concentrated in the stellar disk, which is typically described as a two-component thin and thick disk system (Gilmore & Reid 1983). The Milky Way shows a boxy bulge morphology (e.g. Ness & Lang 2016), which is consistent with almost entirely being formed from the stellar disk. A small fraction ($\approx 1$ percent) of stellar-mass resides in the halo. The stellar halo represents the earliest relic of the Milky Way's formation and includes sub-structure and streaming systems (Belokurov et al. 2006). Detailed characterizations of the empirical properties of the Galactic disk, where most of the stellar mass resides, is important to understanding the processes of spiral galaxy formation, including both dynamical and chemical evolution (Freeman & Bland-Hawthorn 2002). The properties of the Milky Way are summarized in more detail in Bland-Hawthorn & Gerhard

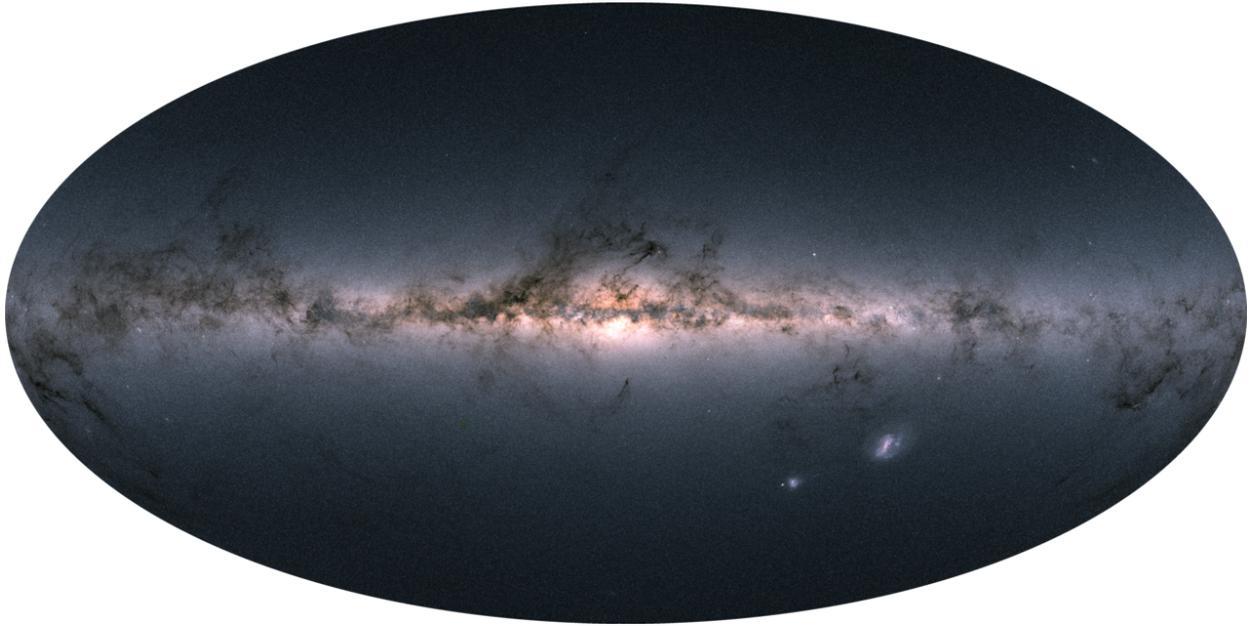

Figure 1.7: The sky map of the observed flux of the $\approx$ 1.7 billion sources in *Gaia* DR2. Image credit: ESA/Gaia/DPAC.

(2016).

Until a few years ago, precision chemical abundances and ages were only measured for a few thousand stars in the solar vicinity (e.g. Rix & Bovy 2013). Now, we have hundreds of thousands to millions of stars with chemical compositions and age estimates, as well as distances and orbital properties, from the bulge to the outer disk (e.g. Ness et al. 2019a). This has opened up many opportunities with respect to advancing our knowledge and understanding of both stellar physics and galactic formation. The *Gaia* all-sky mission has been pivotal in revealing a new view of our Galaxy and its stars, as shown in Figure 1.7. From its astrometric measurements, *Gaia* has delivered stellar velocities and positions for a few million stars, which have been used to characterize stellar orbits via action-angle variables (Trick et al. 2019). Furthermore, time domain missions like *Kepler* and TESS (and in the future, LSST) are contributing time-variability information for a subset of these stars.

A primary challenge in the era of data-rich Milky Way science is how to parse and inter-pret the high volume of stellar data we now have access to. In response to this, data-driven and machine learning methods, that can handle and extract insight from large ensembles of stellar data sets, are being developed. This have led to improved chemical abundance measurements (e.g. Ness et al. 2015; Ting & Rix 2019; Leung & Bovy 2019a), precise distance estimates (e.g. Eilers et al. 2019; Leung & Bovy 2019b) and stellar age estimates (e.g. Martig et al. 2016; Ness et al. 2016), for stars across the Milky Way's disk and bulge. As discussed in Chapter 4, data-driven methods are also being implemented to derive the properties of Milky Way stars from time domain data.

Our understanding of the Milky Way is also being placed in the context of observations of other galaxies (Gadotti et al. 2019), as well as simulated galaxies. Simulations of galaxy formation and evolution can be used to interpret the origin of observed features and structure as well as detailed chemical abundance trends in the Milky Way (e.g. Mackereth et al. 2018; Clarke et al. 2019; Vincenzo & Kobayashi 2020). Some simulations have been sampled with the *Gaia* selection function (Sanderson et al. 2020), and so effectively constructed with this pursuit in mind. Large-volume cosmological simulations (e.g. Illustris and IllustrisTNG) can be used to understand the Milky Way in a broader cosmological context, while also permitting a statistical study of the properties of Milky Way-like galaxies formed in a given simulation volume (e.g. Naiman et al. 2018; Elias et al. 2018).

However, with simulations of galaxy evolution, there is typically a computational trade-off between simulation volume and resolution. "Zoom-in" simulations of Milky Way-like galaxies often permit more detailed studies of internal galactic structure and processes. These simulations, including the FIRE (Hopkins et al. 2014), Auriga (Grand et al. 2017), and Latte (Wetzel et al. 2016) simulations, identify a few to tens of galaxies of interest to resolve at a high mass resolution, while the remainder of the simulation volume is modeled at a lower resolution. Pushing the envelope in terms of both simulation volume and resolution is the TNG50 simulation. With a baryonic mass resolution of $\sim 10^5$ $M_\odot$, the TNG50 volume contains more than $10^3$ galaxies with masses greater than $10^9$ $M_\odot$, about 100 of which are the mass of the Milky Way.

There is now a great opportunity to further our understanding of the Milky Way, both in terms of its detailed formation and evolution and its broader cosmological context. The increasing computational capability described Section 1.4 will continually improve our ability to parse and interpret large collections of stellar observations, as well as advance our simulations of galaxy formation and evolution. This will enable an unprecedented study of the Milky Way by allowing us to connect detailed stellar observations to galaxy simulations with resolved internal structure, dynamics, and chemistry.

## 1.5  Outline of Dissertation contributions

In this Dissertation, we examine the intersection of astronomy and data-driven methods using a diverse collection of data sets, methodologies, and modeling approaches. With these tools, some of which are novel since the pioneering work of early astronomer Annie Jump Cannon, we can use the properties of stars to probe astrophysical objects on multiple scales, ranging from the interiors of stars to the Milky Way galaxy itself, and to the population of massive galaxies in the Universe. Each of the following chapters makes a contribution to a different subfield of astronomy, with the common approach of using large and complex data sets to link measurements of stars to a physical interpretation and its consequences.

Starting at the largest scales in Chapter 2 (published as Blancato et al. (2017)), we examine the connection between the local properties of star formation at the time of stellar birth and the global properties of galaxies that we observe today. Specifically, motivated by the growing evidence for a variable IMF both among and within elliptical galaxies, we use the Illustris cosmological simulation to explore the implications of a variable IMF in the context of the hierarchical buildup of massive galaxies through cosmic time. We find that we are unable to reproduce the observed IMF trends at $z = 0$, which has implications for the reliability of these observations, for the ability of cosmological simulations to capture the build-up of galaxy stellar populations, and for theories of star-formation aiming to explain the physical origin of IMF variations.

In Chapter 3 (published as Blancato et al. (2019)) we narrow our focus to the Milky Way and examine the link between the [Mg/Si] abundance ratio, the structure of the Galactic disk, and models for the chemical enrichment history of the Galaxy. Inspired by the expected differences in the nucleosynthetic origin of Mg and Si, we use the APOGEE and *Gaia* datasets to characterize how [Mg/Si] varies throughout the Milky Way, examining trends with age, [Fe/H], $\alpha$-enhancement, location in the Galaxy, and orbital actions. In an attempt to connect these variations to a physical origin, we fit chemical evolution models to a handful of stars and their Mg, Si, and Fe abundances, finding that we are unable to reproduce the abundance patterns for an ensemble of stars. This result highlights tensions between the data, yield tables, and chemical evolution models, and suggests that a more data-driven approach to nucleosynthetic yield tables and chemical evolution modeling is necessary to maximize insights from large spectroscopic surveys.

At the finest spatial and temporal resolutions we consider, in Chapter 4 (submitted as Blancato et al. (2020)) we explore the prospect of data-driven methods for deriving stellar properties from photometric time series data. If successful, this would enable the properties of millions of stars to be mapped throughout the Galaxy as ongoing and upcoming missions, such as TESS and LSST, plan to deliver light curves for $10^5$ and $10^8$ stars, respectively. In this paper we investigate this potential by training convolutional neural network models on *Kepler* Q9 light curves and high-quality stellar property catalogs to predict a set of stellar parameters ($\nu_{\max}$, $\Delta\nu$, log $g$, $T_{\text{eff}}$) for main sequence and red giant branch stars. With a particular focus on deriving stellar rotation periods from short baseline light curves (such as the TESS 27-day data), we demonstrate how the recovery quality degrades with the baseline of the light curves, finding that the decrease in performance is only marginal down to baselines of 14 days, with the model performing well for stars with rotation periods up to 35 days. This work demonstrates the potential to derive properties from light curve data alone for many stars across the Milky Way.

Finally, in Chapter 5 we summarize the contributions of this Dissertation, and situate it in the context of a complex and changing understanding of the past, present, and future.

# Chapter 2

# Implications of galaxy buildup for putative IMF variations in massive galaxies

## 2.1 Abstract

Recent observational evidence for initial mass function (IMF) variations in massive quiescent galaxies at $z = 0$ challenges the long-established paradigm of a universal IMF. While a few theoretical models relate the IMF to birth cloud conditions, the physical driver underlying these putative IMF variations is still largely unclear. Here we use post-processing analysis of the Illustris cosmological hydrodynamical simulation to investigate possible physical origins of IMF variability with galactic properties. We do so by tagging stellar particles in the simulation (each representing a stellar population of $\approx 10^6$ M$_\odot$) with individual IMFs that depend on various physical conditions, such as velocity dispersion, metallicity, or SFR, at

---
This chapter is a reproduction of a paper by Blancato, Genel, & Bryan (2017) published in The Astrophysical Journal.

the time and place the stars are formed. We then follow the assembly of these populations throughout cosmic time, and reconstruct the overall IMF of each $z = 0$ galaxy from the many distinct IMFs it is comprised of. Our main result is that applying the observed relations between IMF and galactic properties to the conditions at the star-formation sites does not result in strong enough IMF variations between $z = 0$ galaxies. Steeper physical IMF relations are required for reproducing the observed IMF trends, and some stellar populations must form with more extreme IMFs than those observed. The origin of this result is the hierarchical nature of massive galaxy assembly, and it has implications for the reliability of the strong observed trends, for the ability of cosmological simulations to capture certain physical conditions in galaxies, and for theories of star-formation aiming to explain the physical origin of a variable IMF.

## 2.2 Introduction

The stellar initial mass function (IMF) has long been thought a universal feature of star formation. Although observational support for a universal IMF has been mounting for more than half a century, the physical theory behind the IMF still remains an unsolved problem in star formation physics (e.g. Bastian et al. 2010). The elusive origin of the IMF stems from deficient knowledge of the physics governing how molecular clouds collapse and fragment to form protostars and the subsequent accretion of gas onto these protostars (e.g. Offner et al. 2014). Observational measurements of the IMF have also proved to be particularly challenging, both for resolved stellar populations where individual stars can be counted, and for distant (e.g. Chabrier 2003a), unresolved stellar populations where the IMF must be inferred more indirectly (e.g. Tang & Worthey 2015). Recent observations suggesting a variable IMF, as measured within massive elliptical galaxies, serve to further compound the elusive origin of the IMF (e.g. Cappellari et al. 2012; Conroy & van Dokkum 2012b; Spiniello et al. 2014; Ferreras et al. 2010). While understanding the shape of the IMF would certainly provide insight into the physical processes that control star formation, the IMF also has vast implications for the study of galaxy populations as the IMF influences nearly all observable galaxy properties including stellar mass, luminosity, metal content, and star formation history.

Observations of stellar populations within and near the Milky Way gave rise to the conception of a universal IMF. The first determination of the IMF was by Salpeter (1955) who, using field stars in the solar neighborhood, found the IMF to follow a unimodal power law with a slope of $x = 2.35$. Since Salpeter's initial measurement the IMF has been extensively measured in nearby, resolved stellar populations. These measurements have revealed the IMF to turn over at lower masses, following a Chabrier (2003c) log-normal or a Kroupa (2001) segmented power law IMF below $\sim 1$ $M_\odot$ and the original Salpeter IMF at stellar masses greater than $\sim 1$ $M_\odot$. The shape of the IMF has been measured to be largely consistent across a variety of stellar populations within and near the Milky Way, including young clusters (e.g. Hillenbrand 1997; Weights et al. 2009), open clusters (e.g. Massey et al. 1995; Carraro et al. 2005), globular clusters (e.g. McLaughlin & van der Marel 2005; Kruijssen 2008), the Large Magellanic Cloud (LMC) (e.g. Kerber & Santiago 2006; Da Rio et al. 2009) and Small Magellanic Cloud (SMC) (e.g. Sirianni et al. 2002; Sabbi et al. 2008), and M31 and M32 (e.g. Zieleniewski et al. 2015). For a detailed discussion of the nuances of local IMF observations we direct the reader to Chabrier (2003a) and Bastian et al. (2010).

Surprisingly, observations over the past few years of more distant stellar populations suggest deviations from the universal IMF inferred in the Local Group. Several independent methods have been implemented to study the IMF in unresolved stellar populations, includ-ing: (1) dynamical studies where stellar population synthesis (SPS) mass-to-light ratios are compared to dynamically derived mass-to-light ratios (e.g. Dutton et al. 2011; Cappellari et al. 2012, 2013; Conroy & van Dokkum 2012b; Conroy et al. 2013; Martín-Navarro et al. 2015c), (2) absorption line studies where spectral features either sensitive or anti-sensitive to dwarf stars provide the constraints (e.g. Cenarro et al. 2003; van Dokkum & Conroy 2010; Ferreras et al. 2013; La Barbera et al. 2013; Spiniello et al. 2014; McConnell et al. 2016), and (3) lensing studies where SPS masses are compared to masses derived from gravitational lensing (e.g. Ferreras et al. 2008, 2010; Auger et al. 2010; Treu et al. 2010; Thomas et al. 2011; Posacki et al. 2015; Leier et al. 2016).

These measurements predominately infer that the IMF of nearby early-type galaxies (ETGs) becomes more bottom-heavy, i.e. having increasingly numerous low-mass stars with respect to high-mass stars, at higher values of galaxy properties such as velocity dispersion ($\sigma$), metallicity ([M/H]), and metal abundance ratio ([$\alpha$/Fe]). For example, Conroy et al. (2013) find IMF variations with velocity dispersion for a sample of compact ETGs by using mass-to-light ratio as a proxy for the fraction of low-mass stars, as these systems are believed to be stellar dominated at their centers. They find low velocity dispersion galaxies ($\sigma \sim 100$ km s$^{-1}$) to be best described by a Milky Way-like IMF, galaxies with intermediate velocity dispersions ($\sigma \sim 160$ km s$^{-1}$) best fit by a Salpeter IMF, and galaxies with $\sigma \sim 250$ - 300 km s$^{-1}$ best described by an IMF even more bottom-heavy than the Salpeter IMF. Similar IMF trends have also been observed to scale with metallicity. For example, using a sample of ETGs from the CALIFA survey, Martín-Navarro et al. (2015c) find the most metal-poor ETGs in their sample ([M/H] $\sim$ -0.2) to be best described by an IMF slope of $x \sim 2$ and the most metal-rich ETGs ([M/H] $\sim$ 0.2) best described by an IMF slope of $x \sim 2.9$.

IMF variations have also been observed *within* galaxies, highlighting the complications of systematics like aperture radius when comparing IMF measurements across studies (e.g.

Martín-Navarro et al. 2015a; La Barbera et al. 2016; van Dokkum et al. 2017) (however, see Vaughan et al. (2018) who find a constant IMF at all radii for two ETGs with $\sigma = 410$ and 260 km s$^{-1}$). In particular, using deep spectroscopic data taken at various fractions of the effective radius ($R_e$), Martín-Navarro et al. (2015a) find significant radial IMF trends for the highest velocity dispersion galaxies in their sample ($\sigma \sim 300$ km s$^{-1}$), starting with an IMF slope of $x \sim 3$ at galaxy centers, down to an IMF slope of $x \sim 1.9$ at $r = 0.7$ R$_e$. However, for the lower velocity dispersion ETG in their sample ($\sigma \sim 100$ km s$^{-1}$) the IMF is found to be constant with galactocentric distance. These trends may be understood in the context of the 'minor mergers' scenario (Naab et al. 2014; Oser et al. 2012; Rodriguez-Gomez et al. 2016), according to which the most massive ETGs accrete a large number of small galaxies in particular in their outer regions.

Similar IMF variations have also been recently reported for high redshift ETGs. Comparing dynamical to SSP masses, Shetty & Cappellari (2014, 2015) find a Salpeter IMF, rather than the 'universal' Chabrier, for their sample of massive ($> 10^{11}$ M$_\odot$) ETGs at $z \sim 0.8$. For a sample of ETGs at $z \sim 1.4$ with both dynamical and photometric mass estimates, Gargiulo et al. (2015a) report an IMF-$\sigma$ relation consistent with the trends observed at $z \sim 0$. They further posit that the IMF of dense ($> 2500$ M$_\odot$ pc$^{-2}$) ETGs is independent of redshift over the past $\sim$9 Gyr. Additionally, based on the dwarf-sensitive TiO$_2$ feature, Martín-Navarro et al. (2015b) find similar IMF variations for massive ETGs from $0.9 < z < 1.5$ that are consistent with a constant IMF over the past $\sim$8 Gyr.

Though the evidence for IMF variations is mounting, a consensus has yet to be reached, as several studies report discrepant results. For example, Conroy & van Dokkum (2012b) find for their sample of ETGs the strongest IMF correlation to be with [Mg/Fe], with weaker IMF-$\sigma$ and IMF-[M/H] correlations. La Barbera et al. (2015), on the other hand, report negligible IMF correlation with [Mg/Fe]. Additionally, based on studies of the low-mass

X-ray binaries (LMXB) of ETGs, whose number is expected to scale with the IMF, Peacock et al. (2014) argue for an invariant IMF reporting a LMXB population per mass that is constant across a range of galaxy velocity dispersions. Studies based on lensed galaxies have also yielded discrepant results. Smith et al. (2015b) found that for two strong lens ETGs in the SINFONI Nearby Elliptical Lens Locator Survey (SNELLS) with $\sigma \sim 300$ km s$^{-1}$ a bottom-heavy IMF is ruled out in favor of a Kroupa IMF, but find their 1.14 Na I $\mu$m index strengths to suggest they have bottom-heavy IMFs (Smith et al. 2015a). Recently Newman et al. (2017) compared lensing, dynamical, and SPS techniques for inferring the IMF of these SNELLS galaxies, finding that the SPS stellar mass-to-light ratios exceed the total lens mass-to-light ratio, and that there is even a significant discrepancy between the lensing and dynamical masses. Newman et al. (2017) discusses several possibilities for the origin of these tensions, but this study suggests there could be systematic errors in at least one of the techniques used to probe the IMF of ETGs.

These discrepancies highlight the importance of understanding the uncertainties in inferring IMF variations. Emphasizing the difficulties of inferring the IMF from integrated light, Tang & Worthey (2015) find that the degeneracy between a bottom-heavy IMF and decreasing AGB (asymptotic giant branch) strength is only confidently broken for old, metal-rich galaxies with a combination of accurate spectra and photometric observations at the .02 mag level. Clauwens et al. (2015) examines the influence of measurement error and selection bias on IMF variations using a sample of galaxies from the ATLAS-3D project. They find that $\sim$30% gaussian errors on kinematic measurements of mass-to-light ratios lead to similar IMF variations as reported in Cappellari et al. (2012), emphasizing the importance of correctly modeling measurement errors. Clauwens et al. (2015) also find that galaxy selection can significantly influence the inferred IMF trend. Placing a cut on star-formation (as most studies reporting IMF variations do) removes low velocity dispersion galaxies with IMFs comparable to quiescent, high velocity dispersion galaxies. Additionally, excluding galaxies with kinematic masses below the ATLAS-3D mass completeness threshold ($2\times10^{10}$ M$_\odot$) removes the IMF trend with velocity dispersion.

If however proven robust, a variable IMF will have significant implications for our current understanding of galaxy formation and evolution: an understanding that has largely been developed under the assumption of a universal IMF. In particular, a variable IMF will affect the derived properties of galaxy populations. For example, Clauwens et al. (2016) quantifies the effect of the metallicity dependent IMF relation in Martín-Navarro et al. (2015c) on the derived quantities of nearly $2 \times 10^5$ SDSS galaxies. Inferred star formation rates increase by up to two orders of magnitude and stellar mass densities increase by a factor of 2.3 compared to what is inferred with the Chabrier IMF. Adopting an IMF relation dependent on velocity dispersion, McGee et al. (2014) find similar effects on the derived properties of SDSS galaxies, with the shape of the high-mass end of the galaxy stellar mass function shifting from the familiar exponential (using a Chabrier IMF) to a power-law. From a more theoretical standpoint, Vincenzo et al. (2016) use a numerical chemical evolution code to quantify the effect of the IMF shape on metal yields of galaxies. Exploring both the upper mass cut off and the slope of the IMF, they find that the metal yield can vary by up to a full order of magnitude.

Against this observational background, several star-formation models have been recently developed that predict IMF variations. Some models find that the larger density fluctuations associated with higher Mach numbers in star-forming disks cause the low-mass turnover of the pre-stellar core mass function (CMF) to shift to lower masses, implying a more bottom-heavy IMF (e.g. Hopkins 2013; Chabrier et al. 2014; Guszejnov et al. 2016; but see Bertelli Motta et al. 2016a). These models are able to reconcile the universal IMF found across a range of Milky Way stellar populations with a bottom-heavy IMF in more extreme star-formation environments such as starbursts. Also potentially able to account for differences between such environments is the Krumholz (2011) derivation of the low-mass turnover as a function of fundamental constants and a weak dependence on interstellar pressure and metallicity. Another theory is IGIMF (Integrated Galaxy-wide stellar Initial Mass Function), which formulates the shape of the overall IMF of a galaxy based on the properties of its individual molecular clouds, which are in turn controlled by the total galaxy SFR (Weidner et al. 2013b). In this theory, high SFR environments, such as starbursts, are predicted to undergo star-formation which follows a top-heavy IMF, so the excess mass inferred for massive elliptical galaxies is predicted to be in part due to stellar remnants.

Semi-analytical and numerical models have also recently been employed to study IMF variations. The impact of IMF variations on the chemical abundances of galaxies has been quantified using both `SAG` (Gargiulo et al. 2015b) and `GAEA` (Fontanot et al. 2017). Fontanot (2014) uses `MORGANA` to investigate the effect of IMF variations on both stellar mass and star-formation rate, reporting that top-heavy IMF in high star-formation rate environments produces the largest deviations from properties derived under a standard IMF. A variable IMF has also been explored in the context of simulations of individual galaxies, finding that chemical evolution is dependent on the assumed IMF (Bekki 2013; Few et al. 2014). Most recently, Sonnenfeld et al. (2017) developed a simple model of merger-driven galaxy evolution to predict the evolution of IMF trends with mass and velocity dispersion from $z = 2$ to $z = 0$, predicting that the IMF slope of a galaxy is steeper at earlier times.

In this study, we use the Illustris cosmological hydrodynamical simulation to connect the physical conditions in which stars form to the global properties of galaxies at $z = 0$. We construct the IMF for a sample of $z = 0$ Illustris galaxies by using prescribed IMF relations applied to the birth properties of the individual stellar populations that comprise each galaxy. By attempting to reproduce observed relations between the overall IMF of a galaxy (or of its central parts) and its $z = 0$ properties, we are able to provide constraints on relations between IMF and physical conditions at the time of stellar birth.

This paper is organized as follows. In Section 2.3 we describe the Illustris simulation, our galaxy selection, and the method for constructing the IMF. In Section 2.4 we present the primary results: global IMF trends with galactic properties at $z = 0$. In Section 2.5 we explore additional constraints: radial trends, scatter, and redshift evolution. In Section 2.6, we test the robustness of our results to both resolution and variations in the simulation. Finally, in Section 2.7 we discuss our findings, and summarize in Section 2.8.

## 2.3 Methods

### 2.3.1 Simulation suite

To investigate possible physical origins of the observed IMF variations we primarily use cosmological simulations from the Illustris Project (Vogelsberger et al. 2014a,b; Genel et al. 2014), and in particular the highest resolution hydrodynamical simulation in the suite, the Illustris simulation. These simulations evolve down to $z = 0$ a volume large enough to contain statistically significant galaxy populations, and incorporate crucial physics resulting in many realistic galaxy properties. Illustris has been used to study a diverse range of topics in galaxy evolution including, in particular, topics directly relevant to this work, such as the formation of massive, compact ETGs (Wellons et al. 2015, 2016) and the stellar mass assembly of galaxies (Rodriguez-Gomez et al. 2016).

The Illustris simulation treats hydrodynamical calculations using the moving-mesh code `AREPO` (Springel 2010), which has proven advantages over both adaptive mesh refinement (AMR) and smoothed particle hydrodynamics (SPH) techniques (Sijacki et al. 2012; Kereš et al. 2012; Vogelsberger et al. 2012). Gravitational forces are computed using a Tree-PM technique (Xu 1995) that calculates short-range forces using the tree algorithm and long-range forces using the particle mesh (PM) method. A $\Lambda$CDM cosmology with $\Omega_m = 0.2726$, $\Omega_\Lambda = 0.7274$, $\Omega_b = 0.0456$, and $h = 0.704$ from WMAP9 (Hinshaw et al. 2013) is adopted for all simulations used in this study. The galaxy formation physics implemented in Illustris includes radiative cooling, star formation and evolution, including chemical enrichment, black hole seeding and accretion, stellar feedback in the forms of ISM pressure and galactic winds, as well as AGN feedback. Since our study focuses on the stellar populations in Illustris, below we provide a short description of the stellar formation and evolution model. For an in depth discussion of all of the physical models included in Illustris the reader is referred to Vogelsberger et al. (2013a).

Stellar particles form according to the Kennicutt-Schmidt relation (Kennicutt 1989) from dense ISM gas with a time scale of 2.2 Gyr at the density threshold of $n \approx 0.13$ cm$^{-3}$. This gas is pressurized following an effective equation of state for a two-phase medium (Springel & Hernquist 2003). Each stellar particle represents a simple stellar population (SSP) consisting of stars formed at the same time with the same metallicity. Stars in each stellar particle SSP return mass and metals to surrounding gas cells following their expected lifetimes with post-main sequence evolution occurring instantaneously, where low mass stars return mass through AGB winds and more massive stars return most of their mass to the ISM via supernovae. Hence, the mass loss and metal production of each stellar particle as a function of its age are calculated using tables in accordance with the particle's initial mass, metallicity, and assumed IMF.

The SSP of each stellar particle in Illustris is assumed to have a stellar mass distribution described by a Chabrier IMF. The IMF affects the mass and metal return via the evolution of high mass stars, as well as the energy available for galactic wind feedback, which however has a pre-factor that is a tunable parameter of the model. In Section 2.6 we explore simulations not included in the Illustris suite that are evolved with different IMFs, such as the Salpeter IMF as well as a variable IMF. In addition, we study simulations that adopt different degrees of feedback.

In post-processing, structure is identified in each snapshot first using the `FoF` (friends-of-friends) algorithm (Davis et al. 1985) and then an updated version of the `SUBFIND` algorithm (Springel et al. 2001a; Dolag et al. 2009). The `FoF` algorithm identifies dark matter halos using a linking length of one-fifth the mean separation between dark matter particles, with the baryonic particles (gas, stars, and black holes) assigned to the FoF group of their closest dark matter particle if it is close enough by the same separation criterion. The `SUBFIND` algorithm identifies gravitationally-bound substructure (subhalos) within each parent `FoF` group. The dark matter and baryonic components of each subhalo constitute what we refer to as a galaxy.

The publicly released suite of hydrodynamical Illustris simulations[1] includes three runs of the same volume at increasing resolution levels: Illustris-3,-2, and -1. Illustris-1 includes $1820^3$ dark matter particles with masses $m_{\rm DM} = 6.26 \times 10^6$ $M_\odot$ and $\sim 1820^3$ baryonic resolution elements with an average mass of $\overline{m_b} = 1.26 \times 10^6$ $M_\odot$, evolved within a $(106.5 \text{ Mpc})^3$ cube. More details about Illustris-1 and the other simulations used in this study are given in Table 2.1.

### 2.3.2 Galaxy selection

In selecting galaxies in Illustris, we aim to mimic the typical properties of the ETGs examined in the observational IMF studies. We therefore first select galaxies with stellar masses greater than $10^{10}$ $M_\odot$ and specific star formation rates $< 10^{-11}$ yr$^{-1}$. The total stellar mass of each galaxy is calculated as the sum of the stellar particles assigned to it by `SUBFIND` and the

---

[1] http://www.illustris-project.org/data/ (Nelson et al. 2015)

Table 2.1: Simulation parameters

| Simulation | Volume [(Mpc/h)³] | $N_{\text{snapshots}}$ | $N_{\text{DM}}$ | $m_{\text{DM}}$ [M$_\odot$] | $\epsilon_{\text{b}}$ [kpc] | $\epsilon_{\text{DM}}$ [ckpc] | $N_{\text{galaxies}}$ [at $z=0$] | selected $N_{\text{galaxies}}$ [at $z=0$] |
|---|---|---|---|---|---|---|---|---|
| (1) | (2) | (3) | (4) | (5) | (6) | (7) | (8) | (9) |
| Illustris-1 | $75^3$ | 134 | $1820^3$ | $6.3 \times 10^6$ | 0.7 | 1.4 | 4366546 | 371 |
| Illustris-2 | $75^3$ | 136 | $910^3$ | $5.0 \times 10^7$ | 1.4 | 2.8 | 689785 | 229 |
| Illustris-3 | $75^3$ | 136 | $455^3$ | $4.0 \times 10^8$ | 2.8 | 5.7 | 121209 | 103 |
| No feedback | $40^3$ | 31 | $320^3$ | $1.7 \times 10^8$ | 1.8 | 3.6 | 38363 | 178[†] |
| Winds only | $40^3$ | 31 | $320^3$ | $1.7 \times 10^8$ | 1.8 | 3.6 | 37121 | 41[†] |
| IMF-Salpeter | $40^3$ | 31 | $320^3$ | $1.7 \times 10^8$ | 1.8 | 3.6 | 38375 | 23 |
| IMF-Spiniello | $40^3$ | 31 | $320^3$ | $1.7 \times 10^8$ | 1.8 | 3.6 | 37638 | 31 |

(1) Simulation name; (2) Volume of box where the hubble constant is $h = 0.704$ [100 km/s/Mpc];
(3) Number of snapshots produced; (4) Number of dark matter particles; (5) Mass of each dark matter particle;
(6) Baryonic gravitational softening length at $z = 0$; (7) Dark matter gravitational softening length in comoving kpc;
(8) Number of galaxies identified at $z = 0$; (9) Number of massive ($M_* > 10^{10}$ M$_\odot$, $\sigma_* > 150$ km s$^{-1}$),
quiescent (sSFR $< 10^{-11}$ yr$^{-1}$) galaxies at $z = 0$, [†] no sSFR cut specific star formation rate is calculated as the sum of the instantaneous SFRs of its gas cells, divided by the stellar mass. Both the stellar mass and sSFR are calculated within two times the stellar half-mass radius of each galaxy. For Illustris-1 at $z = 0$ these selection criteria result in a sample of 1160 galaxies. The mean stellar mass of this sample is $M_* = 10^{10.88}$ M$_\odot$ and the mean specific star formation rate is sSFR $= 2.45 \times 10^{-12}$ yr$^{-1}$.

Additionally, as done in Spiniello et al. (2014), we limit our sample to galaxies with stellar velocity dispersions greater than 150 km s$^{-1}$. The stellar velocity dispersion, $\sigma_*$, for each galaxy is calculated as the one-dimensional, $r$-band luminosity-weighted velocity dispersion of the stellar particles falling within one-half the projected[2] stellar half-mass radius ($0.5 R^p_{1/2}$). The velocity dispersion criterion reduces our sample from 1160 to 371 galaxies, with $\overline{M_*} = 10^{11.4}$ M$_\odot$, $\overline{\text{sSFR}} = 1.22 \times 10^{-12}$ yr$^{-1}$, and $\overline{\sigma_*} = 198$ km s$^{-1}$.

### 2.3.3 IMF construction

Since the IMF is set at the time of stellar birth, we probe the birth conditions of the stellar particles belonging to our selection of $z = 0$ galaxies. To do this we trace each of the 196335880 stellar particles belonging to the 371 $z = 0$ selected galaxies back to the snapshot in which it first appears and compute several physical quantities (discussed in Section 2.4) that represent its birth conditions.

---
[2]The projected half-mass radius $R^p_{1/2}$ is calculated as the radius containing half the total stellar mass of the galaxy, including stellar particles that fall within this projected radius as viewed from the $z$-direction.

Following Spiniello et al. (2014); Treu et al. (2010) we assign an IMF mismatch parameter, $\alpha_{\rm IMF}$, to each stellar particle:

$$\alpha_{\rm IMF} = \frac{(M_*/L)}{(M_*/L)_{\rm Salp}}, \tag{2.1}$$

where $(M_*/L)_{\rm Salp}$ is the mass-to-light ratio expected assuming a Salpeter IMF and $(M_*/L)$ is the actual mass-to-light ratio assumed for the particle. For stellar populations with an IMF more 'bottom-heavy' compared to the Salpeter IMF, $\alpha_{\rm IMF} > 1$, while stellar populations 'bottom-light' compared to the Salpeter IMF have $\alpha_{\rm IMF} < 1$. A Chabrier IMF is described by $\alpha_{\rm IMF} = 0.6$.

To construct the overall $\alpha_{\rm IMF}$ of each galaxy described in Section 2.3.2 we mass-weight $\alpha_{\rm IMF}^{-1}$ using the birth mass of all (or the innermost subset of) the stellar particles comprising a galaxy. This is equivalent to summing up the light, $L/L_{x=2.35}$, assigned to the stellar particles belonging to each galaxy.

Beginning our exploration, we are inspired by the observations when assigning an $\alpha_{\rm IMF}$ to the stellar particles. For example, we use the $\alpha_{\rm IMF}$-$\sigma_*$ relation presented in Spiniello et al. (2014) as an input relation applied to the local velocity dispersions of the stellar particles. In general though, we have the freedom to construct input $\alpha_{\rm IMF}$ relations that scale with various physical quantities at the star-formation sites. In Section 2.4 we present the Illustris $\log(\alpha_{\rm IMF})$-$\sigma_*$ relations constructed using five different physical quantities associated with the birth conditions of each stellar particle: global stellar velocity dispersion ($\sigma_*$), local dark matter velocity dispersion ($\sigma_{\rm birth}$), local metallicity ([M/H]), global star-forming gas velocity dispersion ($\sigma_{\rm gas}$), and global star-formation rate (SFR).

## 2.4 Investigations of IMF physical drivers

### 2.4.1 Global velocity dispersion

We begin our investigation by first constructing the overall $\log(\alpha_{\text{IMF}})$ of each galaxy in the Illustris sample based on the global stellar velocity dispersion, $\sigma_*$. We trace each stellar particle belonging to a $z = 0$ galaxy back to the progenitor galaxy it was formed in, and compute that galaxy's stellar velocity dispersion in exactly the same way $\sigma_*$ was computed for the $z = 0$ galaxy sample.

With the global $\sigma_*$ associated with each star particle, we construct the overall IMF mismatch parameter ($\log(\alpha_{\text{IMF}})$) according to the prescription outlined in Section 2.3.3. As a starting point, we apply the observed relation presented in Spiniello et al. (2014),

$$\log(\alpha_{\text{IMF}}) = (1.05 \pm .2)\log(\sigma_*) - (2.5 \pm .4), \tag{2.2}$$

which is derived over a range of SDSS ETGs ($z \leq 0.05$) with velocity dispersions between $\sigma_*$ = 150 km s$^{-1}$ and $\sigma_*$ = 310 km s$^{-1}$, by comparing spectral lines sensitive to low-mass stars to the corresponding index strengths in the Conroy & van Dokkum (2012a) SSP models. The mismatch value corresponding to a Chabrier IMF, $\log(\alpha_{\text{IMF}})$ = -0.22, is adopted for all stellar particles with $\sigma_*$ less than 150 km s$^{-1}$.

The red curve in Figure 2.1 shows the resulting overall $\log(\alpha_{\text{IMF}})$-$\sigma_*$ relation for the $z = 0$ Illustris galaxy sample. This relation was constructed based on the innermost parts of each galaxy, using only the stellar particles residing within $0.5R^p_{1/2}$ from the center of the galaxy, to approximately match Spiniello et al. (2014). The red curve in the inset shows the input relation used to construct $\log(\alpha_{\text{IMF}})$, which is the same as the observed Spiniello et al. (2014) relation. The latter is repeated in the main panel as the black curve, to guide the eye. As evident in Figure 2.1, the Spiniello et al. (2014) relation applied at the time of stellar birth is not conserved through the assembly history of massive galaxies. The overall $\log(\alpha_{\rm IMF})$-$\sigma_*$ relation is $\sim 2.5\times$ too shallow compared to the observed $\log(\alpha_{\rm IMF})$-$\sigma_*$ relation. While the observed overall $\log(\alpha_{\rm IMF})$ is reproduced for the lowest velocity dispersion galaxies in the sample, the constructed $\log(\alpha_{\rm IMF})$ of the higher velocity dispersion galaxies becomes increasingly too low. To reproduce the observed $\log(\alpha_{\rm IMF})$-$\sigma_*$ relation, we construct input relations steeper than Equation 2.2. By minimizing the residuals between the resulting Illustris output relation and the Spiniello et al. (2014) relation, we determine input relation that produces the best-fit. The blue curves in Figure 2.1 show that with an input relation $3.5\times$ steeper than the observed relation (shown in the inset), the resulting overall $\log(\alpha_{\rm IMF})$-$\sigma_*$ relation (shown in the main panel) *is* able to reproduce both the slope and normalization of the observed trend within the reported errors.

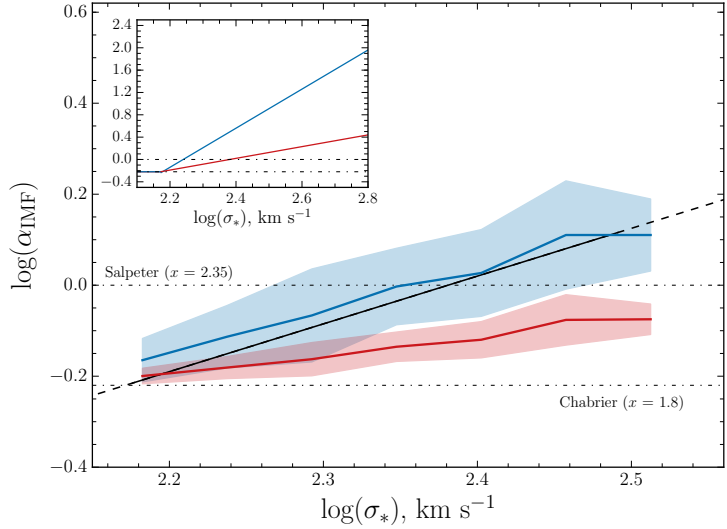

Figure 2.1: Main panel: IMF mismatch parameter, $\log(\alpha_{\rm IMF})$, as a function of $z=0$ global stellar velocity dispersion, $\sigma_*$. Inset: the $\log(\alpha_{\rm IMF})$-$\sigma_*$ relations used as input physical laws that produce the curves in the main panel. The solid black curve shows the observed relation from Spiniello et al. 2014 and the black dashed line shows an extrapolation of the relation. The resulting $z=0$ relations are always shallower than the input relations. Red: input relation as in Spiniello et al. 2014, blue: an input relation that is $3.5\times$ steeper.

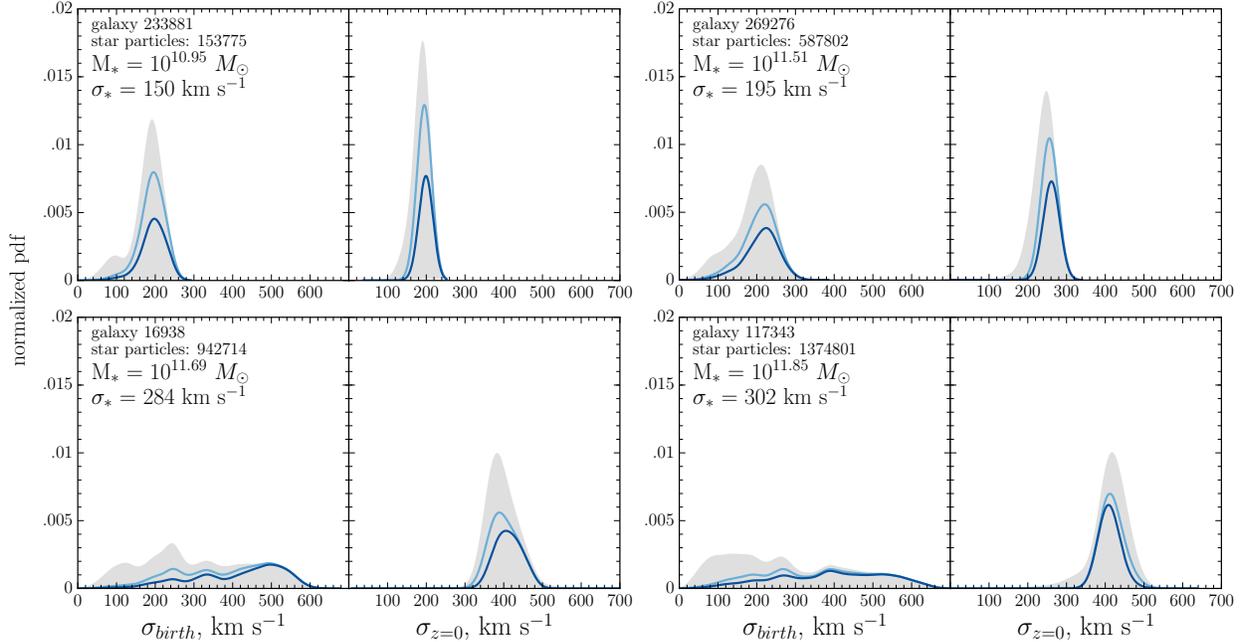

Figure 2.2: Velocity dispersion distributions of the stars in four Illustris galaxies both today (right panels) and at birth, namely in the first snapshot they appear in (left panels). Each pair of plots represents one galaxy, increasing in $z = 0$ stellar mass from top-left to bottom-right. The grey curves correspond to all stars belonging to the $z = 0$ galaxy, while the light (dark) blue curve corresponds to just the stellar particles within one (half) projected stellar half-mass radius $R_{1/2}^p$ ($.5 R_{1/2}^p$) from the center of the $z = 0$ galaxy. Evidently, the velocity dispersions of stars can change dramatically between their birth and $z = 0$, especially in massive galaxies, which present very broad distributions of $\sigma_{\text{birth}}$.

This result suggests that the observed IMF variations with $z = 0$ galactic velocity dispersion cannot be a correlation that exists in the galaxies in which the stars are actually born and where the IMF is set. This observed relation is hence emergent rather than fundamental. It is a manifestation of the complexities of $\Lambda$CDM galaxy formation through hierarchical assembly, where massive galaxies are composed of stellar populations that form in a plethora of progenitor galaxies with varied and evolving properties such as velocity dispersion. Since galaxies with high $\sigma_*$ contain stars that were formed inside galaxies with low $\sigma_*$ and hence have relatively bottom-light IMFs, it is necessary that stars forming in-situ in galaxies with high $\sigma_*$ have extremely bottom-heavy IMFs in order to combine together in their $z = 0$ host galaxies and produce the observed relation.

In this section, we have used the same quantity on the horizontal axes of both the inset and the main panel, namely the assumed physical driver was the same as the independent variable of the $z = 0$ relation. We have shown that the input relation is not preserved through galaxy assembly. For the remainder of this paper, we focus on connecting the global properties of $z = 0$ galaxies to physical properties that are more closely associated with star-formation. We examine the local velocity dispersion and metallicity of the stellar particles, as well global quantities, star-formation rate and gas velocity dispersion, which have been suggested as drivers of IMF variations on the star-formation scale.

### 2.4.2 Local velocity dispersion

Still motivated by the observed IMF trends with velocity dispersion, but aiming for a more physically relevant IMF driver, we construct the overall $\log(\alpha_{\mathrm{IMF}})$ of each Illustris galaxy based on the local velocity dispersion around each individual stellar particle at the time of formation. In particular, we use the one-dimensional velocity dispersion of the 64±1 dark matter particles nearest to each stellar particle in the earliest snapshot where it exists. This quantity, denoted as $\sigma_{\mathrm{birth}}$, probes the local gravitational potential at the time of stellar birth.

Figure 2.2 includes four example velocity dispersion distributions, comparing the local velocity dispersions the stars have in their $z = 0$ host galaxy (right panels) to the velocity dispersions those same stars had at their individual formation times (left panels). We also distinguish between the distributions of local velocity dispersions for stellar particles enclosed within different radii with respect to the center of the $z = 0$ galaxy, which is defined as the position of the most bound particle belonging to the galaxy. Figure 2.4 shows four additional $\sigma_{\mathrm{birth}}$ distributions for comparison to the various other stellar birth properties that will be discussed in following sections.

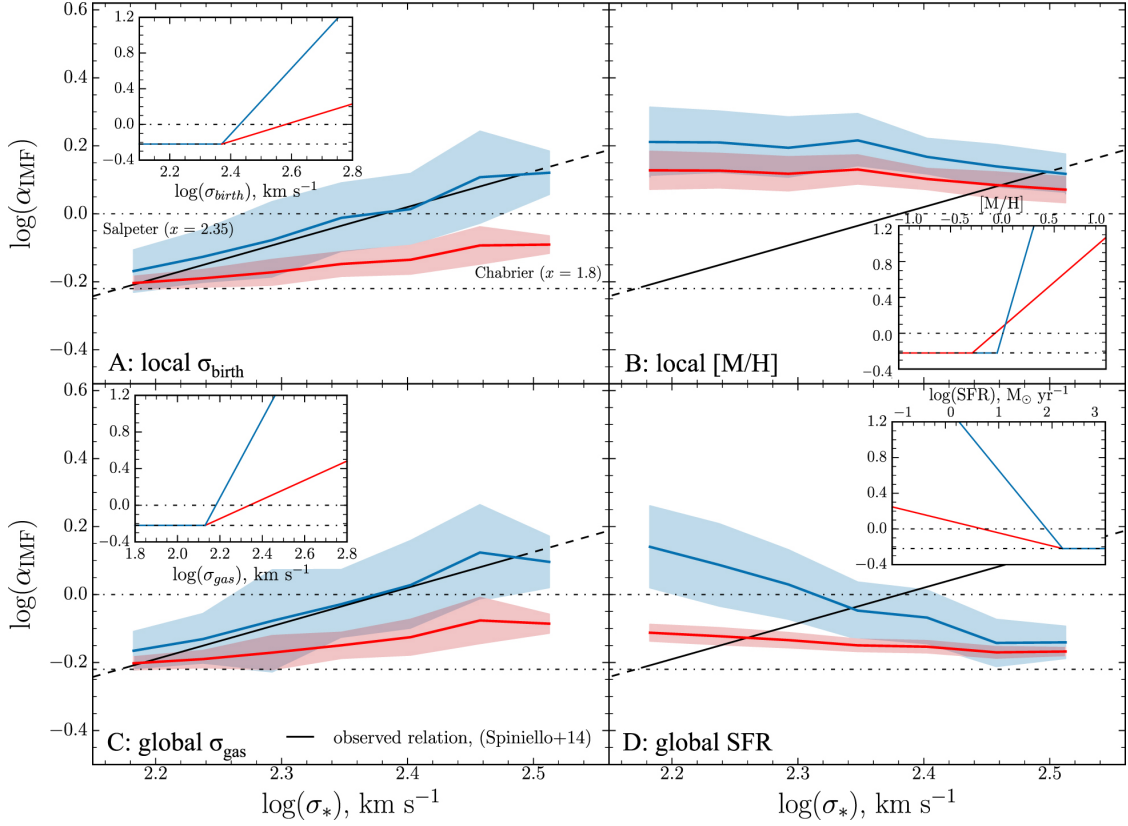

Figure 2.3: Relations between the IMF mismatch parameter $\log(\alpha_{\mathrm{IMF}})$ and various galaxy properties. The insets show the input relations used to construct $\log(\alpha_{\mathrm{IMF}})$, in each panel using a different physical quantity at star-formation time (**Panel A**: local velocity dispersion $\sigma_{\mathrm{birth}}$, **Panel B**: local metallicity [M/H], **Panel C**: global star-forming gas velocity dispersion $\sigma_{\mathrm{gas}}$, **Panel D**: global star-formation rate SFR). The main panels sfhow the resulting constructed relations between $\log(\alpha_{\mathrm{IMF}})$ within $0.5 R_{1/2}^p$ and $z = 0$ global stellar velocity dispersion, $\sigma_*$. The red curves correspond to shallower input relations while the blue curves to steeper ones. In each panel the solid black line shows the observed relation from Spiniello et al. 2014 and the black dashed line shows an extrapolation of the relation. The dot-dashed lines at $\log(\alpha_{\mathrm{IMF}}) = 0$ and $\log(\alpha_{\mathrm{IMF}}) = -0.22$ indicate the Salpeter and Chabrier IMF mismatch parameters respectively. All input relations and fits to the output relations are listed in Table A1. *Note:* In Panel D, the blue $\log(\alpha_{\mathrm{IMF}})$-$\sigma_*$ relation has been shifted by -0.1 dex.

As evident in Figure 2.2 the velocity dispersions that the stellar particles have in their host galaxy at $z = 0$ can be substantially different from the velocity dispersions they had at their time of birth. Qualitatively, the $\sigma_{z=0}$ distributions for galaxies of all masses in our selected sample are singly peaked and best described as a gaussian or a gaussian with a low velocity dispersion tail. In most cases the low velocity dispersion tail is largely built up by stellar particles residing outside the projected stellar half-mass radius of the galaxies. Unlike the $z = 0$ distributions, the birth velocity dispersion distributions are quite varied. Most lower mass galaxies in our sample ($M_* \lesssim 10^{11}\ M_\odot$) have near singly peaked birth distributions, spread out over a broader range of velocity dispersions than their $z = 0$ distributions. Some of these lower mass galaxies end up with a higher mean velocity dispersion at $z = 0$ than a birth, and others vice versa. Gas inflows, outflows, internal dynamical processes, as well as mergers, are all expected to play a role in shifting the velocity dispersion of a galaxy over time either to lower or higher values.

Generally, the range of birth velocity dispersions becomes larger for galaxies with a higher $z = 0$ stellar mass. In particular, the birth velocity dispersion distributions of $M_* \gtrsim 10^{11}\ M_\odot$ galaxies in our sample are usually multi-peaked and spread across a large range of velocity dispersions. This is reflective of the rich merger histories of these high mass galaxies, with their stellar particles being formed in numerous progenitor galaxies with varying masses and velocity dispersions. For example, the $M_* = 10^{11.69}\ M_\odot$ galaxy shown in the bottom left panel of Figure 2.2 underwent 4 major mergers ($\mu > 1/4$), 7 minor mergers ($1/4 > \mu > 1/10$), and 529 very minor mergers ($\mu < 1/10$) throughout its history. On the other hand, the lower mass galaxy shown in the top left panel, with $M_* = 10^{10.95}\ M_\odot$, only underwent 3 major mergers ($\mu > 1/4$), no minor mergers ($1/4 > \mu > 1/10$), and 96 very minor mergers ($\mu < 1/10$) throughout its history.

Radial trends are also present in the birth velocity dispersion distributions of massive galaxies, with the stellar particles closer to the center of each galaxy having, on average, higher birth velocity dispersions. This is due to the spatial distribution of stellar particles inside galaxies set up by mergers. As shown in Rodriguez-Gomez et al. (2016), higher mass galaxies in Illustris consist of a larger fraction of stellar particles formed ex-situ, i.e. not on the main progenitor branch. Galaxies with stellar masses greater than $10^{12}$ M$_\odot$ can have up to 80% of their stellar particles formed ex-situ and later accreted onto the main galaxy via merging. Rodriguez-Gomez et al. (2016) finds that stellar particles formed in-situ tend to reside in the innermost regions of galaxies whereas stars formed ex-situ tend to lie in the outer regions at larger galactocentric distances.

With the birth velocity dispersion and mass of each stellar particle belonging to a galaxy, we construct the IMF mismatch parameter according to the prescription in Section 2.3.3. To start, we shift the observed relation presented by Spiniello et al. (2014) towards higher velocity dispersions,

$$\log(\alpha_{\rm IMF}) = 1.05\log(\sigma_{\rm birth}) - 2.71, \quad (2.3)$$

so that the minimum Chabrier IMF value of $\alpha_{\rm IMF} = 0.6$ is adopted for all stellar particles with $\sigma_{\rm birth}$ less than 235 km s$^{-1}$. This shift in the input relation is applied because the $\sigma_{\rm birth}$ distributions, on average, cover higher values than the $\sigma_*$ distributions. A shift in the relation is needed to place the $\log(\alpha_{\rm IMF})$ of low velocity dispersion galaxies on the observed $\log(\alpha_{\rm IMF})$-$\sigma_*$ relation.[3]

Panel A of Figure 2.3 shows the resulting $\log(\alpha_{\rm IMF})$-$\sigma_*$ relation for $z = 0$ Illustris galaxies constructed based on $\sigma_{\rm birth}$. We show only the relations constructed using the stellar particles residing within $0.5R_{1/2}^p$ of each $z = 0$ galaxy, as observations of IMF variations are mainly constructed using the innermost regions of galaxies, and the same holds for the other panels in Figure 2.3 as well. For the output relations constructed using all the stellar particles belonging to each galaxy, the reader is referred to Table A1. The inset figure in each panel of Figure 2.3 shows the input relations we used to construct $\log(\alpha_{\text{IMF}})$ based on the indicated stellar particle property. The red curve in Panel A shows the $\log(\alpha_{\text{IMF}})$-$\sigma_*$ trend resulting from using Equation 2.3 as the prescribed IMF relation applied to the birth velocity dispersions of each galaxy's stellar particles. As evident in the figure, the output $\log(\alpha_{\text{IMF}})$-$\sigma_*$ is $\sim 2.8\times$ too shallow compared to the observed $\log(\alpha_{\text{IMF}})$-$\sigma_*$ relation (black curve). This can be understood in terms of the birth velocity dispersion distributions shown in Figure 2.2. Using Equation 2.3 as the input relation, there is not enough of a differentiation created between the overall $\log(\alpha_{\text{IMF}})$ of galaxies of different global $\sigma_*$ values. For example, the $\sigma_* = 150$ km s$^{-1}$ and $\sigma_* = 195$ km s$^{-1}$ galaxies shown in Figure 2.2 have birth velocity dispersion distributions that cover a similar range of values. So the $\log(\alpha_{\text{IMF}})$ difference between the two galaxies, where the $\sigma_* = 150$ km s$^{-1}$ galaxy is found to have an overall $\log(\alpha_{\text{IMF}}) = $ -0.219 and the $\sigma_* = 195$ km s$^{-1}$ galaxy is found to have $\log(\alpha_{\text{IMF}}) = $ -0.210, is too small compared to the observed difference of $\Delta\log(\alpha_{\text{IMF}}) = .125$.

---

[3]Using the original Spiniello et al. (2014) relation as input in this case is still unable to reproduce the slope of the observed relation.

To reproduce the observed $\log(\alpha_{\text{IMF}})$-$\sigma_*$ relation, we construct various input $\log(\alpha_{\text{IMF}})$-$\sigma_{\text{birth}}$ relations with steeper slopes. As before, we minimize the residuals between the resulting Illustris output relations and the Spiniello et al. (2014) relation, and select the input relation that produces the best-fit. The blue curve in Panel A of Figure 2.3 shows the resulting $\log(\alpha_{\text{IMF}})$-$\sigma_*$ relation constructed using an input relation that is $\sim 3.5\times$ steeper than Equation 2.3 and a minimum Chabrier $\alpha_{\text{IMF}}$ applied to stellar particles with $\sigma_{\text{birth}} < 235$ km s$^{-1}$. As seen in Panel A of Figure 2.3, this steeper input relation is able to reproduce both the slope and normalization of the observed trend with global velocity dispersion. The increase in the slope of the input relation creates more of a differentiation between galaxies of different $\sigma_*$.

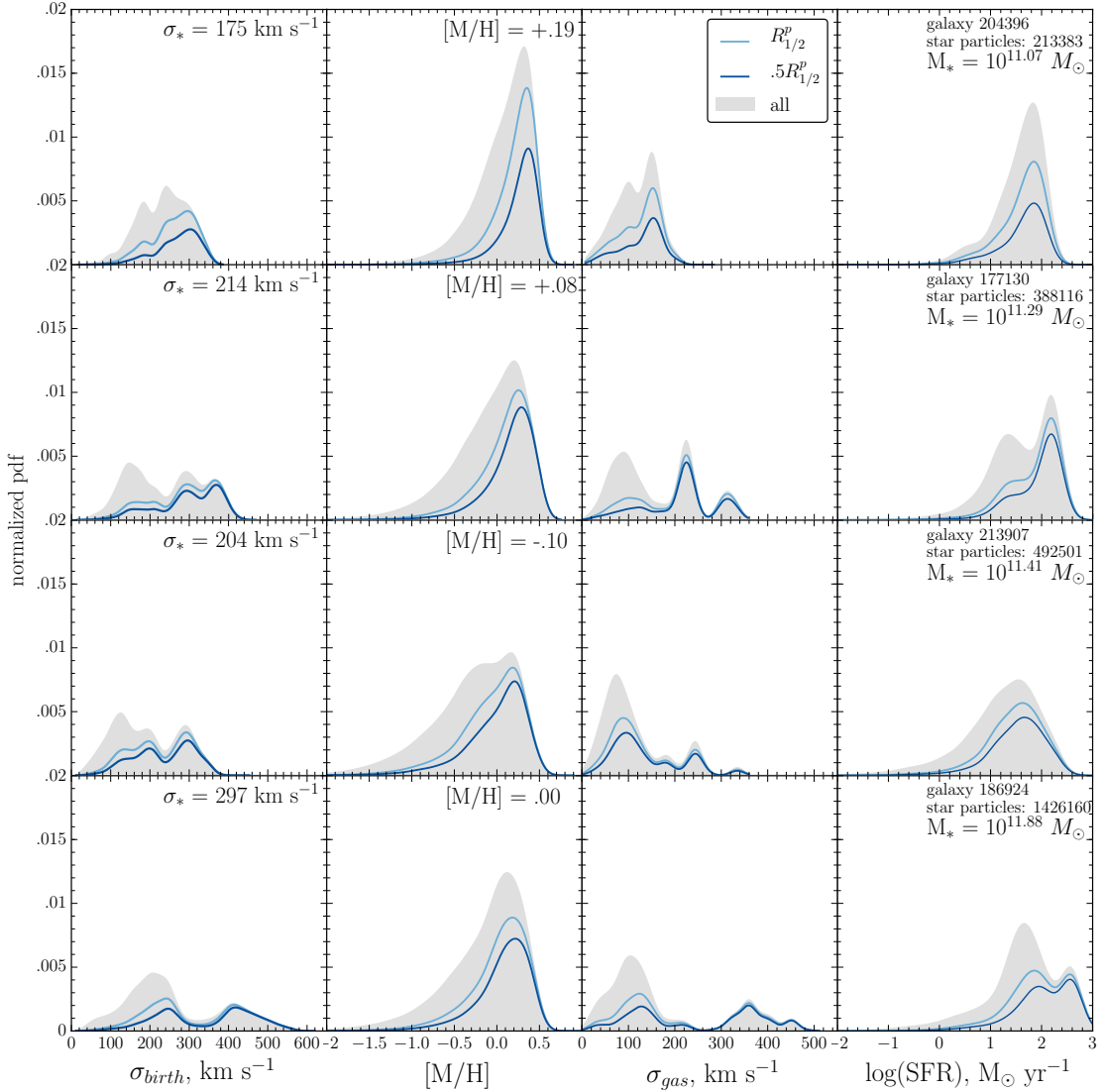

Figure 2.4: Birth properties of stellar particles belonging to four $z = 0$ galaxies, showing how various stellar birth property distributions vary with galaxy mass. Each row shows the properties of one galaxy, increasing in stellar mass from top to bottom where **Column 1**: local, birth velocity dispersion ($\sigma_{\rm birth}$), **Column 2**: birth metallicity ([M/H]), **Column 3**: gas velocity dispersion ($\sigma_{\rm gas}$), **Column 4**: star-formation rate. The grey distributions show the properties of all the stellar particles belonging to the $z = 0$ galaxy while the the light blue and dark blue distributions show the properties of the stellar particles within one and one-half the projected stellar half-mass radius, respectively.

For example, the $\sigma_* = 150$ km s$^{-1}$ galaxy in Figure 2.2 is found to have a similar overall $\log(\alpha_{\rm IMF})$ as before ($\log(\alpha_{\rm IMF})$ = -0.217), but the $\sigma_* = 195$ km s$^{-1}$ overall $\log(\alpha_{\rm IMF})$ increased by $\Delta\log(\alpha_{\rm IMF})$ = .025, placing it closer to the observed $\log(\alpha_{\rm IMF})$-$\sigma_*$ relation.

In addition to increasing the slope of the output $\log(\alpha_{\rm IMF})$-$\sigma_*$ relation, a steeper input relation acts to increase the scatter of the Illustris $\log(\alpha_{\rm IMF})$-$\sigma_*$ relation. As seen in Panel A of Figure 2.3, the scatter of the blue relation is nearly $2\times$ the scatter of the red relation. The increase in scatter with the steeper input relation can also be understood with an example from Figure 2.2. Although the $10^{11.69}$ $M_\odot$ and $10^{11.85}$ $M_\odot$ galaxies shown in Figure 2.2 have similar masses and velocity dispersions, their birth velocity dispersion distributions are quite different. Using Equation 2.3 as the input relation, the difference between their overall $\log(\alpha_{\rm IMF})$ values is $\Delta\log(\alpha_{\rm IMF}) = .025$. But, using the steeper input relation the difference becomes $\Delta\log(\alpha_{\rm IMF}) = .10$. While a steeper input relation is able to create more of an overall differentiation between galaxies of different $\sigma_*$ values, it also creates a larger scatter among galaxies of similar $\sigma_*$ but different formation histories.

### 2.4.3 Local metallicity

The second physical quantity of star-formation we examine is metallicity. Aside from velocity dispersion, observational studies also report that the IMF scales with galaxy metallicity. While both velocity dispersion and metallicity correlate with mass, the quantity that is more fundamentally associated with IMF variations is still unclear.

A metallicity-IMF correlation can be easily imagined through a reversed causal relationship. Simply put, the IMF is expected to influence the metallicity because the number of high to low mass stars will directly affect the chemical evolution of a galaxy. The more top-heavy the IMF, the more metals are injected into the ISM, which increases the metallicity of the stars born in subsequent star formation bursts.

So one might naively expect the overall metallicity of a galaxy by $z = 0$ to be higher with a more top-heavy its IMF. This scenario is however in tension with observational IMF-metallicity relations, which infer a more *bottom*-heavy IMF for the most metal-rich galaxies (e.g. Conroy & van Dokkum 2012b; Martín-Navarro et al. 2015c). One way this tension may be reconciled is by invoking a time-dependent IMF: earlier star formation follows a flatter IMF to build up the metallicity of the ISM and star formation occurring later follows a bottom-heavy IMF to build up the population of low-mass stars (Martín-Navarro 2016; Weidner et al. 2013a; Narayanan & Davé 2013). Particularly, Weidner et al. (2013a) propose that the ISM, enriched by episodes of high star-formation with a flat IMF, is exceptionally turbulent leading to increased fragmentation on lower mass scales. Therefore, proceeding star-formation occurs with a steeper IMF slope. In this scenario, a higher metallicity environment at the time of stellar birth is expected to correspond to a more bottom-heavy IMF.

Our current empirical approach for constructing $\log(\alpha_{\mathrm{IMF}})$ does not take into account how the IMF may influence metallicity. Instead, motivated by Weidner et al. (2013a), we assign an IMF based on the local metallicity of each stellar particle at the time of stellar birth with the idea that stellar particles born into higher metallicity environments form with a steeper IMF slope. Thus, we construct the overall $\log(\alpha_{\mathrm{IMF}})$ of each Illustris galaxy based on the metallicity of each stellar particle at the time of formation. As discussed in Section 2.3.1, the total mass in metals of a stellar particle in Illustris is inherited from the parent gas cell at the time of star formation. The ratio of the total mass in metals heavier than helium to the total mass of the stellar particles at the time of birth, $Z$, is output for each star in the snapshot files. For the stellar particles in our sample, we convert the metallicity mass fraction to the metal abundance [M/H] by assuming each stellar particle to have a primordial hydrogen mass fraction of $X = 0.76$ and scaling to solar units using $Z_\odot = 0.02$ and $X_\odot = 0.70$.

The second column of Figure 2.4 shows the distribution of stellar particle metallicities for four Illustris galaxies, increasing in $z = 0$ stellar mass from top to bottom. The light grey histogram shows the distribution of metallicities of all stellar particles belonging to the galaxy while the light and dark blue distributions show the metallicities of the stellar particles within $R^p_{1/2}$ and $0.5R^p_{1/2}$, respectively. The distributions are similarly shaped, mostly described as a gaussian with a significant low metallicity tail. Similar to the velocity dispersion distributions in Figure 2.2, the stellar particles residing closer to the center of the galaxy have a higher average metallicity than the stellar particles residing in the outer reaches of the galaxy. The global $z = 0$ metallicity of each galaxy, shown in the upper right of each panel, is calculated by taking the mass-weighted average of the metal abundance of the stellar particles within $0.5R^p_{1/2}$ of each galaxy.

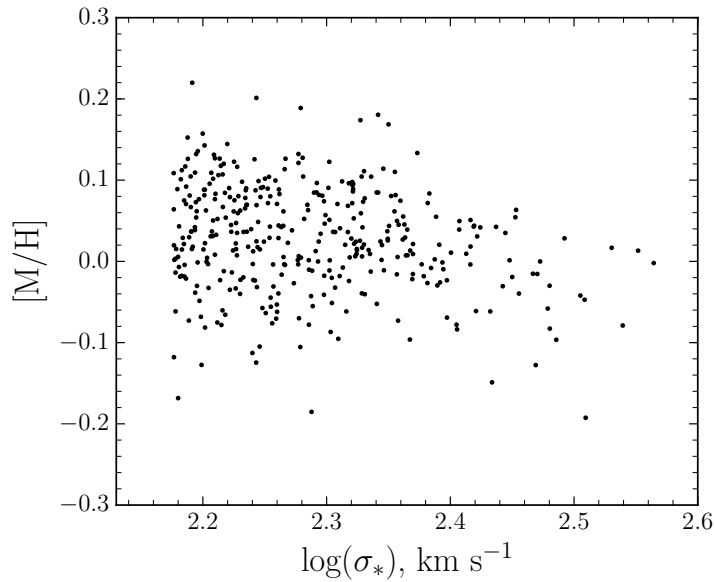

Figure 2.5: Global metallicity, [M/H], as a function of $z = 0$ stellar velocity dispersion, $\sigma_*$ for the 371 selected Illustris-1 galaxies.

With the birth metallicities and mass of each stellar particle belonging to a galaxy, we construct the IMF mismatch parameter as before. As a starting point, we are inspired by Martín-Navarro et al. (2015a) to construct an input $\log(\alpha_{\mathrm{IMF}})$-[M/H] relation that is defined to have a Chabrier $\alpha_{\mathrm{IMF}}$ value at [M/H] = -0.29 and a Salpeter $\alpha$ value at [M/H] = -0.07. This results in the relation $\log(\alpha_{\rm IMF}) = $ [M/H] $+ 0.07$, where stellar particles with metallicities less than -0.29 are assigned $\alpha_{\rm IMF} = 0.6$. Panel B of Figure 2.3 shows the resulting Illustris $\log(\alpha_{\rm IMF})$-$\sigma_*$ relation (red curve) constructed based on the individual [M/H] values that each stellar particle belonging to a galaxy are formed with. Again, we only show the $\log(\alpha_{\rm IMF})$-$\sigma_*$ relations constructed using only the stellar particles residing within $0.5R_{1/2}^p$ of each $z = 0$ galaxy. Evidently, our initial input relation is unable to reproduce the observed $\log(\alpha_{\rm IMF})$-$\sigma_*$ relation, with the slope of the output Illustris relation being nearly flat and in fact slightly negative, showing the opposite trend to the observed one.

To try to reproduce the observed relation, we construct an input $\log(\alpha_{\rm IMF})$-[M/H] that is 4× as steep and defined to have a minimum Chabrier $\alpha$ at [M/H] = -0.05. The increase in the Chabrier minimum is in attempt to decrease the overall $\log(\alpha_{\rm IMF})$ values of lower velocity dispersion galaxies. The blue curve in Panel B of Figure 2.3 shows the $\log(\alpha_{\rm IMF})$-$\sigma_*$ relation constructed using the steeper $\log(\alpha_{\rm IMF})$-[M/H] described. The result is that the normalization of the output relation shifts to higher $\log(\alpha_{\rm IMF})$ values, but its slope is still negative. Additionally, increasing the Chabrier minimum to [M/H] = -0.05 did not act to decrease the overall $\log(\alpha_{\rm IMF})$ of lower velocity dispersion galaxies, as increasing the Chabrier minimum in Section 2.4.2 was able to do.

The inability to reproduce the observed IMF trend with $z = 0$ velocity dispersion using [M/H] as the physical driver of the IMF can be understood from the global [M/H]-$\sigma_*$ relation. As seen in Figure 2.5, this relation is almost flat for our sample comprised of massive $M_*$ > $\sim 10^{10}$ $M_\odot$ galaxies[4] In fact, where it is not *completely* flat, at $2.4 < \log \sigma_* < 2.6$, the output $\log(\alpha_{\rm IMF})$-$\sigma_*$ also has some slope. The negative slope of the [M/H]-$\sigma_*$ relation could either be due to intracluster light contamination or recycling of low metallicity gas. However, even there, the mild slope of the [M/H]-$\sigma_*$ relation combined with its large scatter result in galaxies of similar velocity dispersions having a wide range of global [M/H] values. Since the global [M/H] of each galaxy is related to the distribution of the individual stellar particle [M/H] values (as seen in Figure 2.4) *and* the widths of the [M/H] distributions are broad compared to the galaxy-to-galaxy differences in global [M/H], a differential effect in overall log($\alpha_{\rm IMF}$) between galaxies of different velocity dispersions is not produced using an input relation based on [M/H].

---

[4]The flatness of the [M/H]-$\sigma_*$ relation is generally in agreement with observations of the stellar mass-metallicity relation, such as in Gallazzi et al. (2005), where the SDSS mass-metallicity becomes flat at high stellar masses with a scatter of $\sim$0.3 dex. Figure 5 is also in agreement with the observed velocity dispersion-mass relation over the appropriate velocity dispersion range (Spolaor et al. 2010).

### 2.4.4 Global star-forming gas velocity dispersion

As mentioned in Section 2.2, a few analytical studies have focused on connecting the physics governing star formation to an environment dependent IMF. In particular, Hopkins (2013) developed an analytical formulation where star-forming disks with higher Mach numbers cause the low-mass turnover of the pre-stellar core mass function (CMF) to be shifted to lower masses leading to a more bottom-heavy CMF which implies a more bottom-heavy IMF. Physically, as discussed in Hopkins (2013), a higher star-forming disk Mach number leads to larger density fluctuations which causes more fragmentation on smaller mass scales.

Motivated by a Mach number dependent CMF, we construct the IMF of our Illustris galaxies using the one-dimensional, global star-forming gas velocity dispersion ($\sigma_{\rm gas}$) of the progenitor galaxies in which stellar particles are born. We do not take into account differences in sound speed, but simply use $\sigma_{\rm gas}$ as a proxy for Mach number (Chabrier et al. 2014). To calculate $\sigma_{\rm gas}$ we trace each stellar particle back to the progenitor galaxy in which it was born and consider only the gas cells in that galaxy with non-zero instantaneous star formation rates. We remove net rotation by calculating the total angular momentum vector of each galaxy's star-forming gas component and calculate $\sigma_{\rm gas}$ as the mass-weighted standard deviation of the cell velocities parallel to that angular momentum vector.

The third column of Figure 2.4 shows the $\sigma_{\rm gas}$ distributions for four galaxies in our sample. The birth $\sigma_{\rm gas}$ distributions are similarly multi-peak and spread across a broad range of values as the $\sigma_{\rm birth}$ distributions shown in the first column of Figure 2.4. However, there are a few qualitative differences between the two distributions. First, the $\sigma_{\rm gas}$ distributions are less continuous than the corresponding $\sigma_{\rm birth}$ distributions, reflective of the fact that multiple stellar particles are often born in the same progenitor galaxy and therefore have the same $\sigma_{\rm gas}$ value. The $\sigma_{\rm gas}$ distributions are also shifted to lower velocity dispersion values compared to their $\sigma_{\rm birth}$ counterparts due to the removal of rotation. But similar to the $\sigma_{\rm birth}$ distributions, there is a radial trend in $\sigma_{\rm gas}$ especially for higher mass galaxies, with a larger fraction of stars born in high $\sigma_{\rm gas}$ galaxies residing closer to the center of their $z = 0$ host galaxy.

We construct the overall $\log(\alpha_{\rm IMF})$ of each galaxy based on the $\sigma_{\rm gas}$ distributions. We first construct a $\log(\alpha_{\rm IMF})$-$\sigma_{\rm gas}$ relation inspired by Equation 2.3, but shift the Chabrier minimum of the relation to occur at $\sigma_{\rm gas} = 135$ km s$^{-1}$. This results in the relation $\log(\alpha_{\rm IMF})$ = 1.05log($\sigma_{\rm gas}$) - 2.46, where stellar particles with $\sigma_{\rm gas}$ less than 135 km s$^{-1}$ are assigned $\alpha_{\rm IMF}$ = 0.6. Panel C of Figure 2.3 shows the resulting Illustris $\log(\alpha_{\rm IMF})$-$\sigma_*$ relation (red curve) constructed based on the star-forming gas velocity dispersion of the progenitor galaxy that each stellar particle belonging to a $z = 0$ was formed in, showing just the relation constructed using the stellar particles residing within $0.5R^p_{1/2}$. As with $\sigma_{\rm birth}$, the initial input relation is unable to reproduce the observed $\log(\alpha_{\rm IMF})$-$\sigma_*$ relation, with the slope of the Illustris $\log(\alpha_{\rm IMF})$-$\sigma_*$ $\sim 2.6\times$ shallower than the observed relation.

To reproduce the observed relation, we construct a steeper $\log(\alpha_{\rm IMF})$-$\sigma_{\rm gas}$ relation. The blue curve in Panel C of Figure 2.3 shows the $\log(\alpha_{\rm IMF})$-$\sigma_*$ relation constructed using a $\log(\alpha_{\rm IMF})$-$\sigma_{\rm gas}$ input relation that is 4.1× steeper than the Spiniello et al. (2014) relation. This steeper input relation does, within the reported uncertainty, reproduce both the slope and normalization of the observed trend with global velocity dispersion $\sigma_*$. As with $\sigma_{\text{birth}}$ as a physical driver, the increase in the slope of the input relation creates more of a differentiation between galaxies of different $\sigma_*$, which allows the observed relation to be reproduced. Also similar to the Illustris log($\alpha_{\text{IMF}}$)-$\sigma_*$ relation constructed based on $\sigma_{\text{birth}}$, increasing the slope of the input relation results in a larger scatter in the output relation. This is because galaxies of similar $z = 0$ velocity dispersions can have a range of $\sigma_{\text{gas}}$ distributions.

### 2.4.5 Global star-formation rate

The last physical quantity we consider in constructing the overall log($\alpha_{\text{IMF}}$) of each Illustris galaxy is star-formation rate. Observationally, studies focusing on constraining the high mass end of the IMF suggest that the IMF correlates with galaxy SFR. For example, Gunawardhana et al. (2011) find that for a range of galaxies at $z < 0.35$ with SFRs covering $10^{-3}$ to 100 $M_\odot$ yr$^{-1}$, the most quiescent galaxies are best described by steeper IMF slopes ($x$~2.4), whereas highly star-forming galaxies exhibit shallower IMFs ($x$~1.8). They translate their IMF-SFR to a relation between IMF and SFR surface density, finding that galaxies with higher SFR densities prefer flatter IMF slopes. This result is consistent within the context of IGIMF theory (Weidner et al. 2013b), which connects the global SFR of a galaxy to the formation of stars within individual molecular clouds throughout the galaxy, where galaxies with higher SFR are expected to have a top-heavy galaxy-wide IMF.

On the other hand, Conroy & van Dokkum (2012b) find their strongest IMF trend to be with [Mg/Fe], where galaxies with greater Mg enhancement have more bottom-heavy IMFs. These galaxies with increased Mg abundances are interpreted as having shorter star-formation timescales. Based on the inferred star-formation time scales of massive galaxies with enhanced Mg abundances, Conroy & van Dokkum (2012b) infer that galaxies with high SFR surface densities are described by a more bottom-heavy IMF. Physically, as pointed out by Conroy & van Dokkum (2012b), in the context of the Hopkins (2013) analytical theory for IMF variations, high SFR surface densities promotes turbulence which leads to a more bottom-heavy IMF. In regards to observations reporting that higher SFRs correspond to shallower IMFs and to IGIMF theory which predicts the same, it is suggested that the high $M_*/L$ ratio inferred for these massive elliptical galaxies is at least in part due to an excess of high-mass stellar remnants and not only an excess of low-mass stars. In this scenario, high SFR starbursts induced by mergers form with a top-heavy IMF, and it is the remnants of these high mass stars that produce an excess of mass as measured by $z = 0$.

For each stellar particle that comprises a $z = 0$ galaxy, we record the instantaneous star-formation rate of the progenitor galaxy in which the stellar particle is formed. The fourth column of Figure 2.4 shows the distribution of birth SFRs for the stellar particles comprising four galaxies of various masses. As seen in the figure, the birth SFR distribution becomes multi-peaked and/or broader for galaxies with higher stellar mass, and includes more stellar particles with higher birth SFRs. For lower mass galaxies, where a majority of their stellar populations are formed in-situ, the shape of the birth SFR distribution can be understood as the evolution of the star-formation rate of the main progenitor branch. The low SFR tails of these distributions correspond to the formation of stellar particles before and after the period of peak star-formation where most of the stellar mass is formed. Additionally, for these lower mass galaxies, there is little difference in the birth SFR distributions of all the stellar particles versus just the stellar particles residing within $0.5R_{1/2}^p$ of each galaxy.

The higher mass galaxies shown in the figure ($M_* > 10^{11.2}\ M_\odot$) have stellar particles that, on average, formed in progenitor galaxies with higher SFRs and also cover a broader range of SFRs. The stellar particles formed in progenitor galaxies with SFRs $\sim 100\ M_\odot\ \mathrm{yr}^{-1}$ likely formed during merger-induced nuclear starbursts, whereas the stellar particles making up the lower SFR part of the distributions were formed either before or after the peak star-formation period of these merger events, or in lower SFR galaxies that are later accreted onto the main progenitor. The SFR for the $10^{11.88}$ $M_\odot$ and $10^{11.29}$ $M_\odot$ galaxies particularly show that stellar particles residing closer to the center of the $z = 0$ galaxy are formed in galaxies with higher SFRs than stellar particles residing in the outer edges of the galaxy. This is consistent with the $\sigma_{\text{birth}}$ distribution of massive galaxies, where stellar particles closer to the center of a massive galaxy are formed in high velocity dispersion environments during periods of high SFR nuclear starbursts.

We construct the $\log(\alpha_{\text{IMF}})$-$\sigma_*$ relation for our sample of Illustris galaxies based on the birth SFR distributions described above. First, we use a $\log(\alpha_{\text{IMF}})$-SFR relation that is defined to correspond to a Chabrier IMF at $\log(\text{SFR}) = 2.2$ and a Salpeter IMF at $\log(\text{SFR}) = 0.7$. This starting point is inspired by Gunawardhana et al. (2011) who for their sample of galaxies from GAMA find an IMF-SFR relation -x $\approx$ 0.36 log(SFR) - 2.6. Stellar particles born into galaxies with SFR = 0 $M_\odot$ yr$^{-1}$, where the SFR is an unresolved small value, are assigned zero light, i.e. $\alpha_{\text{IMF}}^{-1} = 0$. The red curve in Panel D of Figure 2.3 shows the $\log(\alpha_{\text{IMF}})$-$\sigma_*$ relation resulting from this initial input relation. As seen in the figure, using SFR as the physical quantity of star-formation, where high SFR environments are expected to correspond to a shallower IMF, to build the Illustris $\log(\alpha_{\text{IMF}})$-$\sigma_*$ results in a trend opposite to that of the observations. Inputting a $\sim$4.7$\times$ steeper $\log(\alpha_{\text{IMF}})$-SFR relation, we are best able to reproduce the steepness of the observed $\log(\alpha_{\text{IMF}})$-$\sigma_*$ trend, but in the opposite direction (blue curve). This is because higher velocity dispersion galaxies in our sample have stellar particles that, on average, formed in progenitors with higher SFRs. This is compared to the stellar particles belonging to low velocity dispersion galaxies, which generally form in progenitors with lower SFRs.

## 2.5 Results beyond the global $z = 0$ trends

### 2.5.1 Radial trends

As suggested by Figures 2.2 and 2.4, the physical quantities that we investigate display various amounts of radial variation depending on the mass of the galaxy. For the more massive galaxies in our sample, stellar particles residing closer to the centers of their $z = 0$ galaxy tend to have higher $\sigma_{\rm birth}$, [M/H], $\sigma_{\rm gas}$, or SFRs compared to the stellar particles residing in the outskirts of the galaxy. Such a radial trend is weaker or non-existent for the lower mass galaxies we examine. This is reflective of lower mass galaxies being composed of a smaller fraction of stellar particles formed ex-situ compared to high mass galaxies. So in constructing the overall $\log(\alpha_{\rm IMF})$ using only stellar particles within $0.5R^p_{1/2}$, the $\log(\alpha_{\rm IMF})$-$\sigma_*$ relation is steeper than when constructing $\log(\alpha_{\rm IMF})$ based on all the stellar particles belonging to each galaxy. Table 2.2 lists the comparison between the output $\log(\alpha_{\rm IMF})$-$\sigma_*$ relations constructed using all stellar particles versus just the innermost stellar particles.

Here we examine in more detail radial trends of $\log(\alpha_{\rm IMF})$ for a sample of galaxies with varying global velocity dispersions. To construct the overall IMF mismatch parameter we use the steep $\log(\alpha_{\rm IMF})$-$\sigma_{\rm birth}$ relation, $\log(\alpha_{\rm IMF}) = 3.7\log(\sigma_{\rm birth}) - 8.99$, for which we were able to reproduce the observed $\log(\alpha_{\rm IMF})$-$\sigma_*$ trend. Figure 2.6 shows $\log(\alpha_{\rm IMF})$ as a function of projected radius for 11 Illustris galaxies, starting at a radius of 1/7 the projected stellar half-mass radius, $(1/7)R^p_{1/2}$, out to radius of 4 times the projected stellar half-mass radius, $4R^p_{1/2}$. The top panel shows $\log(\alpha_{\rm IMF})$ constructed in cylinders (i.e. spheres projected along the line of sight), including all stellar particles falling within the indicated radius. The bottom panel shows $\log(\alpha_{\rm IMF})$ constructed in hollow cylinders, where each indicated radius shows $\log(\alpha_{\rm IMF})$ constructed with just the stellar particles falling within two consecutive radii. The $\log(\alpha_{\text{IMF}})$ value for $(1/7)R^p_{1/2}$ is constructed using stellar particles residing between $(1/8)R^p_{1/2}$ and $(1/7)R^p_{1/2}$.

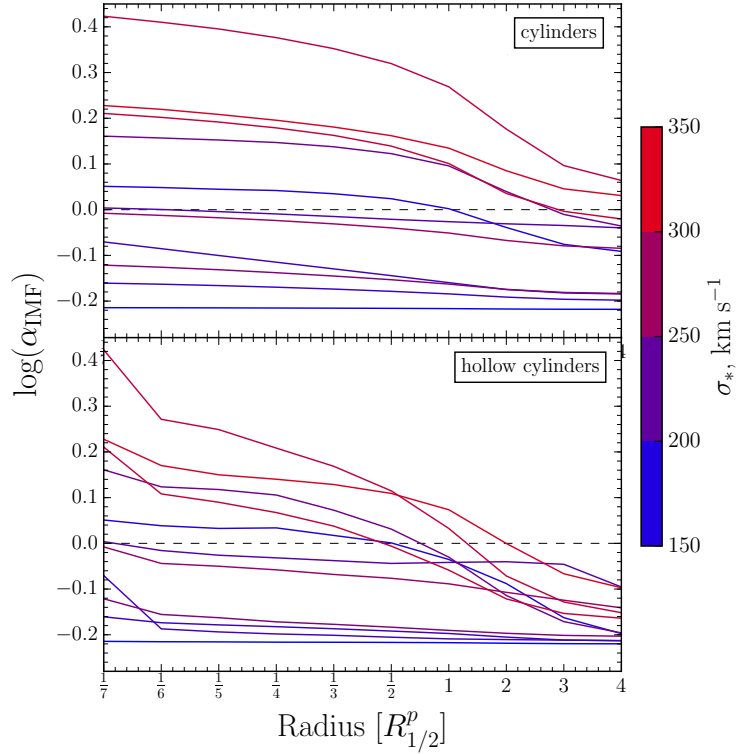

Figure 2.6: IMF slope as a function of galactocentric distance for 11 galaxies in our sample, using the steep $\log(\alpha_{\mathrm{IMF}})$-$\sigma_{\mathrm{birth}}$ relation (see text). Each line represents a galaxy colored by its $z = 0$ velocity dispersion. The top panel shows the IMF slope constructed in cylinders, including all stars within each radius and the bottom panel shows the IMF slope constructed in hollow cylinders, including the stellar particles between two consecutive radii. These constructed Illustris $\log(\alpha_{\mathrm{IMF}})$ radial gradients are in qualitative agreement with observations of IMF gradients for galaxies of both low and high velocity dispersion.

As evident in Figure 2.6, the highest velocity dispersion galaxies ($\sigma \sim 250$ - $350$ km s$^{-1}$) exhibit the greatest decrement in $\log(\alpha_{\text{IMF}})$ towards larger radii, whereas $\log(\alpha_{\text{IMF}})$ for lower velocity dispersion galaxies ($\sigma \sim 150$ - $250$ km s$^{-1}$) stays more constant with radius. Qualitatively, this trend of higher $\sigma_*$ galaxies displaying the largest radial IMF trends is in agreement with observations such as Martín-Navarro et al. (2015a). Though, it is difficult to directly compare our results to observations by radius due to the differences in how effective radius ($R_e$) is measured and how the stellar half-mass radius is calculated for Illustris galaxies.

The decrement in IMF mismatch parameter, $\Delta\log(\alpha_{\text{IMF}})$, and the maximum $\log(\alpha_{\text{IMF}})$ of our highest velocity dispersion Illustris galaxies is similar to what is reported in Martín-Navarro et al. (2015a). The decrement in $\log(\alpha_{\text{IMF}})$ of the most bottom-heavy galaxy shown in Figure 2.6, with $\sigma_* = 302$ km s$^{-1}$, is $\Delta\log(\alpha_{\text{IMF}}) = 0.36$ in cylinders from a galactocentric radius of $(1/7)R^p_{1/2}$ to $4R^p_{1/2}$. Considering hollow cylinders in which $\log(\alpha_{\text{IMF}})$ is calculated, from $(1/7)R^p_{1/2}$ to $4R^p_{1/2}$ there is a larger decrement of $\Delta\log(\alpha_{\text{IMF}}) = 0.57$. Martín-Navarro et al. (2015a) reports for their high velocity dispersion galaxy ($\sigma \sim 300$ km s$^{-1}$) a decrement of $\Delta\text{x} = 1.15$, which roughly corresponds to $\Delta\log(\alpha_{\text{IMF}}) = 0.35$, from the center of the galaxy (r = 0 $R_e$) to 0.7 $R_e$.

### 2.5.2 Scatter

As mentioned, increasing the slope of the input $\alpha_{\text{IMF}}$ relation increases the scatter of the resulting overall $\log(\alpha_{\text{IMF}})$-$\sigma_*$ relations (Figure 2.1 and Figure 2.3). This is consistent with observational studies that also show substantial scatter in the reported IMF-$\sigma$ relations (Posacki et al. 2015; Conroy & van Dokkum 2012b; Cappellari et al. 2012). For example, based on dynamical modeling of ETGs in the ATLAS$^{\text{3D}}$ project, Cappellari et al. (2013)

reports a 1$\sigma$ scatter of ≈0.12 dex (or 32%) in their derived relation between IMF mismatch parameter and velocity dispersion.

Cappellari et al. (2013)'s reported scatter is comparable to the scatter seen in our Illustris log($\alpha_{\rm IMF}$)-$\sigma_*$ relations. Constructing the overall log($\alpha_{\rm IMF}$) of each galaxy based on the stellar velocity dispersion of each stellar particle's progenitor galaxy (Section 2.4.1), we produce a 1$\sigma$ scatter of 0.079 dex (20%) using the original Spiniello et al. (2014) relation as input and a 1$\sigma$ scatter of 0.123 dex (32.7%) using the 3.6× steeper input relation. Similarly, constructing the overall log($\alpha_{\rm IMF}$) using the local velocity dispersion of each stellar particle at the time of birth (Section 2.4.2), we produce a 1$\sigma$ scatter of 0.045 dex (11%) using the shallow input relation and a 1$\sigma$ scatter of 0.125 dex (33%) using the steeper input relation that is able to reproduce the overall log($\alpha_{\rm IMF}$)-$\sigma_*$ relation. As discussed in Section 2.4.2, the scatter in our Illustris IMF relations is due to galaxies of similar global $z = 0$ velocity dispersions having varying stellar birth property distributions like $\sigma_{\rm birth}$ or $\sigma_*$. These differences in the physical conditions of star-formation reflect an intrinsic scatter in the formation histories of galaxies with the same global $z = 0$ properties.

### 2.5.3 Redshift evolution

Finally, we examine the redshift evolution of the log($\alpha_{\rm IMF}$)-$\sigma_*$ relation by repeating the analysis described in Section 2.4.2 but now with a sample of galaxies at a higher redshift. In Illustris-1 at $z = 2$, we select the 311 galaxies with stellar masses greater than $10^{10}$ M$_\odot$ and stellar velocity dispersions greater than 150 km s$^{-1}$. We do not place a cut on star-formation as we did for the $z = 0$ sample since fewer galaxies meet the sSFR $< 10^{-11}$ yr$^{-1}$ criterion at $z = 2$. The average stellar mass, stellar velocity dispersion, and specific star formation rate of the $z = 2$ sample is $\overline{M_*} = 10^{10.89}$, $\overline{\sigma_*} = 197$ km s$^{-1}$, and $\overline{\rm sSFR} = 7.60 \times 10^{-10}$ yr$^{-1}$.

We construct the overall log($\alpha_{\rm IMF}$) of each $z = 2$ galaxy based on the local, birth velocity dispersion of the stellar particles following the same procedure outlined in Section 2.4.2. Figure 2.7 shows the resulting $z = 2$ $\log(\alpha_{\rm IMF})$-$\sigma_*$ relation using the same steep input relation that was able to reproduce the observed $\log(\alpha_{\rm IMF})$-$\sigma_*$ relation for the $z = 0$ galaxy sample. As seen in the figure, the overall $\log(\alpha_{\rm IMF})$ values of the $z = 2$ galaxies are, on average, higher than the $z = 0$ relation. Furthermore, the more quiescent galaxies generally have higher $\log(\alpha_{\rm IMF})$ values than the more star-forming galaxies. To more directly compare to the $z = 0$ relation, which only includes quiescent galaxies, we fit the $z = 2$ $\log(\alpha_{\rm IMF})$-$\sigma_*$ relation only for galaxies with sSFR $< (3t_h)^{-1}$ yr$^{-1}$ where $t_h$ is the age of the Universe at a given redshift (Damen et al. 2009). In Figure 2.7 the red line shows the fit to the 29 quiescent $z = 2$ galaxies, which is $\sim 1.4\times$ steeper and offset by $\sim 0.17$ dex compared to the $z = 0$ relation.

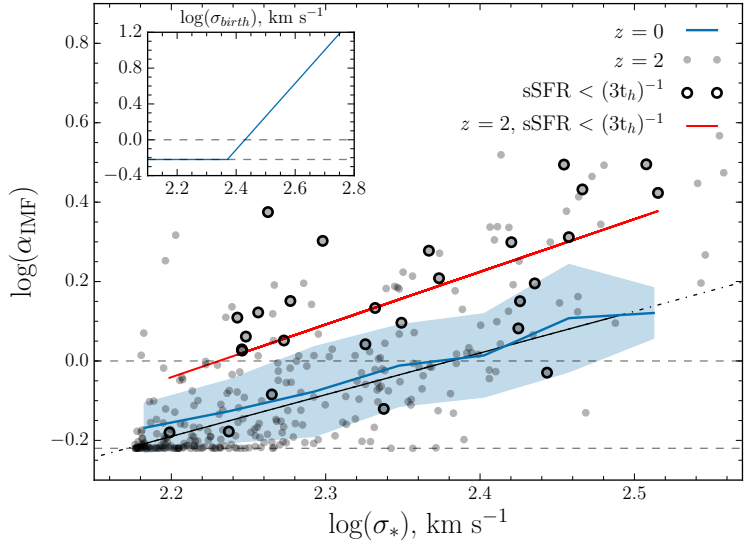

Figure 2.7: IMF mismatch parameter versus global velocity dispersion for the $z = 2$ massive galaxies. The grey points show the $z = 2$ galaxies, while the blue relation shows the $z = 0$ $\log(\alpha_{\rm IMF})$-$\sigma_*$ relation constructed using the steep input relation (shown in the inset as in Figure 3). The red line shows the fit to the $z = 2$ galaxies defined as quiescent with sSFR $< 6.82 \times 10^{-11}$ yr$^{-1}$ (outlined in black). At fixed velocity dispersion, quiescent galaxies at $z = 2$ are *more* bottom-heavy than their $z = 0$ counterparts.

The offset of the $z = 2$ $\log(\alpha_{\rm IMF})$-$\sigma_*$ relation towards higher $\log(\alpha_{\rm IMF})$ values compared to the $z = 0$ relation is due to the assembly history of massive galaxies. In ΛCDM, massive galaxies are thought to first build up their in-situ stellar populations and then at later redshifts accrete smaller systems and build up their ex-situ stellar populations (Naab et al. 2009; Oser et al. 2010). In Illustris, quiescent galaxies at $z = 2$ have already formed a significant portion of their in-situ stellar particles, but have yet to accumulate a majority of their ex-situ stellar particles. The higher $\log(\alpha_{\rm IMF})$ values of the $z = 2$ galaxies suggests the stellar particles already belonging to galaxies by $z = 2$ are formed in higher velocity dispersion environments than the stellar particles that will be added to the galaxies at later times. To go from the $z = 2$ to the $z = 0$ $\log(\alpha_{\rm IMF})$-$\sigma_*$ relation, stellar particles added to galaxies after $z = 2$ decrease the overall $\log(\alpha_{\rm IMF})$ values of the galaxies.

As will be discussed in Section 2.7, our $z = 2$ $\log(\alpha_{\rm IMF})$-$\sigma_*$ relation is seemingly in tension with IMF observations beyond $z = 0$, which suggest that the relation has remained constant over the past ∼8 Gyrs. Robust IMF determinations out to $z = 2$ will be needed to fully assess the implications of our high redshift results.

## 2.6 Dependence on resolution & physics variations

### 2.6.1 Convergence with resolution

First we confirm the convergence of our results to degradation in simulation resolution. We repeat the same analysis as described in Section 2.4.2 for Illustris-1 on the two lower resolution simulations, Illustris-2 and Illustris-3. All three simulations have the same box size of (106.5 Mpc)$^3$, but Illustris-1 contains ∼2 × 1820$^3$ resolution elements while Illustris-2 and Illustris-3 contain ∼2 × 910$^3$ and ∼2 × 455$^3$ resolution elements respectively. In Illustris-2 the average baryonic particle mass is $\overline{m_b} = 1.0 \times 10^7$ M$_\odot$ and in Illustris-3 it is $\overline{m_b} = 8.05 \times 10^7$ M$_\odot$. Refer to Table 2.1 for more Illustris-2 and Illustris-3 simulation parameters.

Implementing the same galaxy selection outlined in Section 2.3.2 results in 229 galaxies in Illustris-2 and 103 galaxies in Illustris-3 that meet the three criteria at $z = 0$ of stellar mass, specific star formation rate, and velocity dispersion. The average stellar mass, specific star formation rate, and stellar velocity dispersion of the Illustris-2 sample is $\overline{M_*} = 10^{11.44}$, $\overline{\text{sSFR}} = 2.52\times10^{-12}$ yr$^{-1}$, and $\overline{\sigma_*} = 199$ km s$^{-1}$ while for the Illustris-3 sample they are $\overline{M_*} = 10^{11.48}$, $\overline{\text{sSFR}} = 2.64\times10^{-12}$ yr$^{-1}$, and $\overline{\sigma_*} = 204$ km s$^{-1}$. To determine the $\log(\alpha_{\text{IMF}})$-$\sigma_*$ relation we use both the observed relation as input (Equation 2.2) and the steeper $\log(\alpha_{\text{IMF}})$-$\sigma_{\text{birth}}$ relation we constructed which was found to reproduce the observed trend with global velocity dispersion.

The leftmost panels of Figure 2.8 show the resulting $\log(\alpha_{\text{IMF}})$-$\sigma_*$ trends for Illustris-2 and Illustris-3, constructed using just the birth velocity dispersions of the stellar particles within $0.5R^p_{1/2}$ from the center of each galaxy. As with Illustris-1, using the observed relation to set the IMF of stellar particles at their birth times results in a $z = 0$ $\log(\alpha_{\text{IMF}})$-$\sigma_*$ relation that is shallower than observed (red curve). The Illustris-2 relation is $\sim$2.5$\times$ shallower than the observed relation while the Illustris-3 relation is $\sim$2.8$\times$ shallower than the observed relation. We construct the $\log(\alpha_{\text{IMF}})$-$\sigma_*$ using the same, steeper input relation we found to reproduce the global trend with Illustris-1, $\log(\alpha_{\text{IMF}}) = 3.7\log(\sigma_{\text{birth}}) - 8.99$. The blue curve in the figure shows that the same steep input relation that was able to match the global $\sigma_*$ trend in Illustris-1 is also able to reproduce the observed trend, within the uncertainties, in Illustris-2 and Illustris-3. Thus, we conclude our main results to be robust to resolution degradation.

We also consider the effect of resolution on the radial IMF trends we explored in 2.5.1. Figure 2.9 shows the average IMF profile (measured in cylinders) in four velocity dispersion bins for Illustris-1, Illustris-2, and Illustris-3. For the two lowest velocity dispersion bins, $\sigma_*$=150-200 km s$^{-1}$ and $\sigma_*$=200-250 km s$^{-1}$, the radial profiles for the three resolution levels are similar, although in the $\sigma_*$=200-250 km s$^{-1}$ bin the Illustris-1 profile is slightly steeper at $\mathrm{R}^p_{1/2} < 1$ compared to Illustris-2 and -3. For the $\sigma_*$=250-300 km s$^{-1}$ bin, the Illustris-2 and -3 average profiles are at higher $\log(\alpha_{\mathrm{IMF}})$ values compared to Illustris-1. This is also seen comparing Figure 2.8 to Panel A of Figure 2.3. In this velocity dispersion bin, the Illustris-3 radial profile is significantly shallower than the Illustris-1 and -2 profiles, which is due to the larger smoothing length of the Illustris-3 simulation. For the Illustris-2 and -3 simulations, the velocity dispersion bin with the highest $\log(\alpha_{\mathrm{IMF}})$ values is not the $\sigma_*$=300-350 km s$^{-1}$ bin, but the $\sigma_*$=250-300 km s$^{-1}$ bin. This is likely due to the most massive galaxies in the lower resolution simulations being more affected by intracluster light, which acts to reduce the overall $\log(\alpha_{\mathrm{IMF}})$. While the most relevant radial profile comparison would be between Illustris-1 and higher resolution simulations, as we discuss next, these high resolution simulations at the appropriate mass scale are not currently available.

Additionally, we attempt to test the robustness of our results to higher resolutions. This is motivated by Sparre & Springel (2016), who present zoom-in simulations of Illustris galaxies with mass resolution up to 40 times better than that of Illustris-1. While galaxies in Illustris-1 do undergo nuclear starbursts (Wellons et al. 2015), in some of these zoom-in simulations the merger-driven nuclear starbursts are stronger. This could potentially influence the constructed IMF for two reasons. First, these starburst episodes produce more stellar mass. Second, these extreme star-formation environments display larger velocity dispersions. Hence they have the potential to be the sites where the bottom-heavy IMF of ETGs is built up.

These galaxies were selected by Sparre & Springel (2016) based on their $z = 0$ quiescence and that they undergo a major merger between $z = .5$ and $z = 1$. Each is run at three resolution levels: 1 - $m_{\mathrm{dm}} = 4.42 \times 10^6$ $M_\odot$, 2 - $m_{\mathrm{dm}} = 5.53 \times 10^5$ $M_\odot$, 3 - $m_{\mathrm{dm}} = 1.64 \times 10^5$ $M_\odot$. We focus on the two galaxies that exhibit the largest increase in star-formation rate at their respective merger times (galaxies 1349 and 1605). For the IMF analysis, we construct the overall log($\alpha_{\rm IMF}$) based on the $\sigma_{\rm birth}$ distributions of the stellar particles residing within $0.5R_{1/2}^p$ of each respective galaxy. We find that the overall velocity dispersion of both galaxies increases with resolution level: $\sigma_* = 115$, 123, and 134 km s$^{-1}$ for galaxy 1349 and $\sigma_* = 113$, 123, and 139 km s$^{-1}$ for galaxy 1605. However, using the steep input relation as discussed in Section 2.4.2, log($\alpha_{\rm IMF}$) hardly changes: -0.22, -0.22, -0.217 for galaxy 1349 and -0.22, -0.22, -0.22 for galaxy 1605, for zoom levels 1, 2, and 3, respectively. The reason is that the corresponding $\sigma_{\rm birth}$ distributions lie mostly below the Chabrier minimum of log($\alpha_{\rm IMF}$) = -0.22 set at $\sigma_{\rm birth} = 235$ km s$^{-1}$. Using a shifted input relation so that fewer stellar particles are assigned the minimum Chabrier value, we do see more significant increases in $\log(\alpha_{\rm IMF})$ with increasing resolution level. This suggests the possibility that for zooms of higher velocity dispersion galaxies of at least $\sigma_* = 300$ km s$^{-1}$ a shallower input relation might suffice to reproduce the observed $\log(\alpha_{\rm IMF})$-$\sigma_*$ relation. However, such simulations would require a very significant investment of computing time and are currently not available. Hence, our zoom analysis is at this time inconclusive.

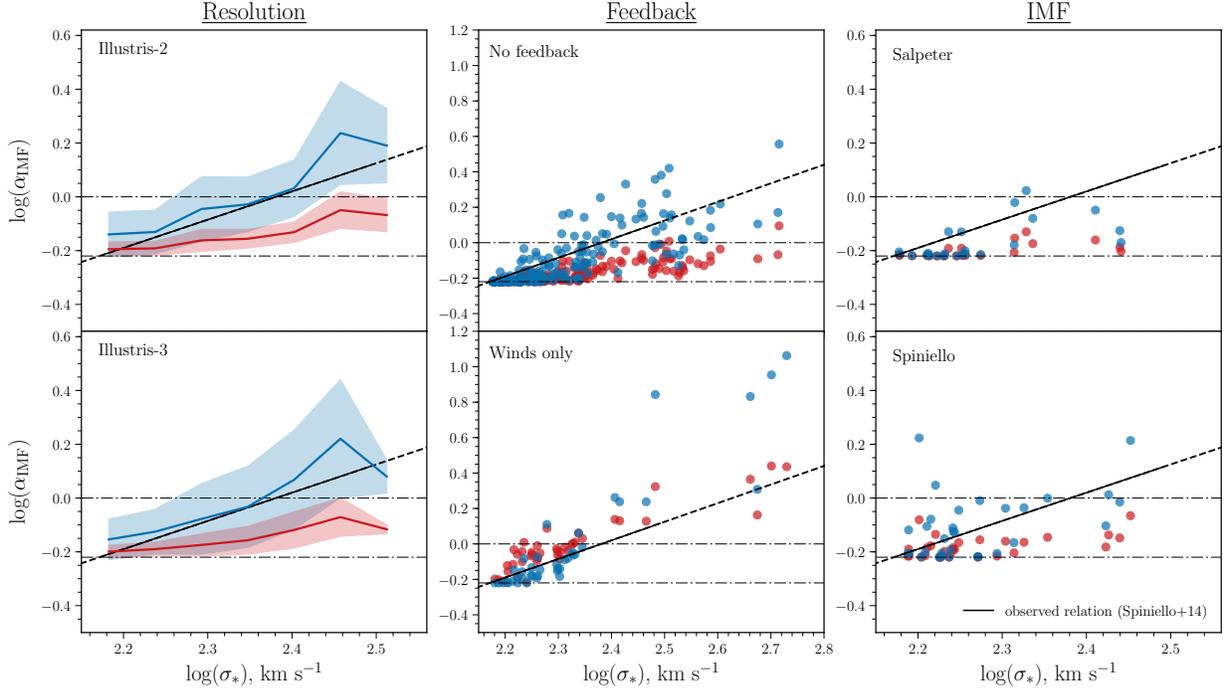

Figure 2.8: The log($\alpha_{\rm IMF}$)-$\sigma_*$ Illustris relations, constructed based on the birth velocity dispersion distributions of each galaxy, for the two lower resolution Illustris simulations (Illustris-2 and Illustris-3), the two simulations with varying feedback (no feedback and winds only), and the two simulations with varying IMFs (IMF-Salpeter and IMF-Spiniello). In each panel, the red curve or points show the observed relation (Equation 2.3) used as input to construct the overall log($\alpha_{\rm IMF}$) of each galaxy and the blue curve shows the resulting relation using the steeper log($\alpha_{\rm IMF}$)-$\sigma_{\rm birth}$ relation. In each panel, the log($\alpha_{\rm IMF}$)-$\sigma_*$ constructed using just the stellar particles within $0.5R_{1/2}^p$ is shown.

Lastly, we test the dependence of our results to time resolution by diluting the snapshots by a factor of 3 in our Illustris-2 analysis. We find our main result to be unaffected by this decrease in time resolution, justifying our choice for the number of snapshots produced in modified physics runs which will be discussed in the following section.

### 2.6.2 Variations in simulation physics

We also test the robustness of our results to variations in simulation physics, first considering variations in feedback. We ran a box of 40 Mpc/h on a side with $2 \times 320^3$ resolution elements (with a Chabrier IMF) once with no feedback and once with galactic winds but no AGN feedback. In each simulation we select galaxies with stellar masses greater than $10^{10}$ M$_\odot$ and stellar velocity dispersions greater than 150 km s$^{-1}$. Since no galaxies in the winds only simulation meet the sSFR criterion, we do not cut on sSFR. This selection results in 178 $z = 0$ galaxies in the no feedback simulation and 41 $z = 0$ galaxies in the winds only simulation. The average stellar mass, specific star formation rate, and stellar velocity dispersion of the no feedback sample is $\overline{M_*} = 10^{11.51}$, $\overline{\rm sSFR} = 2.68 \times 10^{-11}$ yr$^{-1}$, and $\overline{\sigma_*} = 216$ km s$^{-1}$, while for the winds only sample $\overline{M_*} = 10^{11.43}$, $\overline{\rm sSFR} = 1.74 \times 10^{-10}$ yr$^{-1}$, and $\overline{\sigma_*} = 223$ km s$^{-1}$.

The middle panels of Figure 2.8 show the resulting $\log(\alpha_{\rm IMF})$-$\sigma_*$ relations constructed within $0.5R^p_{1/2}$ for the no feedback simulation (top) and winds only simulation (bottom), where the overall $\log(\alpha_{\rm IMF})$ of each galaxy is calculated based on the local, birth velocity dispersions of the stellar particles. For the no feedback simulation, the shallow input relation (as used in Section 2.4.2) produces a $\log(\alpha_{\rm IMF})$-$\sigma_*$ relation that is shallower than the observed relation, similar to the corresponding Illustris-1 relation. We find that the same steep input relation that was necessary to reproduce the observed $\log(\alpha_{\rm IMF})$-$\sigma_*$ trend for Illustris-1 is also able to reproduce the observed trend in the no feedback simulation. The winds only simulation, on the other hand, varies from the Illustris and no feedback results. As seen in bottom, middle panel of Figure 2.8, the shallow input relation *does* reproduce the observed $\log(\alpha_{\rm IMF})$-$\sigma_*$ relation while the steeper input relation results in a $\log(\alpha_{\rm IMF})$-$\sigma_*$ relation that is $\sim 2\times$ too steep.

The impact of varying the simulation feedback on the overall $\log(\alpha_{\rm IMF})$ calculated for each galaxy can be understood by considering what star-formation is suppressed. AGN feedback suppresses star-formation in massive galaxies through both quasar and radio mode, while galactic winds suppress star-formation in lower mass galaxies with shallower potentials. In the no feedback simulation, without AGN feedback or galactic winds, star-formation is not suppressed either in low- or high-mass galaxies. The fraction of high to low velocity dispersion stellar particles in the no feedback simulation ends up being similar to the fraction in the full physics Illustris simulations, and a steep input relation is needed to reproduce the observed $\log(\alpha_{\rm IMF})$-$\sigma_*$ trend. In the winds only simulation, star-formation in low-mass galaxies is suppressed but not in high-mass galaxies. The fraction of high to low velocity dispersion stellar particles in enhanced, leading to galaxies having higher $\log(\alpha_{\rm IMF})$ values. Since more massive galaxies have an increasing fraction of stellar particles formed ex-situ in lower-mass galaxies, an increasing fraction of low velocity dispersion stellar particles are suppressed, allowing the shallow input relation to reproduce the slope of the $\log(\alpha_{\rm IMF})$-$\sigma_*$ trend. Of the physical quantities and simulation variations explored in this paper, a stellar mass function tilted in favor of star-formation in high-mass galaxies is able preserve the overall IMF trend without requiring a steep, physical IMF relation. Although, indicated by the lack of massive quiescent galaxies, the galaxy population in the winds only simulation is unrealistic. As shown in Vogelsberger et al. (2013a), not including AGN feedback results in a $z = 0$ stellar mass function and stellar mass to halo mass relation that are too high compared to observations and the fiducial Illustris model, as well as a star-formation rate density and a stellar mass density functions with redshift also being too high.

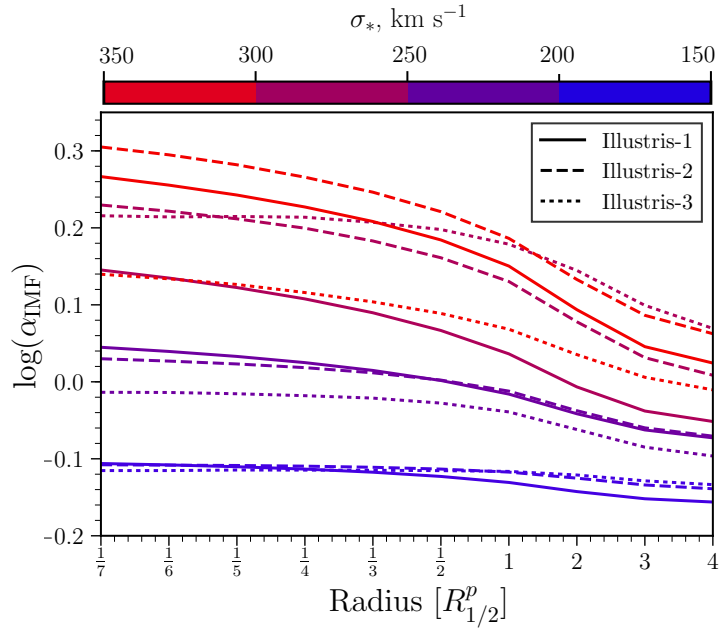

Figure 2.9: The average radial IMF profiles in four velocity dispersion bins for Illustris-1, -2, and -3. Each curve represents the average IMF mismatch parameter as a function of galactocentric radius, colored by velocity dispersion bin. For higher velocity dispersions, Illustris-3 exhibits shallower radial profiles at $R^p_{1/2}$ < 1 compared to Illustris-1 and -2.

Lastly, we begin to explore the effect of varying the IMF with which the simulation is run. As discussed in Section 2.3.1, a Chabrier IMF law is used to govern mass and metal return from stellar particles in Illustris, as well as to calculate the mass-loading factors of galactic winds. Our empirical approach to constructing the IMF mismatch parameter of Illustris galaxies is not expected to depend directly on the IMF used to run the simulation.

But, different mass returns and feedback may affect the local velocity dispersions which stellar particles are born into. To test if the IMF that the simulation is run with alters our main result, we run simulations with different IMF laws but otherwise with the same physics models as in Illustris, using smaller boxes at lower resolutions. We run a box of 40 Mpc/h on a side with $2 \times 320^3$ resolution elements incorporating two IMF laws: 1) a pure Salpeter IMF law with a slope of $x = 2.35$ and 2) the variable IMF law presented by Spiniello et al. (2014) dependent on the local dark matter velocity dispersion $\sigma_{\rm birth}$. More details about these additional simulations are listed in Table 2.1.

Using the same galaxy selection criteria outlined in Section 2.3.2, 23 galaxies in the Salpeter simulation at $z = 0$ and 31 galaxies in the Spiniello simulation at $z = 0$ meet the stellar mass, specific star formation rate, and velocity dispersion criteria. The average stellar mass, specific star formation rate, and stellar velocity dispersion of the Salpeter sample is $\overline{M_*} = 10^{11.34}$, $\overline{\rm sSFR} = 1.27 \times 10^{-12}$ yr$^{-1}$, and $\overline{\sigma_*} = 190$ km s$^{-1}$, while for the Spiniello sample $\overline{M_*} = 10^{11.38}$, $\overline{\rm sSFR} = 3.94 \times 10^{-13}$ yr$^{-1}$, and $\overline{\sigma_*} = 189$ km s$^{-1}$. Again, we determine the $\log(\alpha_{\rm IMF})$-$\sigma_*$ relation using both the observed relation as input (Equation 2.3) and the steeper $\log(\alpha_{\rm IMF})$-$\sigma_{\rm birth}$ relation we constructed which was found to reproduce the observed trend with global $\sigma_*$ in the Illustris-1 analysis.

The rightmost panels of Figure 2.8 show the resulting $\log(\alpha_{\rm IMF})$-$\sigma_*$ relations constructed within $0.5R_{1/2}^p$ for our Salpeter and Spiniello IMF runs. Consistent with the results in Section 2.4.2, for both modified IMF simulations the output $\log(\alpha_{\rm IMF})$ values constructed using the steeper $\log(\alpha_{\rm IMF})$-$\sigma_{\rm birth}$ input relation lie closer to the observed relation than the $\log(\alpha_{\rm IMF})$ values constructed with the shallow input relation. While there are a few outliers and the number of high velocity dispersion galaxies in each simulation is small, incorporating a different and even variable IMF in the Illustris galaxy formation model does not seem to significantly modify the results of this paper. Since the formal fits of the Salpeter and Spiniello output log($\alpha_{\text{IMF}}$)-$\sigma_*$ relations are shallower than the observed relation, this serves to strengthen our claim that a steeper input relation is required to produce the observations of IMF variations. Lastly, while the galaxy population in these modified IMF simulations undoubtedly differs from the galaxy population in Illustris, the investigation of these differences and the implications of simulations which self-consistently include a variable IMF is the topic of future work.

## 2.7 Discussion

### 2.7.1 Simulation limitations

Before we discuss the possible implications of this work, it is important to reiterate the extent to which the results are dependent on the simulation models and resolution. While the Illustris galaxy formation and evolution models reproduce many key galaxy properties and scaling relations, there is certainly room for improvement (Vogelsberger et al. 2014a,b; Genel et al. 2014). For example, Illustris produced massive galaxies with too high of a stellar mass (Vogelsberger et al. 2014b), and galaxy sizes that are too large (Pillepich et al. 2018). In particular, relevant to studying IMF variations through the hierarchical build-up of galaxies, is the merger rate. Rodriguez-Gomez et al. (2015) show the Illustris merger rate to match some observations well, though there remain qualitative differences among the observations. If the merger rate in Illustris is too high, this might explain why a steep physical IMF relation is needed to conserve the global IMF relation to $z = 0$.

Another simulation parameter that could be influencing our results is resolution. Observational IMF studies are starting to reveal that IMF variations are confined to the most inner parts of galaxies, at radii typically $< 0.3 R_e$ (Martín-Navarro et al. 2015a; van Dokkum et al. 2017). The resolution of the Illustris simulation, with the highest resolution simulation having a baryonic smoothing length of 0.7 kpc, prevents us from gaining a realistic understanding of IMF gradients at small radii. As considered in Section 2.6.1, higher resolution simulations could yield more powerful nuclear starbursts, which in turn could reduce the steepness of the physical IMF laws we currently find necessary. At this time though, there are no simulations at the relevant mass scale to test this hypothesis.

Given the above dependence on simulation details, it is advised that the results of this work be interpreted in a more qualitative sense. The exact input relation we find necessary to reproduce the observed IMF trends are certainly sensitive to the galaxy formation and evolution model used, as well as resolution. While the quantitative results may change with new and improved simulation, the qualitative results are more robust.

### 2.7.2 Implications for IMF observations

Observations suggest that with increasing galaxy velocity dispersion, galaxies also have increasingly bottom-heavy IMFs where the most massive $z = 0$ galaxies are characterized by super-Salpeter IMF slopes. Since these massive galaxies are believed to be primarily composed of ex-situ stellar populations, which formed in smaller systems with lower velocity dispersions and only later accreted onto the main galaxy, then the physical explanation for the steepness of the correlation between $z = 0$ global galaxy properties and the IMF is unclear. The result that even steeper physical IMF relations are needed to preserve the observed IMF variations through the assembly of massive galaxies has implications for observations of IMF variations and theoretical work predicting IMF variations.

One consequence of our analysis is that massive, quiescent galaxies at higher redshifts would have global IMFs even more bottom heavy than their $z = 0$ counterparts. This is because massive galaxies in Illustris first build up their in-situ stellar populations, which are mainly formed in high velocity dispersion environments, and only later accrete smaller systems with stellar populations formed in low velocity dispersion environments. For massive galaxies, these smaller systems reside even within an effective radius where global IMF measurements are typically made (Rodriguez-Gomez et al. 2016), reducing the overall IMF of the galaxy.

Based on studies of ETG populations at $z \sim 1$ and $z \sim 1.4$, observations beyond the local Universe are beginning to suggest that the IMF-$\sigma$ relation remains roughly constant over the last 8 Gyrs (Martín-Navarro et al. 2015b; Gargiulo et al. 2015a). Contrary to these studies, Sonnenfeld et al. (2015) find that the overall $\log(\alpha_{\mathrm{IMF}})$ at fixed velocity dispersion decreases from the $z \sim 0$ value out to $z \sim 0.8$. In Sonnenfeld et al. (2017), they postulate that a possible source of this apparent evolution could be their assumption of fixed dark matter density profile. A robust determination of the evolution of IMF-$\sigma$ relation from $z \sim 0$ to $z \sim 2$ will require careful determination of the dark matter fraction of galaxies to break the dark matter-IMF degeneracy. To reconcile our prediction for $z \sim 2$ with observations that suggest a constant IMF-$\sigma$ evolution, one option is to invoke a time-dependent physical IMF 'law'. In the context of our analysis, this would mean applying a shallower $\log(\alpha_{\mathrm{IMF}})$-$\sigma_{\mathrm{birth}}$ relation to higher-redshift stellar population.

A further prediction of our analysis is that satellite galaxies, at fixed velocity dispersion, should have more bottom-heavy IMFs than central galaxies. Since satellites are expected to undergo fewer minor mergers than centrals, they should better preserve to $z = 0$ their bottom-heavy IMFs that are in place at higher redshift before they become satellites. To demonstrate this prediction, we examine the satellite fraction for $z = 0$ Illustris-1 galaxies in four velocity dispersion bins, with widths of 25 km s$^{-1}$ except for the highest $\sigma_*$ bin that includes all 78 galaxies with $\sigma_* > 225$ km s$^{-1}$. For each velocity dispersion bin, we determine $\log(\alpha_{\mathrm{IMF}})$ quintiles and calculate the fraction of satellite galaxies within each quintile. Figure 2.10 shows the satellite fraction, normalized by the total satellite fraction in the velocity dispersion bins, for each $\log(\alpha_{\mathrm{IMF}})$ quintile. We note that the formal errors associated with the normalized satellite fractions are large, especially when comparing different velocity dispersion bins within each quintile. However, some individual bins are above or below the total satellite fraction with statistical significance.

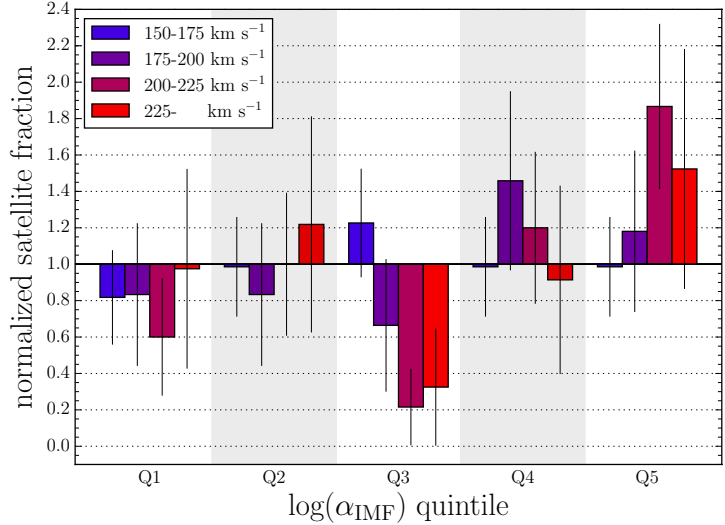

Figure 2.10: The satellite fraction of $z = 0$ Illustris-1 galaxies in 4 velocity dispersion bins, normalized by the total satellite fraction in the bin, for each $\log(\alpha_{\mathrm{IMF}})$ quintile. The bars are colored according to velocity dispersion, and each group of bars shows one quintile. The errors are calculated assuming a binomial distribution, where we add in quadrature the error associated with the total satellite fraction and the error associated with the satellite fraction in each quintile. In the 200-225 km s$^{-1}$ bin, satellite galaxies make up 65% of the most extreme $\log(\alpha_{\mathrm{IMF}})$ galaxies, which is $\sim 1.8\times$ larger than their total fraction in the bin.

The fifth quintile represents the galaxies with the most bottom-heavy IMFs in each velocity dispersion bin. For the lowest velocity dispersion bin, from 150 - 175 km s$^{-1}$, the satellite fraction is the same as the total satellite fraction in the bin, indicating that for these low velocity dispersion galaxies, satellites do not have higher $\log(\alpha_{\mathrm{IMF}})$ values than centrals. This can be understood as lower velocity dispersion galaxies are more dominated by in-situ evolution and less affected by minor mergers. In higher velocity dispersion bins, the fraction of satellites composing the most extreme $\log(\alpha_{\mathrm{IMF}})$ galaxies is higher than the total satellite fraction. In particular, in the 200-225 km s$^{-1}$ bin $\sim 65\%$ of the highest $\log(\alpha_{\mathrm{IMF}})$ galaxies are satellites, which is significantly enhanced with respect to their total fraction in the bin. For these galaxies with 200-225 km s$^{-1}$, the highest IMF values of individual galaxies are log($\alpha_{\rm IMF}$) $\sim$ 0.32 (compared with the typical log($\alpha_{\rm IMF}$) $\sim$ -0.05 for this $\sigma_*$ bin), and a majority of these 'extreme IMF' galaxies are satellites. For galaxies with $\sigma_* >$ 225 km s$^{-1}$, of the 8 galaxies with log($\alpha_{\rm IMF}$) > 0.20, 2 are satellites and 6 are centrals. The most extreme log($\alpha_{\rm IMF}$) value for a central in this bin is log($\alpha_{\rm IMF}$) = 0.32 and for a satellite it is log($\alpha_{\rm IMF}$) = 0.36.

### 2.7.3 Implications for IMF theory

In our post-processing analysis of Illustris, we find that steep physical IMF relations, as applied to the birth properties of stellar particles, are required to reproduce the observed $z = 0$ IMF trend with global velocity dispersion. Input relations more than 3$\times$ steeper than the observed relation are needed, which means that some individual stellar populations must be formed with mass-to-light ratios up to $\sim$20$\times$ greater than the Salpeter mass-to-light ratio. These required extreme mass-to-light ratios are $\sim$10$\times$ greater than the overall mass-to-light ratios measured in observations of massive galaxies.

To gain an idea of what IMF slope could give rise to an $M_*/L$ ratio excess this large, we calculate mass-to-light ratios with the FSPS (Flexible Stellar Population Synthesis) library (Conroy & Gunn 2010; Conroy et al. 2009b) and the Python FSPS package[5]. We model a single burst of star-formation with solar metallicity and an exponentially declining star-formation history truncated at 4 Gyrs, and calculate the $r$-band $M_*/L$ ratio at an age of 10 Gyr for several IMF slopes. We find that unimodal IMF slopes greater than $x = 4$ are required to produce a $M_*/L$ ratio that is $\sim$20$\times$ greater than the Salpeter $M_*/L$ ratio, where an IMF slope of $x = 4$ results in a $M_*/L$ excess of $\sim$11.5 and an IMF slope of $x = 4.5$ results

---
[5]http://dan.iel.fm/python-fsps/

in an $M_*/L$ excess of $\sim$24.

While IMF slopes this steep have yet to be robustly observed, an IMF slope this extreme could have more of an immediate implication for analytical IMF variation theories. For example, the functional form of the CMF could be mapped to a unimodal IMF slope to determine how high of a Mach number would be required to produce these extreme $M_*/L$ ratios in theories that predict that high Mach number environments promote a bottom-heavy IMF (Hopkins 2013; Chabrier et al. 2014; Guszejnov et al. 2016).

Additionally, in Section 2.4.5 we determined the overall $\log(\alpha_{\mathrm{IMF}})$-$\sigma_*$ relation with an input relation based on star-formation rate. Motivated by IGIMF theory, we assigned stellar particles born into low SFR galaxies a higher $\log(\alpha_{\mathrm{IMF}})$ value and stellar particles born into high SFR lower $\log(\alpha_{\mathrm{IMF}})$ values. With this input relation, we are able to reproduce the slope and normalization of the observed $\log(\alpha_{\mathrm{IMF}})$-$\sigma_*$ relation, but in the opposite direction. Combined with our ability to reproduce the slope of the observed relation and its sign with an input relation based on $\sigma_{\mathrm{birth}}$, this suggests that stellar particles formed in high SFR environments also form in high velocity dispersion environments. If this is the case, then there is an apparent tension between the analytical IMF theories which predict that high Mach number (and therefore high velocity dispersion) environments promote a bottom-heavy IMF (Hopkins 2013; Chabrier et al. 2014; Guszejnov et al. 2016), and IGIMF theory Weidner et al. (2013b) which predicts that high SFR environments lead to a top-heavy IMF.

While this tension seems to exist, the main driver of IMF variations in analytical and simulation work is far from settled. For example, recent SPH simulations of star-formation Bertelli Motta et al. (2016b) show no correlation between Mach number and peak of the IMF in high density environments. Moreover, in low density environments they actually find that a higher Mach number shifts the peak of the IMF to higher masses, implying a more *top-heavy* IMF – opposite of what Hopkins (2013) predicts. In general, if observations prove to be robust, any theory predicting IMF variations will have to accommodate both the correlation of overall IMF with global $z = 0$ velocity dispersion and with metallicity. One possibility is an IMF 'law' which depends on a combination of local velocity dispersion and metallicity. Combined, this could be an additional probe of Mach number, with metallicity controlling cooling and therefore the sound speed.

As discussed in Weidner et al. (2013b), to resolve the tension between IGIMF theory and the observed mass excess of ETGs it is suggested that at least part of the inferred mass is due to an increased number of stellar remnants. IMF variation studies focused on measuring dwarf sensitive spectral lines do break the degeneracy between low-mass stars and stellar remnants (e.g. van Dokkum & Conroy 2010; La Barbera et al. 2013), suggesting that the inferred, excess mass is low-mass stars. However, the exact parameterization of the IMF of massive ETGs, especially below 1 $M_\odot$, is still unconstrained. Determining the physical driver of IMF variations will require both theoretical work predicting the shape of the IMF as a function of environment and the ability to observationally constrain the shape of the IMF. Recent work to constrain the shape of low-mass end of the IMF from high-quality spectra appears promising (Conroy et al. 2017).

### 2.7.4 Comparison to other work

The closest approach in the existing literature to the post-processing IMF analysis of Illustris presented in this paper is that of Sonnenfeld et al. (2017). For a sample of galaxies at $z = 2$, Sonnenfeld et al. (2017) assume that all subsequent stellar mass growth occurs via dry mergers, based on a toy merging model. By using empirical $\log(\alpha_{\rm IMF})$-$\sigma$ or $\log(\alpha_{\rm IMF})$-$M_*$ relations, a $\log(\alpha_{\rm IMF})$ value is assigned to each central galaxy at $z = 2$. The overall $\log(\alpha_{\rm IMF})$ of each galaxy at later redshifts is determined by the addition of smaller systems, with their own IMFs assigned by the same empirical relations.

While Sonnenfeld et al. (2017) explore the mixing between IMF dependence on stellar mass and velocity dispersion, the most directly comparable result to this work is the $\log(\alpha_{\mathrm{IMF}})$-$\sigma$ relation based on the velocity dispersion model. Here they find that the slope and normalization of the $\log(\alpha_{\mathrm{IMF}})$-$\sigma$ relation is preserved from $z = 2$ to $z = 0$, such that galaxies at fixed velocity dispersion have the same $\log(\alpha_{\mathrm{IMF}})$ at different times. As shown in Section 2.5.3, our analysis predicts that at fixed velocity dispersion $\log(\alpha_{\mathrm{IMF}})$ is higher at $z = 2$ than at $z = 0$. These qualitatively discrepant results could be due to a number of factors stemming from significant methodology differences.

First, we assign a $\log(\alpha_{\mathrm{IMF}})$ value to stellar particles based on the *birth* velocity dispersion, whereas in Sonnenfeld et al. (2017) $\log(\alpha_{\mathrm{IMF}})$ values are assigned based on the $z = 2$ velocity dispersion for the central galaxies and based on later redshift values for the smaller systems. This difference could be why our analysis required steeper input relations to reproduce the observed relation in the first place, as birth velocity dispersion values are generally lower than at $z = 2$ and later. Another difference between the works is that the Sonnenfeld et al. (2017) model does not take into account the addition of newly quenched galaxies to the population after $z = 2$. This certainly affects the comparison to our $z = 0$ IMF-$\sigma$ relation, which does include more recently quenched galaxies. Finally, as recognized by Sonnenfeld et al. (2017), the pure dry merger model does not correctly reproduce the redshift evolution of the velocity dispersion of galaxies, suggesting that a dissipational component is missing. The addition of such a component to their model would also break the constancy of the $\log(\alpha_{\mathrm{IMF}})$-$\sigma$ relation. On the other hand, our analysis is based on the innermost region of each galaxy, following observational constraints, while Sonnenfeld et al. (2017) do not address radial gradients. When we consider the full galaxy extent, we find an even stronger shallowing of the $\log(\alpha_{\mathrm{IMF}})$-$\sigma$ relation with redshift, suggesting the differences between the two studies are even larger than at face value.

As mentioned in the Introduction, IMF variations have been studied in the context of semi-analytical models. In particular, Fontanot et al. (2017) incorporates IGIMF theory into the `GAEA` (GAlaxy Evolution and Assembly) model to study the implications of IMF variations on the chemical evolution and dynamical properties of galaxies. There, a broken power-law IMF is employed where for stellar masses less than $\sim 1 M_\odot$ the standard Kroupa IMF is adopted, and for masses greater than that the slope of the IMF is determined by the instantaneous star-formation rate. In accordance with IGIMF theory, higher star-formation rates correspond to shallower (more top-heavy) IMF slopes at the high-mass end.

For $z = 0$ galaxies formed under the IGIMF model, Fontanot et al. (2017) compares the stellar mass-to-light ratio ($M_*/L$) and stellar mass excesses to the corresponding Chabrier equivalents. Overall, their results are in qualitative agreement with observations of IMF variations, where higher mass galaxies exhibit a greater mass (and $M_*/L$) excess compared to the mass (or $M_*/L$) derived assuming a Chabrier IMF. But for $10^9$ $M_\odot < M_* < 10^{10.8}$ $M_\odot$, there is a negative slope to the 'true' versus Chabrier stellar mass relation, which only turns over and becomes positive for stellar masses $10^{10.8}$ $M_\odot < M_* < 10^{12.2}$ $M_\odot$. In contrast, the overall IMF-$\sigma_*$ found in this study (Section 2.4.5), assuming a SFR dependent IMF relation where higher SFRs correspond to a less bottom-heavy IMF, exhibits an overall negative slope from the lowest to highest velocity dispersion galaxies. While this tension could possibly be highlighting an interesting difference between the Illustris and `GAEA` galaxy formation models, there are several factors that make the comparison difficult including the post-processing nature of our analysis and our different parameterization of the IMF-SFR relation.

Concerning the chemical enrichment of galaxies, Fontanot et al. (2017) find that the IMF variations they include are actually able to reproduce the observed $\alpha$-enhancement of massive galaxies ($M_* > 10^{11}$ $M_\odot$), though they do note that IMF variations are not the only solution to produce the observed metallicity of galaxies. Adopting a standard universal IMF, the [O/Fe] ratio of galaxies in `GAEA` with $M_* = 10^{12}$ $M_\odot$ is ∼0.2 dex too low compared to observations of massive galaxies. At the time, the post-processing nature of our analysis prevents us from making a meaningful comparison to this result. In future work, as discussed in the next section, we will self-consistently incorporate different IMF laws into the Illustris galaxy formation models, and by tracking the build-up of individual metals we will be able to study how the chemical enrichment of galaxies is altered.

### 2.7.5 Prospects for future work

Future work will include validating our results with both higher resolution simulations, and improved galaxy formation models (Pillepich et al. 2018; Weinberger et al. 2017). High-resolution simulations of individual galaxies, at the relevant mass scales, will allow for a more realistic study of radial gradients in the IMF and fully capture the impact of star-formation occurring in nuclear starbursts. Additionally, improved galaxy formation models that can better match observed galaxy properties than the current fiducial Illustris model can, such as the $z = 0$ stellar mass function, will be a crucial test of the robustness of our main result.

In addition to validating our results with improved simulations, future work will include expanding on our analysis. Inspired by observations and theoretical work, in this study we examined five physical quantities associated with star-formation and/or IMF variations. There are other quantities not considered here that have also been advocated to influence the IMF, including pressure and redshift (Krumholz 2011; Munshi et al. 2014). Combinations of quantities can also be considered, such as metallicity and velocity dispersion.

Expanding on the analysis of the variable IMF simulations presented in Section 2.6.2, further work is also required for studying the impact of a variable IMF that is incorporated self-consistently on the evolution and properties of galaxy populations. Since the number of low- to high-mass stars in a galaxy determines the amount of metals injected into the ISM, the energy available for supernovae feedback, and the mass of baryons trapped in low-mass stars, the inclusion of a variable IMF could significantly alter our current picture of galaxy evolution. While a variable IMF remains controversial because measuring the IMF is observationally difficult, the numerous studies reporting IMF variations and the analytical work predicting IMF variations in extreme star-formation environments necessitates that the implications of a variable IMF be investigated.

## 2.8 Summary & Conclusion

In this study, we present an investigation of the physical origin of IMF variations using the cosmological simulation Illustris in post-processing. For a sample of massive ($M_* > 10^{10}$ $M_\odot$) and quiescent (sSFR $< 10^{-11}$ yr$^{-1}$) galaxies we connect the physical conditions in which stellar particles form to the properties of the galaxies they reside in at $z = 0$. We do this by constructing the overall IMF mismatch parameter, $\alpha_{\text{IMF}}$, of each galaxy based on various formation conditions associated with the individual stellar particles that comprise it. By attempting to reproduce the observed relations between overall IMF and global velocity dispersion, we are able to gain insight into how galaxy-wide quantities at $z = 0$ are related to the IMF – a local property defined at the time of star-formation. Our findings are summarized as follows:

- A much steeper than observed physical IMF relation is needed to reproduce the reported IMF trends with global $z = 0$ velocity dispersion under the hierarchical assembly of massive galaxies. This result ties observations of an IMF that varies with $z = 0$ galactic properties to a physical origin, but requires some individual stellar populations to be formed with super-Salpeter IMFs that are even steeper than the IMFs reported by observational studies. These extreme IMFs are up to ∼20× in excess of the Salpeter $M_*/L$, which could imply a unimodal IMF slope of $x > 4$.

- Of the five physical quantities we consider, we are able to reproduce the observed IMF trend with $z = 0$ velocity dispersion by constructing the overall $\alpha_{\rm IMF}$ of each galaxy based on the global and local birth velocity dispersion of the stellar particles, and the global star-forming gas velocity dispersion of the progenitor galaxy in which each stellar particle was formed. All of these quantities are roughly related to Mach number, which in analytical models of IMF variations corresponds to increased fragmentation on lower mass scales in more extreme star-formation environments.

- We are unable to reproduce the observed IMF trend with $z = 0$ velocity dispersion when constructing the overall $\alpha_{\rm IMF}$ of each galaxy based on the metallicities of individual stellar particles. The relations obtained in this way are too shallow. This is due to the scatter and near flatness of the [M/H]-$\sigma_*$ relation, and to the fact that the global metallicity of each galaxy is composed of a broad distribution of individual stellar particle metallicities.

- Using the star-formation rate of the progenitor galaxy in which each individual stellar particle was formed to construct the overall $\alpha_{\rm IMF}$, we obtain steep relations between IMF and $z = 0$ velocity dispersion, but in the opposite direction to what is observed. Inspired by IGIMF theory, we construct a $\log(\alpha_{\rm IMF})$-SFR relation in which a low SFR corresponds to a bottom-heavy IMF. For the $z = 0$ massive quiescent galaxies we focus on in this study, this simple relation does not reproduce the direction of the observed IMF trend. If stellar particles which form in merger induced starbursts have both high velocity dispersions and high SFRs, there is tension between IMF theories as to whether these stellar populations are expected to form with a bottom-heavy or top-heavy IMF.

- We find radial gradients in the constructed $\log(\alpha_{\mathrm{IMF}})$ for massive galaxies due to the different formation conditions of stellar particles that reside at the center of galaxies versus in the outer regions. This result reinforces the need for consistent comparisons of IMF variations across observational studies that may be probing the IMF at different radii. It also further supports the idea that the IMF is a local property of galaxies, which are composites of numerous stellar populations formed in a diverse range of physical conditions throughout cosmic time.

- The scatter in the constructed Illustris IMF relations is reflective of the diverse formation histories of galaxies that have similar $z = 0$ velocity dispersions. In other words, galaxies with similar galactic quantities can be comprised of stellar populations which formed in different physical environments. This is in agreement with, and can provide a possible explanation for, the scatter in the observed IMF-$\sigma*$ relations.

- Based on our analysis we make two predictions for observations: (1) the $\log(\alpha_{\mathrm{IMF}})$-$\sigma_*$ relation at high redshift would be more bottom-heavy than the $z = 0$ relation, and (2) at high velocity dispersions galaxies with extreme bottom-heavy IMFs are preferentially satellite galaxies.

*Acknowledgements.* We thank Richard Bower and Romain Teyssier for useful discussions, and Martin Sparre for sharing with us the outputs of his zoom-in simulations. We also thank the anonymous referee for a helpful report. KB is supported by the NSF Graduate Research Fellowship under grant number DGE 16-44869. SG acknowledges support provided by NASA through Hubble Fellowship grant HST-HF2-51341.001-A awarded by the STScI, which is operated by the Association of Universities for Research in Astronomy, Inc., for NASA, under contract NAS5-26555. GB acknowledges financial support from NASA grant NNX15AB20G and NSF grant AST-1615955. The Flatiron Institute is supported by the Simons Foundation. The Illustris simulation was run on the CURIE supercomputer at CEA/France as part of PRACE project RA0844, and the SuperMUC computer at the Leibniz Computing Centre, Germany, as part of project pr85je. Modified simulations used in this study were run on Columbia University's High Performance Computing cluster Yeti and on the Odyssey cluster supported by the FAS Division of Science, Research Computing Group at Harvard University. Post-processing analysis of simulation data was run on the Comet cluster hosted at SDSC, making use of the Extreme Science and Engineering Discovery Environment (XSEDE), which is supported by National Science Foundation grant number ACI-1053575.

## 2.A  Input & output IMF relations

For each simulation studied, Table 2.2 shows the input relations constructed based on the indicated star-formation quantity and the fit to the resulting $\log(\alpha_{\text{IMF}})$-$\log(\sigma_*)$ output relation. The output relations listed under $0.5R_{1/2}^p$ correspond to the relations shown in Figure 2.3 and Figure 2.8.

Table 2.2: IMF Relations

| Simulation | Quantity | Input relation, $\log(\alpha_{\rm IMF}) =$ | Output relation, $\log(\alpha_{\rm IMF}) =$ | |
|---|---|---|---|---|
| | | | $R_{all}$ | $0.5R^p_{1/2}$ |
| (1) | (2) | (3) | (4) | (5) |
| Illustris-1 | $\sigma_*$ | $1.05\times\log(\sigma_*)$ - 2.71 | $0.25\times\log(\sigma_*)$ - 0.76 | $0.42\times\log(\sigma_*)$ - 1.12 |
| | | $3.5\times\log(\sigma_*)$ - 7.84 | $0.52\times\log(\sigma_*)$ - 1.31 | $0.96\times\log(\sigma_*)$ - 2.26 |
| | $\sigma_{\rm birth}$ | $1.05\times\log(\sigma_{\rm birth})$ - 2.71 | $0.21\times\log(\sigma_*)$ - 0.68 | $0.38\times\log(\sigma_*)$ - 1.03 |
| | | $3.7\times\log(\sigma_{\rm birth})$ - 8.99 | $0.51\times\log(\sigma_*)$ - 1.30 | $0.97\times\log(\sigma_*)$ - 2.29 |
| | [M/H] | $1\times$[M/H] + 0.07 | $-0.12\times\log(\sigma_*) + 0.26$ | $-0.12\times\log(\sigma_*) + 0.39$ |
| | | $4\times$[M/H] - 0.02 | $-0.13\times\log(\sigma_*) + 0.32$ | $-0.20\times\log(\sigma_*) + 0.65$ |
| | $\sigma_{\rm gas}$ | $1.05\times\log(\sigma_{\rm gas})$ - 2.46 | $0.24\times\log(\sigma_*)$ - 0.73 | $0.41\times\log(\sigma_*)$ - 1.10 |
| | | $4.3\times\log(\sigma_{\rm gas})$ - 9.38 | $0.51\times\log(\sigma_*)$ - 1.31 | $0.96\times\log(\sigma_*)$ - 2.27 |
| | SFR | $-0.15\times\log({\rm SFR}) + 0.10$ | $-0.12\times\log(\sigma_*) + 0.18$ | $-0.21\times\log(\sigma_*) + 0.34$ |
| | | $-0.7\times\log({\rm SFR}) + 1.32$ | $-0.78\times\log(\sigma_*) + 2.01$ | $-1.03\times\log(\sigma_*) + 2.49$ |
| Illustris-2 | $\sigma_{\rm birth}$ | $1.05\times\log(\sigma_{\rm birth})$ - 2.71 | $0.22\times\log(\sigma_*)$ - 0.69 | $0.42\times\log(\sigma_*)$ - 1.12 |
| | | $3.7\times\log(\sigma_{\rm birth})$ - 8.99 | $0.49\times\log(\sigma_*)$ - 1.25 | $1.09\times\log(\sigma_*)$ - 2.54 |
| Illustris-3 | $\sigma_{\rm birth}$ | $1.05\times\log(\sigma_{\rm birth})$ - 2.71 | $0.21\times\log(\sigma_*)$ - 0.66 | $0.38\times\log(\sigma_*)$ - 1.05 |
| | | $3.7\times\log(\sigma_{\rm birth})$ - 8.99 | $0.47\times\log(\sigma_*)$ - 1.22 | $1.05\times\log(\sigma_*)$ - 2.48 |
| No feedback | $\sigma_{\rm birth}$ | $1.05\times\log(\sigma_{\rm birth})$ - 2.71 | $0.20\times\log(\sigma_*)$ - 0.65 | $0.37\times\log(\sigma_*)$ - 1.04 |
| | | $3.7\times\log(\sigma_{\rm birth})$ - 8.99 | $0.47\times\log(\sigma_*)$ - 1.25 | $1.01\times\log(\sigma_*)$ - 2.44 |
| Winds only | $\sigma_{\rm birth}$ | $1.05\times\log(\sigma_{\rm birth})$ - 2.50 | $0.72\times\log(\sigma_*)$ - 1.71 | $1.04\times\log(\sigma_*)$ - 2.42 |
| | | $3.7\times\log(\sigma_{\rm birth})$ - 8.99 | $1.17\times\log(\sigma_*)$ - 2.77 | $2.19\times\log(\sigma_*)$ - 5.09 |
| IMF-Salpeter | $\sigma_{\rm birth}$ | $1.05\times\log(\sigma_{\rm birth})$ - 2.71 | $0.13\times\log(\sigma_*)$ - 0.50 | $0.20\times\log(\sigma_*)$ - 0.65 |
| | | $3.7\times\log(\sigma_{\rm birth})$ - 8.99 | $0.34\times\log(\sigma_*)$ - 0.95 | $0.56\times\log(\sigma_*)$ - 1.44 |
| IMF-Spiniello | $\sigma_{\rm birth}$ | $1.05\times\log(\sigma_{\rm birth})$ - 2.71 | $0.14\times\log(\sigma_*)$ - 0.52 | $0.26\times\log(\sigma_*)$ - 0.77 |
| | | $3.7\times\log(\sigma_{\rm birth})$ - 8.99 | $0.34\times\log(\sigma_*)$ - 0.91 | $0.70\times\log(\sigma_*)$ - 1.69 |

(1) Simulation name; (2) Star-formation quantity; (3) Applied input relation;

(4) Fit to output relation with input relation applied to all star particles belonging to each galaxy;

(5) Fit to output relation with input relation applied to all star particles within $0.5R^p_{1/2}$.

# Chapter 3

# Variations in $\alpha$-element ratios trace the chemical evolution of the disk

## 3.1 Abstract

It is well established that the chemical structure of the Milky Way exhibits a bimodality with respect to the $\alpha$-enhancement of stars at a given [Fe/H]. This has been studied largely based on a bulk $\alpha$ abundance, computed as a summary of several individual $\alpha$-elements. Inspired by the expected subtle differences in their nucleosynthetic origins, here we probe the higher level of granularity encoded in the inter-family [Mg/Si] abundance ratio. Using a large sample of stars with `APOGEE` abundance measurements, we first demonstrate that there is additional information in this ratio beyond what is already apparent in [$\alpha$/Fe] and [Fe/H] alone. We then consider *Gaia* astrometry and stellar age estimates to empirically characterize the relationships between [Mg/Si] and various stellar properties. We find small but significant trends between this ratio and $\alpha$-enhancement, age, [Fe/H], location in the

---

Galaxy, and orbital actions. To connect these observed [Mg/Si] variations to a physical origin, we attempt to predict the Mg and Si abundances of stars with the galactic chemical evolution model *Chempy*. We find that we are unable to reproduce abundances for the stars that we fit, which highlights tensions between the yield tables, the chemical evolution model, and the data. We conclude that a more data-driven approach to nucleosynthetic yield tables and chemical evolution modeling is necessary to maximize insights from large spectroscopic surveys.

## 3.2 Introduction

Alpha ($\alpha$) elements (such as Mg, Ti, Si, Ca) are primarily produced through the successive fusion of helium nuclei in high-mass ($M_* > 8\ M_\odot$) stars, and are released to the interstellar medium (ISM) when these stars explode as core collapse supernovae (CC-SN). In contrast, iron-peak elements (such as Fe, Mn, Cr, Ni) are produced in both CC-SN and SN Ia supernovae (SN Ia). The time delay between production of yields from CC-SN and SN Ia leads to informative contrasts between different families of elements. For example, the relative abundance of $\alpha$-elements to iron (Fe) has been of longstanding interest as an indicator of the star-formation history of a galaxy, as well as the contribution of yields from CC-SN versus SN Ia at the site of star-formation (Venn et al. 2004; Tinsley 1979; Pagel 1998).

Fundamental chemical properties of the Milky Way have been revealed by investigating $\alpha$ and Fe abundances. In particular, it has become well established that the chemical structure of the Galaxy exhibits a bimodality with respect to the $\alpha$-enhancement of stars at a given [Fe/H]. This bimodality was first observed locally in the solar neighborhood (Fuhrmann 1998; Prochaska et al. 2000; Reddy et al. 2006; Adibekyan et al. 2012b; Bensby et al. 2014), and was also apparent within the first year of the Apache Point Observatory Galactic Evolution

Experiment (APOGEE) survey, where stars within the solar circle ($d < 1$ kpc) were observed to follow two "sequences" in the [$\alpha$/Fe] vs. [Fe/H] plane. As shown in Anders et al. (2014), the "low-$\alpha$ sequence" is focused near solar [$\alpha$/Fe] spanning a range of metallicities from [Fe/H] $\sim$-0.8 to 0.4 dex, while the "high-$\alpha$ sequence" covers a range of enriched $\alpha$ abundances, from $\sim$0.3 dex at [Fe/H] $\sim$-1.0 to $\sim$0.1 dex at [Fe/H] $\sim$ 0. At [Fe/H] $\sim$ 0.1, the high- and low-$\alpha$ sequences appear to merge.

Beyond the solar neighborhood, large surveys such as APOGEE and *Gaia* have enabled the empirical characterization of the bimodal $\alpha$ sequence throughout the Milky Way disk (Bovy et al. 2012b; Nidever et al. 2014; Hayden et al. 2015; Mackereth et al. 2017). For instance, using SDSS/SEGUE spectra Bovy et al. (2012b) examine how the $\alpha$ abundances of stars vary with location from disk midplane ($|z| \sim$ 0.3 - 3 kpc), as well as with Galactocentric radius (R = 5 - 12 kpc). They find that, compared to stars with low-$\alpha$ abundances, the population of stars with the highest [$\alpha$/Fe] enrichment are more vertically extended, yet radially concentrated. Using a larger sample of stars from APOGEE, Hayden et al. (2015) confirm this result and further characterize how the ratio of stars with low versus high-$\alpha$ abundances varies throughout the disk. They find that the inner disk (R $\lesssim$ 9 kpc) is comprised of both low- and high-$\alpha$ sequence stars; however, low-$\alpha$ sequence stars are primarily confined close to the disk midplane ($|z| \lesssim 1$ kpc) and high-$\alpha$ sequence stars are primarily located at $|z| \sim$ 0.5 - 2 kpc. Contrary to the inner disk, at all distances from the midplane the outer disk (R $\gtrsim$ 9 kpc) is markedly devoid of stars with high-$\alpha$ abundances. This observed chemical structure of the Milky Way has been cited as evidence for "inside-out" (Larson 1976) and "upside-down" (Bournaud et al. 2009) formation scenarios of the disk components of our Galaxy where, radially, the central disk was formed before the outer disk and, vertically, the thick disk was formed before the thin disk.

Although the $\alpha$-enriched component of the Milky Way has come to be associated with a "hotter disk" and the solar-$\alpha$ component often associated with a "cooler disk" (Bensby et al. 2003; Navarro et al. 2011; Bovy et al. 2012b), the origin of these two populations, and whether they are unique, is still debated (Haywood et al. 2016; Toyouchi & Chiba 2016; Bovy et al. 2012a). Note that these hotter and cooler populations do not necessarily follow the same morphology as have previously been identified as the "thick" and "thin" disks (Gilmore & Reid 1983). However, there has been growing evidence that, in addition to their chemical differences, the two sequences are also dynamically distinct. For example, using a sample of stars from APOGEE and *Gaia*, Mackereth et al. (2019) find that at fixed age and [Fe/H], the low- and high-$\alpha$ sequences display different age-velocity dispersion relationships with respect to both radial and vertical velocity dispersion. As discussed in Mackereth et al. (2019), these kinematic differences between the low- and high-$\alpha$ sequence are suggestive of disparate heating mechanisms contributing to the formation of the two populations. Similarly, Gandhi & Ness (2019) report differences between the low- and high-$\alpha$ sequence in terms of their orbital actions ($J_\phi$, $J_R$, $J_z$) at all stellar ages.

While differences between the high- and low-$\alpha$ sequence have been characterized empirically, the origin of the bimodality remains an open question. Nonetheless, recent simulation work modeling Milky Way-like galaxies has made substantial progress (Grand et al. 2018; Mackereth et al. 2019; Clarke et al. 2019). For example, analyzing the large-volume EAGLE cosmological simulation, Mackereth et al. (2018) find that a bimodal $\alpha$ sequence occurs in galaxies that experience an early phase of anomalously rapid mass accretion. Consequently, they report that only $\sim$5% of Milky Way-like galaxies in the EAGLE volume exhibit a bimodal $\alpha$ sequence similar to our Galaxy's, implying that this chemical structure is rare. More recently, Clarke et al. (2019) propose that clumpy star-formation is responsible for the observed bimodal $\alpha$ sequence. Using GASOLINE to perform high-resolution simulations, Clarke et al. (2019) reproduce the Milky Way's chemical bimodality in the [O/Fe]-[Fe/H] plane. Investigating the birth sites of stars in the two sequences, they find that stars in the high-$\alpha$ sequence are formed in clumps that start off with low-$\alpha$ abundances, but rapidly self-enrich due to high star formation rates (SFRs). Meanwhile, low-$\alpha$ sequence stars are the product of a more extended star-formation mode that occurs with a substantially lower SFR. Since the incidence of clumps are common in high-redshift galaxies, Clarke et al. (2019) conclude that chemical bimodality should be prevalent among Milky Way-mass galaxies.

Typically, the $\alpha$ abundances investigated in observational and theoretical studies are computed as an average of many individual $\alpha$-elements. However, there is additional information in the relative abundance of different $\alpha$-elements themselves. We know from stellar nucleosynthesis theory that the different $\alpha$-elements vary in the details of their production mechanisms. Given the precision of recent spectroscopic surveys like `APOGEE`, `GALAH`, and `LAMOST`, and the many abundance measurements now available for hundreds of thousands of stars, we can begin to go beyond considering a mean $\alpha$ abundance and examine the information encoded by individual $\alpha$-elements. Full exploitation of the information contained in these multi-element abundance vectors, and how this vector varies with dynamical properties, is still underway. Recently, Weinberg et al. (2019) has mapped multiple `APOGEE` abundances from R = 3 - 15 kpc and $|z|$ = 0 - 2 kpc. This includes the comparison of various elements to Mg, including $\alpha$-elements (like S, Si, O, and Ca), light odd-$Z$ elements (like Al, P, K, and Z), as well as iron-peak elements (like Cr, Mn, Fe, V, Co, and Ni). Through this broad exploration, they find small variations among the different abundance ratios throughout the disk. By considering these subtle differences between inter-family element combinations in depth, more details of the Galaxy's formation history, and the physics of nucleosynthesis, can be gleaned.

For example, a detailed examination of an inter-family abundance ratio has been carried out for the Sagittarius dwarf galaxy, where stars are observed to be deficient in magnesium (Mg) compared to silicon (Si) (McWilliam et al. 2013; Hasselquist et al. 2017; Carlin et al. 2018). The discrepancy between the enrichment of Mg and Si has been interpreted as suggesting the Sagittarius galaxy was formed with a "top-light" stellar initial mass function (IMF), meaning an IMF with fewer high-mass stars compared to the canonical IMF. This is argued to be a consequence of varying yield dependencies on stellar mass between $\alpha$-elements like Mg and O and $\alpha$-elements like Si and Ca. However, this argument is complicated by the fact that some $\alpha$-elements (like Si) are also produced by SN Ia (Tinsley 1979).

The interpretation of Sagittarius [Mg/Si] observations are supported by theoretically motivated expected differences in how various $\alpha$-elements are produced. As discussed in Hasselquist et al. (2017), hydrostatic $\alpha$-elements, such as Mg and O, are produced in massive stars during the hydrostatic burning phase, and these elements get ejected into the ISM during CC-SN explosions. In constrast, explosive $\alpha$-elements, such as Si and Ca, are produced in massive stars during the explosive nucleosynthesis leading up to the CC-SN explosion. While both hydrostatic and explosive $\alpha$-elements are produced in massive stars and released via CC-SN, explosive $\alpha$-elements are produced in shells that lie closer to the cores of massive stars, whereas hydrostatic $\alpha$-elements are produced in the outermost shells. This makes the yields of hydrostatic $\alpha$-elements more dependent on the mass of the star, while explosive $\alpha$-element yields are relatively independent of stellar mass (Woosley & Weaver 1995). In this regard, different $\alpha$-elements can be used to probe the population of massive stars at a given epoch of star-formation.

Inspired by the Sagittarius results, as well as the increasing precision and wealth of multi-abundance measurements of stars located throughout the Milky Way's disk, in this paper we perform a focused investigation of the magnesium to silicon abundance ratio. By considering this specific inter-family element ratio, we are able to attain a more resolved understanding of how $\alpha$-elements vary in the Galaxy. We are also able to isolate particular enrichment events associated only with the production of these inter-family elements. We do this by examining how the ratio of Mg to Si varies with age, as well as dynamics, and find that the low- and high-$\alpha$ sequences exhibit markedly different behavior. To put these results in the context of the Galaxy's chemical evolution, we further attempt chemical evolution modeling to extract both ISM and stellar population parameters at the time of star-formation. While in this study we concentrate only on the ratio of Mg to Si, our methods are generalizable. An in-depth analysis of many inter-family abundance ratios, and even a full matrix of element ratios, appears to be a promising avenue for extracting the most information from large spectroscopic surveys. These detailed characterizations of particular stellar abundance ratios (including [Mg/Si]) are also of interest to studies of planet formation and occurrence (Adibekyan et al. 2015).

This paper is organized as follows. In Section 3.3 we use yield tables to explore theoretical Mg and Si contributions from different nucleosynthetic sources. Then in Section 3.4 we introduce the data and methods of the paper including: the `APOGEE` sub-sample, clustering of the low- and high-$\alpha$ sequences, and empirical motivation for examining the ratio of Mg to Si. In Section 3.5 we present the main empirical results of the paper, showing how [Mg/Si] varies: between the low- and high-$\alpha$ sequences, with age and metallicity, throughout the Galactic disk, and with orbital parameters. In the second half of the paper we focus on interpreting these empirical results within the context of Galactic chemical evolution (GCE). To do this, in Section 3.6 we use *Chempy* to fit chemical evolution models to a sample of `APOGEE` stars and examine variations with the IMF slope, number of SN Ia, and ISM parameters. In this section we discuss the limitations we encounter when attempting to fit these models to a diverse set of stars. Finally, in Section 3.7 we distill the main takeaways of both the empirical results and attempted chemical evolution modeling, and discuss potential paths forward.

## 3.3 Expected Mg and Si yields

Before empirically characterizing the ratio of magnesium to silicon throughout the Milky Way, we further motivate examining this abundance ratio by exploring theoretical Mg and Si yields as expected from yield tables. We do this using the GCE code *Chempy* (Rybizki et al. 2017), which is discussed in more detail in Section 3.6. In short, using *Chempy* we initialize a simple stellar population (SSP) with a Chabrier IMF (Chabrier 2003b) and evolve the SSP from 0 to 13.5 Gyr in 1350 linear-spaced steps, keeping track of enrichment from CC-SN, SN Ia, and asymptotic giant branch (AGB) stars. The yields from CC-SN and AGB stars both depend on the mass of dying stars at each time step, whereas yields from SN Ia, parameterized as a power-law delay time distribution (DTD) (Maoz et al. 2010), are independent of stellar mass. The yields we report throughout this paper are the net yields, which is the newly synthesized material from these nucleosynthetic channels.

We consider two initial SSP metallicities. This includes an SSP with a solar-like metallicity ($Z = 0.01$), as well as a metal-poor SSP with $Z = 0.0001$. We also compute SSPs assuming two sets of yield tables. The first is the *Chempy* default yield tables which include CC-SN yields from Nomoto et al. (2013), SN Ia yields from Seitenzahl et al. (2013), and AGB yields from Karakas (2010). The alternative *Chempy* yield set instead uses CC-SN yields from Chieffi & Limongi (2004), SN Ia yields from Thielemann et al. (2003), and AGB yields from Ventura et al. (2013). In the following figures we only show the Mg and Si yields from the default yield set at $Z = 0.0001$ and $Z = 0.01$. The alternative yield set exhibits similar trends as the default yield set; however, for the alternative yield set there is less of a difference in the yields produced at the two metallicities.

The *upper two panels* of Figure 3.1 show the cumulative Mg and Si yields as a function of time. The total net yields summing the contributions from the three nucleosynthetic channels, as well as the individual yields from CC-SN and SN Ia, are shown. The AGB yields are negligible compared to the SN yields and scale of the plots, and thus are not shown. For both magnesium and silicon, there is an early phase of element production from CC-SN, which ends after ∼40 Myr. After this time, both the Mg and Si yields remain constant until ∼150 Myr, at which point SN Ia ejecta begins to enrich the ISM. As seen in the figure, SN Ia produce a negligible amount of Mg compared to Si, which results in the Mg yield remaining relatively constant until present day, while the Si yield continues to increase. To more closely examine the differences between the two elements, the *bottom panel* of Figure 3.1 shows how the ratio of Mg to Si evolves over time. Here we see that Mg/Si decreases with time, first due to more Si ejecta from CC-SN, then due to the Si ejecta from SN Ia. While there is a normalization offset and small differences comparing the $Z = 0.01$ and $Z = 0.0001$ SSPs, the overall trend is persistent at both metallicities.

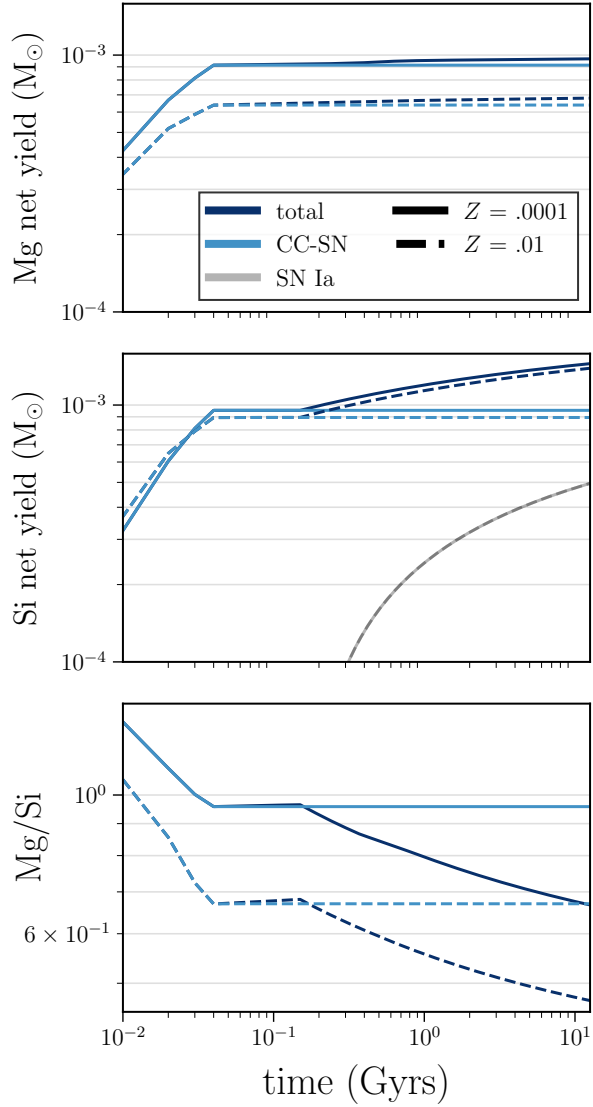

Figure 3.1: ***Upper two panels:*** expected magnesium (Mg) and (Si) net yields for a 1 M$_\odot$ SSP assuming a Chabrier IMF and metallicities of $Z = 0.0001$ (dashed) and $Z = 0.01$ (solid). At each time step, the cumulative net yields from CC-SN (Nomoto et al. 2013) and SN Ia (Seitenzahl et al. 2013) are shown. The AGB yields (not shown) are negligible compared to the SN yields. ***Bottom panel:*** expected ratio of Mg to Si net yields throughout time from the same SSPs. The Si produced in SN Ia reduces the total Mg/Si ratio after 100 Myr.

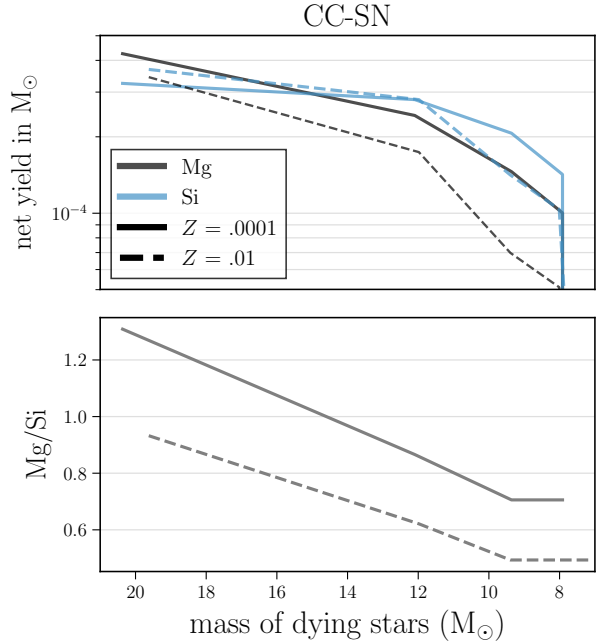

Figure 3.2: ***Top panel:*** expected magnesium and silicon yields from CC-SN assuming the same SSP parameters and yield tables as in Figure 3.1. Shown are the yields generated from dying stars with masses from $M_* = 8 - 20\ M_\odot$. The yields of Si are more constant with stellar mass than those of Mg, especially for the $Z = 0.0001$ SSP. ***Bottom panel:*** expected ratio of Mg to Si net yields as a function stellar mass for both the $Z = 0.1$ (dashed) and $Z = 0.0001$ (solid) SSPs. As stellar mass decreases, more Si is produced compared to Mg.

Focusing on the CC-SN yields, Figure 3.2 shows the Mg and Si yields per mass of dying star. For the metal-poor SSPs, the higher-mass stars ($> 15$ M$_\odot$) produce more Mg than Si, whereas lower-mass stars ($< 15$ M$_\odot$) produce more Si than Mg. This results in an ISM that is Mg/Si rich at early times when the massive stars are exploding, and then becomes Mg/Si deficient once the lower mass CC-SN begin to explode at later times. For the solar-metallicity SSP we see that, while Si and Mg also exhibit different dependencies on stellar mass, the Si yields are greater than the Mg yields at every mass. So while the ratio of Mg to Si yields for the $Z = 0.01$ SSP also decreases with time, the population is always deficient in Mg compared to Si.

From this exploration of the magnesium and silicon yields generated throughout the lifetime of a SSP, we confirm the expected behavior of these yields discussed in Section 3.2 and build intuition for how the abundance ratio of [Mg/Si] might vary in the simplest case. Considering the theoretical expectations, Figure 3.1 shows that the yield tables predict Si, but not Mg, to be produced in SN Ia, which causes Mg/Si to continually decrease once SN Ia enrichment commences. Additionally, Figure 3.2 confirms that the amount of Mg (hydrostatic) and Si (explosive) generated in CC-SN exhibits different dependencies on stellar mass, making Mg/Si sensitive to the high-mass end slope of the IMF.

While these trends in Mg to Si are easily understood for the case of a SSP, in reality the interpretation of [Mg/Si] will not be as straightforward. Stars can form from an ISM that is enriched by several generations of stellar populations, at different starting metallicities, and exhibit incomplete mixing, all which will alter the [Mg/Si] abundance. To begin to disentangle what can be learned from variations in the ratio of Mg to Si, these variations must first be empirically quantified. With this goal, in the next section we describe the data we use to carry out the characterization of [Mg/Si] throughout the Milky Way disk.

## 3.4 Data and Methods

### 3.4.1 Main data sample

To investigate how the Mg to Si abundance ratio varies in the Milky Way, we use data from the publicly available `APOGEE` Data Release 14 (DR14) (Abolfathi et al. 2018). `APOGEE` DR14 is a main SDSS-IV (Blanton et al. 2017) campaign carried out with the `APOGEE` near-IR spectrograph (Wilson et al. 2010) on the 2.5 meter SDSS telescope (Gunn et al. 1998) located at Apache Point Observatory. The `APOGEE` Stellar Parameter and Chemical Abundances Pipeline (ASPCAP) (García Pérez et al. 2016) is used to derive stellar parameters and chemical abundances by $\chi^2$ fitting to 1D local thermal equilibrium models. Produced is a data catalog consisting of stellar atmosphere parameters (e.g. $T_{\text{eff}}$, log $g$), as well as chemical abundances including a global metallicity, [M/H], an $\alpha$ abundance, [$\alpha$/M] (which is a combination of O, Mg, Si, S, Ca, and Ti), and measurements of 19 individual elemental abundances: C/C I, N, O, Na, Mg, Al, Si, P, S, K, Ca, Ti/Ti II, V, Cr, Mn, Fe, Co, Ni, and Rb.

The entire `APOGEE` DR14 ASPCAP catalog contains 277,371 stars. To obtain stellar age estimates, we match this catalog with the Ness et al. (2016) catalog, which provides stellar ages for 73,151 `APOGEE` stars included in DR14. Ness et al. (2016) derive these ages using *The Cannon* (Ness et al. 2015) and infer stellar masses to a precision of ∼0.07 dex based on the `APOGEE` spectra. These mass estimates translate to stellar age estimates with ∼40% errors, and are found to be primarily based on CN absorption features.

To examine trends with structural and orbital properties, we further match the `APOGEE` sample with a *Gaia* product catalog. Sanders & Das (2018) include distances, Galactocentric coordinates, and actions for a majority of stars in our sample. Distances are derived with a Bayesian approach using spectroscopic, photometric, and astrometric properties of each star. From these distances, Sanders & Das (2018) compute the Galactocentric radius (R) of each star, as well as the distance from the disk midplane ($z$). The angular momenta ($J_\phi$), vertical actions ($J_z$), and radial actions ($J_R$) are computed in the McMillan et al. (2018) potential using the Stäckel Fudge method (Sanders & Binney 2016).

Finally, we discard stars with STAR_BAD indicated in the APOGEE ASPCAPFLAG and also remove stars without [Fe/H], [Mg/Fe], [Si/Fe], and [$\alpha$/Fe] measurements. This results in 72,125 stars with which to carry out our study. The median (min, max) of the relevant chemical abundances are -0.13 dex (-1.17 < [Fe/H] < 0.61 dex) in [Fe/H], 0.034 dex (-0.28 < [$\alpha$/Fe] < 0.42 dex) in [$\alpha$/Fe], 0.050 dex (-0.48 < [Mg/Fe] < 0.61 dex) in [Mg/Fe], and 0.027 dex (-0.61 < [Si/Fe] < 0.59 dex) in [Si/Fe]. The median (min, max) of the stellar atmosphere parameters are 2.46 (1.05 < log $g$ < 3.72) in log $g$ and 4780 K (3980 < $T_{\rm eff}$ < 5800 K) in $T_{\rm eff}$. The median Galactocentric radius and distance from the disk midplane of the sample are R = 9.47 kpc and $|z|$ = 0.37 kpc, and the 3$\sigma$ spatial extent spanned is 0.14 < R < 44.6 kpc and 0.00 < $|z|$ < 10.7 kpc. Lastly, the median (3$\sigma$ range) of the actions are 2050 kpc km s$^{-1}$ (-390 < $J_\phi$ < 3570 kpc km s$^{-1}$) in $J_\phi$, 1.51 (-0.81 < log($J_R$) < 2.76) in log($J_R$), and 0.94 (-1.67 < log($J_z$) < 2.50) in log($J_z$).

### 3.4.2 Additional datasets

We assume that the measured [Mg/Fe] and [Si/Fe] abundances of stars reflect their abundances at the time of birth. While stochastic effects, like binary interactions and planetary engulfment, might change abundances over time, we expect these affects to be negligible. Additionally, given the narrow temperature range of our star sample, we also assume that dust does not differentially impact the measured abundances.

To test that our empirical results are robust, we corroborate our findings with two other samples: a sample of red clump (RC) stars (which are constrained in $T_{\text{eff}}$ and $\log g$) and a sample of main-sequence stars with chemical abundance measurements from the High Accuracy Radial velocity Planet Searcher (HARPS). In the following sections we describe and motivate the use of each of these datasets.

### 3.4.2.1   APOGEE red clump stars

RC stars are low-mass, core helium-burning stars. As discussed in Girardi (2016), the constancy of the core mass for these ~1.5 $M_\odot$ stars at the start of the core helium-burning phase is what leads these stars to "clump" to the same luminosity in the color-magnitude diagram. For similar reasons, RC stars also span narrower ranges in stellar parameters such as $T_{\text{eff}}$ and $\log g$ compared to red giant branch (RGB) stars. This property makes RC stars a good check of our results, and if what we find is dependent on stellar atmosphere parameters.

The sample of RC stars we use is from the `APOGEE` Red-Clump (RC) Catalog, which is derived from the main `APOGEE` stellar catalog based on the procedure developed in Bovy et al. (2014). The selection makes use of both photometric and spectroscopic data, and identifies probable RC stars based on their metallicity, color, effective temperature, and surface gravity. This selection results in minimal contamination from RGB stars. The DR14 RC catalog contains 29,502 RC stars with distances accurate to 5-10%, and 18,357 of these stars have age estimates from Ness et al. (2016). The median value (min, max) of the chemical abundances for the RC sample are -0.13 dex (-0.90 < [Fe/H] < 0.51 dex) in [Fe/H], 0.029 dex (-0.14 < [$\alpha$/Fe] < 0.40 dex) in [$\alpha$/Fe], 0.038 dex (-0.31 < [Mg/Fe] < 0.53 dex) in [Mg/Fe], and 0.032 dex (-0.41 < [Si/Fe] < 0.46 dex) in [Si/Fe]. The median (min, max) of the stellar atmosphere parameters are 2.45 (1.80 < $\log g$ < 3.13) in $\log g$ and 4880 K (4190 < $T_{\text{eff}}$ < 5440 K) in $T_{\text{eff}}$. As expected, the RC stars are in a narrower range in $\log g$ and $T_{\text{eff}}$ than the main sample described in Section 3.4.1.

### 3.4.2.2 HARPS solar twins

Solar twin stars are main-sequence G dwarfs that are spectroscopically similar to the Sun. A typical solar twin has an effective temperature within 100 K, surface gravity log $g$ within 0.1 dex, and bulk metallicity or iron abundance [Fe/H] within 0.1 dex of the solar values (e.g. Ramírez et al. 2014; Nissen 2015). As a result, the spectra of solar twins may be compared differentially to the solar spectrum with minimal reliance on stellar atmospheric models, yielding extremely high-precision (0.01 dex level) abundance measurements (Bedell et al. 2014). Similarly precise spectroscopic parameters and therefore isochronal ages are also possible for these stars (Ramírez et al. 2014; Spina et al. 2018).

We use a set of 79 solar twin stars observed at high signal-to-noise and high resolution with the HARPS spectrograph. These stars have highly precise ages derived from isochrone fits to their spectroscopic parameters with an estimated uncertainty of 0.4 Gyr in Spina et al. (2018). Using the same spectra and parameters, abundances for several $\alpha$-elements including Mg and Si were derived in Bedell et al. (2018) through a differential equivalent width technique.

The stars in this dataset are limited in both number and scope: all are located within $\sim$ 100 pc of the Sun and, by definition, all have approximately solar metallicity. Despite these limitations, the precision of age and abundance measurements in this sample makes it a valuable source of information. This sample also serves as a test for the generalizability of our results to main-sequence stars.

### 3.4.3 Soft clustering of low and high [$\alpha$/Fe] stars

To quantify differences in [Mg/Si] between stars with solar $\alpha$ abundances and those with enriched $\alpha$ abundances, we first separate the full APOGEE sample into a 'low-$\alpha$' sequence and a 'high-$\alpha$' sequence. We cluster the subsample of APOGEE stars described in Section 3.4.1 based on two input features, $\vec{X} = \{[\alpha/\text{Fe}], [\text{Fe/H}]\}$, which is the parameter space in which the two sequences are typically identified. Visually, the two proposed clusters in this feature space are distributed anisotropically and not cleanly separable, so we decide to achieve a soft clustering through fitting a mixture model.

Fixing the number of clusters to two, we fit the data with a Gaussian mixture model. We find the results of the clustering to be sensitive to the initialization of the component means, so we compute a 2D kernel density estimate (KDE) of the data and use the highest-density locations in each of the two sequences to initialize the means. After convergence, for each star we record the component assignment ($z$) probability that the star belongs to the high-$\alpha$ sequence $P(z = \text{high-}\alpha)$. As seen in Figure 3.3, stars with high-$\alpha$ abundances are assigned with high probability to the same mixture component, and stars with low-$\alpha$ abundances are assigned with high probability to the other mixture component. As expected, stars with intermediate-$\alpha$ abundances at a fixed [Fe/H] are assigned a lower probability of belonging to either the low or high-$\alpha$ sequence.

For the results throughout this paper, we use the assignment probabilities to weight the stars when making comparisons between the low- and high-$\alpha$ sequence. Since 92% of the stars in our sample are assigned with high probability to either component (i.e. $P(z = \text{high-}\alpha.) = $ 0-5% or 95-100%), only a small fraction of stars with more ambiguous component assignments are given less weight.

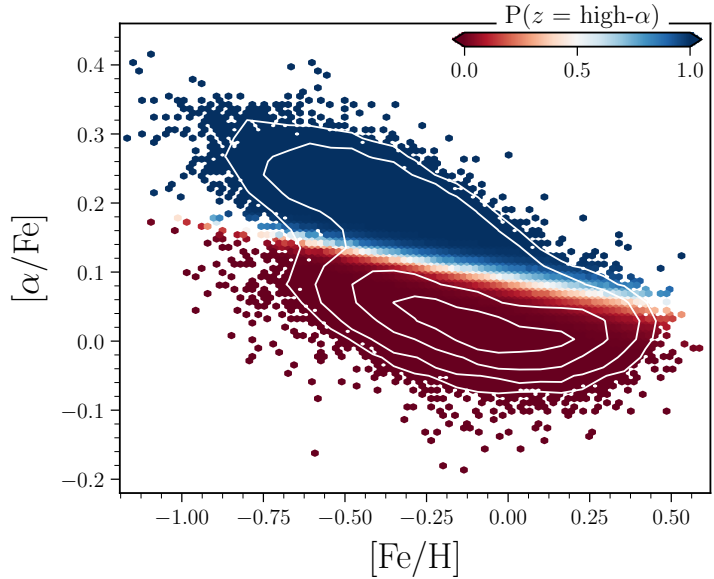

Figure 3.3: Soft clustering of the APOGEE stars in the $[\alpha/\text{Fe}]$-$[\text{Fe/H}]$ plane. Each bin is colored by the mean assignment ($z$) probability of the stars contained within the bin, where blue indicates a high probability of being assigned to the high-$\alpha$ sequence and red indicates a high probability of being assigned to the low-$\alpha$ sequence. The contours indicate the density of stars (with levels indicated at 30, 100, 500, and 1500 stars).

### 3.4.4 Demonstration that [Mg/Si] does not simply trace [$\alpha$/Fe]

In Section 3.2 we described the physical motivation for looking at the ratio of magnesium to silicon. We now demonstrate, in a quantifiable way, that there is additional information encoded in the ratio of Mg to Si beyond its relation to a global $\alpha$ abundance. We do this by building a model to predict the $[\alpha/\text{Fe}]$ abundances of stars using two sets of input features. In the first instance we use $\vec{\mathbf{X}} = ([\text{Fe/H}], [\text{Mg/Fe}], [\text{Si/Fe}])$, and in the second instance we use $\vec{\mathbf{X}} = ([\text{Fe/H}], [\text{Mg/Si}])$. The former set contains the individual Mg and Si abundances, and the latter contains just the ratio of the two.

We split the APOGEE sample described in Section 3.4.1 into a training set (70%) and a hold-out set (30%). The training set is used for model training and hyperparameter tuning. The hold-out set, which is data not seen during training or model selection, is used to evaluate the performance of the final model. To select the best model, with the training set we perform a grid search over model hyperparameters with a 10-fold cross-validation, and ultimately select the model that results in the best average r$^2$ score. The r$^2$ score is computed as the standard coefficient of determination, r$^2$ = 1 - $\frac{1}{N\sigma^2}\sum_i(y_{\text{true},i} - y_{\text{pred},i})^2$, where $y_{\text{true}}$ and $y_{\text{pred}}$ are the true and model-predicted values of the dependent variable, $N$ is the number of observations, and $\sigma^2$ is the variance of $\vec{y_{\text{true}}}$. An r$^2$ score closer to 1 indicates that the model predicts the variation in $y_{\text{true}}$ well, whereas an r$^2$ score of 0 indicates that the model does not capture any of the variation.

First we consider a linear model to predict [$\alpha$/Fe]. Since the dimensionality of the input feature space is low, we first implement unregularized ordinary least squares (OLS). The top row of Figure 3.4 shows the results of the OLS model applied to the hold-out set. Using the individual Mg and Si abundances, we find that even this simple linear model predicts [$\alpha$/Fe] quite well, with an r$^2$ = 0.95. However, the results are considerably worse when using the input feature vector of $\vec{X}$ = ([Fe/H], [Mg/Si]). The r$^2$ is reduced to 0.59 and, as seen in the figure, the model tends to over-predict the $\alpha$ abundances of low-$\alpha$ sequence stars, and under-predict those high-$\alpha$ sequence stars.

One possible reason why the [Mg/Si] abundance does not predict [$\alpha$/Fe] well is that the relationship is not captured by a linear model. To test this, we also train a vanilla feed-forward neural network (also referred to as a multilayer perceptron (MLP)) using both sets of input features. We create a sequential MLP model and perform a small grid search over several hyperparameters including: the hidden layer size (5, 10, 25, 100), the number of hidden layers (1, 2), the activation function (ReLU, tanh), and the regularization $\alpha$ (5 values from $10^{-3}$ to $10^3$). As seen in the bottom row of Figure 3.4, we find that an MLP model only performs marginally better (r$^2$ = 0.63) than the OLS model at predicting the [$\alpha$/Fe] abundances of the hold-out set from $\vec{X}$ = ([Fe/H], [Mg/Si]).

To test if the dimensionality of the input feature vector is driving the difference in perfor-mance between the models trained with the two sets of features described above, we trained models with three additional sets of input features: (1) $\vec{X}$ = ([Fe/H], [Mg/Fe]), (2) $\vec{X}$ = ([Fe/H], [Si/Fe]), and (3) $\vec{X}$ = ([Fe/H], average([Mg/Fe], [Si/Fe])). Training with (1) we find the OLS $r^2$ to be 0.93 and the MLP $r^2$ to be 0.93. Training with (2) we find the OLS $r^2$ to be 0.74 and the MLP $r^2$ to be 0.77. And training with (3) we find the OLS $r^2$ to be 0.94 and the MLP $r^2$ to be 0.94. From this experiment we find that for this specific problem, the difference in the dimensionality of the input feature vector only marginally effects the prediction performance because [Mg/Fe] more strongly traces [α/Fe] than [Si/Fe].

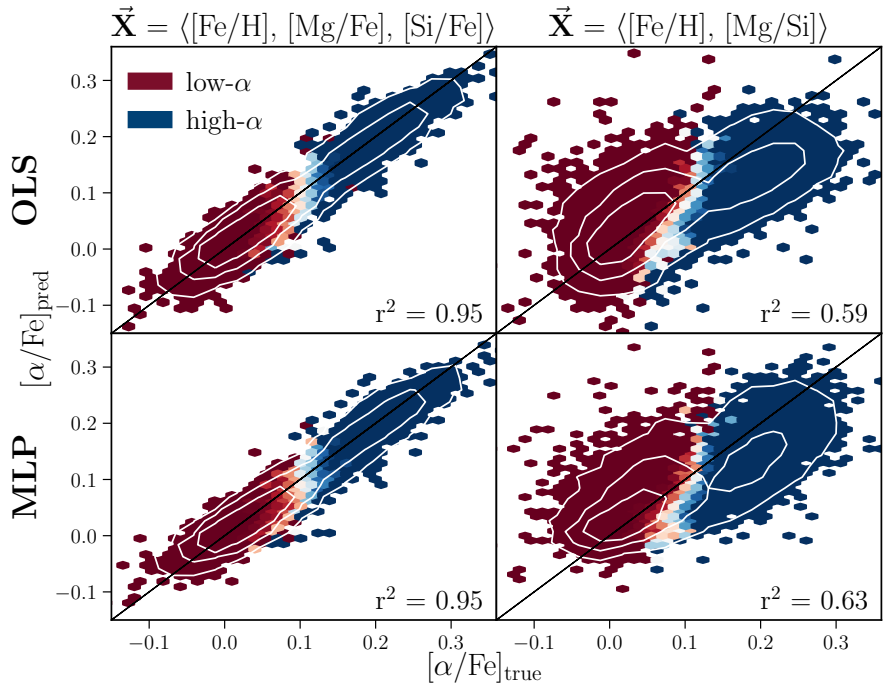

Figure 3.4: Prediction of [α/Fe] using two sets of input features, $\vec{X}$ = ([Fe/H], [Mg/Fe], [Si/Fe]) (left) and $\vec{X}$ = ([Fe/H], [Mg/Si]) (right). In each panel, the predicted vs. true [α/Fe] abundances of stars in the hold-out set are shown, which are data not used during training and model selection. The binning color indicates the dominant α class, and the contours show the density of stars. The input vector $\vec{X}$ = ([Fe/H], [Mg/Si]) fails to predict the α-enhancement of stars assuming both a linear model (OLS, top-row), and a more flexible model (MLP, bottom-row).

In the end, using both a simple linear model and a more flexible nonlinear model, the individual [Mg/Fe] and [Si/Fe] abundances predict the $\alpha$-enhancements of stars very well. This is expected, because the global $\alpha$ abundance is computed from information about the individual $\alpha$-element abundances. However, the predictive ability of the ratio of two $\alpha$-elements, [Mg/Si], is significantly worse. Even with a more flexible nonlinear model, [Mg/Si] does not predict the global $\alpha$-enhancement of stars as well as the individual [Mg/Fe] and [Si/Fe] abundances do. What this suggests is that there is additional information contained in the residuals of the prediction when using [Mg/Si] that is distinct from the two elements' relationship to a bulk $\alpha$ abundance. However, this demonstration does not tell us what the residuals from the [Mg/Si] prediction do trace. In this paper we investigate this information and seek to understand what [Mg/Si] reveals about the chemical evolution of the Milky Way.

## 3.5 Empirical characterization of [Mg/Si]

### 3.5.1 The [Mg/Si] abundance of the low- and high-$\alpha$ sequences

We begin our investigation by first characterizing the ratio of Mg to Si for both the low- and high-$\alpha$ sequences. The *left panel* of Figure 3.5 shows the [$\alpha$/Fe]-[Fe/H] distribution for the sample of $\sim$70,000 `APOGEE` stars described in Section 3.4, where the bins are colored by the mean [Mg/Si] of the stars that fall within each bin. For the entire sample, the inner $90^{\text{th}}$ percentile of the [Mg/Si] abundances spans nearly 0.2 dex from [Mg/Si]= -0.05 - 0.14 dex, indicating that the stars have varying [Mg/Fe] and [Si/Fe] abundances. The median [Mg/Fe] and [Si/Fe] values for the high-$\alpha$ sequence are 0.25 dex and 0.1 dex, respectively. The median [Mg/Fe] and [Si/Fe] values for the low-$\alpha$ sequence are 0.035 dex and 0.015 dex, respectively. These differences in Mg and Si abundances lead to a mean [Mg/Si] $\sim$ 0.096 dex for the high-$\alpha$ sequence and mean [Mg/Si] $\sim$ 0.02 dex for the low-$\alpha$ sequence.

To quantify how the ratio of Mg to Si varies for stars *within* the low- and high-$\alpha$ sequences,

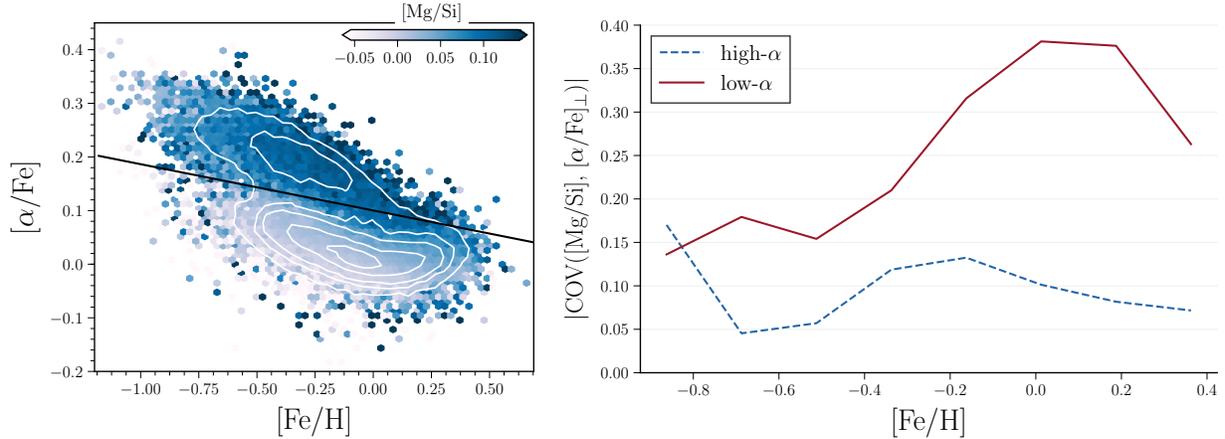

Figure 3.5: **Left:** [α/Fe]-[Fe/H] distribution for the ~70,000 stars in our APOGEE subsample. The bins are colored by the [Mg/Si] abundance, while the contours represent the density of stars. Here, we see that the high-α sequence stars have higher [Mg/Si] abundances than low-α sequence stars. **Right:** absolute value of the covariance between [Mg/Si] and [α/Fe]$_\perp$ as a function of metallicity for the low-α (red) and high-α (blue) sequences. In the low-α sequence, [Mg/Si] and [α/Fe]$_\perp$ jointly vary more strongly than in the high-α sequence.

the *right panel* of Figure 3.5 shows how [Mg/Si] varies with [α/Fe] across the [α/Fe]-[Fe/H] plane. Here, we compute the absolute value of the covariance between [Mg/Si] and [α/Fe]$_\perp$ in eight [Fe/H] bins from -1.0 ≤ [Fe/H] ≤ +0.5 dex, where [α/Fe]$_\perp$ is measured perpendicular to a linear class boundary separating the low- and high-α sequences. In computing the covariance for the high- and low-α sequence separately, we weight the stars according to their cluster assignment probabilities as described in Section 3.4.3. As seen in Figure 3.5, [Mg/Si] and [α/Fe] jointly vary more strongly for low-α sequence stars than for high-α sequence stars across nearly the entire range of metallicities. The absolute value of the covariance for the high-α sequence never surpasses ~0.2, whereas for the low-α sequence it reaches ~0.4 at [Fe/H] ~ 0.2 dex. This shows that the ratio [Mg/Si] behaves differently in the low- and high-α sequences, which suggests that this ratio could be probing differences in the chemical enrichment histories of the two sequences.

Before continuing, we corroborate the finding that the high-α sequence has an excess of Mg relative to Si compared to the low-α sequence, with both the APOGEE RC sample described in Section 3.4.2.1 and the HARPS sample described in Section 3.4.2.2. First, to obtain cluster membership probabilities for both of these datasets, we apply the same Gaussian mixture model trained on the full `APOGEE` sample. For the RC stars, the high-$\alpha$ sequence is found to have a mean [Mg/Si] of 0.07 dex, while the low-$\alpha$ sequence has a mean [Mg/Si] of 0.003 dex. For the HARPS stars, the mean [Mg/Si] of the high-$\alpha$ sequence is 0.07 dex and the mean [Mg/Si] of the low-$\alpha$ sequence is 0.01 dex. These differences are similar to what is found for the full `APOGEE` sample. After verifying that the [Mg/Si] trends are robust to log $g$ and $T_{\text{eff}}$ variations, and are also present in a sample of stars with higher-quality spectra, in the following section we explore how [Mg/Si] evolves with stellar age and metallicity.

### 3.5.2 [Mg/Si] trends with age and metallicity

To further examine differences in the [Mg/Si] abundance between the low- and high-$\alpha$ sequences, we now consider variations with other stellar properties available to us, including ages and [Fe/H]. Investigating correlations with these additional parameters will allow us to build intuition for how the production of Mg relative to Si varies both through time and in different star-formation environments.

As described in Section 3.4.1, we use stellar age estimates from Ness et al. (2016), which are derived from CN absorption lines and are accurate to ∼40%. Considering these uncertainties, we divide our sample of `APOGEE` stars into three wide-age bins, roughly separating young stars born 0 - 3 Gyr ago, intermediate-aged stars born 3 - 7 Gyr ago, and old stars born 7 - 14 Gyr ago. This binning results in 561 young stars, 3,981 intermediate-aged stars, and 8,254 old stars in the high-$\alpha$ sequence (P($z$ = high-$\alpha$) > 0.95), and 19,957 young stars, 23,685 intermediate-aged stars, and 8,747 old stars in the low-$\alpha$ sequence (P($z$ = high-$\alpha$) < 0.05). The median age of stars in the high-$\alpha$ sequence is 8.26 Gyr with a standard de-viation of 3.0 Gyr, while the median age of stars in the low-$\alpha$ sequence is 3.7 Gyr with a standard deviation of 2.8 Gyr. There is only a relatively small fraction of high-$\alpha$ stars that are young, and old low-$\alpha$ stars that are old (e.g. see Silva Aguirre et al. (2018)). These have corresponding abundance and age errors that indicate these measurements are robust. As a consequence of our large sample, we have a sufficient numbers of stars to examine the trends of the high- and low-$\alpha$ sequence to high significance, across all ages.

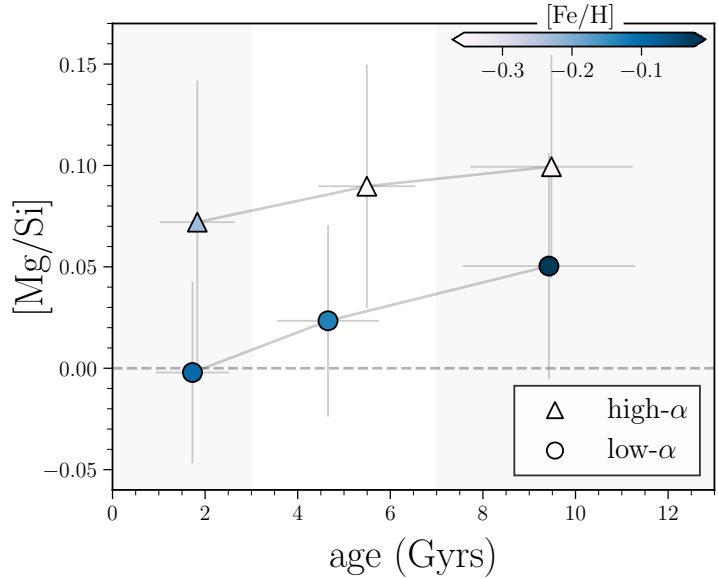

Figure 3.6: The [Mg/Si] abundance of the low-$\alpha$ (circles) and high-$\alpha$ (triangles) sequences as a function of stellar age. The gray error bars indicate the standard deviations of [Mg/Si] and age in each stellar age bin, while the errors on the mean are smaller than the size of the symbols.

Figure 3.6 shows the mean [Mg/Si] abundance in each age bin, for both the low- and high-$\alpha$ sequences. As expected from Section 3.5.1, in each bin the high-$\alpha$ sequence has a higher mean [Mg/Si] than the low-$\alpha$ sequence. It should be noted that while the differences in the mean [Mg/Si] values are significant, specifically the standard errors on the means are smaller than the size of the symbols in the figure, the $1\sigma$ dispersions around the mean of the distributions (indicated in gray) do overlap. So, while the distributions peak at different values, they are not completely disparate. The takeaway of Figure 3.6 is the trends we observe with age and the mean [Mg/Si]. Without yet taking metallicity into consideration, we notice two things. The first is that stars born earlier in time (older ages) are more enhanced in Mg relative to Si than stars born at later times (younger ages). This is consistent with the relative theoretical Mg/Si yields discussed in Section 3.3. The second observation is the mean [Mg/Si] abundance varies more with age for the low-$\alpha$ sequence stars than for the high-$\alpha$ sequence stars. The low-$\alpha$ sequence spans $\sim$0.049 dex in [Mg/Si] from the young age bin to the old age bin, while the high-$\alpha$ sequence only spans $\sim$0.026 dex. We also verify that this trend with age is present in the RC sample. Considering the same age bins, the [Mg/Si] of low-$\alpha$ sequence stars decreases by 0.038 dex from the oldest to youngest stars. For the high-$\alpha$ sequence stars, this decrease in [Mg/Si] with age is only 0.023 dex. We cannot confirm this [Mg/Si] age trend with the HARPS data because our sample only includes eight $\alpha$-enriched stars, all with ages between 8.1 and 9 Gyr.

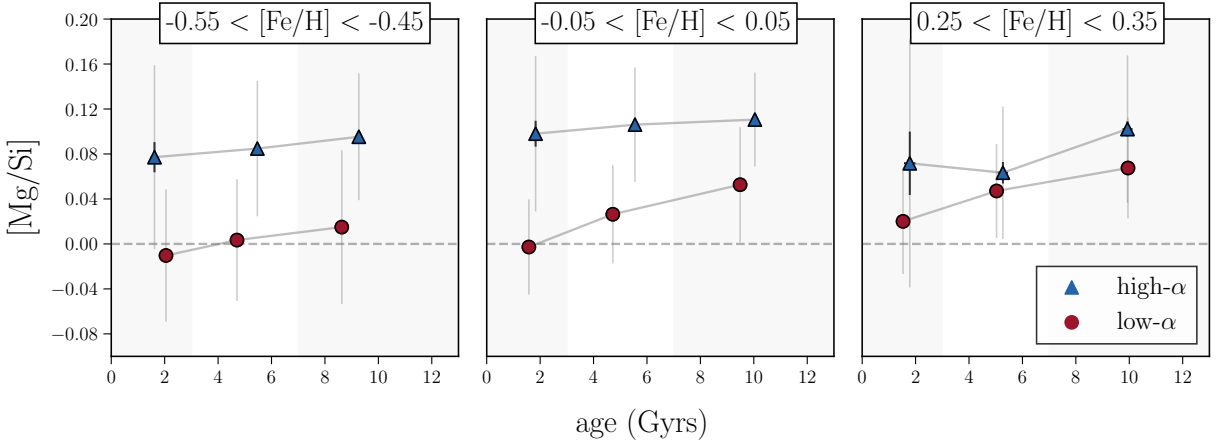

Figure 3.7: [Mg/Si] abundance as a function of stellar age in three narrow metallicity bins of 0.1 dex. The left panel shows a metal-poor bin, the middle panel shows a solar-metallicity bin, and the right panel shows a metal-rich bin. In each panel, the high-$\alpha$ sequence is shown in blue, and the low-$\alpha$ is shown in red. The light gray error bars represent the 1$\sigma$ standard deviation of [Mg/Si] in each age bin, while the black error bars represent the error on the mean.

Figure 3.6 also reveals that metallicity varies with age and [Mg/Si]. For high-$\alpha$ sequence stars, younger stars are more metal-rich than older stars, with the average [Fe/H] decreasing by ∼0.12 dex with increasing stellar age. However, for low-$\alpha$ sequence stars, younger stars are instead more metal-poor than older stars, with the average [Fe/H] increasing by ∼0.07 dex with increasing stellar age. The opposite [Fe/H] age trend between the low- and high-$\alpha$ sequences is a further line of evidence that the chemical enrichment histories of the two sequences is different. To understand the variation of [Mg/Si] with age independently of metallicity, we condition on [Fe/H] by examining how the [Mg/Si] age trend varies is narrow metallicity bins. Figure 3.7 shows three 0.1 dex wide [Fe/H] bins: a metal-poor bin (-0.55 < [Fe/H] < -0.45 dex), a solar metallicity bin (-0.05 < [Fe/H] < 0.05 dex), and a metal-rich bin (0.25 < [Fe/H] < 0.35 dex).

In each metallicity bin, we see that the [Mg/Si] abundance of the low-$\alpha$ sequence increases with stellar age. This suggests that the general trend of the relationship is independent of variations in [Fe/H]. However, the normalization of the trend does vary with metallicity. Stars in the metal-rich bin have overall higher [Mg/Si] abundances than stars in the metal-poor bin. In contrast to the low-$\alpha$ sequence, the [Mg/Si] abundance of the high-$\alpha$ sequence remains relatively constant with age in the [Fe/H] bins. This is the case in the metal-poor and solar-metallicity bin, where similar [Mg/Si] values are found at each age. However, in the metal-rich bin there is some variation with the oldest high-$\alpha$ sequence stars having larger ratios of Mg to Si than the intermediate-age and young high-$\alpha$ sequence stars. Part of this trend can be attributed to the lack of separation between the two sequences at higher metallicities. As seen in Figure 3.3, at metallicies higher than [Fe/H] ∼ 0.0 dex, the high-$\alpha$ sequence appears to merge into the low-$\alpha$ sequence. This makes the definition of the two sequences more ambiguous at these high metallicities, which is reflected in assignment probabilities discussed in Section 3.4.3.

Based on Figure 3.7, we can begin to hypothesize about the origin of these [Mg/Si] abundance trends by considering the nucleosynthetic channels of the two elements. As discussed in Section 3.2, CC-SN produce both Mg and Si. However, Si is also produced in SN Ia. Given this, one explanation for the decrease in [Mg/Si] with age for the low-$\alpha$ sequence is that this is an imprint of the time-dependent yield contributions from SN Ia, which we expect from Figure 3.1. Since CC-SN enrichment occurs instantaneously compared to SN Ia enrichment, the Mg abundance is relatively constant in time. Therefore, for the low-$\alpha$ sequence a possible reason for the decrease in [Mg/Si] with age is the steady increase in Si over time. What this could imply about the high-$\alpha$ sequence, where [Mg/Si] is constant with stellar age, is that the environment in which these stars form is unpolluted by SN Ia ejecta at all times. While we expect the oldest (and majority) of the high-$\alpha$ sequence stars to be unaffected by SN Ia, what Figure 3.7 suggests is that even the younger high-$\alpha$ sequence stars are unpolluted by Si from SN Ia. This could indicate that at all star-formation epochs, the formation of low- and high-$\alpha$ sequence stars occurs distinctly. Nonetheless, there could be numerous alternative explanations for the trends we find in Figure 3.7. One is that the Mg enrichment of the gas from which high-$\alpha$ sequence stars are formed could steadily increase over time to match the increasing Si enrichment. This would result in a flat [Mg/Si] age trend without requiring isolation from SN Ia pollution. A possible mechanism for this increase in Mg enrichment is if high-$\alpha$ sequence stars formed according to an IMF with a high-mass end slope that became flatter over time. This would produce relatively more high-mass stars which, as discussed in Section 3.2, would yield more Mg than Si. We attempt to explore these possibilities in Section 3.6 through GCE modeling.

### 3.5.3 Spatial and orbital trends with [Mg/Si]

#### 3.5.3.1 Trends with Galactic location

We now turn toward empirically characterizing the relationship between [Mg/Si] and disk structure and dynamics. We do this by establishing how the Mg to Si ratio varies with location throughout the disk, and by investigating how the stellar actions are related to [Mg/Si]. By making this connection between the chemistry and the structural and orbital properties of stars, we hope to understand how [Mg/Si] might encode unique information regarding the formation and build-up of the Milky Way disk.

To start, we consider how [Mg/Si] varies with Galactocentric radius (R) and distance from the disk midplane ($|z|$). As discussed in Section 3.4.1, we match our sample of `APOGEE` stars with the Sanders & Das (2018) catalog to obtain the coordinates of each star. Inspired by Figure 4 of Hayden et al. (2015), we divide the sample into three $|z|$ bins: 0 - 0.5 kpc, 0.5 - 1.0 kpc, and 1.0 - 2.0 kpc. Each of these three bins in $|z|$ is further divided into six radius bins, ranging from 3 to 15 kpc in 2 kpc wide annuli. To quantify the impact of each star's distance measurement uncertainty on the ($|z|$, R) binning, we perform a Monte Carlo simulation. In short, during each Monte Carlo iteration every star's coordinates are re-sampled from:

$$z_i \sim \mathcal{N}(|z|, z_{\text{err}}) \tag{3.1}$$

$$R_i \sim \mathcal{N}(\text{R}, \text{R}_{\text{err}}), \tag{3.2}$$

where the coordinates are described as normal distributions centered at the Sanders & Das (2018) derived values, with a variance equal to the reported errors. After sampling $z_i$ and $R_i$ for each star, the stars are re-binned by ($|z|$, R) and the mean [Mg/Si] and [Fe/H] values of each bin are recorded. Repeating this procedure N = $10^3$ times, we obtain the median and range of the mean [Mg/Si] and [Fe/H] values in each bin.

The *top panel* of Figure 3.8 shows the low- and high-$\alpha$ sequence [Mg/Si]-[Fe/H] distributions in the different ($|z|$, R) bins, as well as the results of the Monte Carlo sampling simulation described above. In the Figure, the displayed contours are based on the mean posterior distances to each star, while the error bars represent the results of the Monte Carlo sampling. First we examine the high-$\alpha$ sequence. At all distances from the midplane, the [Mg/Si]-[Fe/H] distribution evolves similarly from the inner disk to the outer disk, where the peak of the distribution appears to increase marginally in [Mg/Si] abundance from the inner disk (3 kpc) to the mid-disk (7 - 9 kpc), and then decreases toward the outer disk (15 kpc). However, in the outermost region of the disk considered (13 - 15 kpc) the distance errors combined with the limited sample size significantly impact the binning.

In contrast to the high-$\alpha$ sequence, the low-$\alpha$ sequence displays more significant spatial [Mg/Si]-[Fe/H] trends. As seen in bottom row of the *top panel* of Figure 3.8, close to the disk midplane ($|z|$ = 0 - 0.5 kpc) the peak of the [Mg/Si] distribution decreases from $\sim$0.02 dex in the inner part of the Galaxy to $\sim$-0.01 dex in the outer regions. While this gradient in [Mg/Si] with radius is weak, it is significant compared to the stellar distance uncertainties and sample sizes of each bin. A similar trend with radius is seen farther from the midplane. This figure shows that the low-$\alpha$ sequence stars currently residing in the inner region were formed from gas that was more enriched in Mg relative to Si compared to the stars currently residing in the outer disk.

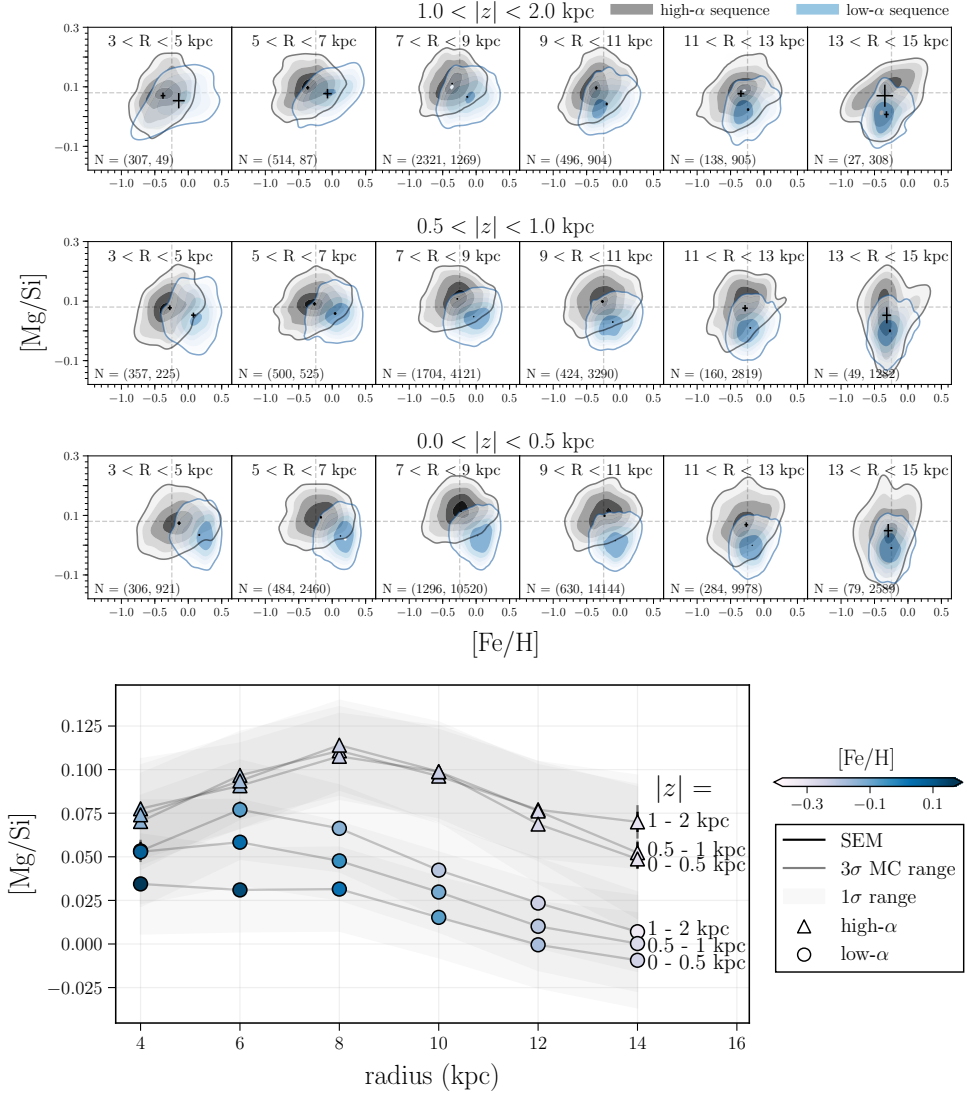

Figure 3.8: **Top:** distribution of high-$\alpha$ (gray) and low-$\alpha$ (blue) sequence stars in the [Mg/Si]-[Fe/H] plane as a function of Galactocentric radius (R) and distance from the disk midplane ($|z|$). Each row shows the distribution in 2 kpc wide radius bins, from the inner disk (left) to the outer disk (right). The bottom row shows the stars closest to the disk midplane from $|z| = 0 - 0.5$ kpc, the middle row shows stars from $|z| = 0.5 - 1$ kpc, and the top row shows stars farthest from the midplane at $|z| = 1 - 2$ kpc. In each panel the error bars represent the range of the mean [Mg/Si]-[Fe/H] values based on $10^3$ Monte Carlo samplings of R and $|z|$. The number of stars in each bin is indicated as N = (# high-$\alpha$ stars, # low-$\alpha$ stars). **Bottom:** this figure summarizes the trends shown in the panel above, showing the mean [Mg/Si] as a function of radius for each R,$|z|$ bin. The black error bars represent the standard error on the mean, the gray error bars represent the $3\sigma$ range from the Monte Carlo simulation, and the gray bands represent the $1\sigma$ range of the [Mg/Si] distribution. Here, trends with [Mg/Si], [Fe/H], and position within the Galaxy are more apparent. For high-$\alpha$ sequence stars (triangles) the peak [Mg/Si] occurs at an intermediate disk radius of $\sim 8$ kpc, regardless of height from the disk midplane. For low-$\alpha$ sequence stars (circles), the peak [Mg/Si] occurs more toward the inner disk at $< 6$ kpc, and in each radius bin stars at larger distances from the disk midplane are more enhanced in [Mg/Si] than stars residing closer to the disk midplane.

The *bottom panel* of Figure 3.8 summarizes the mean trends seen in the *top panel* of Figure 3.8, for both the low- and high-$\alpha$ sequences. The [Mg/Si]-radius relationship is shown for each bin in $|z|$ and the marker color indicates the mean [Fe/H]. The shaded gray regions indicate the 1$\sigma$ distribution, the black error bars indicate the standard error on the mean, and the gray error bars indicate the 3$\sigma$ range from the Monte Carlo simulation described above. While the information shown in this figure is the same as that in the *top panel* of Figure 3.8, this representation of the data better lends itself to visualizing the relevant trends. First consider the high-$\alpha$ sequence. For these stars we see that the mean [Fe/H] decreases from the inner to the outer disk, and that the trend with mean [Mg/Si] and radius is similar regardless of distance from the disk midplane. Here, we see more clearly that high-$\alpha$ sequence stars in the intermediate disk (at $\sim$8 kpc) have the highest [Mg/Si] abundances on average ($\sim$ 0.11 dex) compared to the average [Mg/Si] abundances of stars in the inner disk ($\sim$0.07 dex) and outer disk ($\sim$0.06 dex). This trend of an average [Mg/Si] which first increases from $\sim$3 kpc to $\sim$8 kpc, and then decreases from $\sim$8 kpc to $\sim$15 kpc, is robust to the standard errors on the mean (which span lengths smaller than the height of the markers in the figure).

The low-$\alpha$ sequence stars exhibit markedly different behavior with mean [Mg/Si] and ($|z|$, R). As seen in the *bottom panel* of Figure 3.8, while the shape of the [Mg/Si]-radius curve is similar for all bins in $|z|$, at each radius stars that reside farther from the disk midplane have higher [Mg/Si] abundances compared to those that reside closer to the disk midplane. For instance, in the R = 5 - 7 kpc bin, low-$\alpha$ stars residing at $|z|$ = 0 - 0.05 kpc have an average [Mg/Si] abundance of 0.03 dex, while the stars residing at $|z|$ = 1 - 2 kpc have an average [Mg/Si] abundance of 0.076 dex. Compared to the high-$\alpha$ sequence stars, the shape of the mean [Mg/Si]-radius trends are also different. For $|z|$ > 0.5 kpc, the stars with the highest [Mg/Si] abundances on average are located at R $\sim$ 6 kpc, and then the mean [Mg/Si] decreases with radius until R $\sim$ 15 kpc. For the stars closest to the disk midplane ($|z| <$ 0.5 kpc), the mean [Mg/Si] is nearly constant from R $\sim$ 3 - 9 kpc, and then decreases until R $\sim$ 15 kpc. To test the robustness of these observed trends to variations in stellar properties, namely in $T_{\text{eff}}$, we examined the [Mg/Si]-radius relationship considering stars with temperatures constrained to 200 K wide bins across 4300 K to 5100 K. We find that the general trends presented in the bottom panel of Figure 3.8 hold across these narrower temperature ranges.

The main takeaways from this exploration of how [Mg/Si] varies throughout the disk are summarized as follows. First, from the *top panel* of Figure 3.8 we learn that, similar to how the distribution of [$\alpha$/Fe]-[Fe/H] varies with location in the disk (as in Figure 4 of Hayden et al. (2015)), the distribution of [Mg/Si]-[Fe/H] also varies with Galactic coordinates. We examine the high- and low-$\alpha$ sequence stars separately, and observe that, while their [Mg/Si]-[Fe/H] distributions significantly overlap, there are small differences between how the distributions of the two sequences change with R and $|z|$. From the *bottom panel* of Figure 3.8 we learn that the large number of stars in each bin affirms that the differences between the two $\alpha$ sequences in their mean [Mg/Si]-spatial trends are significant.

The trends seen with mean [Mg/Si] abundance and $|z|$ for the low-$\alpha$ sequence stars could suggest that the low-$\alpha$ sequence part of the disk was built both up and out from layers of gas that had varying overall [Mg/Si] normalizations that scale with $|z|$, but similar gradients with respect to radius. On the other hand, the high-$\alpha$ sequence trends suggest that the gas that formed these stars was similarly enriched in Mg and Si at all distances from the disk midplane. These observations are consistent with how the ages of stars vary throughout the disk for both the low- and high-$\alpha$ sequences. Considering the same spatial bins defined in Figure 3.8, at every $|z|$, low-$\alpha$ sequence stars that reside closer to the Galactic center are older than their counterparts at larger radii. And at fixed R, low-$\alpha$ sequence stars that reside closer to the disk midplane are younger than the stars that at greater heights from the midplane. However, for the high-$\alpha$ sequence stars, we find little to no trend with stellar age and Galactic coordinates.

Aside from their different trends with $|z|$, the mean [Mg/Si] of the high- and low-$\alpha$ sequences also peak at different radii, with the low-$\alpha$ sequence stars having higher [Mg/Si] abundances more toward the inner disk and the high-$\alpha$ sequence stars having higher [Mg/Si] abundances in the intermediate disk. These differences must be reflective of the time sequence of star-formation, and of radial migration effects. Low-$\alpha$ sequence stars largely form in the disk midplane where there are gradients in [Mg/Si] and [Fe/H]. As time goes on, these stars form at progressively larger radii. On the other hand, high-$\alpha$ sequence stars form more toward the central regions of the Galaxy with higher and clumpier efficiencies. Since high-$\alpha$ sequence stars are generally older, they presumably have had more time to migrate both vertically and radially. Distinct radial migration patterns, for example caused by different dynamical effects of the Galactic bar on the low- versus high-$\alpha$ sequence populations, could also potentially result in the trends we observe. These broad speculations should be revisited in the context of a more rigorous modeling approach.

#### 3.5.3.2 Trends with orbital properties

Aside from considering location within the Galaxy, we also characterize the relationship between [Mg/Si] and stellar orbital properties. We do this by examining how [Mg/Si] varies with the three actions, $J_\phi$, $J_R$, and $J_z$. As mentioned in Section 3.4.1, we obtain actions for our `APOGEE` sample from Sanders & Das (2018), who adopt a McWilliam et al. (2013) Milky Way potential to compute $J_\phi$, $J_R$, and $J_z$ from *Gaia* astrometry using the Stäckel Fudge method (Sanders & Binney 2016). Assuming an axisymmetric and relatively time-independent potential, the three actions uniquely define a stellar orbit. Concise physical explanations of the three actions are given in Trick et al. (2019). The azimuthal action, $J_\phi$, quantifies a star's rotation about the center of the Galaxy. This quantity is the same as angular momentum in the vertical direction, and throughout this paper we refer to azimuthal action of a star as its angular momentum. The radial action, $J_R$, describes the amount of oscillation a star exhibits in the radial direction, which is related to the eccentricity of the orbit. Stars with $J_R = 0$ are on circular orbits. Finally, the vertical action, $J_z$, measures the excursion of a star in the vertical direction, where $J_z = 0$ indicates that a star is confined to the disk midplane.

To begin, the top row of Figure 3.9 shows how the actions of high- and low-$\alpha$ sequence stars differ across stellar ages. Similar to what Gandhi & Ness (2019) report for stars with LAMOST abundances, we find that the low- and high-$\alpha$ sequences are distinct in their actions at all ages. First considering angular momentum, we see that low-$\alpha$ sequence stars have higher $J_\phi$ values than high-$\alpha$ sequence stars at each age. This separation between the two sequences is robust to the standard errors on the mean $J_\phi$ in each age bin, and at most ages the distributions are separated by at least one standard deviation. The separation in $J_\phi$ between the two sequences is consistent with the notion that the low-$\alpha$ sequence comprises a more radially extended disk, while the high-$\alpha$ sequence is confined closer to the Galactic center. Since $J_\phi$ traces radial location, these trends are also reflective of a gas disk where star-formation proceeded outward over time with a radial gradient in chemical abundances. This is inside-out formation of the disk, where stars with lower $J_\phi$ values (residing in the inner disk) formed at early times, and since then the radial size of the disk was built up from progressively younger stars.

We also find the high- and low-$\alpha$ sequences to be distinct in radial action, $J_R$, and vertical action, $J_z$. As seen in the middle panel of the top row of Figure 3.9, in every age bin the high-$\alpha$ sequence stars exhibit larger $J_R$ values than the low-$\alpha$ sequence stars. This trend is robust to the standard errors on the mean; however, the distributions do overlap within $1\sigma$. What these trends with $J_R$ suggest is that, at all ages, high-$\alpha$ sequence stars are on more eccentric orbits than low-$\alpha$ sequence stars, and that for both sequences older stars are described by more eccentric orbits than younger stars. One explanation for the relationships between radial action and age is that older stars have had more time for their orbits to be perturbed to more eccentric orbits, and this perturbation occurs regardless of a star's $\alpha$-enhancement. The observed trends with $J_z$ are similar to those with $J_R$. As seen in the figure, at every age the high-$\alpha$ sequence stars are described by larger vertical actions than the low-$\alpha$ sequence stars. This trend also evolves with age for both sequences, where older stars have higher $J_z$ values than younger stars. These trends are robust to the standard errors on the mean, and the distributions between the two sequences are distinct by at least $1\sigma$. The differences in the $J_z$ values of the low- and high-$\alpha$ sequence stars is reflective of their general location within the Galaxy, where low-$\alpha$ sequence stars are more confined to the disk midplane and high-$\alpha$ sequence stars mostly comprise the "thick", vertically extended disk of the Milky Way.

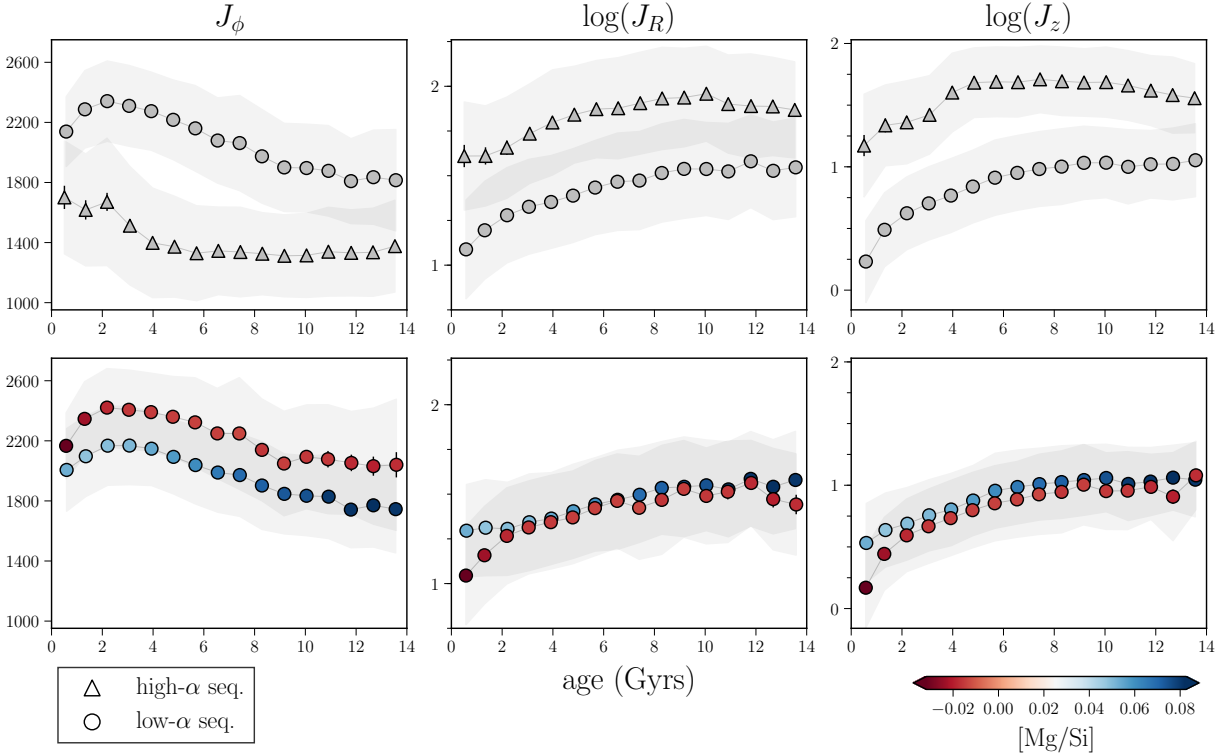

Figure 3.9: **Top row**: evolution of orbital actions ($J_\phi$, $J_R$, $J_z$ [kpc km s$^{-1}$]) with stellar age for low- and high-$\alpha$ sequence stars. For 16 age bins, the average action value is shown and in gray the $1\sigma$ standard deviation is indicated. At every age, the actions of high- and low-$\alpha$ sequence stars are distinct. **Bottom row**: low-$\alpha$ sequence action-age relations now divided into stars with lower vs. higher [Mg/Si] abundances. For $J_\phi$, part of the scatter in the relation with age is described by a gradient in [Mg/Si], where stars with lower [Mg/Si] values have higher angular momentum than those with higher [Mg/Si] values.

We speculate why high-$\alpha$ sequence exhibit higher $J_z$ and $J_R$ values than low-$\alpha$ sequence stars at every age. One possibility is that because the low- and high-$\alpha$ sequences form in different spatial locations (and presumably from different reservoirs of gas), they subsequently experience distinct modes of star-formation that result in different initial orbital properties. An additional possibility is that heating processes may be more active in the thicker, inner disk where high-$\alpha$ sequence stars reside, which perturbs these stars to higher radial and vertical excursions. Lastly, dynamical times are shorter in the inner disk. So the dynamical ages of high-$\alpha$ stars are generally older than their stellar ages, which means they would have more time to be perturbed to higher $J_z$ and $J_R$ values.

Considering the evolution with age, as with the radial action, older stars have had more time to be perturbed to larger vertical excursions. However, while the $J_z$ values of low-$\alpha$ sequence stars increase steadily with stellar age, those of high-$\alpha$ sequence stars remain nearly constant from 4 - 14 Gyr. This suggests that $\alpha$-enhancement could be a signature of initial orbital properties, as well as the dynamical processes that perturb orbits over time. The behavior of the low-$\alpha$ sequence can be broadly understood in the context of recent work by Ting & Rix (2019), who explore the observed age dependence of $J_z$ for low-$\alpha$ `APOGEE` RC stars with ages < 8 Gyr. Ting & Rix (2019) find that a simple analytic model of vertical heating can describe the trends in the data. They posit that heating is dominated by orbit scattering, presumably from giant molecular clouds (GMCs), with $J_z \propto t^{1/2}$. They also take into account how the exponentially declining SFR of the Galaxy results in the decline of GMC occurrence over time, and that the disk mass density decreases with time as well. However, they do not consider high-$\alpha$ sequence stars. We find in Figure 3.9 that, unlike the age-$J_z$ relationship for the low-$\alpha$ sequence, the high-$\alpha$ age-$J_z$ relationship saturates at some stellar age. This saturation matches our physical expectation that, at a certain point, already dynamically hot high-$\alpha$ sequence stars spend such a small fraction of their time near the disk midplane that GMC scattering cannot further heat these stars.

Given the relationships revealed in Section 3.5.2 between [Mg/Si] and stellar age, the bottom row of Figure 3.9 shows the [Mg/Si] abundance of low-$\alpha$ sequence stars separated into two bins. Divided based on the mean [Mg/Si] value of all the low-$\alpha$ sequence stars, one bin contains the stars with lower [Mg/Si] abundances and the other contains those with higher [Mg/Si] abundances. As seen in the figure, for the radial and vertical actions, at every age there is little to no dynamical separation between stars with low versus high [Mg/Si] enrichment. However for angular momentum, a gradient with [Mg/Si] partly constitutes the scatter in the $J_\phi$-age relationship. At every age, stars with lower [Mg/Si] abundances have higher angular momenta than those with higher [Mg/Si] abundances. Due to the large number of stars in the sample, this trend in the mean [Mg/Si] is robust to the standard errors. For the high-$\alpha$ sequence, we find no separation in any action-age relationship with [Mg/Si] abundance. The trends with $J_\phi$-[Mg/Si] and age for the low-$\alpha$ sequence reflect the chemical segregation of the star-forming thin disk, at a given time. The absence of a trend with $J_R$ and $J_z$ suggests that heating and/or migration effects act similarly on mono-age stellar populations at the radii where they reside. For the high-$\alpha$ sequence, the additional absence of a trend with $J_\phi$ is reflective of the lack of chemical segregation in the thicker disk in which high-$\alpha$ stars were generally born. These observations support the notion that the

low- and high-α sequences experience different modes of enrichment.

As we have done throughout Section 3.5, large samples of stars with high-quality spectra and astrometric measurements allow us to empirically characterize the relationship between chemical abundances, stellar ages, and the dynamical properties of stars. However, the physical origins of these connections between chemistry, age, and dynamics remain an open question in Milky Way science. Nonetheless, empirical investigations can place constraints on theoretical expectations, trends observed in simulations of Milky Way galaxies, and on GCE models. For the remainder of this paper, we now shift our focus and attempt to understand the [Mg/Si] abundance trends we report by connecting the observed abundance patterns to underlying stellar physics.

## 3.6 Chemical evolution modeling

### 3.6.1 Motivation

In Section 3.3 we established a theoretical motivation for examining the ratio of magnesium to silicon. In Section 3.4.4 we used a large sample of stars with abundance measurements from `APOGEE` to show that the [Mg/Si] abundance contains different information than the individual [Mg/Fe] and [Si/Fe] abundances. Then, in Section 3.5 we characterized how the [Mg/Si] abundance varies between the low- and high-α sequences, with age and [Fe/H], with location in the Galaxy, and with the orbital actions. After establishing these empirical [Mg/Si] trends, we now seek to understand these trends by investigating the physical origin of the observed magnesium and silicon abundances.

One way observed abundances can be linked to a physical origin is through GCE modeling. Broadly, the goal of GCE modeling is to employ physically motivated models of star-formation and stellar evolution together with a parameterization of galactic ISM physics to predict observed abundance patterns of stars throughout cosmic time. The simplest form of GCEs involves a single "zone" that instantiates a site of star-formation surrounded by a reservoir of gas. Through inflow and outflow processes, primordial gas, and gas enriched from the products of star-formation, is exchanged between the gas reservoir and the star-formation site.

As briefly described in Section 3.3, *Chempy* is a recently developed code that allows for Bayesian inference of GCE model parameters. The publicly available *Chempy* model presented in Rybizki et al. (2017) is a one-zone GCE with seven free parameters. This includes three parameters related to stellar physics, and four to the describe the ISM. The stellar physics parameters include the high-mass slope of the IMF, a normalization constant for the number of exploding SN Ia, and the SN Ia enrichment time delay. The ISM parameters include a parameter which sets the star-formation efficiency, the peak of the SFR, the fraction of stellar yields which outflow to the surrounding gas reservoir, and the initial mass of the gas reservoir. With these parameters specifying the GCE, *Chempy* computes the enrichment of the ISM through time based on a set of yield tables. These yield tables describe stellar feedback from three nucleosynthetic channels: CC-SN, SN Ia, and AGB stars. The CC-SN and AGB yields are mass and metallicity dependent, and in the model these yields are ejected immediately following stellar death. SN Ia feedback is independent of mass and metallicity, with the yields being deposited into the ISM according to the Maoz et al. (2010) DTD. As we will describe in Section 3.6.2, with a list of observed stellar abundances, a stellar age estimate, and choice of yield tables, *Chempy* can be used to obtain posterior distributions for the GCE model parameters.

We present the results of our *Chempy* modeling in the following sections. Our approach is to fit each star's abundance pattern with its own one-zone GCE model, which is in contrast with approaches that fit a single model to either the abundance pattern of multiple stars, or a mean abundance pattern of many stars. However, we caution that ultimately we encounter problems in reproducing the `APOGEE` abundances, which limits the scope and interpretation of the fits. The challenges we come across are discussed in detail in Section 3.6.4, but in short we find that the inferred *Chempy* model parameters result in predicted abundances that are unable to match the observed abundance patterns, particularly the [Si/Fe] abundance. Presumably, this failure is due to the inability of current yield tables to describe the abundance patterns of stars that exhibit a diversity of enrichment histories.

### 3.6.2 Fitting `APOGEE` stellar abundances

Given the limitations we encounter when using *Chempy* to model observed `APOGEE` abundance patterns, we decide to narrow the scope of this study by focusing on just a handful of mono-age, mono-abundance stars. First, we select stars from two narrow [Fe/H] bins: a solar metallicity bin from -0.02 < [Fe/H] < 0.02 dex, and a metal-poor bin from -0.32 < [Fe/H] < -0.28 dex. To consider trends with age, we further select stars in three age bins: 1.5 - 2.5 Gyr, 4.5 - 5.5 Gyr, and 8.5 - 9.5 Gyr. We then randomly select 10 stars in each [Fe/H] and age bin, with five stars having a high probability of belonging to the low-$\alpha$ sequence, $P(z =$ high-$\alpha) < 0.5$, and the other five stars having a high probability of belonging to the high-$\alpha$ sequence, $P(z =$ high-$\alpha) > 0.5$.

We use the default *Chempy* yield tables which include CC-SN yields from Nomoto et al. (2013), SN Ia yields from Seitenzahl et al. (2013), and AGB yields from Karakas (2010). Based on these yield tables, the *Chempy* inference routine can fit a list of 22 chemical abundances. However, since the focus in this paper is on quantifying the information encoded in magnesium and silicon abundances, we fit just these two abundances, as well as [Fe/H]. We also pass the age estimates from Ness et al. (2016) to *Chempy* to be used during inference.

While the *Chempy* GCE model is described by seven parameters, we decide to only infer four of them. This is so the posterior is not overwhelmingly dominated by the priors, as the likelihood is only being computed from three abundances. The three we decide not to infer are the SN Ia delay time, the fraction of stellar feedback that returns to the surrounding gas reservoir, and the initial mass of gas reservoir. Through empirical experimentation, we notice our data to be relatively uninformative regarding these parameters. Additionally, as described in Rybizki et al. (2017), observational constraints on these three parameters are less certain. During the Markov chain Monte Carlo (MCMC) routine, the parameters that we do not infer are set to default priors values as given in Rybizki et al. (2017).

The four *Chempy* parameters we infer are described in detail as follows. Two of them, $\alpha_{\mathrm{IMF}}$ and $N_{\mathrm{Ia}}$, are typical SSP parameters. The IMF parameter is defined as the high-mass slope of a Chabrier IMF (Chabrier 2003b), which in the *Chempy* model is defined over the stellar mass range of 0.08 - 100 $M_\odot$. The IMF parameter determines the ratio of low- versus high-mass stars formed in a burst of star-formation, which alters the relative yields produced. The SN Ia parameter, $N_{\mathrm{Ia}}$, is the normalization constant of the Maoz et al. (2010) DTD, in which SN Ia enrichment is modeled as a power law in time. In *Chempy* specifically, $N_{\mathrm{Ia}}$ is defined as the number of SNIa explosions per solar mass over a 15 Gyr time period. The remaining two parameters we infer are ISM parameters, star formation efficiency (SFE) and $\mathrm{SFR}_{\mathrm{peak}}$. The SFE governs the amount of gas infalling from the gas reservoir that is needed to sustain the conversion of ISM gas to stars. This is defined as $m_{\mathrm{SFR}}/m_{\mathrm{ISM}}$. As illustrated in Figure 7 of Rybizki et al. (2017), low SFE values result in more gas being required for the star formation. That leads to a bigger gas reservoir in the ISM, which dilutes the stellar enrichment and keeps the ISM metallicity lower compared to a higher SFE value. The fourth parameter we infer is the peak time of the SFR, $\mathrm{SFR}_{\mathrm{peak}}$. In *Chempy* this is parameterized as a gamma distribution with $\mathrm{SFR}_{\mathrm{peak}}$ as the scale parameter, and the shape parameter fixed at $k = 2$. As seen in Figure 6 of Rybizki et al. (2017), an early SFR peak results in star-formation that is concentrated at early times, whereas a later SFR peak results in a smoother SFR distribution.

Table 3.1: *Chempy* parameter priors

| parameter name | *Chempy* default | | | this work | | |
|---|---|---|---|---|---|---|
| | $\mu$ | | $\sigma$ | $\mu$ | | $\sigma$ |
| $\alpha_{\text{IMF}}$ | -2.29 | $\pm$ | 0.2 | -2.29 | $\pm$ | 0.3 |
| $\log_{10}(\text{N}_{\text{Ia}})$ | -2.75 | $\pm$ | 0.3 | -2.75 | $\pm$ | 0.5 |
| $\log_{10}(\text{SFE})$ | -0.3 | $\pm$ | 0.3 | -0.3 | $\pm$ | 0.5 |
| $\log_{10}(\text{SFR}_{\text{peak}})^\dagger$ | 0.55 | $\pm$ | 0.1 | 0.45 | $\pm$ | 0.3 |

$^\dagger$ In Rybizki et al. (2017), a Gaussian prior in linear space with $\mu = 3.5$ Gyrs and $\sigma = 1.5$ Gyrs is placed on the SFR$_{\text{peak}}$ parameter.

We adopt the same priors for all of the stars that we fit. Each prior is defined as a normal distribution with mean $\mu$ and width $\sigma$, as listed in Table 3.1. Here we report both the priors used in Rybizki et al. (2017), and the priors we choose to adopt in this work. The priors set forth in Rybizki et al. (2017) are motivated by recent work related to observational estimates of Milky Way-like GCE parameters. While adopting these priors in the *Chempy* model results in agreement with the Sun's chemistry, for our work we choose to broaden the priors. This is because we are fitting stars of various ages that exhibit a range of chemical abundance patterns, which is tentatively reflective of a multitude of enrichment histories. We increase the width of the priors to better capture the uncertainty in our knowledge regarding the parameter values that may describe this diversity of stars.

As described in Rybizki et al. (2017), inference of the *Chempy* GCE parameters is achieved through posterior sampling, specifically utilizing the `emcee` implementation of MCMC (Foreman-Mackey et al. 2013); `emcee` is an affine-invariant ensemble sampler, which initializes a collection of walkers in a defined parameter space. Each walker evaluates the posterior at every step of its random walk through the space, and by comparing the current evaluation to the evaluation at the previous step, the proposed step is either accepted or rejected with some probability. In the context of *Chempy*, the sampling routine is summarized as follows. During each walker's steps, the current parameter configuration is used to evaluate the log prior. Then, keeping track of the chemical abundances at each timestep, the *Chempy* GCE model is run with these parameter values from $t = 0$ to the provided estimated time of stellar birth. The log likelihood is then computed by summing the square difference between the observed and model abundances, weighted by the observational uncertainties. The posterior is then evaluated by summing the log prior and log likelihood.

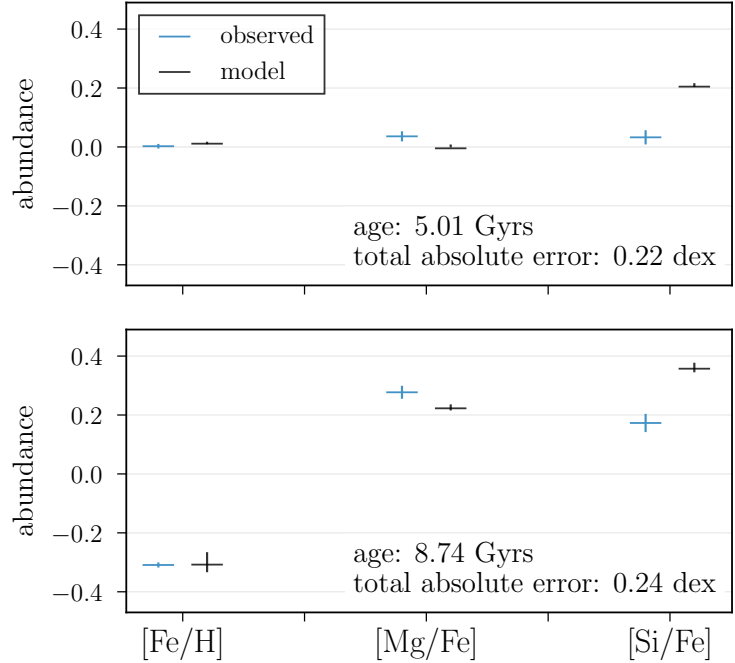

Figure 3.10: Comparison of the observed abundances to the *Chempy* predicted abundances for two example stars. The uncertainties on the observed abundances are indicated, and for the predicted abundances the range of possible abundance values obtained from the $1\sigma$ posterior distribution is shown.

For each of the stars we fit, we initialize the MCMC routine with 28 walkers. The starting locations of the walkers are typically set to be confined to a tight 4-dimensional ball centered on the prior means, and after several iterations the walkers explore the larger, relevant parameter space. Instead of running each MCMC instance for a pre-defined number of steps, we monitor the convergence of the chains in real time and terminate the runs once some convergence criterion is satisfied. The commonly used Gelman-Rubin statistic is not appropriate for chains produced with `emcee` since the chains are not independent of one another. Instead, Foreman-Mackey et al. (2013) suggest assessing convergence by computing the integrated autocorrelation time of the chains, $\tau$, which is an estimate of the number of steps the walkers need to draw an independent sample from the posterior distribution. Chains with longer estimated $\tau$'s require more steps to generate a given number of independent posterior samples. We modify the *Chempy* inference routine to compute an estimate of $\tau$ for each chain every 300 steps. We do this using the `emcee` implementation of Goodman & Weare (2010)'s method for estimating the integrated autocorrelation time. Once the chains reach a length of $30\times\tau$, they are terminated. Figure 3.12 in the Appendix shows the distribution of integrated autocorrelation times for each of the runs, as well as the estimated number of independent posterior samples generated. Often, more than 30 independent samples are obtained because longer chains are needed to generate a reliable estimate of $\tau$. In addition to being used to monitor convergence, we utilize the integrated autocorrelation times to remove the burn-in phase of the chains. For each star, we determine the maximum integrated autocorrelation time across all parameters and chains, and remove three times this value before computing the posterior percentiles.

### 3.6.3 Results

We now present the results of the *Chempy* GCE modeling for the mono-age, mono-abundance stellar samples described above. As an example of the full posterior distributions obtained, Figure 3.13 in the Appendix shows the trace plots and joint posterior distributions for two stars that we fit: a solar-metallicity, 5 Gyr-aged star in the low-$\alpha$ sequence, and a metal-poor, 9 Gyr-aged star in the high-$\alpha$ sequence. Details regarding these particular fits are discussed in the Appendix. While there are potentially interesting differences between the posteriors of the two stars, these differences must be interpreted in the context of how well the models describe the data. To examine this we use the inferred parameter values to run instances of the *Chempy* GCE model and predict the stellar abundance patterns at the time of stellar birth. We then compare these predicted abundances to the observed abundances that were used to infer the posteriors. This approach to model evaluation is similar to posterior predictive checking (Gelman & Stern 1996). Figure 3.10 shows the observed and predicted abundances for both of the stars included in Figure 3.13. The observed abundances are the [Fe/H], [Mg/Fe], and [Si/Fe] measurements from `APOGEE` that were passed to the likelihood calculation, and the predicted abundances are based on a *Chempy* model run using parameter values from each star's posterior distributions. The horizontal markers indicate predicted abundances based on the $50^{th}$ percentile posterior values, while the vertical range indicates the predicted abundance range considering *Chempy* models with the $16^{th}$ and $84^{th}$ percentile values of the parameter posteriors.

As seen in the figure, the [Fe/H] and [Mg/Fe] abundances of the 5 Gyr-aged solar-metallicity star are well predicted by the model, falling nearly within the observational uncertainty ranges. However, the [Si/Fe] abundance is not reproduced as well. The model over-predicts the silicon abundance by nearly 0.2 dex, and the observed abundance is not within the model's $1\sigma$ posterior range. The predicted abundances for the 9 Gyr-old metal-poor star exhibit similar discrepancies with respect to the observed abundances. As seen in the *bottom panel* of Figure 3.10, the *Chempy* model run with the $50^{th}$ percentile posterior values results in predicted abundances that agree or nearly agree with the observed [Fe/H] and [Mg/Si] abundances. But, similar to the solar-metallicity star, the range of predicted [Si/Fe] abundance is disparate from the observed [Si/Fe] abundance, and is over-predicted by about 0.2 dex.

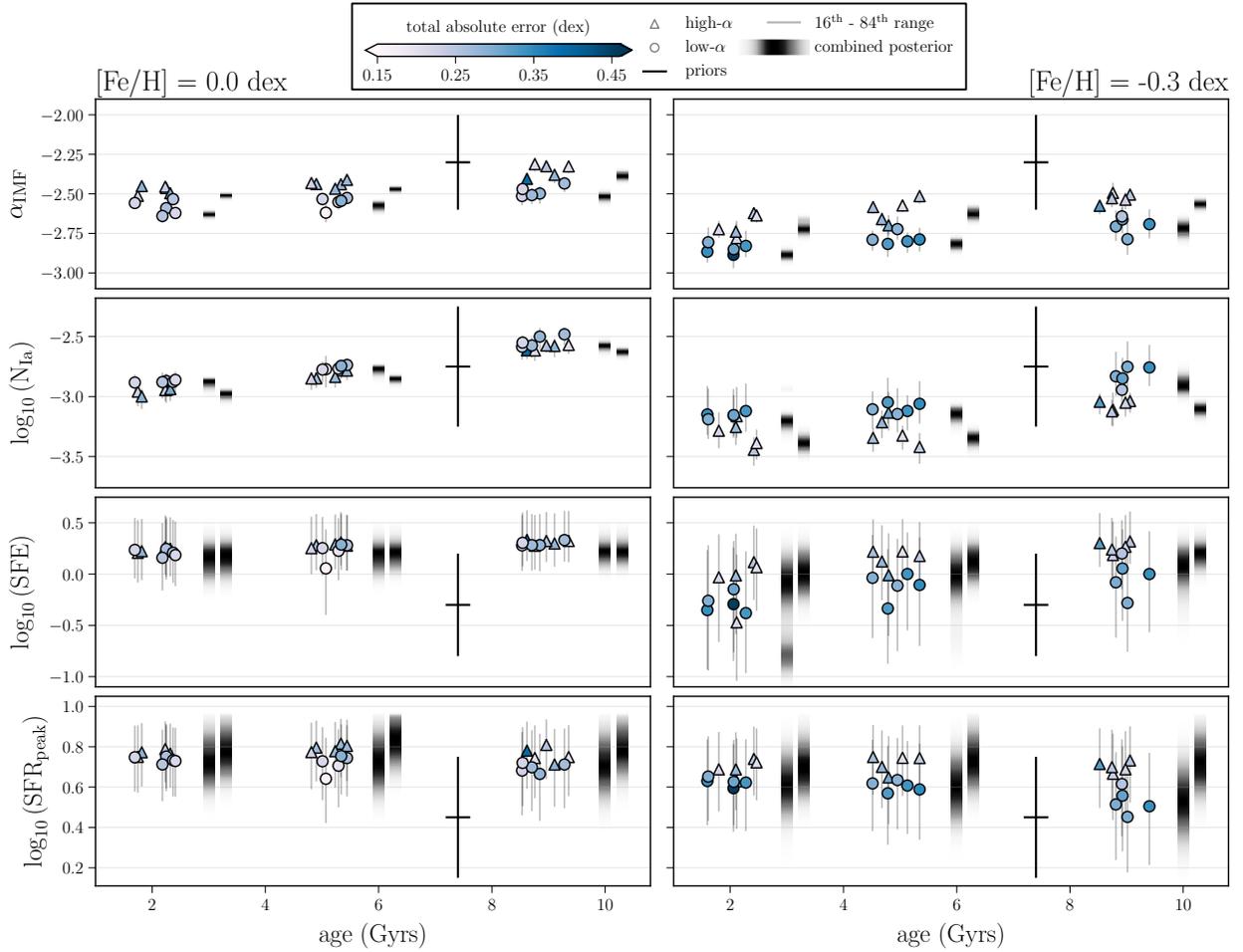

Figure 3.11: *Chempy* posterior distributions for mono-age, mono-abundance stars with abundance measurements from `APOGEE`. Each panel shows the values of the inferred GCE parameters ($\alpha_{\rm IMF}$, $N_{\rm Ia}$, SFE, and $\rm SFR_{peak}$) for stars with ages of 2, 5, and 9 Gyr. The plots on the left show stars with solar [Fe/H] abundances, while the plots on the right show stars with [Fe/H] abundances of -0.3 dex. For each star, the $50^{\rm th}$ percentile of the posterior distribution is indicated by the markers, where high-$\alpha$ sequence stars are indicated by triangles and low-$\alpha$ sequence are indicated by circles. The error bars on the markers span the $1\sigma$ (or $16^{\rm th}$ - $84^{\rm th}$) posterior percentile range, and the markers are colored by the total absolute error between the observed and predicted abundances. The prior on each parameter is shown for comparison, with the variance of the distribution indicated by the solid black line. The combined posteriors, binning stars with the same age, [Fe/H] abundance, and $\alpha$-enhancement, are shown in grayscale, with darker shades indicating a higher posterior probability.

The main takeaway from Figure 3.10 is as follows. While we are relatively confident that the chains are not unconverged, we find that the inferred parameters do not always result in *Chempy* models that are able to reproduce the observed abundances, particularly when it comes to [Si/Fe]. Possible reasons for this will be discussed further in Section 3.6.4. While we next present how the inferred GCE parameters vary for stars of different ages and abundances, the quality of the model fits cautions any significant physical interpretation of these trends.

Moving beyond the two example stars discussed up to this point, we now present the posteriors for the full sample of mono-age, mono-abundance stars that we fit. As discussed in Section 3.6.2, we fit a collection of 60 stars that fall into two [Fe/H] bins, three age bins, and have $\alpha$-enhancements characteristic of both the low- and high-$\alpha$ sequences. Figure 3.11, which displays summary statistics of the posterior distributions at each age and [Fe/H], summarizes the results of these fits. In each panel, the individual posteriors for each star are described by the markers placed at the $50^{\text{th}}$ percentile value, and the error bars which span the $1\sigma$ (or $16^{\text{th}}$ - $84^{\text{th}}$) percentile range. The markers are colored by the "total absolute error", which is the sum of the absolute difference between the observed abundances and predicted abundances based on the $50^{\text{th}}$ posterior values. We use this error as a goodness-of-fit measure, which we discuss further in Section 3.6.4. We also show the combined posterior, which is indicated by the grayscale colorbar. These combined posteriors are determined by multiplying the individual posterior distributions of all the stars that fall within the same [Fe/H], age, and $\alpha$ sequence bin, which assumes that individual posterior probability distributions are independent of one another. Given that the stars have similar properties modulo any differences in [Mg/Fe] and [Si/Fe], the combined posteriors more strongly constrain the parameters than the individual posteriors.

First consider the $\alpha_{\text{IMF}}$ parameter, which is the GCE parameter we find to be most constrained by the data. For all the stars that we fit, the posteriors appear to disfavor IMF slopes that are more top-heavy than $\alpha_{\text{IMF}} > $ -2.25. Additionally, for both the solar-metallicity and metal-poor stars, we find a weak trend between the inferred $\alpha_{\text{IMF}}$ and stellar age. Younger stars are found to have IMF slopes described by a more bottom-heavy IMF, whereas older stars are described by more Chabrier-like IMF slopes. At each age, there is a further trend with $\alpha_{\rm IMF}$ and [Fe/H]. Metal-poor stars are found to be described by a more bottom-heavy IMF slope than their solar-metallicity counterparts of the same age. There are also smaller differences between the inferred IMF slope of low- and high-$\alpha$ sequence stars, which is most apparent when considering the combined posteriors. For both metallicities and at all ages, high-$\alpha$ sequence stars are found to have more top-heavy IMF slopes than low-$\alpha$ sequence stars.

Aside from the IMF parameter, $N_{\rm Ia}$ is the other parameter we find to be tightly constrained by the `APOGEE` abundances. As seen in the second row of Figure 3.11, the $\sim 1\sigma$ posterior ranges appear to disfavor different regions of the $N_{\rm Ia}$ parameter space, with dependence on stellar age and metallicity. First we notice that at every age, the metal-poor stars are found to be fit by fewer SN Ia than their solar-metallicity counterparts of the same age. This difference is $\sim 0.25$ dex on average. In addition to the variations with metallicity, there are also trends with age at fixed [Fe/H]. For the solar-metallicity stars, younger stars are found to be described by fewer SN Ia compared to older stars, spanning values from $\log(N_{\rm Ia})$ = -3 to -2.5. For the metal-poor stars, a similar trend is observed with age, except these values span $\log(N_{\rm Ia})$ = -3.5 to -2.75. Lastly, similar to the IMF parameter, there are small differences between the $N_{\rm Ia}$ posteriors of high-$\alpha$ sequence stars and low-$\alpha$ sequence stars. As apparent in the combined posteriors, low-$\alpha$ sequence stars are described by fewer SN Ia than high-$\alpha$ sequence stars, at every age and metallicity.

Unlike the IMF and SN Ia parameters, the two ISM parameters are not as tightly constrained by the data. As seen in the third row of Figure 3.11, the $16^{\rm th}$ - $84^{\rm th}$ ranges of the SFE posteriors are more comparable to the range of the prior, especially for the metal-poor stars. The binned posteriors of metal-poor stars are more constraining, and show the SFE values for these stars to be > -0.25 dex, with weak trends seen with age and $\alpha$-enhancement. For the 2 Gyr-aged metal-poor stars, we see a discrepancy between the 50$^{\text{th}}$ percentile values of the individual posteriors and the combined posterior distributions, and bimodality in the combined posterior for the high-$\alpha$ sequence. This is a result of these posteriors being non-Gaussian, exhibiting a one-sided tail in addition to a peak. For the solar-metallicity stars, the posteriors are more constrained. For these stars, SFE values less than $\sim$0 dex are mostly disfavored, with posterior medians at $\sim$0.25 dex for stars of each age and $\alpha$ enhancement. Compared to stars of the same age that are metal-poor, the solar-metallicity stars exhibit slightly higher SFE values.

Finally, the fourth row of Figure 3.11 shows the posteriors for SFR$_{\text{peak}}$. As with the SFE parameter, the ranges of the SFR$_{\text{peak}}$ posteriors are more comparable to the range of the prior. Despite this, some regions of the parameter space are still disfavored. For the solar-metallicity stars, their posteriors suggest a low probability of an SFR peak occurring before $\sim$3 Gyr, with a median posterior value of $\sim$5.5 Gyr for all of the stars, with only a slight difference between the low-$\alpha$ and high-$\alpha$ sequence. For the metal-poor stars, their posteriors suggest a low probability of an SFR peak occurring before $\sim$2 Gyr, with the median posterior value of $\sim$4 Gyr, with a slight trend with age and $\alpha$ enhancement.

### 3.6.4 Limitations

As emphasized throughout Section 3.6.3, we find that the predicted *Chempy* abundances are often in tension with the observed abundances, particularly with regard to [Si/Fe]. We now quantify these discrepancies, and discuss possible reasons for them that ultimately limit the interpretation of our chemical evolution modeling results. In the end, we advocate that, given the current quality and quantity of abundance measurements for stars located all throughout the Milky Way, a more data-driven approach to GCE modeling is a possible path forward.

Before examining how well the *Chempy* models are able to reproduce the abundance patterns of stars with varying metallicities, ages, and $\alpha$-enhancements, we first compare the default and alternative *Chempy* yield tables. As discussed in Rybizki et al. (2017), the choice of yield tables is a hyperparameter of the *Chempy* model, and ideally the observed abundances can be used to determine which yield tables are most probable. Consider the posteriors inferred under the default yield tables (based on Nomoto et al. (2013), Seitenzahl et al. (2013), and Karakas (2010)), which are the tables used for the results shown in Section 3.6.3.

For the solar-metallicity stars, the average absolute difference between the predicted and observed [Fe/H] abundances is 0.007 dex, with all of the stars having their abundances over-predicted. For [Mg/Fe] the predicted abundances match the observed abundances somewhat worse, with the average absolute difference being 0.056 dex (100% under-predicted). However, the *Chempy* models are unable to reproduce the observed silicon abundances. The average absolute difference between the predicted and observed [Si/Fe] abundances is 0.187 dex (100% over-predicted). For the metal-poor stars, the predicted abundances are less consistent with the observed abundances. On average, the [Fe/H] predictions are 0.035 dex off from the observed abundances (100% over-predicted), the [Mg/Fe] predictions are 0.073 dex off from the observed abundances (100% under-predicted), and the [Si/Fe] predictions are 0.178 dex off from the observed abundances (100% over-predicted).

The differences between the [Fe/H] and [Mg/Fe] predictions are at least partly due to the smaller measurement uncertainties on the [Fe/H] abundances. These smaller [Fe/H] uncertainties result in larger penalties in the likelihood evaluation for discrepant [Fe/H] predictions compared to the penalties for discrepant [Mg/Fe] predictions. The observational uncertainties on the [Mg/Fe] and [Si/Fe] abundances are comparable, so the observational uncertainties are not the primary driver of the [Si/Fe] over-prediction.

To test if perhaps another set of yield tables could result in better [Si/Fe] predictions, we also infer the parameter posteriors assuming the alternative *Chempy* yield tables (based on Chieffi & Limongi (2004), Thielemann et al. (2003), and Ventura et al. (2013)). For the solar-metallicity stars, the predicted [Fe/H] abundances are off by 0.06 dex (96% under-predicted, 4% over-predicted), the predicted [Mg/Fe] abundances are off by 0.11 dex (100% over-predicted), and the predicted [Si/Fe] abundances are off by 0.32 dex (100% over-predicted). Compared to the default yield tables, the alternative yields result in poorer abundance matches. Considering the metal-poor stars, the predicted [Fe/H] abundances are off by 0.09 dex (100% under-predicted), the predicted [Mg/Fe] abundances are off by 0.14 dex (100% over-predicted), and the predicted [Si/Fe] abundances are off by 0.34 dex (100% over-predicted). Again, the alternative yield tables result in worse matches to the observed abundances.

We make several observations from this comparison of the predicted and observed abundances for both the default and alternative yield tables. First is that the default yield tables result in better matches to the observed abundances for all elements, for both the solar-metallicity and metal-poor stars. As found in Philcox et al. (2018) who present a scoring system for comparing yield tables, the optimal choice of tables even when fitting only proto-solar abundances is dependent on what specific abundances are being fit. Presumably the optimal yield tables will also differ for stars with varying abundance patterns. A detailed ranking of the various yield tables for a collection of stars with different abundance properties may reveal informative trends regarding what tables are preferred by the data. The other observation we make is that, regardless of yield table choice, the silicon abundance is the most difficult to reproduce and is systematically over-predicted by $\sim$0.2 - 0.4 dex. Because of this, despite the [Mg/Si] trends that the default *Chempy* yield tables exhibit in Section 3.3, the inferred parameter values are not able to correctly reproduce the relative amounts of magnesium and silicon. As found in Philcox et al. (2018) there seems to be an Mg to Si offset for all tested CC-SN yield tables, which include those used in this work and more recent ones from the literature; see their Figure 4. It is unlikely that other nucleosynthetic channels or the GCE model parameterization are causing this offset. Instead, it is found that the explosion energies of the CC-SN could be the reason and lowering them could remedy the discrepancy (Heger & Woosley 2010; Fryer et al. 2018).

While the default yield tables are preferred by the data, these yield tables result in some systematic trends between stellar properties and how well the *Chempy* models can reproduce the abundance patterns of stars (as seen in Figure 3.11). While we find no significant trends with stellar age, there are some differences in how well the models describe metal-poor versus solar-metallicity stars, and low-$\alpha$ versus high-$\alpha$ sequence stars. Considering metallicity, we find that the abundances of the solar metallicity stars are reproduced slightly better than those of the metal-poor stars. To compare the fits, we compute the "total absolute error" between the predicted and observed abundances by summing the absolute values of the individual [Fe/H], [Mg/Fe], and [Si/Fe] differences. The average total error of the solar-metallicity stars is 0.25 dex, while that of the metal-poor stars is 0.29 dex. As mentioned previously, a majority of this error is from the model's over-prediction of silicon. The worse model predictions for metal-poor stars compared to solar-metallicity stars could be due to a number of factors, and we note that a similar finding is reported in Rybizki et al. (2017), where Arcturus with an [Fe/H] $\sim$ -0.5 dex is fit more poorly than the Sun. Generally, stellar models are gauged to solar abundances, which could result in the yields preferentially producing solar abundance patterns.

We also find differences in the ability of the models to reproduce the abundances of low- versus high-$\alpha$ sequence stars. First considering solar-metallicity stars of all ages, we find that on average the high-$\alpha$ stars have marginally higher total errors than low-$\alpha$ stars (0.26 dex compared to 0.24 dex). This small difference between the two sequences is mostly due to worse [Mg/Fe] predictions for the high-$\alpha$ stars. For the metal-poor stars, we find a larger discrepancy in the accuracy of the predicted abundances between the low- versus high-$\alpha$ sequences. On average the high-$\alpha$ stars have a total error of 0.25 dex, while the low-$\alpha$ stars have an average total error of 0.32 dex. This difference is mainly from both the [Fe/H] and [Si/Fe] abundance predictions, which are each $\sim$0.03 dex worse for the low-$\alpha$ sequence stars. The more significant difference between the low- and high-$\alpha$ sequence fits at low metallicities could be a consequence of the high-$\alpha$ sequence abundance pattern resulting directly from CC-SN yields. Fine-tuning of the abundances with Fe from SN Ia is additionally needed to produce low-$\alpha$ sequence abundance patterns.

In summary, while we find the yield tables of Nomoto et al. (2013), Seitenzahl et al. (2013), and Karakas (2010) to be preferred by the data, ultimately we are not able to describe the abundance patterns of these `APOGEE` stars due to the systematic over-prediction of [Si/Fe]. The reasons for this fall into three categories, which include shortcomings with the yield tables, the model, or the data. Considering the yield tables, one possibility is that there are nucleosynthetic contributions missing from the yield tables that are necessary to reproduce the Si abundances of these stars. The yield tables might also not be descriptive enough to accurately characterize stars with abundance patterns arising from a variety of initial environmental conditions and that have experienced a multitude of different evolutionary pathways. On the model side, while here we fit each star with its own unique one-zone model, the ISM parameterization might be too simplistic to reproduce the abundance patterns of these stars. Parameterizations that allow for things like a bursty star-formation history, an adaptive coronal gas mass and volume, or different mixing of the ISM will better describe the abundances. Finally, shortcomings may lie with the data. In this paper, our primary focus was to empirically characterize the ratio of magnesium to silicon, and then understand the physical origin of these variations. With this goal, we fit *Chempy* with just three abundances ([Fe/H], [Mg/Fe], and [Si/Fe]) in an attempt to generate the simplest versions of the model that would be able to reproduce the data. However, GCE models could potentially be better constrained by different or more abundances, or even the entire chemical abundance vector available from `APOGEE`. That said, two abundances are highly constraining in the sense that they highlight deepened tensions between the data, yield tables, and GCE models, even in a simple application.

## 3.7  Summary & Conclusion

In this paper we present a detailed investigation of the [Mg/Si] abundance in the Milky Way disk. We do this using a large sample of stars with `APOGEE` abundance measurements, estimated stellar ages, and *Gaia* astrometry. Inspired by the increasing precision and sheer number of stars with available abundances, our primary goal is to go beyond the information contained in a bulk $\alpha$ abundance and examine the higher level of granularity encoded in an inter-family abundance ratio. The specific choice of magnesium and silicon is motivated by the expected subtle differences in their nucleosynthetic origins, which has been previously used to interpret the [Mg/Si] abundances of stars in the Sattitargius dwarf galaxy. Our endeavor for Milky Way stars includes both an empirical characterization of the [Mg/Si] abundance throughout the Galaxy, as well an attempt to link [Mg/Si] variations to a physical origin through galactic chemical evolution modeling.

After gaining intuition in Section 3.3 for how Mg and Si yields evolve through time in a single burst of star-formation, we then focus on establishing how the observed [Mg/Si] abundance varies in the Galaxy. We make connections between [Mg/Si] and various stellar properties, including stellar age, [Fe/H], location, and stellar orbital actions. With the goal of better understanding the origin of the Milky Way's bimodal $\alpha$ sequence, we identify differences in [Mg/Si] between low- and high-$\alpha$ sequence stars. Our findings are summarized as follows.

- High-$\alpha$ sequence stars are more enhanced in [Mg/Si] than low-$\alpha$ sequence stars, with the difference in the average [Mg/Si] between the two sequences being $\sim$0.08 dex. The two sequences also exhibit distinct behavior in the variation of their [Mg/Si] abundances across the [$\alpha$/Fe]-[Fe/H] plane, where [Mg/Si] varies more strongly with [$\alpha$/Fe] and [Fe/H] in the low-$\alpha$ sequence than it does in the high-$\alpha$ sequence. Given the expected theoretical Mg and Si yields (Figure 3.1), observed variations in [Mg/Si] at early times could potentially be used to discriminate between different IMFs. This is tentatively supported by the *Chempy* inferences in Section 3.6, where high-$\alpha$ sequence stars are found to be described by a slightly more top-heavy IMF than low-$\alpha$ sequence stars.

- The enhanced [Mg/Si] abundance of the high-$\alpha$ sequence compared to the low-$\alpha$ sequence is persistent at all stellar ages. Considering the evolution of [Mg/Si] with stellar age, from old to young ages the [Mg/Si] abundance of the low-$\alpha$ sequence decreases nearly 2$\times$ more than the [Mg/Si] abundance of the high-$\alpha$ sequence. This difference is persistent for solar-metallicity stars, as well as metal-poor stars, but is minimized at higher [Fe/H] abundances where the two $\alpha$ sequences become less distinct.

- Inspired by Hayden et al. (2015)'s characterization of the [$\alpha$/Fe]-[Fe/H] distribution in the disk, we examine how the [Mg/Si]-[Fe/H] distribution varies with location in the Galaxy, from R = 3 - 15 kpc and $|z|$ = 0 - 2 kpc. We find that the [Mg/Si]-[Fe/H] distributions of the high- and low-$\alpha$ sequences considerably overlap at each R and $|z|$, but that there are significant differences in how the mean [Mg/Si] of the two sequences varies throughout the Galaxy. For the high-$\alpha$ sequence, the trend with mean [Mg/Si] and radius is the same at all heights from the disk midplane, with the highest [Mg/Si] occurring at R $\sim$ 8 kpc. However, for the low-$\alpha$ sequence the highest [Mg/Si] abundances occur at R $\sim$ 6 kpc and the mean [Mg/Si] at each radius scales with $|z|$. Since the [Mg/Si] abundance is correlated with age, these trends seemingly reflect the age gradients with R and $|z|$ across the disk, which are distinct for the low- and high-$\alpha$ sequences.

- Considering trends with orbital actions, our findings confirm the results of Gandhi & Ness (2019) that the high- and low-$\alpha$ sequences are distinct in $J_\phi, J_r,$ and $J_z$ at all stellar ages. Examining how [Mg/Si] varies with actions, we find that at all ages low-$\alpha$ sequence stars with lower [Mg/Si] abundances have higher angular momenta than their enriched [Mg/Si] counterparts. A similar trend is not found for the radial or vertical actions, or the high-$\alpha$ sequence and any action. Presumably, this reflects the radial metallicity gradient in the disk at a given age, and a bulk chemical composition of star-forming gas that varies with time and Galactocentric radius, but not with distance from the disk midplane.

These observed trends with [Mg/Si] abundance support established and reveal new connections between chemistry and orbital dynamics. Given the expected differences in the chemical enrichment processes responsible for generating Mg and Si yields, the varying [Mg/Si] abundance relations between low versus high-$\alpha$ sequence stars bolsters the notion that the two sequences are distinct in their origin and formation. Detailed matching of these observational trends between [Mg/Si], age, and dynamics to the properties of simulated Milky Way-like galaxies can place powerful constraints on disk formation and chemical evolution mechanisms.

In the second half of this paper we focus our attention on understanding the physical origin of the observed [Mg/Si] variations. Specifically, we perform GCE modeling, which combines physically motivated models of star-formation, the ISM, and galactic evolution to predict the chemical content of the ISM through time. Here, we employ the recently developed *Chempy* code which, given a set of observed abundances and estimated stellar age, allows for inference of GCE model parameters.

We infer GCE parameters for a set of mono-age, mono-metallicity stars with various [Mg/Fe] and [Si/Fe] abundances. For each star, we obtain posterior distributions for four parameters: the slope of the high-mass IMF, the number of SN Ia explosions per solar mass over a 15 Gyr time period, the SFE, and the peak of the SFR. While there are potentially interesting relationships between these parameters and stellar properties, our interpretation of them is limited. This is because the *Chempy* models are unable to reproduce the `APOGEE` abundances for the diversity of stars that we fit. A majority of the discrepancy between the predicted and observed abundances comes from the consistent over-prediction of [Si/Fe], which has also been reported by Philcox et al. (2018) for a number of tested CC-SN yield tables. We also find systematic trends with metallicity and $\alpha$ enhancement, and the predictive quality of the *Chempy* models. This reveals tensions between the GCE models, yield tables, and observed abundance measurements.

The main conclusions of this paper are as follows. Given the large number of stars with high-quality abundance measurements available, small variations in these abundances can be characterized by averaging stars with similar properties, such as $\alpha$-enhancement, location, age, or actions. In this way, we were specifically motivated to examine the inter-family ratio of two $\alpha$-elements, Mg and Si, because of expected differences in how they are produced in CC-SN and SN Ia events. This approach we take in dissecting the [Mg/Si] abundance can be generalized to isolate other particular enrichment channels. In theory, other inter-family or intra-family abundances can be used to probe specific nucleosynthesis or disk formation mechanisms, and these empirical relationships can serve as detailed constraints for simulations of Milky Way-like galaxies. However, in practice we encountered challenges in connecting variations in Mg and Si to underlying stellar physics. Specifically, we found that a flexible model of GCE is unable to predict even just three stellar abundances that represent a diversity of enrichment histories. This failure highlights tensions between the chemical evolution models, the yield tables, and the data. As we have now entered a regime of rich stellar abundance information, a more data-driven approach to chemical evolution models and nucleosynthetic yield tables may be a way forward.

*Acknowledgements.* We would like to thank Gail Zasowski, Dan Foreman-Mackey, Victor Debattista, and Adam Wheeler for helpful conversations. We are also grateful to Nick Carriero for his assistance in planning the *Chempy* runs on the Flatiron Institute computing cluster.

KB is supported by the NSF Graduate Research Fellowship under grant number DGE 16-44869. KB thanks the LSSTC Data Science Fellowship Program, her time as a Fellow has benefited this work. MN is supported by the Alfred P. Sloan Foundation. KVJ's contributions were supported in part by the National Science Foundation under grants NSF PHY-1748958 and NSF AST-1715582. Her work was performed in part during the Gaia19 workshop and the 2019 Santa Barbara *Gaia* Sprint (also supported by the Heising-Simons Foundation), both hosted by the Kavli Institute for Theoretical Physics at the University of California, Santa Barbara.

Funding for the Sloan Digital Sky Survey IV has been provided by the Alfred P. Sloan Foundation, the U.S. Department of Energy Office of Science, and the Participating Institutions. SDSS-IV acknowledges support and resources from the Center for High-Performance Computing at the University of Utah. The SDSS website is www.sdss.org. SDSS-IV is managed by the Astrophysical Research Consortium for the Participating Institutions of the

## 3.A  Convergence

Figure 3.12 shows the distribution of the integrated autocorrelation times for each *Chempy* sampled parameter. The average integrated autocorrelation time is computed across the chains for each star that we run. Also shown is the effective number of independent posterior samples, based on the estimated integrated autocorrelation times and chain lengths.

As an example, Figure 3.13 shows the trace plots and joint posterior distributions for two stars that we fit with *Chempy*: a solar-metallicity, $\sim$5 Gyr-aged star in the low-$\alpha$ sequence, and a metal-poor, ∼9 Gyr-aged star in the high-$\alpha$ sequence. As seen in the figure, the chains appear well-mixed and stationary. This, in addition to the integrated autocorrelation times, is suggestive of convergence. Examining the posteriors of the 5 Gyr-old star, we see that for the parameter describing the slope of the IMF, $\alpha_{\rm IMF}$, the posterior distribution peaks at lower values than the prior. This suggests that stellar populations with fewer high-mass to low-mass stars, compared to the typical Chabrier IMF, is necessary to produce the [Fe/H], [Mg/Fe], and [Si/Fe] abundances of this star. For the remaining parameters, the deviations between the posterior and prior distributions are less significant. The peak of the SN Ia parameter posterior is similar to the mean of the prior, but compared to the priors higher SFE and later $SFR_{peak}$ values appear to be preferred. Considering the relationship between posterior distributions, the strongest correlations are between the IMF slope and number of SN Ia, the IMF slope and the $SFR_{peak}$, and the number of SN Ia and the $SFR_{peak}$. As described in Rybizki et al. (2017), the correlation between the $\alpha_{IMF}$ and $N_{Ia}$ parameters is because an increased metal production from more CC-SN requires more Fe from SN Ia to generate the correct balance of $\alpha$-element and Fe enrichment. The $SFR_{peak}$ is additionally correlated with $\alpha_{IMF}$ and $N_{Ia}$ parameters because a later $SFR_{peak}$ increases the number of metal-rich stars, which eject more metals.

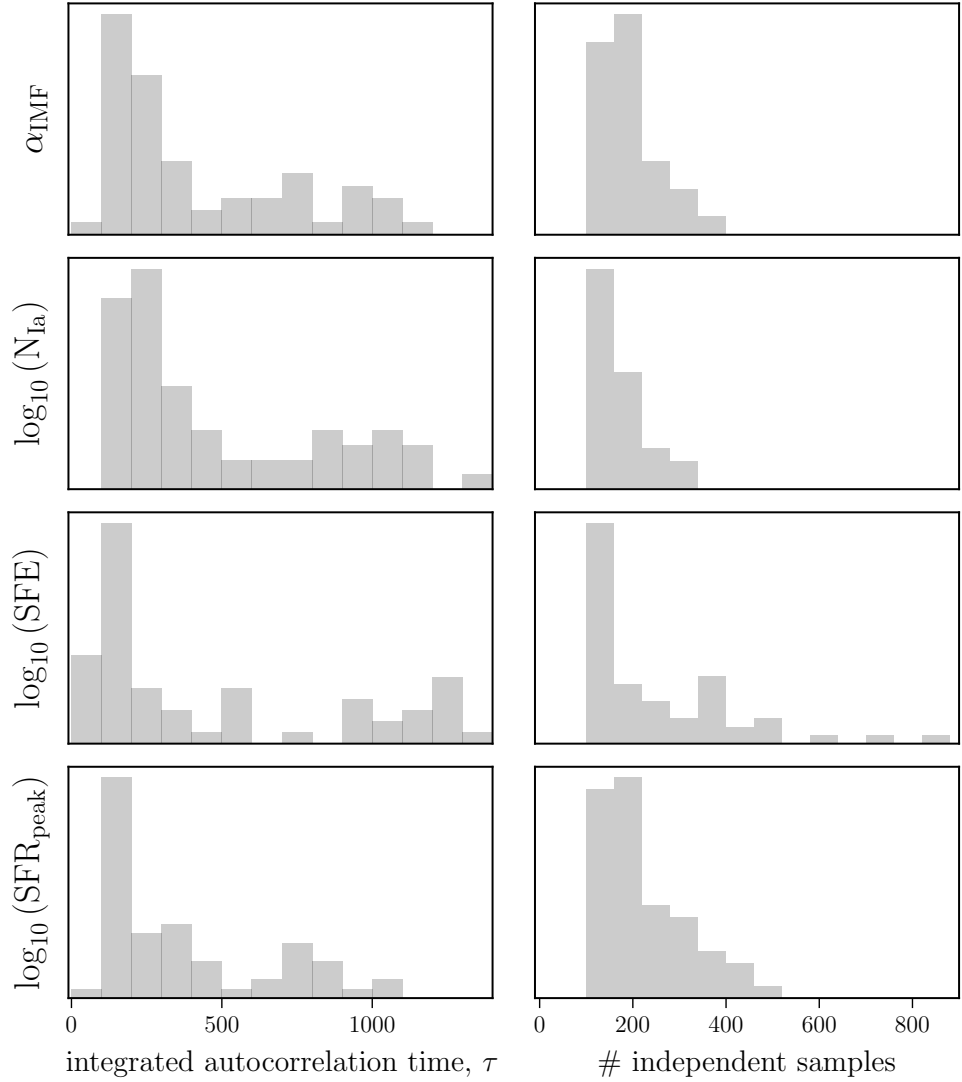

Figure 3.12: Distribution of average integrated autocorrelation times (left), and the average number of estimated independent posterior samples (right) for each inferred *Chempy* parameter.

The posteriors of the ∼9 Gyr-old metal-poor star exhibit behavior mostly similar to the posteriors of the ∼5 Gyr-old solar-metallicity star. The $\alpha_{IMF}$ posterior for this star is nearly the same as that of the 5 Gyr-aged star, suggesting that the chemical abundance pattern of this star arose from a stellar population that also formed with fewer high-mass to low-mass stars, compared to the typical Chabrier IMF. The SFE and $SFR_{peak}$ posteriors for this star are also similar to the 5 Gyr-aged star, with a higher efficiency and later SFR peak preferred compared to the priors. The main difference between the posteriors of the two stars is in the $N_{Ia}$ parameter. For the metal-poor ∼9 Gyr-aged star, a case of fewer exploding SN Ia is preferred because this enables the production of the lower [Fe/H] and higher $\alpha$ abundances of this star.

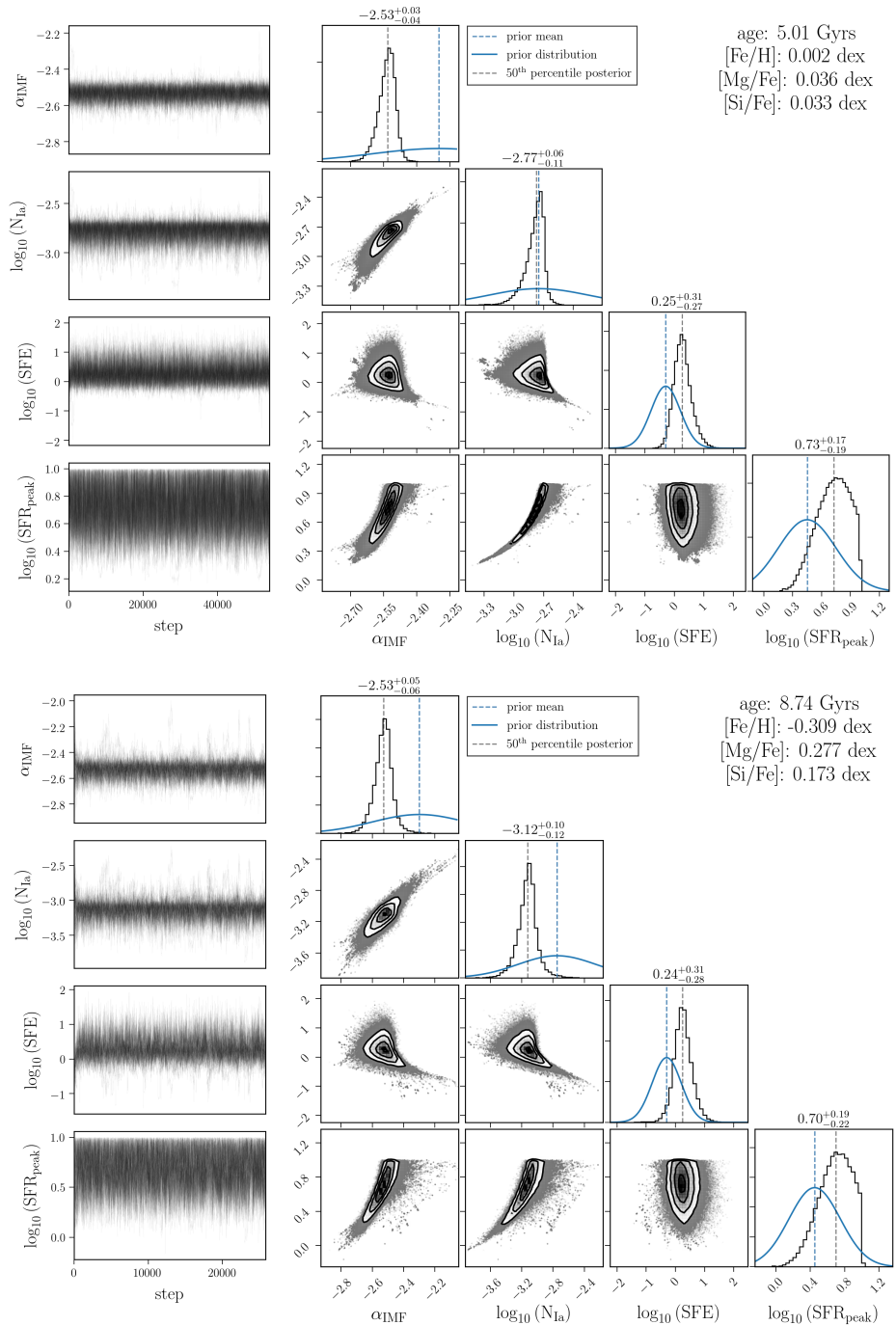

Figure 3.13: Example trace plots (left) and joint posterior distributions (right) of the four inferred *Chempy* parameters. The *top panel* shows the posterior distributions for a solar-metallicity, ∼5 Gyr-aged star in the low-$\alpha$ sequence, and the *bottom panel* shows the posterior distributions for a metal-poor, ∼9 Gyr-aged star in the high-$\alpha$ sequence. The blue curves indicates the prior for each parameter, while the histograms show the marginalized posterior distributions with the 50$^{\text{th}}$ percentile values indicated by the dashed lines. The sharp edges in the SFR$_{\text{peak}}$ posterior distributions are a result of the model being constrained to explore SFR peaks occurring before ∼12.6 Gyr.

# Chapter 4

# Data-driven derivation of stellar properties from photometric time series data using convolutional neural networks

## 4.1 Abstract

Stellar variability is driven by a multitude of internal physical processes that depend on fundamental stellar properties. These properties are our bridge to reconciling stellar observations with stellar physics, and for understanding the distribution of stellar populations within the context of galaxy formation. Numerous ongoing and upcoming missions are charting brightness fluctuations of stars over time, which encode information about physical processes such as rotation period, evolutionary state (such as effective temperature and

---

This chapter is a reproduction of a paper by Blancato, Ness, Huber, Lu , & Angus (2020) submitted to The Astrophysical Journal.

surface gravity), and mass (via asteroseismic parameters). Here, we explore how well we can predict these stellar properties, across different evolutionary states, using only photometric time series data. To do this, we implement a convolutional neural network, and with data-driven modeling we predict stellar properties from light curves of various baselines and cadences. Based on a single quarter of *Kepler* data, we recover stellar properties, including surface gravity for red giant stars (with an uncertainty of $\lesssim 0.06$ dex), and rotation period for main sequence stars (with an uncertainty of $\lesssim 5.2$ days, and unbiased from $\approx 5$ to 40 days). Shortening the *Kepler* data to a 27-day TESS-like baseline, we recover stellar properties with a small decrease in precision, $\sim 0.07$ dex for $\log g$ and $\sim 5.5$ days for $P_{\rm rot}$, unbiased from $\approx 5$ to 35 days. Our flexible data-driven approach leverages the full information content of the data, requires minimal feature engineering, and can be generalized to other surveys and datasets. This has the potential to provide stellar property estimates for many millions of stars in current and future surveys.

## 4.2 Introduction

In the coming years the number of stars with photometric time series observations is projected to increase by several orders of magnitude. The ongoing TESS mission (Ricker et al. 2014) will deliver light curves for the order of $10^5$ stars, while LSST (LSST Science Collaboration et al. 2009) is planned to deliver light curves for an unprecedented number of $\sim 10^8$ stars. The large stellar samples covered by these space- and ground-based surveys will enable further probing of known, and possibly reveal new, empirical connections between time domain variability and stellar physics. In combination with *Gaia* parallaxes (Gaia Collaboration et al. 2016, 2018b), these observations have the additional potential to markedly extend the characterization of stellar properties and populations throughout the Milky Way. However, while high quality light curves have been used to infer a number of stellar properties, fast and automated methods that can be employed on shorter baseline and sparser cadence observations will be crucial to maximize insights from the forthcoming volume of time domain data.

Brightness variability in the time domain encodes information about stellar properties through physical processes including oscillations, convection, and rotation. With high-cadence time domain data from the *Kepler* (Borucki et al. 2008) and CoRoT (Baglin et al. 2006) missions, solar-like oscillations have been detected in a large ensemble of stars. These oscillations are a result of turbulent convection near the stellar surface, which induces acoustic standing waves in the interiors of both main sequence and evolved stars, generating stellar fluctuations across a range of timescales (e.g. Aerts et al. 2010). Solar-like oscillations are typically parameterized through two average parameters, $\nu_{\max}$ and $\Delta\nu$, which can be precisely measured in power spectra computed from high-cadence time series data (e.g. Hekker et al. 2009; De Ridder et al. 2009; Gilliland et al. 2010; Bedding et al. 2010; Mosser et al. 2010; Stello et al. 2013; Yu et al. 2018). The frequency of maximum power, $\nu_{\max}$, is dependent on the temperature and surface gravity of a star (Brown et al. 1991; Belkacem et al. 2011), while the large frequency separation between consecutive overtones, $\Delta\nu$, is dependent on the stellar density (Ulrich 1986). The combination of $\nu_{\max}$ and $\Delta\nu$ thus allows a direct measurement of stellar masses ($M_*$) and radii ($R_*$) (e.g. Kjeldsen & Bedding 1995; Stello et al. 2009a,b; Kallinger et al. 2010; Huber et al. 2011).

In stars with convective envelopes, stellar granulation is also imprinted in a star's photometric variability. The circulation of convective cells produces brightness fluctuations at the stellar surface, where brighter regions correspond to hotter, rising material (granules) and darker regions correspond to cooler, sinking material (intergranule lanes). Because the size of the granules is dependent on the pressure scale height (Freytag & Steffen 1997; Hu-ber et al. 2009; Kjeldsen & Bedding 2011), the variability timescale of granulation has been demonstrated to scale with the surface gravity ($\log g$) of a star (Mathur et al. 2011; Kallinger et al. 2016; Pande et al. 2018). The relationship between granulation timescale and surface gravity has led to the development of the "Flicker method", in which brightness variations on timescales less than 8 hours are used to estimate $\log g$ (Bastien et al. 2013, 2016; Cranmer et al. 2014). With this estimate of $\log g$, and a probe of effective temperature ($T_{\text{eff}}$) (e.g. from spectroscopy, broad-band photometry), a stars relative position on the Hertzsprung-Russell (HR) diagram, and thus evolutionary state, can be determined.

In addition to oscillations and granulation, stellar rotation also contributes to variability in the observed brightness of a star. Star spots on the surface of magnetically active stars quasi-periodically cross the observable stellar face, imprinting semi-regular patterns in the photometric time series (Strassmeier 2009; García et al. 2010). Based on these modulations, stellar rotation periods, ($P_{\text{rot}}$), have been estimated by examining light curves from ground-based surveys (e.g. Irwin et al. 2009), the *Kepler* and CoRoT missions (e.g. Mosser et al. 2009; do Nascimento et al. 2012; Reinhold et al. 2013; Nielsen et al. 2013; García et al. 2014; Santos et al. 2019), as well as more recently the *K2* and TESS missions (e.g. Curtis et al. 2019; Reinhold & Hekker 2020). Since rotation at the surface is linked to processes occurring in the stellar interior (e.g. dynamos, turbulence) (e.g. Zahn 1992; Mathis et al. 2004; Browning et al. 2006; Decressin et al. 2009; Wright et al. 2011), there is a prospect of using rotation period measurements to probe fundamental stellar properties, as well as the magnetic and dynamical evolutionary history of stars.

Of particular value is the connection between rotation period and stellar age. As main sequence stars evolve, stellar winds transport angular momentum away from the star, slowing the rate at which it rotates (Weber & Davis 1967; Kawaler 1988; Bouvier et al. 1997). The empirical relationship between stellar age and rotation period was first realized by Skumanich (1972), which prompted the development of gyrochronology (Barnes 2003) as a tentative tool for estimating stellar ages from rotation and color alone. Recent theoretical work has focused on deriving the gyrochronology relations from stellar physics (e.g. Matt et al. 2012; Reiners & Mohanty 2012; Gallet & Bouvier 2013), open clusters, and other stellar samples, for which precise and independent measurements of both stellar age and rotation period can be made have been used to calibrate these relationships (e.g. Kawaler 1989; Barnes 2003, 2007; Cardini & Cassatella 2007; Meibom et al. 2009; Mamajek & Hillenbrand 2008; Agüeros et al. 2018; Douglas et al. 2016, 2019).

However, as illustrated in Angus et al. (2015), a robust empirical calibration of gyrochronology has proven to be challenging. Based on *Kepler* stars with asteroseismic age estimates, it is found that multiple age-period-color relationships are necessary to describe the properties of the stellar sample, which suggests that the gyrochronology relationship is under-specified. Furthermore, in Angus et al., accepted for publication in AJ, it is demonstrated that empirically calibrated gyrochronology models are not able to sufficiently reproduce the ages of rotating stars, particularly of late K- and early M-type dwarfs, which suggests that the simple gyrochronology relation as proposed by Skumanich (1972) is unable to capture the full complexity of stellar spin-down. Adding additional physics appears necessary, and in the semi-empirical modeling of rotational evolution pursued in Spada & Lanzafame (2020), it is found that including a mass and age-dependent core-envelope coupling timescale is needed to reproduce the rotation periods of stars in old open clusters (e.g. Curtis et al. 2019).

The uncertainty of gyrochronology relations have also been revealed from a theoretical perspective. For instance, van Saders et al. (2016) find that weakened magnetic breaking limits the predictive capability of gyrochronology, specifically for stars in the second half of their main sequence lifetimes. In the context of theoretical stellar rotation models, Claytor et al. (2020) determine the biases associated with the inference of stellar age from rotation period for lower main sequence stars based on current theoretical models of stellar angular momentum spin-down. Furthermore, combining theoretical models of stellar rotation with expected observational biases, van Saders et al. (2019) use forward modeling to probe rotation periods across a population of stars, finding that current models of magnetic braking fail at longer rotation periods, and that particular care is necessary to correctly interpret stellar ages from rotation period distributions.

As a result of the physical processes described above, high cadence stellar photometry contains rich information at multiple timescales about fundamental stellar properties including mass, radius, and age. Careful analysis of light curves from missions like *Kepler* and CoRoT have revealed the potential of this data and enabled the determination of stellar properties for thousands of stars. However, the imminent volume of time domain data that will be delivered by surveys like TESS and LSST necessitates the development of new methods for estimating stellar properties from shorter baseline and sparser cadence data. Automated pipelines to measure the asteroseismology parameters $\nu_{\mathrm{max}}$ and $\Delta\nu$ have been developed and applied to large samples of *Kepler* stars (Huber et al. 2009), and Bayesian methods for inferring these parameters have been tested on small ($<$ 100 stars) samples (Davies et al. 2016; Lund et al. 2017), and on $\sim$13,000 *K2* Campaign 1 stars (Zinn et al. 2019).

Automated methods for extracting rotation periods from *Kepler* photometry have also been put forth. Producing the largest catalog of homogeneously derived rotation periods to date, McQuillan et al. (2014) derive rotation periods for $\sim$30,000 main sequence stars with a peak identification procedure in the autocorrelation function (ACF) domain based on a minimum baseline of $\sim$2 years of observational coverage (see McQuillan et al. (2013)). Instead, taking a probabilistic approach, Angus et al. (2018) infer posterior PDFs (probability distribution functions) for the rotation periods of ∼1,000 stars based on a Gaussian process model. This method has the benefit of not assuming strictly sinusoidal periodicities, and compared to traditional methods it provides more robust credible intervals on the inferred rotation periods. However, Angus et al. (2018) find that the posteriors still underestimate the true uncertainties, and the method relies on computationally expensive posterior sampling. Most recently, Lu et al. in prep. implement a random forest model to predict rotation periods from light curves and *Gaia* data, with a particular focus on deriving the long periods of M-dwarfs from TESS data.

Data-driven techniques have shown promise in their capability to efficiently identify red giant branch (RGB) stars with solar-like oscillations, and to estimate fundamental stellar properties like $T_{\text{eff}}$ and $\log g$ from time domain data. Learning a generative model for RGB stars, Ness et al. (2018) use *The Cannon* (Ness et al. 2015) to model the ACF amplitude at each lag as a polynomial function of stellar properties ($T_{\text{eff}}$, $\log g$, $\nu_{\text{max}}$, $\Delta\nu$). Trained on ∼4-year baseline data, Ness et al. (2018) find the variance of their $\log g$ estimator to be $< 0.1$ dex and the variance of their $T_{\text{eff}}$ estimator to be $< 100$ K, with the information required to learn these properties being contained in ACF lags up to 35 days and 370 days, respectively, for $\log g$ and $T_{\text{eff}}$. Taking a similar approach, Sayeed et al. in prep. learn a local linear regression model between the power density at each frequency of smoothed *Kepler* power spectra and stellar properties. For upper main sequence and RGB stars that do not exhibit rotation, Sayeed et al. in prep. learn a $\log g$ estimator with a variance $< 0.07$ dex based on the 10 nearest neighbors in the frequency domain of the training set. Neural networks have also been implemented for RGB asteroseismology. Training a convolutional neural network (CNN) on an image representation of *Kepler* power spectra, Hon et al. (2017) classify RGB stars versus core helium burning stars to an accuracy of 99%, and in Hon et al. (2018b) their approach predicts $\nu_{\text{max}}$ to an uncertainty of about 5%. In Hon et al. (2018a) it is found that based on power spectra images derived from 4-year, 356-day, 82-day, and 27-day data, that the classification accuracy decreases from ∼98% based on 4-year data to ∼93% based on 27-day data.

In this work, we pursue systematically and consistently estimating a set of stellar properties directly from photometric time series data. We do this by fitting a flexible 1-dimensional (1D) CNN to the data, which is able to capture the structure of the data in the time domain on multiple scales and requires minimal feature engineering. Using a single quarter of *Kepler* data and asteroseismology-quality stellar measurements as our training set, we build models to classify stellar evolutionary state and a set of stellar properties across the RGB (including red giants and red clump stars) and main sequence from light curves of various baselines and cadences, and compare these results to models based on the ACF and frequency domain transformations of the data. The CNN classification model distinguishes RGB stars from main sequence and sub-giant stars to an accuracy of ∼90%, and for RGB stars we demonstrate that the CNN regression model trained on 27-day *Kepler* light curves is able to predict log $g$ to an rms precision of ∼0.07 dex, $\Delta\nu$ to an rms precision of ∼1.1 $\mu$Hz, $\nu_{\max}$ to an rms precision of ∼17 $\mu$Hz, and $T_{\text{eff}}$ to an rms precision of ∼300 K. For main sequence stars, we predict rotation periods up to $P_{\text{rot}}$ ∼35 days based on 27-day and even 14-day data, with an rms precision of ∼6 days. We also find that for observations spaced 1 day apart (over 97 days), we can recover $P_{\text{rot}}$ from ≈5 to 40 days with an rms precision of ∼6.2 days. Our approach, which leverages the full information content of the data, serves as a proof of concept in the pursuit of estimating stellar properties for many millions of stars from variable quality time domain data.

## 4.3 Training data

### 4.3.1 The Kepler data

To predict stellar and asteroseismology parameters from time-domain variability, we build models trained on long-cadence (29.4-minute sampling) *Kepler* data. We download all available Q9 light curves from the *Kepler* mission archive[1], which supplies ∼97 days of time-domain observations for 166,899 stars. To minimize the amount of data-processing, we train our models based on this single quarter of observations.

Light curves are often transformed to different representations, in the frequency and time-lag (i.e. ACF) domains. This is commonly done in order to extract signals that concentrate in these forms: $\nu_{\max}$ and $\Delta\nu$ from the frequency spectrum, and rotation period from the peak in the ACF. In this work, we do not collapse the data to a few measurable signatures. We leverage the entire set of observed fluxes to predict the stellar properties. It is therefore unclear as to whether a particular choice of data representation will be better than another. We examine how well we can derive stellar properties using (i) time, (ii) frequency and (iii) time-lag representations of the data. We report the differences between the approaches in Section 4.6. We note however that any preferential representation, in terms of prediction performance, may simply reflect the compatibility of the data representation, given our modeling choice.

---

[1] https://archive.stsci.edu/pub/kepler/lightcurves

### 4.3.2 Light curve processing

#### 4.3.2.1 The time domain

The flux measurements we use are the Pre-search Data Conditioning Simple Aperture Photometry (PCDSAP) flux values, which have been corrected for systematic errors and anomalies caused by the spacecraft and instrument (Twicken et al. 2010; Jenkins et al. 2010). Before transforming the time series data to different domains, we apply two additional processing steps to each light curve. First, we remove observations with a `SAP_QUALITY` flag greater than 0. We then apply a local sigma clipping algorithm to each light curve, removing observations with flux values more than three standard deviations away from the mean flux computed in a sliding window of 50 consecutive observations. Finally, we transform the light curves to be in units of relative flux, $\Delta f/f$. These three pre-processing steps are applied to the light curves before any further processing and transformation into other data domains.

For the models we build based on the data in the original time domain, we apply the following additional processing steps. First we enforce the light curves to be on a common time grid, so that the structure of the input data is standardized. Since the cadence of the *Kepler* data is mostly regular with flux measurements at every 29.4 minutes, we set any missing flux values in the time grid to zero. In Section 4.8 we discuss alternative imputation choices that can be explored, however our model is successful taking the simplest zero-imputation approach. We then normalize the relative flux values of each light curve by subtracting the mean ($\mu$) and dividing by the standard deviation ($\sigma$) of the relative flux values, so that $f_{scaled} = (f - \mu)/\sigma$. Because the standard deviation of each individual light curve provides useful information in comparing across the collection of light curves, we supply this as an additional feature to the model as discussed in Section 4.4.2.

In Figure 4.1 we illustrate how the time domain data varies for stars across the HR

diagram. In the main panel of this figure, we show (in grey) the distribution of stars in the stellar radius against effective temperature plane ($R_*$-$T_{\text{eff}}$) for the set of ∼150,000 stars from Berger et al. (2018) that have *Kepler* Q9 light curves available. The main sequence is distributed across $\log(R_*) \lessapprox 0.5 \ R_\odot$ and 6500 K $\lessapprox T_{\text{eff}} \lessapprox$ 3000 K, and the RGB is located at $\log(R_*) \gtrapprox 0.5 \ R_\odot$ and 5500 K $\lessapprox T_{\text{eff}} \lessapprox$ 3000 K with the red clump resolved at $\log(R_*) \sim$ 1 $R_\odot$ and $T_{\text{eff}} \sim$ 4800 K. Stars with $T_{\text{eff}} \gtrapprox$ 6500 K have thin convective or entirely radiative envelopes, and thus include "classical" pulsators such as delta Scuti or gamma Doradus stars. To demonstrate how the shape of the light curves vary across the regions of the HR diagram, the middle signal in the inset panels shows the time domain data for two stars with different property values in the main sequence, as well as for two stars in the RGB/red clump with different property values. For the main sequence stars (at about the same stellar radius) their light curves clearly indicate a stellar rotation signal with the hotter, upper main sequence star having a shorter rotation period of 10 days and the cooler, lower main sequence stars having a longer rotation period of 39 days. For the two RGB stars with $T_{\text{eff}} \sim$ 5000 K, we see that unlike the main sequence stars the light curves don't exhibit a rotation signal within the 97-day baseline that is shown, but that the amplitudes of the short timescale variations differs between the stars at different $R_*$. From these example light curves in the main sequence and RGB we see that the time domain data varies across the HR diagram, where stars with different properties exhibit distinctive light curves characteristics. What this suggests is that from the light curves alone we can place stars (to some degree of precision) on the HR diagram and predict other fundamental stellar properties.

#### 4.3.2.2 The ACF

In addition to working with data in the original light curve space, we also test building models based on the ACF of the time series. The ACF describes the strength of periodic signals present in time series data by measuring the similarity of the time series with itself at different lags. The ACF has been shown to be an effective domain for measuring the surface gravity of stars (Kallinger et al. 2016), the rotation periods of main sequence stars (McQuillan et al. 2014), as well as the temperatures, surface gravities, and asteroseismology observables of RGB stars (Ness et al. 2018).

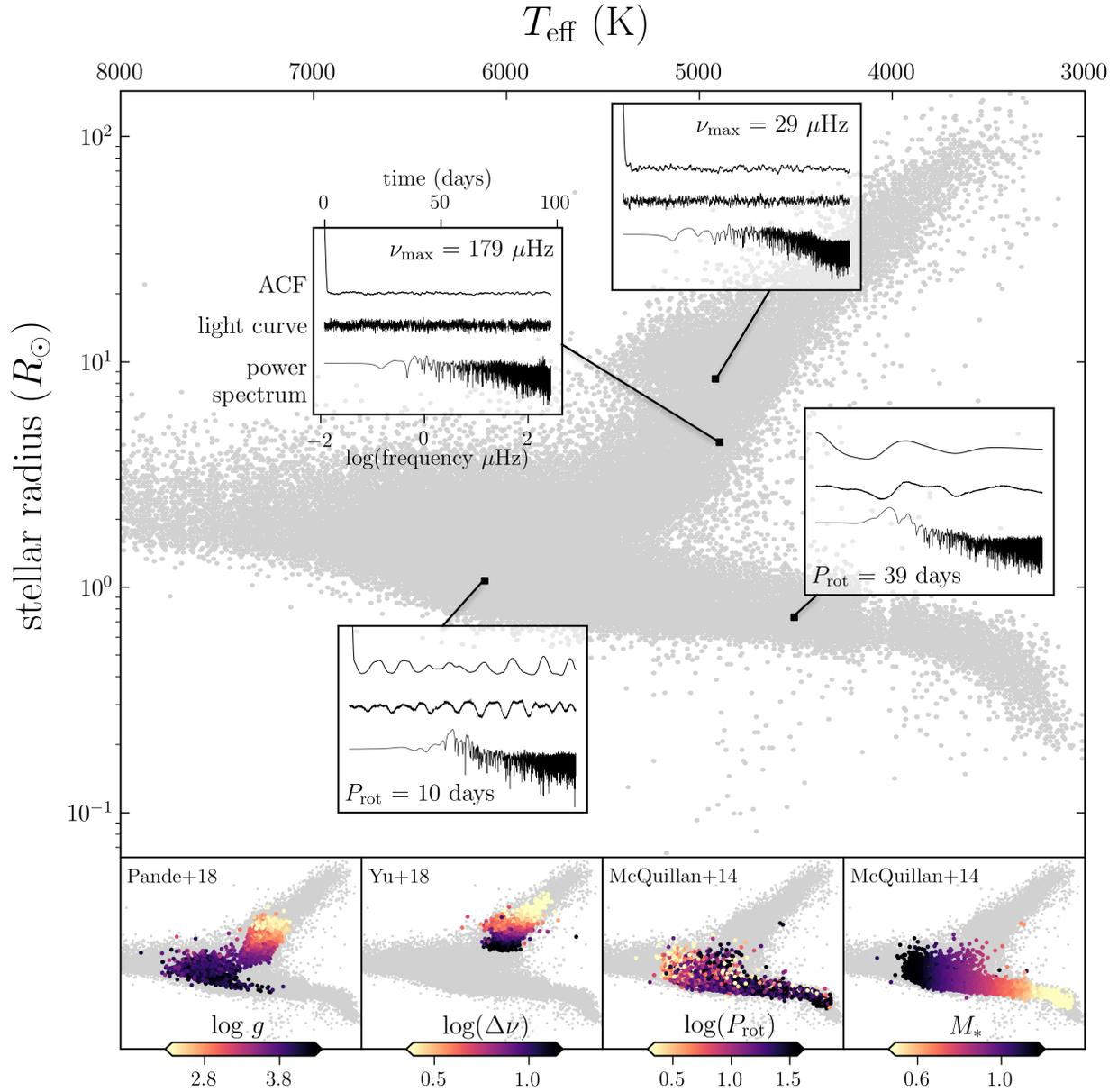

Figure 4.1: Demonstration of how the *Kepler* light curves and stellar properties jointly vary across different regions of the HR diagram, illustrating the potential to learn the fundamental properties of a star from its light curve alone. The HR diagram (shaded in grey) is from the Berger et al. (2018) catalog of derived stellar radii and $T_{\rm eff}$ for ∼150,000 stars. In the **main panel**, the insets show the original light curve, as well as the ACF and power spectrum computed from the light curve, at various positions along the RGB and main sequence. The **lower panels** indicate how various stellar properties ($\log g$, $\Delta\nu$, $P_{\rm rot}$, and $M_*$) from different catalogs also vary across the HR diagram.

For observations evenly space in time, $t_k = (k-1)\Delta t$, the ACF at each lag $k$ is,

$$\text{ACF}_k = \frac{\sum_{i=1}^{N-k}[(x_i - \overline{x})(x_{i+k} - \overline{x})]}{\sum_{i=1}^{N}(x_i - \overline{x})^2}, \qquad (4.1)$$

where the numerator is the co-variance between the time series and itself at lag $k$, and the denominator is the variance of the time series, which normalizes the ACF to be 1 at lag $k = 0$ and defined over the range $[-1, 1]$ (e.g. see Ivezić et al. 2014, Chapter 10). To compute the ACF according to Equation 4.1, we first linearly interpolate the flux of each light curve to a common, evenly spaced time grid defined from 0 to 97.4 days with a $\Delta t = 0.0204$ days (i.e. the long-cadence sampling).

The main panel of Figure 4.1 shows example ACFs for stars in the main sequence and RGB. For the two stars in the main sequence, we see that the second peak of the ACF corresponds to the rotation period of the star, with the peaks at later lags being integer multiples of the period. However, for the two RGB stars, ACF shows less visible structure. For these stars that don't exhibit strong rotation over the baseline of the data, the information contained in the ACF is more subtle. For example, granulation, as a stochastic process, is much less coherent than rotation, which results in a less structured imprint of this signal in the ACF.

### 4.3.2.3 The frequency domain

Another representation of stellar time series data is in the frequency domain. The power spectrum of a star's light curve quantifies the strength of the flux signal across a range of timescales ($T$), represented as a the spectral density ($P$) as a function of frequency ($f = 1/T$). The primary asteroseismology observables, $\nu_{\max}$ and $\Delta\nu$, are defined and identified in the power spectrum representation of stellar light curves (e.g. Bedding et al. 2010; Yu et al. 2018). For discretely sampled data the fast Fourier transform (FFT) algorithm, which represents the light curves as a summation of sinusoidal functions, is typically used to compute the power spectrum of stellar time series. However, the FFT algorithm requires that the time series be regularly sampled over the entire observation window. In the case of unevenly sampled or missing data, an alternative method for generating a frequency domain representation of time series data is to compute a periodogram as an estimate of the true power spectrum. A commonly used algorithm in astronomy is the Lomb-Scargle (LS) periodogram (Lomb 1976; Scargle 1982), which is a least squares method for detecting sinusoidal periodic signals in time series data.

To compute the LS periodogram of the *Kepler* light curve data, we use the implementation provided by the `astropy` package. Following the recommendations of VanderPlas (2018), we compute the periodogram on a frequency grid with a minimum frequency of $f_{\min}$ = 0 Hz, a maximum frequency of $f_{\max} = 1/(2\delta t)$ Hz, and a frequency spacing of $\Delta f = 1/(n_o T)$ Hz, where $T$ is the baseline of the observations (e.g. 97.39 days for Q9) and $n_o$ is the oversampling factor, which we set to $n_o = 10$. The value for the Nyquist frequency, $f_{\max}$, is a pseudo-windowing limit, where we take $\delta t$ to be most frequent spacing of the time series observations (0.0204 days).

In the main panel of Figure 4.1 we show example periodograms for stars at different locations in the HR diagram. Considering the two RGB stars, the frequency of maximum power is a prominent feature of the power spectra. For the star with the larger stellar radius, $\nu_{\max}$ is at a lower frequency of 29 $\mu$Hz while the $\nu_{\max}$ of the star with a smaller stellar radius is at a higher frequency of 179 $\mu$Hz. For the main sequence stars $\nu_{\max}$ is not visible. For these stars, the frequency of maximum power resides at frequencies greater than the range permitted by the Nyquist frequency ($\gtrapprox$ 240 $\mu$Hz). Even though $\nu_{\max}$ lies beyond the frequency grid of the power spectra for these stars, the overall shape and other features of the spectrum contain useful information that can potentially be indicative of the properties of the star.

### 4.3.3 Stellar property catalogs

There are a number of catalogs in the literature providing stellar property estimates for *Kepler* stars. Many of these catalogs have stars in common, but there is no joint database that exists. Here we try to systematically explore the intersection of several important and relevant catalogs for data-driven inference work. Figure 4.2 shows the coverage and set intersection (e.g. catalog 1 $\cap$ catalog 2) of six stellar property catalogs with the stars that have *Kepler* Q9 light curves available. As seen in the figure, the Berger et al. (2018) catalog includes a majority of the *Kepler* stars, delivering estimates of $R_*$ and evolutionary state across the HR diagram. The catalog that provides stellar property estimates for the next greatest number of stars is the McQuillan et al. (2014) rotation period catalog for $\sim$30,000 main sequence stars, and following this the Yu et al. (2018) and Pande et al. (2018) catalogs provide $\nu_{\max}$, $\Delta\nu$, $M_*$, $R_*$, and log $g$ for $\sim$13,000 stars and $\nu_{\max}$, log $g$, and $T_{\text{eff}}$ for $\sim$10,000 stars respectively, primarily for stars on the RGB. The remaining catalogs shown in Figure 4.2 provide stellar properties for fewer stars, with a minimum of $\sim$4,000 to be included in the figure.

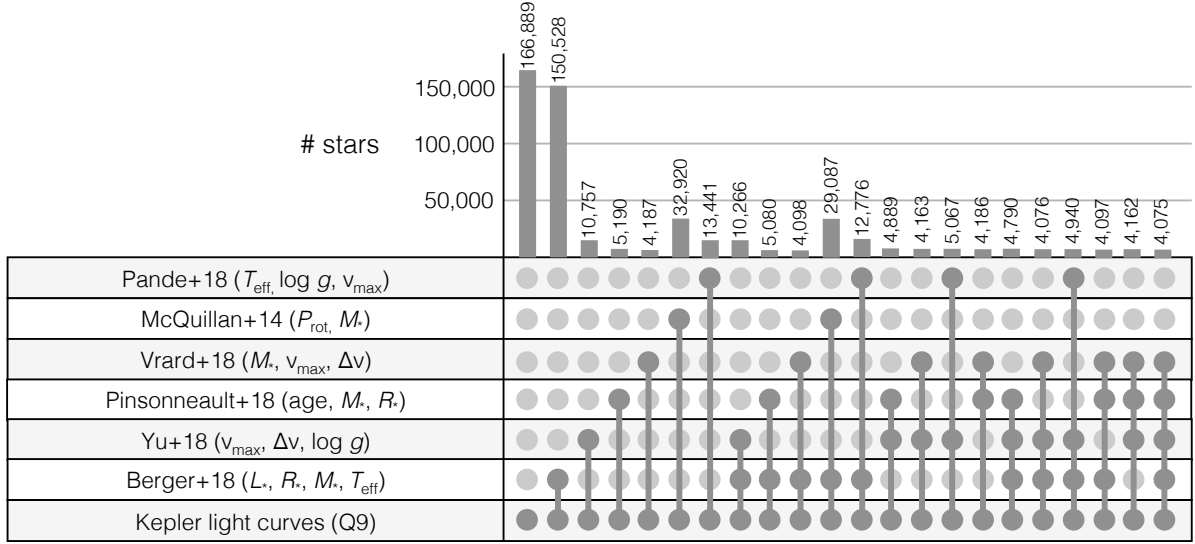

Figure 4.2: UpSet plot (Lex et al. 2014) showing the set intersections of the *Kepler* stars with Quarter 9 light curves and the various stellar property catalogs available in the literature. The histogram indicates the number of stars contained in the set defined in each column, where the shaded circles indicate which catalogs are intersected. For conciseness, only the intersections that contain a minimum of 4,000 stars is displayed. This plot demonstrates the various datasets that can be used to train a model to predict stellar properties from light curves..

As an initial proof of concept of our modeling approach, we focus on the three catalogs covering the greatest number of stars (McQuillan et al. 2014; Yu et al. 2018; Pande et al. 2018) where the stellar properties are homogeneously derived. However, models can certainly be tested on the other catalogs, as well as on a set of stellar properties combining the estimates from multiple catalogs. The stellar property catalogs compiled in Figure 4.2 demonstrates the various datasets that can be constructed and used to train data-driven models of stellar properties.

We now provide a brief description of how the stellar properties we predict in Section 4.6 are derived. For the Yu et al. (2018) sample, which includes RGB stars, we successfully recover the asteroseismology observables, $\Delta\nu$ and $\nu_{\max}$, as well as $\log g$, each which are derived as follows:

- $\Delta\nu$: derived from the *Kepler* 29.4-minute cadence data across available quarters using

the SYD pipeline described in Huber et al. (2009), which consider the light curves in both the frequency and ACF domain of the data (see Huber et al. (2009) for details). The mean of the reported uncertainties on $\Delta\nu$ is $\sim$0.05 $\mu$Hz, and the mean fractional uncertainty is $\sim$1%.

- $\nu_{\mathrm{max}}$: derived with the same pipeline as $\Delta\nu$ (see Huber et al. (2009) for details). The mean of the reported uncertainties on $\nu_{\mathrm{max}}$ is $\sim$0.9 $\mu$Hz, and the mean fractional uncertainty is $\sim$2%.

- log $g$: derived along with mass and radius from scaling relations. The mean of the reported uncertainties on log $g$ is $\sim$0.01 dex, and the mean fractional uncertainty is $\sim$0.5%.

For the Pande et al. (2018) sample, which includes RGB as well as sub-giant stars, we successfully recover $T_{\mathrm{eff}}$ and log $g$, which are derived as follows:

- $T_{\mathrm{eff}}$: taken from Mathur et al. (2017), which compiled temperatures from various sources including spectroscopic and photometric based measurements (see Mathur et al. (2017) for details). The mean of the reported uncertainties on $T_{\mathrm{eff}}$ is $\sim$140 K, and the mean fractional uncertainty is $\sim$2.5%.

- log $g$: determined from *Kepler* 29.4-minute cadence data based on an empirical relationship between log $g$, $T_{\mathrm{eff}}$ and $\nu_{\mathrm{max}}$ which has been established using the Fourier transform of the 1-minute cadence *Kepler* benchmark dataset, consisting of $\sim$500 stars (Huber et al. 2011; Bastien et al. 2013). The mean of the reported uncertainties on log $g$ is $\sim$0.25 dex, and the mean fractional uncertainty is $\sim$8%.

and finally for the McQuillan et al. (2014) sample, which covers main sequence stars, we successfully recover the stellar rotation period, and weakly recover $M_*$, which are derived as

follows:

- $M_*$: derived from the Baraffe et al. (1998) isochrone models taking $T_\text{eff}$ as input, where $T_\text{eff}$ is either from the *Kepler* Input Catalog (KIC) or Dressing & Charbonneau (2013), if available. As reported in McQuillan et al. (2014), given a $\sim$200 K precision for the $T_\text{eff}$ estimates the typical uncertainty on $M_*$ is $\sim$0.1 $M_\odot$. Assuming a 0.1 $M_\odot$ uncertainty across the entire stellar mass range, this translates to a mean fractional uncertainty of $\sim$12%.

- $P_\text{rot}$: derived from a minimum of 8 of the 12 *Kepler* 29.4-minute cadence quarters from Q3 - Q14. The rotation period for each star is identified using an automated peak identification procedure in the ACF domain (see McQuillan et al. (2013)), excluding stars from the sample that are eclipsing binaries, KOIs, and without convective envelopes ($T_\text{eff} > 6500$ K). The mean of the reported uncertainties on $P_\text{rot}$ is $\sim$0.6 days, and the mean fractional uncertainty is $\sim$3%.

## 4.4 Methods

In this section we discuss our modeling approach, as well as outline our training and evaluation procedures. The modeling code is made publicly available on GitHub at: https://github.com/kblancato/theia-net.

### 4.4.1 Modeling approach

As demonstrated in Figure 4.1, the properties of stars and the traits of their light curves vary jointly across the HR diagram. Given these correlations, our goal is to predict the properties of a star based on its light curve alone. To achieve this, the model we choose should capture the time structure of the data, that is, how each flux value is related to other values in the time series. Typically, the time structure of light curves is characterized by transforming the data to either the ACF domain or the frequency domain, described in Sections 4.3.2.2 and 4.3.2.3, respectively. After performing these data transformations, informative features in these domains are identified and used to infer stellar properties that features are known to correlate with. These data transformations require additional computational time and preconceptions of how to transform the data to produce the features of interest. Transformations of data may also result in information loss. Given these considerations, in this paper our goal is to build a model that can learn directly from the time series data itself, requiring minimal pre-processing or handcrafted engineering of the raw data.

To do this, we implement a 1D CNN to accomplish the supervised learning task of mapping light curve data to stellar properties. CNN based models have been very successfully used for many supervised learning tasks, particularly for image classification (e.g. Krizhevsky et al. 2017; He et al. 2015; Simonyan & Zisserman 2014; Goodfellow et al. 2014; Ronneberger et al. 2015). They are built from a hierarchy of artificial neural networks, known as "universal function approximators" (Hornik et al. 1990; Hornik 1991), which learn increasingly abstract representations of the input data, $\vec{X}$, by non-linearly transforming the data through a series of hidden layers that relate $\vec{X}$ to an output prediction $\vec{Y}$. CNNs are a special class of neural network architecture, that differ from fully-connected neural networks, by their inclusion of only partially connected, or so-called convolutional layers, which detect the topological structure of the input data, capturing how neighboring image pixels are related spatially, or how adjacent time series measurements are related temporally. The convolution operation relates elements of the input data to each other through weight sharing. This makes the modeling more efficient and less prone to overfitting than the fully-connected counterpart, by effectively reducing the number of model parameters that need to be learned. CNN models have been successfully used for a variety of tasks in astronomy, including the classification of galaxy morphology and properties based on galaxy images (e.g. Dieleman et al. 2015; Huertas-Company et al. 2018; Domínguez Sánchez et al. 2018), to predict characteristics of stellar feedback in CO2 emission maps (Van Oort et al. 2019; Xu et al. 2020), and to predict the 3D distribution of galaxies from the underlying dark matter distribution in large-volume cosmological simulations (Zhang et al. 2019; Yip et al. 2019).

With the CNN as our model of choice, the modeling approach we take is a so called end-to-end discriminative approach. A model is learned from a set of objects, for which the input data (light curves), and labels (stellar properties) which describe it, are both defined. The model takes the time series light curve data as an input, and through the training process learns an informative set of features from the data which optimize the stellar property predictions. This procedure requires no handcrafted transformations or feature engineering of the data as a separate procedure before model training. For the task that we tackle here of predicting stellar properties from time series data, there are a number of alternative models that can capture the time dependence of the data. We discuss an alternative method that is also suited to this problem, recurrent neural networks (RNNs), in the discussion.

### 4.4.2 Model architecture

The CNN model architecture we implement has two convolutional layers, followed by three fully-connected layers, which together perform the stellar property prediction. Given that the size of our training sets are on the order of $10^4$ examples, we define a relatively small network architecture so as to minimize the number of network parameters that need to be learned and to prevent overfitting. For comparison, AlexNet (Krizhevsky et al. 2017), with 5 convolutional layers and 3 fully-connected layers, had a total of 60 million network parameters and was trained on 1.2 million images.

Figure 4.3 is a visual representation of the model, showing the operations performed to transform the light curve data to a stellar property prediction. The left-most block represents the light curve data itself, which has been pre-processed and scaled as described in Section 4.3.2.1. The first operation applied to the time series data is a 1D convolution with one input channel, i.e. the scaled flux values at each time, and a specified number of output channels, $N_K$, which corresponds to the number of learned kernels each having its own weight matrix and bias. This makes the number of parameters to learn for each convolutional layer $[(K_W \times K_H)+1] \times N_K$, where $K_W$ is the kernel width and $K_H$ is the kernel height (in the 1D case $K_H = 1$). The addition of one accounts for the single bias parameters learned per kernel. The convolution operation takes the input vector, $\vec{X}$, of length $n(X_{\text{in}})$, and transforms it into a new vector of length $n(X_{\text{out}})$, which is computed as:

$$n(\vec{X}_{\text{out}}) = \left[\frac{n(\vec{X}_{\text{in}}) + 2 \times P - D \times (K_W - 1) - 1}{S} + 1\right], \qquad (4.2)$$

where $P$ is number of zeros padded to either side of the time series, $D$ is the dilation factor, and $S$ is the stride over which the convolution is taken. In Figure 4.3, the block to the immediate right of the light curve data represents the output of the first convolution layer, where each of the $N_K$ output channels has a length described by Equation 4.2.

After each convolution, three additional operations are performed on the data before it is passed to the next layer of the model. First, an activation function is applied to introduce non-linearities into the model. This captures the non-linear relationship between the data and the labels that describe it. To do this, we implement the commonly used ReLU (Rectified Linear Unit) activation function, defined as $max(0, \vec{X}_{\text{out}})$. Following the activation function, a pooling operation is applied. Pooling, or "down-sampling", reduces the dimensionality of the data vector that will be passed to the following convolutional layer and aids in the prevention of overfitting. The pooling operation slides over the data vector, and typically takes either the maximum or average of the data values within each window, resulting in an output vector of length $n(\vec{X}_{\text{out}}) = n(\vec{X}_{\text{in}})/K_{\text{pool}}$ when the stride is set equal to the pooling kernel width, $K_{\text{pool}}$. Lastly, batch normalization (Ioffe & Szegedy 2015) is applied to each output channel. Batch normalization solves the problem of "internal covariate shift", in which the distribution of each hidden-layer value changes during training, as the parameters of the previous layers are updated. To enforce that the distribution of the hidden layer values is similar throughout the training process, the batch normalization operation standardizes the values of each hidden layer by subtracting and dividing by the batch mean. This operation adds two new parameters for the model to learn, that weight and shift the normalized vector, but leads to faster and more stable training and also acts to regularize the model.

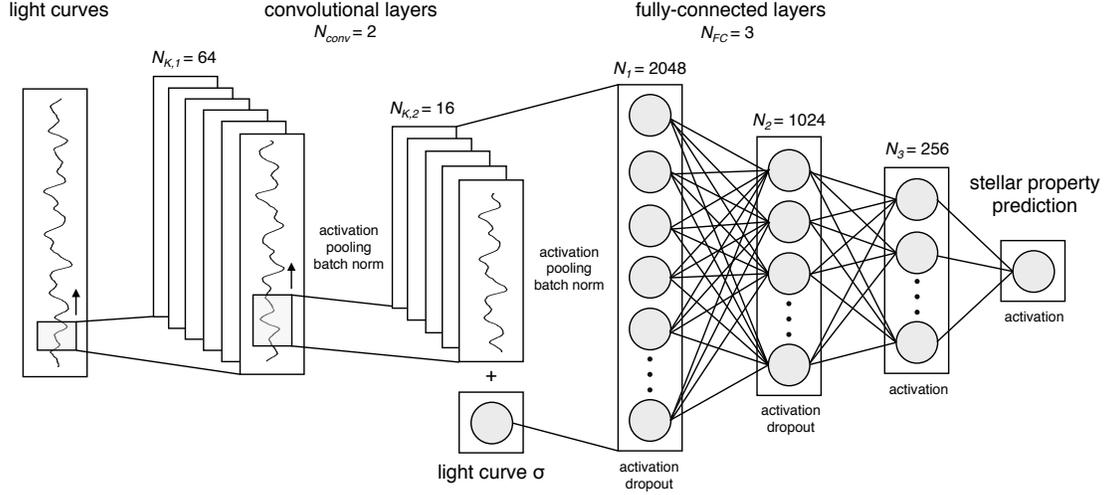

Figure 4.3: Schematic of the CNN architecture implemented to predict stellar properties from light curves. The left-most panel represents the input time series data, the next two panels indicate the two convolutional layers with different kernel widths and output channels, the fourth and fifth layers show the two fully-connected layers where each circle represents a hidden unit, and the last layer is the stellar property prediction. The symbols describing the network architecture are defined in Table 4.1. For the classification of evolutionary state, the last layer is replaced with a prediction of the probability of the star belonging to the RGB, sub-giant branch, and the main sequence.

After the activation function, pooling, and batch normalization operations, convolutions are performed on each of the $N_K$ output channels from the previous layer, which has the same properties as the first convolution, as described above. Following this second convolution operation, an activation function, pooling, and batch normalization are again applied to the data. After the two convolution layers, the output channels produced by the second convolution are flattened to a single dimension, and the data is then passed to the fully-connected part of the network. The fully-connected part of the network, represented in the last four panels of Figure 4.3, is a typical multilayer perceptron (MLP). Each element of the flattened data vector produced by the second convolution layer is mapped to $N_1$ hidden units in the first MLP layer, with each hidden unit having its own learnable weight parameter, $w_1$. With the addition of a bias parameter, $b_1$, the output of the first MLP layer is $\vec{h}(X) = f\left(\sum_{i=0}^{n(\vec{X}_{\text{out}})}(w_{1,i}X_i + b_1)\right)$ where $f$ is the specified activation function and $n(w_{1,i}) = N_1$. To this layer we also pass the scaled standard deviation of the flux values (light curve $\sigma$) for each light curve, to capture how the amplitude of the light curves vary across the sample. The second and third fully-connected layers take the output of the layer immediately preceding it, and perform the same operation with each layer learning its own set of weights and biases.

The last operation of the model architecture, shown as the right-most panel of Figure 4.3, is the prediction of the output stellar property $\vec{Y}$. In the case of regression, $\vec{Y} = \left(\sum_{i=0}^{N_3}(w_{4,i}g(X)_i + b_4)\right)$, where $n(w_{4,i}) = 1$. We experiment with one hyperparameter describing the fully-connected part of the model architecture, which is the dropout probability, $D_{\text{FC}}$, applied to the first and second MLP layers. Dropout is a form of regularization, where, during each training iteration, the values for a number of hidden units are randomly set to zero with a probability of $p$ (Hinton et al. 2012). The two dropout probabilities we consider are $D_{\text{FC}} = [0.0, 0.3]$.

The architecture described above is a smaller capacity, 1D regression version of the "vanilla" end-to-end CNN architectures that are commonly used for the task of image classification. In Table 4.1, we summarize the parameters of the model architecture, and note which (hyper)parameters we experiment with varying, which we will discuss in Section 4.4.5. In the following sections we also describe how we split the data for training, validation, and testing, and we describe our training procedure and model evaluation metrics.

### 4.4.3 Datasets

We split each of the data samples into three sets to form a training set (72%), a validation set (13%), and a test set (15%). This split of the data was chosen to include as many stars as possible in the training sets, while having at least a thousand stars in the validation and test sets to be representative of the entire parameter range. We test a 50%-25%-25% and 90%-5%-5% train-validate-test split for two parameters, $\Delta\nu$ and $P_{\rm rot}$, and find only marginal differences in model performance, with variations in the $r^2$ score of the best models being on the order of a few percent. The full Yu et al. (2018) sample includes 10,755 stars with 7,769 in the training set, 1,372 in the validation set, and 1,614 in the test set. The full Pande et al. (2018) sample includes 13,439 stars with 9,709 in the training set, 1,714 in the validation set, and 2,016 in the test set. The full McQuillan et al. (2014) sample includes 27,001 stars with 19,507 in the training set, 3,443 in the validation set, and 4,051 in the test set. Figure 4.10, in the Appendix, shows the distribution of the stellar properties we learn for each sample, including $\Delta\nu$, $\nu_{\rm max}$, and log $g$ for the Yu et al. (2018) sample, log $g$ and $T_{\rm eff}$ for the Pande et al. (2018) sample, and $P_{\rm rot}$ and $M_*$ for the McQuillan et al. (2014) sample. In forming the training, validation, and test sets, we draw stars evenly from the underlying stellar property distribution.

The training sets are used to train a given model, i.e. learn the optimal weights and biases of the network. The validation sets, which don't contribute to learning the network param-eters, are used to evaluate the performance of the network throughout the training process, as well as perform the hyperparameter selection. To prevent overfitting, we implement early stopping based on monitoring the loss of the validation set, which we describe further in Section 4.4.4. The test sets, which are independent of learning the network parameters, and are used to evaluate the performance of the model after training has been terminated. We describe the model evaluation and selection procedure in more detail in Section 4.4.6

For model training we scale the distribution of each stellar property we predict to the range [0, 1] by computing $\vec{Y}_{\text{scaled}}$ as,

$$\vec{Y}_{\text{scaled}} = \frac{\vec{Y} - \min(\vec{Y})}{\max(\vec{Y}) - \min(\vec{Y})}, \quad (4.3)$$

where this operation is performed separately for the training, validation, and test sets to prevent information leakage. In this context, information leakage refers to when the distribution of one dataset is incorrectly used to inform the scaling of another dataset, making them no longer independent, which often leads to inflated model performance.

### 4.4.4 Training procedure

The models are trained using NVIDIA Tesla GPUs. We implement our model architecture and training procedure in the machine learning library `PyTorch` (Paszke et al. 2017), which includes the `nn` module that can be used to define a variety of network architectures, as well as compute model gradients and perform tensor computations with GPU support.

For our training task, to predict continuous stellar properties, the loss function, $\mathcal{L}$, we optimize is the mean square error (MSE), which is a common choice for regression problems. The mean squared difference between true and predicted target value is computed as,

$$\text{MSE} = \frac{1}{N_{\text{batch}}} \sum_{i-1}^{N} (Y_i - \hat{Y}_i)^2, \tag{4.4}$$

where $N_{\text{batch}}$ is the number of data examples in the batch, $Y_i$ is the true stellar property of interest, and $\hat{Y}_i$ is the predicted stellar property, computed through the series of convolution and fully-connected network operations as described in Section 4.4.2.

For each model we train, the training data is batched into sets of $N_{\text{batch}} = 256$ stars. For the Yu et al. (2018) sample this results in 31 training batches, for the Pande et al. (2018) sample this results in 38 training batches, and for the McQuillan et al. (2014) sample this results in 77 training batches. During each training iteration, which includes the forward and backward pass, one batch of the training data through the network architecture and used to update the model parameters. One epoch of training has been completed once all of the training batches have been passed through the network. Batching the training data reduces the memory requirements during each training iteration, decreases the training time since the weights are updated more frequently, and acts to improve how well the model generalizes to unseen data.

The training procedure, which is typical for neural network models, can be summarized as follows: (1) forward pass of the batch through the network architecture to compute $\vec{\hat{Y}}_{\text{batch}}$, (2) compute $\mathcal{L}$ according to Equation 4.4, (3) backpropagation of $\mathcal{L}$ through each layer of the network architecture, (4) compute $\nabla_{\vec{\theta}} \mathcal{L}$, the gradient of the loss function with respect to each model parameter $\vec{\theta}$, (5) update the value of each model parameter to minimize $\mathcal{L}$. Steps 1 through 5 are repeated for every training batch iteration, and the model is trained for $N_{\text{epochs}}$ = 800 epochs or until an early stopping criterion is met. During training, we also compute the loss function for a validation set described in Section 4.4.3. At the beginning of each training epoch, steps 1 and 2 listed above are carried out on the validation dataset and the loss is monitored as the model trains. Since the validation set is not used to update the model weights, the performance of the model on this dataset is diagnostic of how generalizable the model is to new data. To combat overfitting, we implement an early stopping criterion based on the validation loss as a function of epoch. If the validation loss increases for $N_{\text{stop}}$ = 50 epochs within a tolerance of $N_{\text{tol}} = 10^{-2}$, then training is terminated and the model parameters before the validation loss increased is saved as the final model.

To update the values of the model parameters during training (i.e. step 5 above), we use `PyTorch`'s implementation of the AdamW (Adaptive Moment Estimation) optimization method (Kingma & Ba 2014), with "Decoupled Weight Decay Regularization" (Loshchilov & Hutter 2017). AdamW is an adaptive learning rate optimization method that computes individual learning rates for each model parameter based on the exponential moving average of the first and second moments of the loss function gradient, $\nabla_{\vec{\theta}}\mathcal{L}$, with two parameters $\beta_1$ and $\beta_2$, that set the exponential decay parameters for each moment. For the models we train, we fix the exponential decay parameters to their defaults in the original Adam paper of $\beta_1 = 0.9$ and $\beta_2 = 0.999$. Adam differs from traditional stochastic gradient descent, which uses a single learning rate for all parameters throughout the duration of the training process. Each model parameter, $\theta_i$, is updated at timestep $t$:

$$\theta_i^t = \theta_i^{t-1} - \alpha \frac{\hat{m}_i^{t-1}}{\sqrt{\hat{v}_i^{t-1}} + \epsilon} \tag{4.5}$$

where $\epsilon = 10^{-8}$ is typically added to promote numerical stability, $\hat{m}_i^t$ is the bias corrected exponential average of the first moment of the gradient with respect to parameter $\theta_i^t$ and $\hat{v}_i^t$ is the exponential average of the second moment of the gradient with respect to parameter $\theta_i^t$, both computed as defined in Kingma & Ba (2014).

The initial learning rate, $\alpha$, controls the step size at which the model parameters are updated. The optimal learning rate is problem specific, but is typically set in the range of $[10^{-4}, 10^0]$. Learning rates that are too low can result in training that takes many iterations to find a minimum in the loss function gradient, and without sufficient training time the parameter space may not have been explored sufficiently and a local minimum solution is returned. Learning rates that are too high can overstep the minimum in the loss function gradient and ultimately fail to converge on a desirable solution. As will be described in Section 4.4.5, we experiment with three different initial learning rate values, $\alpha = [10^{-5}, 10^{-4}, 10^{-3}]$, and choose the one that leads to the best results for predicting a given stellar property.

Lastly, to introduce regularization into the optimization routine, we add a weight decay term to the loss function described in Equation 4.4. The weight decay term, $\lambda ||\theta||^2$, is a typical L2 regularization that penalizes model parameters that become too large by a factor of $\lambda$. As will be described in the following section, we test two different weight decays parameters, $\lambda = [10^{-5}, 10^{-1}]$.

### 4.4.5 Model hyperparameters

As is evident in Table 4.1, there are numerous parameters that must be set to define the model architecture as well as the training procedure. These "hyperparameters" are parameters whose values are determined before training begins, and are not updated through the course of the training process. Since the dimensionality of the hyperparameter space is large, it is not feasible to evaluate all possible hyperparameter combinations and the effect each has on model performance. However, as an improvement beyond choosing ad-hoc or values selected empirically, we heuristically choose a small set of hyperparameters that are varied systematically, and perform a grid search over the combinations. We train one model with each hyperparameter combination defined in the grid, and select the preferred hyperparameter values based on how the model performs on the validation dataset. Limiting the search to just five hyperparameters, we test varying $K_W$, $\alpha$, $\lambda$, $\epsilon$, and $D_{\text{FC}}$. We define a grid over the values of these parameters, and train a model with each combination of hyperparameters. For each run, all other architecture and training hyperparameters are set to the values listed in Table 4.1.

Table 4.1: Model and training parameters

| | Parameter | Value(s)/Setting(s) | Description |
|---|---|---|---|
| Architecture | $N_{\text{conv}}$ | 2 | number of convolutional layers |
| | $N_{K,1}$ | 64 | number of output kernels |
| | $K_{W,1}$ | 3, 5, 6, 8, 12, 20 | kernel width |
| | $P_1$ | 4, 5, 2, 3, 1, 5 | padding |
| | $S_1$ | 3, 3, 2, 2, 2, 2 | convolution stride |
| | $T_{pool,1}$ | average | pooling type |
| | $K_{pool,1}$ | 4 | width of pooling kernel |
| | $f_{\text{conv},1}$ | ReLU | convolution activation function |
| | $N_{K,2}$ | 16 | same as above for second convolution |
| | $K_{W,2}$ | 5, 8, 10, 12, 16, 30 | — |
| | $P_2$ | 1, 0, 0, 1, 1, 1 | — |
| | $S_2$ | 1, 2, 2, 2, 1, 1 | — |
| | $T_{pool,2}$ | average | pooling type |
| | $K_{\text{pool},2}$ | 2 | — |
| | $f_{\text{conv},2}$ | ReLU | — |
| | $N_{\text{FC}}$ | 3 | number of fully-connected layers |
| | $N_1$ | 2048 | number of hidden units in fully-connected layer |
| | $f_{\text{FC},1}$ | ReLU | fully-connected activation function |
| | $D_{\text{FC},1}$ | 0.0, 0.3 | dropout probability applied to fully-connected layer |
| | $N_2$ | 1024 | same as above for second fully-connected layer |
| | $f_{\text{FC},2}$ | ReLU | — |
| | $D_{\text{FC},2}$ | 0.0, 0.3 | — |
| | $N_3$ | 256 | same as above for third fully-connected layer |
| | $f_{\text{FC},3}$ | ReLU | — |
| | $D_{\text{FC},3}$ | 0.0 | — |
| Optimization | optimizer | AdamW | — |
| | $\alpha$ | $10^{-5}, 10^{-4}, 10^{-3}$ | learning rate |
| | $\lambda$ | $10^{-5}, 10^{-1}$ | weight decay parameter |
| | $\epsilon$ | $10^{-8}, 10^{-2}$ | numerical stability term |
| | $\mathcal{L}$ | mean squared error (MSE) | loss function |
| Training | $N_{\text{batch}}$ | 256 | training batch size |
| | $N_{\text{epochs}}$ | 800 | maximum number of training epochs |
| | $N_{\text{stop}}$ | 50 | number of epochs to stop training if no improvement |
| | $N_{\text{tol}}$ | $10^{-2}$ | early stopping tolerance |

As described in Section 4.4.4, the optimal learning rate is problem specific and the consequences for choosing too low or high of a rate can result is poor model performance. Therefore, we experiment with three values for the initial learning rate, $\alpha = [10^{-5}, 10^{-4}, 10^{-3}]$. We also experiment with two values for the weight decay parameter, $\lambda = [10^{-5}, 10^{-1}]$. We prioritize varying this parameter because the amount of regularization in the optimization procedure directly impacts the values of the model parameters and controls how well the model generalizes to unseen data. We also experiment with two values of the numerical stability term $\epsilon$, testing values of both $10^{-8}$ and $10^{-2}$.

In addition to the two optimization-related hyperparameters, we also test varying one of the model architecture parameters. Motivated by our physical understanding of how information about different stellar properties are encoded at different timescales in the light curves, we decide to test varying the kernel widths of the convolution layers. Presumably, smaller kernel widths are more sensitive to information encoded on shorter timescales, while larger kernel widths will pick up information imprinted on longer timescales. We choose 8 different kernel widths to test for the first convolution layer, $K_{W,1} = [3, 5, 6, 8, 12, 20]$, which corresponds to convolution over timescales of $t_{\text{conv},1} = [.061, .102, .123, .163, .245, .408]$ days respectively. For the second convolution layer we choose larger kernel widths, with $K_{W,2} = [5, 8, 10, 12, 16, 30]$, where each element in $K_{W,2}$ is paired with its corresponding element in $K_{W,1}$. After the second kernel is applied, these kernel widths result in time series that are convolved over timescales of $t_{\text{conv},2} = [.306, .817, 1.23, 1.96, 3.92, 12.25]$ days respectively. To ensure that each convolution and pooling operation results in an integer number of output data elements, we modify the zero-padding and stride parameters for each layer as necessary. The values for $P_1$, $P_2$, $S_1$, and $S_2$ for each element of $K_{W,1}$ and $K_{W,2}$ is listed in Table 4.1.

### 4.4.6 Model evaluation and selection

As described in Section 4.4.3, we select the best model (over the grid of hyperparameters tested) based on the models performance on the validation set. The validation set is not used to train the model, and thus yields a more realistic report of how the model performs on unseen data. To assess the performance of a given model, we compute three evaluation metrics: the coefficient of determination ($r^2$), the bias ($\Delta$) and the rms. The $r^2$ score is computed as,

$$r^2 = 1 - \frac{1}{N\sigma^2} \sum_i \left(Y_i - \hat{Y}_i\right)^2, \tag{4.6}$$

where $Y$ and $\hat{Y}$ are the true and model predicted values of the dependent variable, $N$ is the number of observations in the validation or test set, and $\sigma^2$ is the variance of $\vec{Y}$. An $r^2$ score closer to 1 indicates that the model predicts the variation in $Y$ well, whereas an $r^2$ score of 0 indicates that the model does not capture any of the variation. The bias and root mean square of the estimator are computed as,

$$\Delta = \frac{1}{N} \sum_i \left(\hat{Y}_i - Y_i\right), \tag{4.7}$$

and,

$$\text{rms} = \left[\frac{1}{N} \sum_i \left(Y_i - \hat{Y}_i\right)^2\right]^{\frac{1}{2}}, \tag{4.8}$$

respectively. Both of these metrics are in the units of the stellar property, $Y$, where less bias and smaller rms values both indicate better model performance.

For each stellar property that we predict for the Yu et al. (2018), Pande et al. (2018), and McQuillan et al. (2014) samples, we train 144 models with the hyperparameter choices described in Section 4.4.5. Each of these models is trained according to the procedure outline in Section 4.4.4, with the model architecture described in Section 4.4.2. Based on the same validation set for each stellar property, we compute the $r^2$, $\Delta$, and rms for each of the 144 models we train to predict the property. For each stellar property, we select the best model according to a two-step procedure. If possible, we first we eliminate all models with bias values greater than 10% of the mean or rms values greater than 50% of the standard deviation of the stellar property validation set distribution, described as follows:

$$\Delta \;\leq\; 0.1 \times \frac{\sum_i Y_i}{N}, \tag{4.9}$$

$$\text{rms} \;\leq\; 0.5 \times \sqrt{\frac{1}{N}\sum_i (Y_i - \overline{Y})^2}, \tag{4.10}$$

after eliminating models that don't meet both of the criteria above, we then rank the models according to their $r^2$ scores. We then visually inspect the performance of the top 10 models trained for each property, and select the model with the highest $r^2$ score that doesn't exhibit structure in the true versus predicted plots for the validation set. As evident by comparing Equations 4.6, 4.7, and 4.8, the three evaluation metrics are closely related, so a high $r^2$ score is correlated with small $\Delta$ and rms values. Depending on the specific use case of the stellar property predictions, this model selection process can be easily modified to emphasize a particular or different evaluation metric. The final performance results we show in the following sections are based on the test set performance, which is data that was not used to train, validate, or select the best model.

## 4.5 Classification of evolutionary state

Before attempting the regression problem described in Section 4.4, we start with the broader task of predicting a star's evolutionary state based on its light curve. By determining a star's general location on the HR diagram, this classification task serves as an initial probe of our modeling capabilities, before we move on to the task of predicting continuous (as opposed to categorical) stellar properties. This classification model also has the utility to be used a front-end to an automated stellar property derivation pipeline.

### 4.5.1 Data

To train the classification model we build a dataset based on the overlap between the stars listed in the Berger et al. (2018) catalog and the stars with *Kepler* Q9 light curves, which includes ∼150,000 stars as shown in Figure 4.2. Stars in the Berger et al. (2018) catalog are classified into three evolutionary states; main sequence, sub-giant, or RGB, based on fitting solar-metallicity evolutionary tracks to the transition between the end of the main sequence and start of the RGB in the temperature-stellar radius plane as shown in Figure 5 of Berger et al. (2018). Of the total catalog, 67% of stars are classified as main sequence stars, 21% as sub-giant stars, and 12% as RGB stars. We randomly sample the same number of stars from the three classes to ensure a balanced classification problem, with the dataset including 13,355 stars from each the main sequence, sub-giant branch, and RGB (which includes red clump stars), totaling 40,065 stars. We split this dataset into three parts as described in Section 4.4.3, which results in 28,945 stars in the training set, 5,109 stars in the validation set, and 6,010 stars in the test set.

### 4.5.2 Methods

For the classification problem we make two main modifications to the model, one to the model architecture described in Section 4.4.2 and one to the training procedure described in Section 4.4.4. First, instead of the output of the final fully-connected layer of the model being a single value (as shown in Figure 4.3), for the classification problem the output of the model is equal to the number of distinct classes $C$ (in this case, $C=3$). To convert the model output to a prediction probability over classes we apply the softmax function, $\sigma(y)_i = e^{y_i}/\sum_{j=1}^{C} e^{y_i}$, and assign each star to the class with the highest probability. The second change we make is to the loss function. Instead of computing the mean square error described by Equation 4.4, we instead compute the cross entropy loss which is appropriate for training multi-class classification problems. The cross entropy over $C$ classes is computed as,

$$CE = \sum_{j=0}^{C}\left[-Y_c + \log\left(\sum_{j=0}^{C} \exp \hat{Y}_j\right)\right], \tag{4.11}$$

where the first term, $Y_c$, is the indicator variable of the star's true class membership, and the second term is the log of the sum of the un-normalized class probabilities $\hat{Y}_j$ over the $C$ classes output from the model. In addition to the above changes to the model architecture and loss function, we also evaluate model performance with metrics that are relevant for classification models, which are different than the metrics used in Section 4.4.6. We focus on three performance metrics; the accuracy, average precision, and the area under the receiver operator curve. The multi-class accuracy, which quantifies the number of correct predictions averaged across $C$ classes, is computed as,

$$\text{accuracy} = \frac{1}{N} \sum_{j=0}^{C} (TP_j + TN_j), \tag{4.12}$$

where $N$ is the total number of stars in the validation or test set, $TP_j$ is the number of stars correctly identified as belonging to class $j$ (i.e. true positives), $TN_j$ is the number of stars correctly identified as not belonging to class $j$ (i.e. true negatives). An accuracy closer to unity indicates better model performance.

We also compute the average precision across classes (in a one-versus-rest manner), which summarizes the precision-recall curve. Given the class probabilities output by the model as described above, different probability thresholds can be placed to define the boundary between the classes. Precision, defined as $P = TP/(TP + FP)$, where $FP$ is the number of false positives, measures how many correct predictions are made for stars belonging to a certain class at a given threshold. Recall, $R = TP/(TP + FN)$, where $FN$ is the number of false negatives, measures how many stars belonging to a classes are recovered from the total population of that class. The precision-recall curve describes the trade-off between precision and recall at different class threshold boundaries, with the best threshold being one that produces both a high precision and high recall. The average precision, computed as,

$$\mathrm{AP} = \sum_n P_n(R_n - R_{n-1}), \qquad (4.13)$$

where $P_n$, $R_n$, and $R_{n-1}$ are the precision and recall values at the $n$th and $n$th-1 probability thresholds, is the weighted mean of precisions at each recall threshold, with AP closer to unity indicating better model performance. In addition to accuracy and average precision, we also measure model performance by computing the area under the receiver operator characteristic (AUROC) curve. At different probability thresholds, the ROC curve shows the true positive rate, $TPR = TP/(TP + FP)$ (i.e. the recall), as a function of the false positive rate, $FPR = FP/(FP + TN)$, which describes the number of stars incorrectly classified as belonging to a class relative to the total number of stars that do not belong to the class. Models with low FPRs and higher TPRs indicate good performance, which corresponds to an area under the ROC curve closer to unity.

While the various classification metrics described above are related, they each emphasize different aspects of the model performance. The accuracy is the most general, measuring the fraction of total correct predictions. While this is a good overall metric of model performance, for more specific use cases of the predictions it is often not detailed enough. The precision captures how often the model is correct when the model predicts a specific class instance, which is relevant when the consequences of a false positive prediction are high. On the other hand the recall, which captures the fraction of a class that is correctly identified, is relevant when the consequences of a false negative prediction are high. Depending on the specific application of the classification model, it can be important to consider these different metrics together, and not only the accuracy alone. For example, if the classifier is used to select targets for follow up spectroscopy of one class, a classifier with high precision, but with low recall, would lead to an inefficient observing program.

For the classification problem, we perform the same hyperparameter grid search as described in Section 4.4.5, training a total of 144 models. In the following section we report all of the metrics described above, however, since our goal here is to demonstrate the general performance of the classification model, we select the best model based on which hyperparameter combination results in the best overall accuracy on the validation set. The model performance reported in the next section is on the independent test set.

### 4.5.3 Results

Figure 4.4 shows the performance of the best model we train, evaluated using the metrics described above, to classify stars as main sequence, sub-giant, or RBG based on their light curves. The middle panel of the figure shows the confusion matrix, with the true class labels along the $y$-axis and the predicted class labels along the $x$-axis. Stars that fall into the diag-onal bins are correctly classified by the model, while the stars that fall into the off-diagonal bins are incorrectly classified. As evident by the confusion matrix, the model performs the best at distinguishing RGB stars from the other evolutionary states, at an accuracy of 91%. For the remaining true RGB stars, 3% are misclassified as main sequence stars and 6% are misclassified as sub-giant stars. Examining the predictions for main sequence stars, 56% are correctly classified, while 40% of main sequence stars are misclassified as sub-giants and 4% as RGB. For the sub-giant stars, only 57% are classified correctly, while 34% are incorrectly classified as main sequence stars and 9% are misclassified as RGB stars.

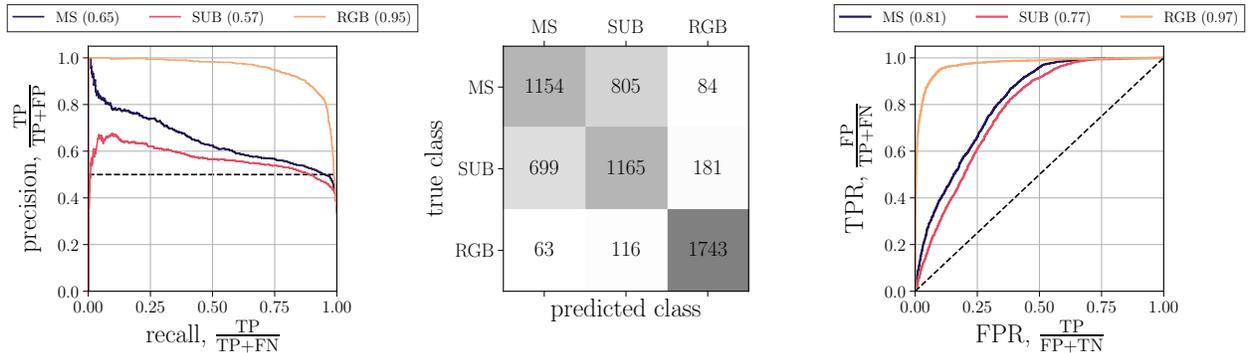

Figure 4.4: Evolutionary state classification performance of the CNN model for a test set of stars. **Left panel:** the precision-recall curve showing the one versus rest classification of stars as belonging to the main sequence (MS), sub-giant branch (SUB), or RGB. **Middle panel:** confusion matrix showing the number of false positives and false negatives for each class on the off-diagonal entries. **Right panel:** one versus rest receiver operating characteristic curve for each evolutionary state.

The precision-recall and ROC curves in Figure 4.4 show how varying the class discrimination threshold based on the prediction probabilities results in classifiers with different performance properties. The precision-recall curve shows that the RGB stars are clearly separable from the main sequence and sub-giant stars, with high precision values maintained at most recall thresholds resulting in a average precision of AP = 0.95. The main sequence and sub-giant stars exhibit worse performance, with average precisions of AP = 0.65 and AP = 0.57, respectively. The ROC curve shows similar behavior with regards to the classifica-tion performance. The TPR for the RGB stars is high across nearly the entire range of FPR thresholds, with an AUROC = 0.97. For the main sequence and sub-giant branch stars, high TPRs are only achieved along with higher FPRs. The TPR of main sequence stars reaches ∼0.95 at FPRs greater than 0.5, with an AUROC = 0.81, and the TPR of sub-giant stars reaches ∼0.95 at FPRs greater than 0.6, with an AUROC = 0.77. This is still better than the performance of a random classifier, characterized by an AUROC = 0.5.

To summarize, we find that the model does well at distinguishing between main sequence and RGB stars, but mixes up the identification of a significant portion of main sequence and sub-giant branch stars. The performance of the classification model we train likely reflects that the light curves of main sequence and RGB stars vary enough to be informative as to these evolutionary states, but that light curves vary across large regions of the HR diagram in a continuous (rather than discrete) manner. Part of the reason for the poorer results may also be related to the quality of classifications. Due to the lack of spectroscopy, Berger et al. (2018) used solar-metallicity isochrones to separate evolutionary stages, which will introduce significant noise since the exact border between main-sequence and sub-giant stars is sensitive to metallicity. In contrast, the sub-giants and red giants are clearly separated by luminosity with relatively small dependence on metallicity, thus yielding more accurate classifications.

## 4.6 Predicting stellar properties

### 4.6.1 Results of CNN stellar property recovery

In the previous section, we demonstrated the potential of using a 1D CNN model in the time domain to classify a star's evolutionary state. We now turn our attention to the main goal of this paper, which is to predict stellar properties from light curve data. As shown in Figure 4.2, there are many possible training sets that can be constructed to predict a variety of stellar properties given the catalogs that are available in the literature. Here, we focus on the three catalogs with the large numbers of stars available: the Yu et al. (2018) catalog which includes 10,757 stars, the Pande et al. (2018) catalog with includes 13,441 stars, and the McQuillan et al. (2014) catalog which includes 32,920 stars. We split each of these stellar samples into a training set, a validation set, and a test set as described in Section 4.4.3, and train individual models to predict each sample and stellar property combination according to the procedure described in Section 4.4.4. We perform the hyperparameter search as described in Section 4.4.5, and for each parameter we present the predictions resulting from the best of the 144 models trained, selected as described in Section 4.4.6.

Figure 4.5 shows the stellar property predictions for each sample's test set derived from the best trained models. For each stellar property, we show in the top panel the true stellar property value versus the model predicted stellar property value, where the one-to-one line indicates a perfect prediction. In the bottom panels, we show the fractional difference between the model predicted and the true stellar property values, as a function of the true values. The bottom panel therefore more clearly highlights the parameter space where the model is biased. In this panel we also indicate the regions of 3 and 5 standard deviations from a perfect prediction (expect for the $P_{\rm rot}$ panel which shows -0.5$\sigma$ to 1$\sigma$), when the fractional difference is equal to zero. For each stellar property the mean ($\mu$) and standard deviation ($\sigma$) of the true test set values are also indicated, as well as the model evaluation metrics, $r^2$, $\Delta$, and rms, and the fractional bias and rms in parenthesis.

First, we examine the predicted stellar properties based on the Yu et al. (2018) RGB stellar sample, showing the $\nu_{\rm max}$, $\Delta\nu$, and log $g$ predictions in the top row of Figure 4.5. As seen in the figure, we find that we recover all three of these stellar properties well, with $r^2$ scores greater than 0.95. Demonstrating the importance of the hyperparameter search, the worst performing models for these three parameters result in $r^2$ values of $\sim$0.8 - 0.85.

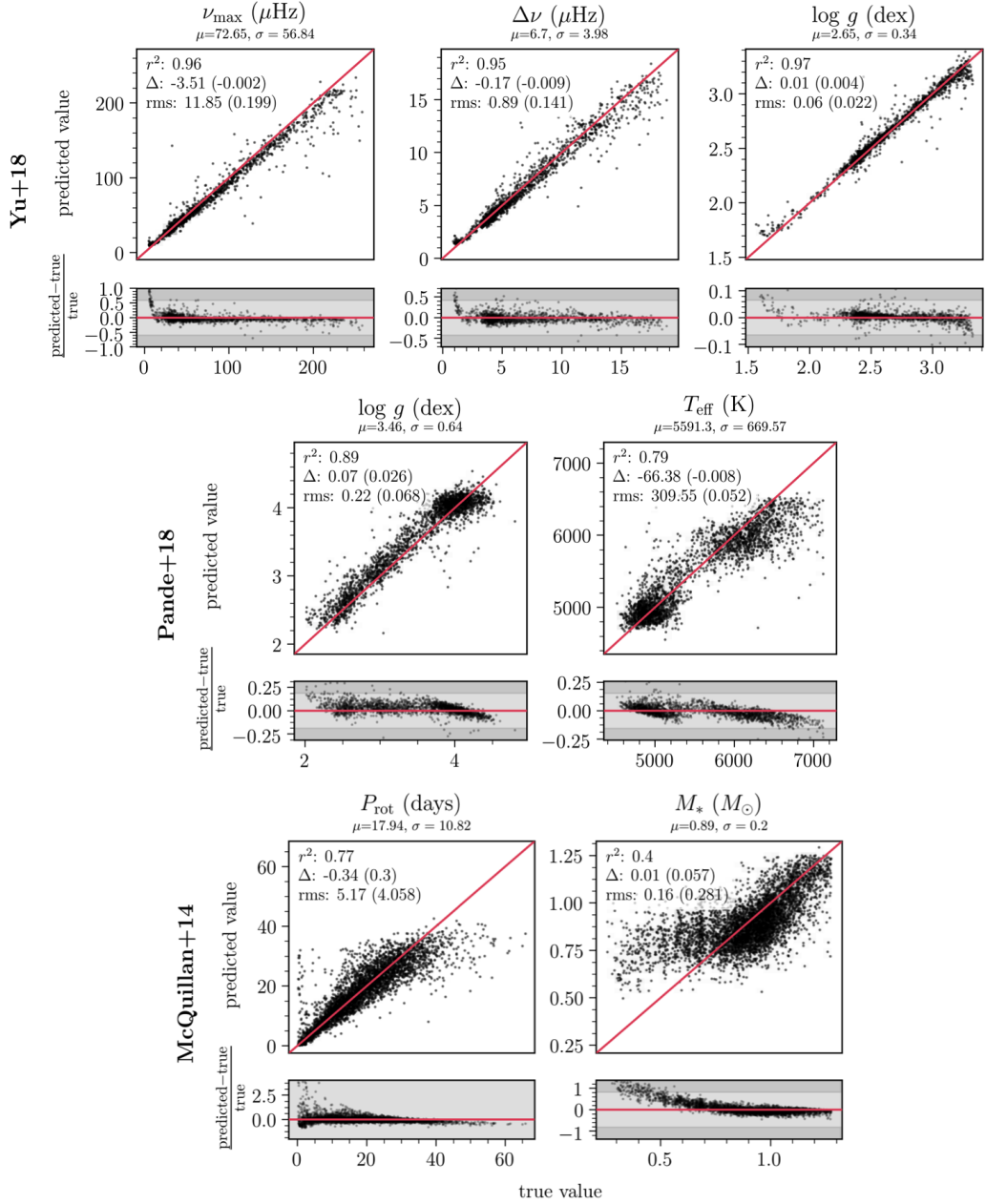

Figure 4.5: Performance of the best CNN model (as selected in Section 4.4.6) in predicting $\nu_{\max}$, $\Delta\nu$ and $\log g$ for the test set of Yu et al. (2018) RGB stars, $\log g$ and $T_{\text{eff}}$ for the test set of Pande et al. (2018) stars, and $P_{\text{rot}}$ and $M_*$ for the test set of McQuillan et al. (2014) stars. For each predicted stellar property, we show both the predicted values (top panels) and the fractional difference between the predicted and true values (bottom panels) as a function of the true stellar property. The $r^2$, $\Delta$, and rms of the predictions are indicated in each panel, as well as the bias and rms of the fractional differences shown in parenthesis. We also report the mean ($\mu$) and standard deviation ($\sigma$) of the distribution of the true property values. The red line indicates a perfect prediction and the shaded regions in the bottom panels indicate the standard deviations of the fractional difference ($3\sigma$ light grey, $5\sigma$ dark grey). For $P_{\text{rot}}$, the scale of the $y$-axis is from $-0.5\sigma$ to $1\sigma$ to show the prediction quality across the entire range of true values.

Examining the best $\nu_{\mathrm{max}}$ model, the overall bias of $\Delta$ = -3.5 $\mu$Hz is $\sim$5% of the mean of the true test set values, while the rms of the predictions is 11.85 $\mu$Hz, over the range of $\nu_{\mathrm{max}}$ values from 5 to 250 $\mu$Hz. Considering the prediction quality as a function of $\nu_{\mathrm{max}}$, we see that the predictions of $\nu_{\mathrm{max}}$ values less than $\sim$10 $\mu$Hz and greater than $\sim$150 $\mu$Hz are more biased. This is seen most clearly in the bottom panel of Figure 4.5 which shows the fractional difference, with some predictions falling in the 5$\sigma$ range (and 8 examples comprising 0.5% of the test set fall outside the plot limits). As seen in Figure 4.10 in the Appendix, the Yu et al. (2018) sample includes far fewer stars with these smaller and larger $\nu_{\mathrm{max}}$ values. This means that there are fewer examples for the model to learn this region of the parameter space well during training.

Similar to the $\nu_{\mathrm{max}}$ prediction, $\Delta\nu$ for the Yu et al. (2018) sample is also recovered well. The overall bias of $\Delta$ = -0.17 $\mu$Hz is $\sim$2.5% of the mean of the true test set values. The rms of the predictions is 0.89 $\mu$Hz over the range of $\nu_{\mathrm{max}}$ values from 0.9 to 18.8 $\mu$Hz. Similarly to $\nu_{\mathrm{max}}$, $\Delta\nu$ is biased for the smallest and largest values. For $\Delta\nu$ values less than $\sim$2 $\mu$Hz and greater than $\sim$12 $\mu$Hz, the predictions are more biased. This is seen most clearly in fractional difference plot, with a few predictions falling in the 5$\sigma$ range (and 8 examples comprising 0.5% of the test set fall outside the plot limits). Again, as with the $\nu_{\mathrm{max}}$, the Yu et al. (2018) sample includes far fewer stars with these smaller and larger $\Delta\nu$ values, which means that there are fewer examples for the model to learn this region of the parameter space well at training time.

The final stellar property we predict for the Yu et al. (2018) stellar sample is log $g$. This property is also recovered well by the best trained model, with an overall bias of $\Delta$ = 0.01 dex and an rms of the predictions of 0.06 dex, over the range of log $g$ values from 1.6 to 3.3 dex. Considering the prediction quality across the range of log $g$ values, we find again find that the predictions are more biased in the parameter space regions with fewer representative stars in the training set, as shown in Figure 4.10 of the Appendix. As evident in the bottom panel of Figure 4.5, there is more bias in the predictions for stars with log $g$ values less than 2 dex and also greater than 3.2 dex (and 11 examples comprising 0.7% of the test set fall outside the plot limits).

We now discuss the property recovery for the Pande et al. (2018) stellar sample, which, as shown in Figure 4.1, includes stars from the RBG as well as the sub-giant branch and the upper main sequence. The first property we consider is log $g$. As seen in Figure 4.5, the best CNN model recovers log $g$ with an $r^2$ score of 0.89, an overall bias of $\Delta = 0.07$ dex, which is ∼2% of the mean of the true test set values, and an rms of 0.22 dex over the range of log $g$ values from 2 to 4.8 dex. In the fractional difference plot (which excludes 6 examples comprising 0.3% of the test set given the axis limits), we see that for log $g$ values less than ∼3.75 dex, there is a systematic positive bias. This is perhaps cause by the model trying to correctly predict the larger number of less evolved stars (log $g \sim 4$) at the expense of biasing the log $g$ predictions for the red giants. Compared to the recovery of log $g$ for the Yu et al. (2018) sample of RGB stars, log $g$ is recovered less precisely for the Pande et al. (2018) sample, as evident by both the difference in $r^2$ scores between the two models, 0.89 for the Pande et al. (2018) versus 0.97 for Yu et al. (2018), as well as the higher rms of the Pande et al. (2018) model at an rms of 0.22 dex, compared to 0.06 dex for the Yu et al. (2018) sample. One reason for the difference in log $g$ prediction quality between these two stellar samples is the precision of the stellar properties used to train the models. As mentioned in Section 4.3.3, Yu et al. (2018) use asteroseismology with an uncertainty of 0.01 dex for the derived log $g$ values, while the reported uncertainty on the Pande et al. (2018) log $g$ values based on granulation is much higher, at ∼0.25 dex.

The other property we successfully predict for the Pande et al. (2018) stellar sample is $T_{\text{eff}}$. With an $r^2$ score of 0.79, the bias of the best $T_{\text{eff}}$ model is $\Delta = $ -66.4 K, which is ∼1% of the mean of the true test set values, and the rms is 310 K over a range of temperature values from 4520 K to 7123 K. As seen in the fractional difference plot of Figure 4.5 (which excludes 2 examples comprising 0.1% of the test set), the prediction quality varies across the range of values for both modes of the $T_{\text{eff}}$ distribution. For the cluster of stars with $T_{\text{eff}} \sim 5000$ K, the bias is larger at both cooler and hotter temperatures, and similarly for the cluster of stars with $T_{\text{eff}} > 5500$ K. Of the properties we've discussed so far, including both the Yu et al. (2018) and Pande et al. (2018), the prediction of $T_{\text{eff}}$ is the least precise, achieving and $r^2$ score of $\sim 0.8$ compared to $r^2$ scores greater than 0.9 achieved for $\log g$, $\nu_{\max}$ and $\Delta\nu$. This is expected due to the more indirect relation of $T_{\text{eff}}$ to the physical processes causing brightness variations. Granulation and oscillation amplitudes are predominantly determined by evolutionary state (such as $\log g$, radius and luminosity), which are only indirectly traced by the effective temperature of a star. This is particularly the case for main sequence and sub-giant stars, which can have a wide range of temperatures for a given $\log g$. This is also consistent with larger spread towards hotter $T_{\text{eff}}$ in Figure 4.5.

Finally, the last stellar sample we make predictions for is the McQuillan et al. (2014) sample, which as shown in Figure 4.1 includes stars from across the main sequence with temperatures ranging from $T_{\text{eff}} = 3500$ - 7000 K. The first stellar property we consider for this sample is rotation period, which, as discussed in Section 4.2, is of particular interest for its potential use as a probe of stellar age. With an $r^2$ score of 0.77, the bias of the best $P_{\text{rot}}$ model is $\Delta$ = -0.34 days, which is $\sim 2\%$ of the mean of the true test set values, and the rms is $\sim 5$ days over the range of periods from 0.2 to 66 days. In the bottom panel of Figure 4.5 we show the fractional difference of the predictions as a function of $P_{\text{rot}}$ spanning -0.5$\sigma$ to 1$\sigma$ from the line of perfect prediction, which excludes 35 stars comprising 0.9% of the test set. These excluded stars are fast rotators, for which we see that the prediction quality for stars with $P_{\text{rot}} \lessapprox 5$ days is the most biased. The large reported fractional metrics are inflated by the short rotation period stars that the model greatly over-predicts. Examining the sample of stars with the highest fractional differences, we find that 49 stars have fractional differences larger than 2.5, all of which have true rotation periods < 6.2 days. These 49 stars comprise ∼8% of the stars in the test set with $P_{\rm rot}$ < 6.2 days. If we remove these stars from the fractional bias and rms calculations, these metrics become 0.016 and 0.30 respectively. We suspect that most of the short period stars that the model over-predicts could be binary systems whose rotation periods, as measured in their light curves, does not reflect the true rotation periods of the stars.

In Figure 4.5 we also see that the predictions of $P_{\rm rot}$ values greater than ∼35 days become increasingly more biased. As with the predicted stellar properties for the Yu et al. (2018) and Pande et al. (2018) catalogs, a potential reason for this behavior of the model is that there are simply fewer examples of stars in the McQuillan et al. (2014) catalog with these longer rotation periods, and therefore examples for the model to learn from and be able to sufficiently learn this region of the parameter space. Since deriving stellar rotation periods is of special interest in light of upcoming photometric surveys, in Section 4.7 we investigate the ability to recover rotation periods from both shorter baseline and longer cadence time series data.

The other property we predict for the McQuillan et al. (2014) sample is $M_*$. As seen in Figure 4.5, the model predicts stellar mass well only at the upper mass range, $M_* > 0.8$ $M_\odot$, resulting in an $r^2$ of 0.7. While the bias of the model is only $\Delta = 0.01$ $M_\odot$, which is ∼1% of the mean of the true test set values, the rms of 0.16 $M_\odot$ is large compared to the range of masses covered, from 0.26 to 1.28 $M_\odot$. The fractional difference plot excludes 18 of the low stellar mass stars, comprising 0.4% of the test set, where the fractional bias of the predictions is large. We note that where the model does poorly, at $M_* < 0.7$ $M_\odot$, the density distribution of this property is underrepresented in the training objects, as seen in Figure 4.10 of the Appendix. Of the properties we present, our recovery of $M_*$ is the least successful. One reason for this could be the high fractional uncertainties associated with the $M_*$ values (∼12%), which were derived without the use of *Gaia* parallaxes. Another factor could be that the light curve data alone is not sufficient to predict this property, and perhaps adding additional information to the model, like *Gaia* distances, may improve the recovery.

### 4.6.2 Comparison to modeling in the ACF and frequency domains

As discussed in Section 4.4.1, when deriving stellar properties from photometric time series data, the light curves are often first transformed to an alternate representation of the original data. Two common representation are the ACF as described in Section 4.3.2.2 and the power spectrum as described in Section 4.3.2.3. Each of these representations highlights different features of the data, which are known to correlate with particular stellar properties. For example, peaks in the ACF are informative to stellar rotation periods, and the asteroseismic parameters of $\nu_{\max}$ and $\Delta\nu$ are defined in the frequency domain. One aim of this paper is to investigate how well a deep learning approach can learn various stellar properties from the time domain data itself, because it requires minimal feature engineering and leverages the full information content of the data.

To test how well we learn stellar properties in the time domain compared to the ACF and frequency domains, for each of the properties we predict in Section 4.6.1 we also train models to predict these properties based on the ACF and the LS periodogram. We do this for the three stellar samples of (Yu et al. (2018), Pande et al. (2018), and McQuillan et al. (2014)), as described in Section 4.3.2.2 and 4.3.2.3, respectively. Since the ACF and frequency domain already capture the time dependence of the data, we build and train fully-connected neural network models to predict stellar properties from these data representations. This means that unlike in the CNN case, which includes weight sharing to capture the time-dependence of the input data, in the fully-connected model a weight term is learned for each element of the input. Appropriately, we scale each $n^{th}$ element of the ACF and periodogram relative to the range of values exhibited by the corresponding $n^{th}$ element across all of the stars in the training set. The last four layers represented in Figure 4.3 show the fully-connected architecture we implement, where each flux measurement of the light curve is passed to its own hidden node in the first model layer, each with its own weight term.

For these models, instead of implementing different kernel widths, as for the CNN, the architecture hyperparameter we search over is the number of hidden layers, as well as the number of hidden units in each layer. We test four different model architectures; two with three hidden layers consisting of [$N_1$=2048, $N_2$=1024, $N_3$=256] and [$N_1$=4096, $N_2$=1024, $N_3$=256] hidden units, as well as two with two hidden layers consisting of [$N_2$=1024, $N_3$=256] and [$N_2$=2048, $N_3$=512] hidden units. The other hyperparameters we optimize over are the same as those in Section 4.6.1. These are the learning rate, weight decay, numerical stability term, and the dropout fraction. With the architecture choices described above, for each stellar property we train 96 models in total and select the best model as outlined in Section 4.4.6.

For the stellar properties we consider in Section 4.6.1, Figure 4.6 compares the test performance of the best model, $r^2$, $\Delta$, and rms, for the three models we train: a CNN based on the time series data, a fully-connected NN based on the ACF, and a fully-connected NN based on the LS periodogram. Compared across the three models for each property, better model performance for the evaluation metrics is indicated by darker shades of its entry in Figure 4.6, where the fractional values of the bias and rms (in parenthesis) are used to indicate model performance. In summary, the CNN model results in the best overall performance. As detailed in Section 4.6.1, taking a CNN approach, the $\nu_{\max}$, $\Delta\nu$, and log $g$ are recovered the most successfully with $r^2 > 0.9$, rotation period and $T_{\text{eff}}$ are recovered to $r^2 \sim 0.8$, and stellar mass is recovered the least successfully with an $r^2 = 0.4$.

|  | | time series (CNN) | | | ACF (FC) | | | LS (FC) | | |
|---|---|---|---|---|---|---|---|---|---|---|
|  | | $r^2$ | $\Delta$ | rms | $r^2$ | $\Delta$ | rms | $r^2$ | $\Delta$ | rms |
| Yu+18 | $\nu_{\max}$ [$\mu$Hz] | 0.957 (0.957) | 3.51 (-0.002) | 11.85 (0.199) | 0.734 (0.734) | 1.952 (0.237) | 29.303 (0.68) | 0.904 (0.904) | 6.455 (0.239) | 17.656 (0.493) |
| | $\Delta\nu$ [$\mu$Hz] | 0.95 (0.95) | 0.166 (-0.009) | 0.887 (0.141) | 0.76 (0.76) | 0.228 (0.137) | 1.952 (0.4) | 0.895 (0.895) | 0.307 (0.115) | 1.291 (0.269) |
| | $\log g$ [dex] | 0.971 (0.971) | 0.008 (0.004) | 0.057 (0.022) | 0.793 (0.793) | 0.017 (0.011) | 0.154 (0.064) | 0.907 (0.907) | 0.036 (-0.012) | 0.103 (0.037) |
| Pande+18 | $\log g$ [dex] | 0.888 (0.888) | 0.072 (0.026) | 0.215 (0.068) | 0.737 (0.737) | 0.026 (0.003) | 0.329 (0.1) | 0.598 (0.598) | 0.327 (0.105) | 0.408 (0.137) |
| | $T_{\text{eff}}$ [K] | 0.786 (0.786) | 66.378 (-0.008) | 309.546 (0.052) | 0.472 (0.472) | 82.018 (-0.008) | 486.498 (0.083) | 0.568 (0.568) | 185.168 (-0.027) | 440.021 (0.071) |
| McQuillan+14 | $P_{\text{rot}}$ [days] | 0.772 (0.772) | 0.343 (0.3) | 5.167 (4.058) | 0.757 (0.757) | 0.247 (0.544) | 5.331 (5.171) | 0.502 (0.502) | 3.985 (0.199) | 7.62 (3.974) |
| | $M_*$ [$M_\odot$] | 0.396 (0.396) | 0.009 (0.057) | 0.155 (0.281) | 0.286 (0.286) | 0.002 (0.051) | 0.169 (0.297) | 0.23 (0.23) | 0.004 (0.052) | 0.177 (0.31) |

Figure 4.6: Model performance comparison based on different transformations of the light curve data, considering the time series, ACF, and LS periodogram from left to right. For each stellar sample, the predicted stellar properties are shown along the rows, while the performance metrics ($r^2$, $\Delta$, rms) are shown along the columns. The fractional bias and fractional rms are indicated in parenthesis. Entries shaded in dark grey indicate better model performance, according to the $r^2$ score, fractional bias, and fractional rms.

### 4.6.3 Short baseline predictions

We now explore the prospect of deriving stellar properties from shorter baseline data using the 1D CNN model in the time domain. As discussed in Section 4.2, ongoing and upcoming photometric missions such as TESS and LSST will observe stars with different baselines of observations, most of which will be shorter than 97 days. In particular, the TESS mission is delivering thousands of stellar light curves with a baseline of 27 days. With the application in mind of being able to estimate stellar properties from this short baseline data, in this section we train CNN models to predict stellar properties using baselines of 62, 27, and 14 days. We do this by truncating the *Kepler* light curves to these shorter baselines.

For the shorter baseline models, we use the same sample of Yu et al. (2018), Pande et al. (2018), and McQuillan et al. (2014) stars, described in Section 4.3.3. We simply truncate the light curves to each baseline length, starting at the first observation of the Q9 *Kepler* data. For each of the baselines we test, we recompute the standard deviation of the light curve fluxes based just on the observations that fall within the specified baseline. This prevents information about the light curves at later times from mistakenly inflating model performance. Since the length of the input data $(n(\vec{X}_{\text{in}}))$ varies with baseline, we redetermine the padding and strides of the two convolutional layers of the model, while keeping the kernel widths tested in the hyperparameter search the same as described in Table 4.1. Other than these changes, the models for the baseline tests have the same architecture and training process as described in Section 4.4. The results we report are for the performance of the best model on the Yu et al. (2018), Pande et al. (2018), and McQuillan et al. (2014) test sets selected from the 144 models trained for each baseline.

Figure 4.7 demonstrates the model performance as a function of baseline for all of the stellar properties examined in Section 4.6, except for rotation period, which is specially considered as a property of interest in Section 4.7. Figure 4.11 in the Appendix reports the full list of performance metrics, $r^2$, $\Delta$, rms, fractional $\Delta$, and fractional rms, for each baseline model. From Figure 4.7 and the full list of metrics in Figure 4.11, we find that all stellar properties are recovered remarkably well using short baseline time series data. First, examining the Yu et al. (2018) stellar properties, we find that the $\Delta\nu$ recovery only degrades slightly with decreasing baseline, with an $r^2$ of 0.95 at 97 days and an $r^2$ score of 0.93 at 14 days. The rms increases slightly from 97 days down to 14 days from 0.9 $\mu$Hz to 1.1 $\mu$Hz, while the bias fluctuates marginally. The log $g$ predictions exhibit similar behavior with baseline. The $r^2$ degrades slightly from 0.97 at 97 days to 0.94 at 14 days, while the rms increases from 0.06 to 0.09 dex and the bias is nearly the same for each of the models. For the $\nu_{\rm max}$ predictions the 97- and 62-day models perform similarly with an $r^2$ score of 0.96, while the 27-day model results in and $r^2$ score of 0.91 and the 14-day model results in an $r^2$ score of 0.94, with the rms and bias of each of the models similarly following the $r^2$ score trends.

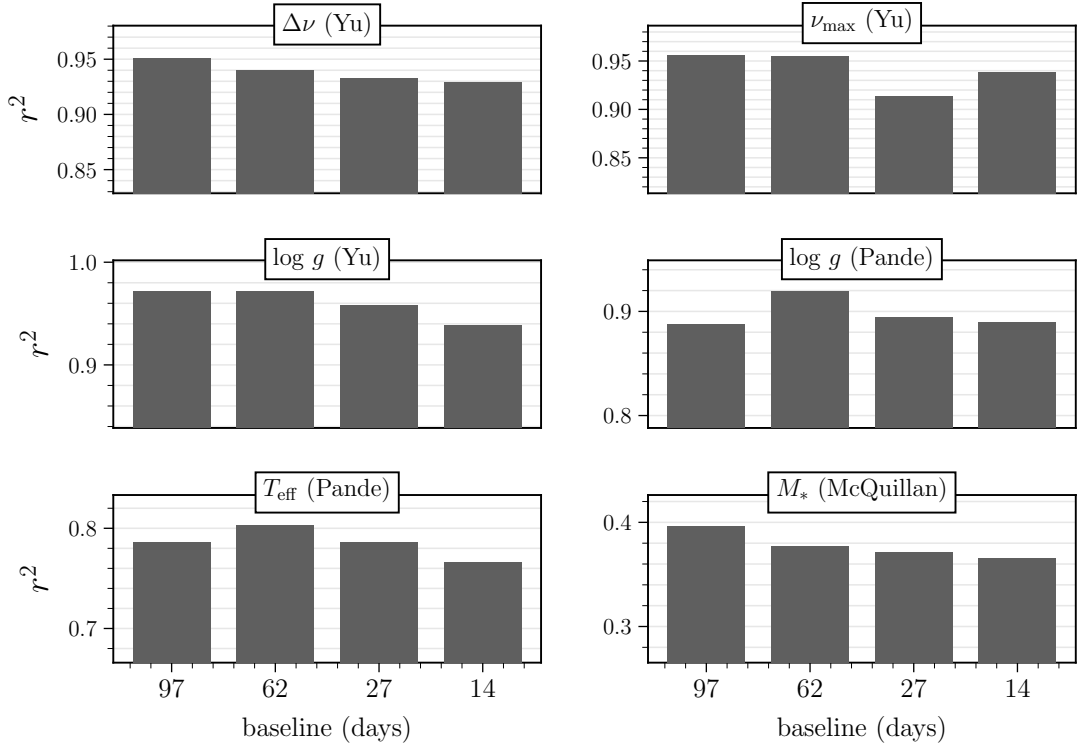

Figure 4.7: Performance of the CNN model as a function of light curve baseline for each of the stellar properties examined in Section 4.6 (except $P_{\rm rot}$, which is discussed in Section 4.7). In each panel the bar chart shows how the $r^2$ score of the best model varies with baselines of 97, 62, 27, and 14 days. The bias, rms, fractional bias, and fractional rms of each of these models is reported in Figure 4.11 in the Appendix.

Considering the Pande et al. (2018) stellar sample we find similar trends, with only a marginal decrease or fluctuations in the model performance across the tested baselines. For log $g$ the $r^2$ score of the best model is 0.92 for the 62-day baseline data, while the 97-day, 27-day, and 14-day models all result in an $r^2$ score of 0.89, with the bias and rms for each of the models being similar. For $T_{\text{eff}}$ the $r^2$ scores follow a similar pattern, with the 62-day model resulting in the highest score of 0.8, while the 97-day and 27-day models have an $r^2$ score of 0.79 and the 14-day model with an $r^2$ score of 0.77, and again the bias and rms of the four models is similar. Lastly, for the $M_*$ predictions for the McQuillan et al. (2014) sample, we see that the $r^2$ score decreases steadily from the 97-day to the 14-day model, from $r^2 = 0.4$ to $r^2 = 0.37$, with the rms slightly increasing with decreasing baseline and the bias fluctuating marginally.

The recovery of stellar properties using short baseline data suggests is that these stellar properties are still sufficiently encoded in the light curve data at these shorter timescales. This results of Figure 4.7 are promising for the prospects of estimating these stellar properties from light curves from surveys such as TESS and LSST. In the following section we explore a similar prospect for the recovery of stellar rotation period, demonstrating how the recovery of $P_{\text{rot}}$ changes with baseline, as well as with the cadence of the observations.

## 4.7 Rotation period of main sequence stars

### 4.7.1 Baseline

We now focus on stellar rotation as a key stellar property, particularly for gyrochronology studies, and examine the prospects of deriving $P_{\text{rot}}$ from light curves with baselines less than 97 days (Section 4.7.1), as well as cadences longer than 29.4 minutes (Section 4.7.2). For the full list of stellar properties examined in Section 4.6, similar cadence tests can also be performed. However, to limit the scope of this paper, we omit this examination.

First, we investigate how well rotation periods can be recovered from light curves as a function of the observation baseline. In addition to the 97-day rotation model trained in Section 4.6, we train three additional CNN models based on light curves with baselines of 62, 27, and 14 days. The 27-day model is of particular interest, as most stars that will be observed by the TESS mission will have observations spanning 27 days. For these shorter baseline models, we prepare the *Kepler* Q9 light curves and modify the CNN padding and stride values for the hyperparameter search the same as described in Section 4.6.3. Further demonstrating the necessity of the hyperparameter search, a number of the short baseline $P_{\rm rot}$ models result in $r^2$ scores of ∼0.5, which is significantly worse than the performance of the best models presented here.

Figure 4.8 demonstrates how the recovery of stellar rotation period degrades as a function of light curve baseline, with the top panel of the figure summarizing the performance of the models by showing how the $r^2$ decreases with decreasing observation lengths. We find that the $r^2$ score changes by less than $\Delta r^2$ = -0.1, from $r^2$ = 0.77 at a baseline of 97-days, to $r^2$ = 0.69 at a baseline of 14-days, with the 27-day "TESS"-baseline model resulting in an $r^2$ of 0.74. Examining the model performance in more detail, for each baseline the bottom half of Figure 4.8 shows how both the light curve predicted (top panel) and the fractional difference between the predicted and true rotation period (bottom panel) vary as a function of the true rotation period for the test set of (McQuillan et al. 2014) stars. For ease of comparison, the 97-day model from Figure 4.5 is also included. Across the four baselines considered, the bias of the models computed across the entire range of true rotation periods remains $|\Delta|$ < 0.6 days without a clear trend with baseline, while the rms of the models increase with decreasing baseline by ∼1 day from the ∼5 days for the 97-day model to ∼6 days for the 14-day model.

Visually inspecting the bias and rms of the models as a function of the true rotation period, we find that for each baseline, the rms increases as rotation period increases, while the fractional rms decreases marginally by 0.05. In general, shorter rotation periods are recovered with less bias, except for the fraction of fast rotators whose rotation periods are over-predicted, as discussed in Section 4.6.1. In the fractional difference plots for the models shown in Figure 4.8, which span -0.5$\sigma$ to 1$\sigma$ along the $y$-axis, 35, 41, 32, and 42 short period stars (comprising just ~1% of the test set) are excluded from the plots for the 97-day, 62-day, 27-day, and 14-day models respectively. As in Section 4.6.1, if we remove the short rotation period stars with the highest fractional differences, the fractional bias and fractional rms metrics decrease significantly. For the 62-day model, removing the over-predicted ~9% of the stars with true rotation periods $< 6.2$ days results in a fractional bias and fractional rms of 0.012 and 0.30 respectively. For the 27-day model, removing the over-predicted ~6% of the stars with true rotation periods < 6.5 days results in a fractional bias and fractional rms of 0.032 and 0.36 respectively. And lastly for the 14-day model, removing the over-predicted ~7% of the stars with true rotation periods < 8 days results in a fractional bias and fractional rms of 0.081 and 0.38 respectively.

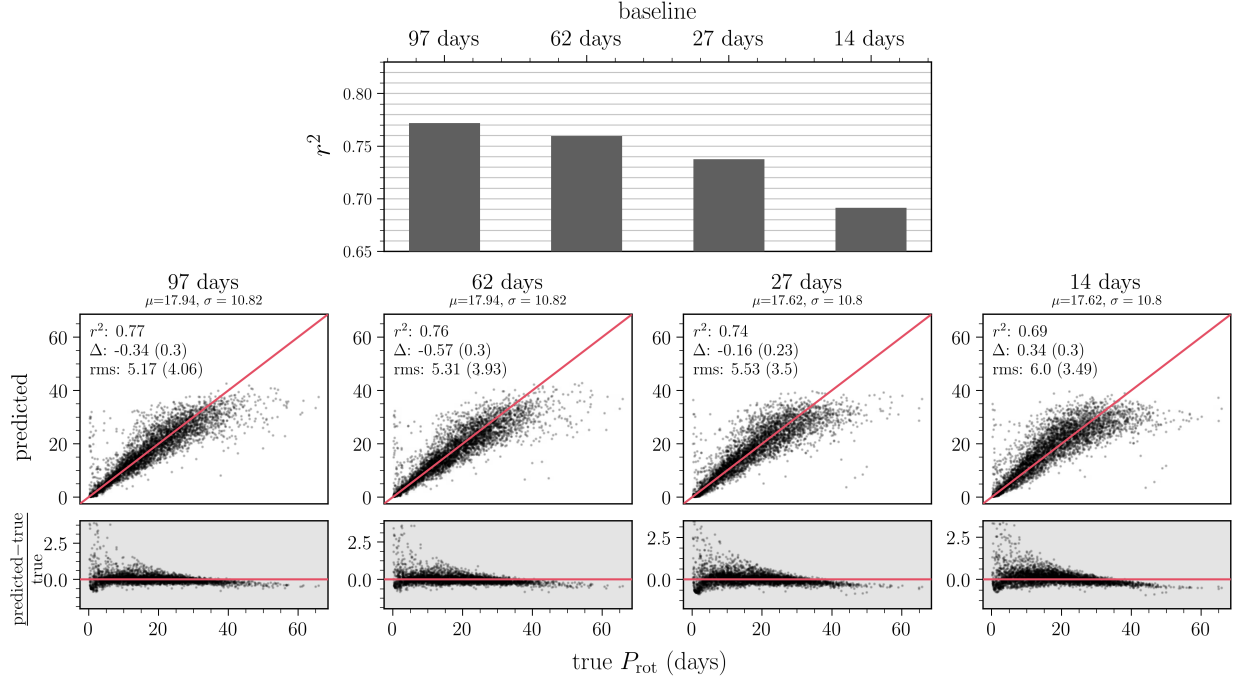

Figure 4.8: Performance of the CNN model based on different light curve baselines for a test set of McQuillan et al. (2014) stars. **Top panel:** summary of model performance showing the $r^2$ score versus the tested baselines of 97, 62, 27, and 14 days. **Bottom panels:** the predicted versus true rotation period, and the fractional difference between the predicted and true values, for models based on each light curve baseline for the test set of stars. The $r^2$, $\Delta$, and rms of the predictions are indicated in each panel, as well as the fractional bias and fractional rms in parenthesis. As discussed in the text, the fractional metrics for the $P_{\rm rot}$ predictions is greatly inflated by the over-prediction of short period stars. If we remove these stars with fractional differences $> 2.5$, the fractional bias and fractional rms decrease significantly.

Another feature of the model performance we notice is that as the baseline decreases, the bias of the predictions at rotation periods > 35 days marginally increases. While all of the models exhibit this behavior to an extent, as evident in the bottom panels of Figure 4.8, the 27-day and 14-day models in particular do not predict rotation periods greater than ~35 days, with slowly rotating stars in the test set having their rotation periods under-predicted. The degradation of the predictions for stars with rotation periods longer than ~35 days could be due to a number of factors. One such factor is that for these more slowly rotating stars, fewer cycles of the rotation period are imprinted in the light curves, and in some cases only a fraction of one full rotation period is present. However, from Figure 4.8, we see that even for rotation periods longer than the baseline, the model can still recover rotation, although at decreasing precision. Another factor that could be impacting the model's ability to precisely recover rotation periods longer than ~35 days is the distribution of the training data. The mean rotation period of the (McQuillan et al. 2014) sample is ~18 days with a standard deviation of ~11 days. As seen in Figure 4.10, there are few stars rotation periods longer than ~35 days, predominantly due to the fact that instrumental systematics in the *Kepler* data become more prominent on monthly timescales. The model could be less effective in predicting long rotation periods also because there are few examples to learn from. Given a more complete rotation period coverage in the training data, the model may be better able to learn the rotation periods of more slowly rotating stars, making unbiased predictions even beyond the baseline of the data.

### 4.7.2 Cadence

In addition to baseline, we also investigate how well we can recover rotation periods from light curves as a function of cadence. We train a CNN to predict rotation periods from light curves with an observation every 2 hours, 10 hours, and 24 hours, all with a baseline of 97 days. Being able to measure rotation periods from less frequently sampled photometric time series is also of interest, as upcoming surveys will observe a substantial number of stars at much sparser cadences than those of the *Kepler* 29.4-minute sampling. In particular, the cadence of LSST observations will be irregular, and the minimum separation between subsequent observations is tentatively ∼3 days.

To modify the light curve data for the cadence models, we again use the same sample of McQuillan et al. (2014) stars described in Section 4.3.3, but instead of using the full light curve we only select every 4th, 21st, and 49th flux observation to achieve light curves with cadences of 2 hours, 10 hours, and 24 hours, respectively. As we did for the baseline models, to prevent information leakage, we compute the standard deviation of the light curves based just on the selected $n$th observations for each tested cadence, which is passed to the first fully-connected layer of the model as shown in Figure 4.3. Since the length of the input data also varies with cadence, we redetermine the padding and strides of the two convolutional layers. For the 2-hour and 10-hour models, we test the same kernel widths as listed in Table 4.1. However, for the 24-hour model, since there are only 98 flux observations for each light curve, instead of testing kernel widths of $K_{W,1} = 12$ and $K_{W,1} = 20$, we test two additional smaller kernel widths of $K_{W,1} = 2$ and $K_{W,1} = 4$. These have a corresponding $K_{W,2} = 3$ and $K_{W,2} = 7$, respectively. For the 10-hour and 24-hour models, we also modify the size of the fully-connected part of the model. For the 10-hour cadence data with an input size of 228 observations, we change the size of the first fully-connected layer to have $N_1 = 512$ hidden neurons, the second fully-connected layer to have $N_2 = 256$ hidden neurons, and the last fully-connected layer to have $N_3 = 128$ hidden neurons. For the 1-day cadence data with an input size of 98 observations, we change the size of the first fully-connected layer to have $N_1$ = 256 hidden neurons, the second fully-connected layer to have $N_2 = 128$ hidden neurons, and the last fully-connected layer to have $N_3 = 64$ hidden neurons. Other than modifying the kernel widths explored and the reduction of the capacity of the fully-connected part of the model, the cadence models have the same architecture and training process as described in Section 4.4. The results we report here are the performance of the best model on the McQuillan et al. (2014) test set selected from the 144 models trained for each cadence.

Figure 4.9 demonstrates how the recovery of rotation period degrades as a function of the light curve cadence, with the top panel of the figure summarizing the performance of the models by showing how the $r^2$ decreases with sparser cadence. As seen in the figure, in terms of the $r^2$ score, the 0.5-hour, 2-hour, and 10-hour cadence models have nearly identical model performance with $r^2$ values of 0.77, 0.78, and 0.76, respectively. However, for the 24-hour cadence light curves, the performance of the model decreases to an $r^2$ of 0.67. Examining the performance of the model on the McQuillan et al. (2014) test set in more detail, for each cadence the bottom half of Figure 4.9 shows how both the light curve predicted (top panel) and the fractional difference between the predicted and true rotation period (bottom panel) vary as a function of the true rotation period. For ease of comparison, the 29.4-minute model from Figure 4.5 is also included. Considering the four cadences tested, the bias of the models computed for the entire range of true rotation periods remains $|\Delta| < 0.7$ days, without a clear trend with cadence. The rms of the 0.5-hour, 2-hour, and 10-hour models are all ∼5.2 days, while the rms of the 24-hour model is somewhat larger, at ∼6.2 days, and the fractional rms increases marginally by ∼0.06 from the 0.5-hour to 24-hour cadence models. If we remove the short rotation period stars with the highest fractional differences, the fractional bias and fractional rms metrics decrease significantly. For the 2-hour model, removing the over-predicted ∼7.5% of the stars with true rotation periods $< 6.2$ days results in a fractional bias and fractional rms of 0.10 and 0.32 respectively. For the 10-hour model, removing the over-predicted ∼9.4% of the stars with true rotation periods $< 5.6$ days results in a fractional bias and fractional rms of -0.02 and 0.27 respectively. And lastly for the 24-hour model, removing the over-predicted ∼25% of the stars with true rotation periods $< 5.2$ days results in a fractional bias and fractional rms of 0.042 and 0.40 respectively.

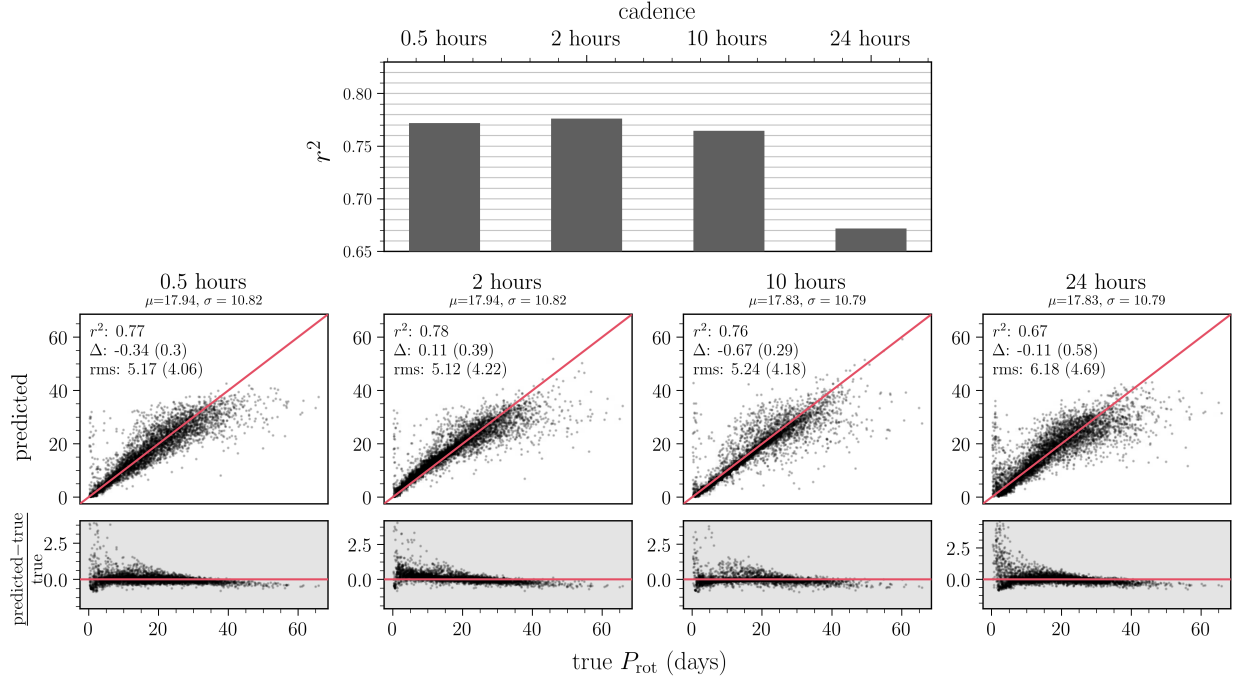

Figure 4.9: Performance of the CNN model based on different light curve cadences for a test set of McQuillan et al. (2014) stars. **Top panel:** summary of model performance showing the $r^2$ score versus the tested cadences of 0.5 hours, 2 hours, 10 hours, and 24 hours. **Bottom panels:** the predicted versus true rotation period, and the fractional difference between the predicted and true values, for models based on each light curve cadence for the test set of stars. The $r^2$, $\Delta$, and rms of the predictions are indicated in each panel, as well as the fractional bias and fractional rms in parenthesis. As discussed in the text, the fractional metrics for the $P_{\rm rot}$ predictions is greatly inflated by the over-prediction of short period stars. If we remove these stars with fractional differences $> 2.5$, the fractional bias and fractional rms decrease significantly.

Inspecting the bias and rms of the models as a function of rotation period, we see that the 0.5-hour, 2-hour, and 10-hour cadence models all perform similarly across the range of rotation periods, with the shorter and longer rotation periods being recovered less precisely than rotation periods spanning ∼5-35 days, as with the baseline models discussed in Section 4.7.1. In the fractional difference plots shown in Figure 4.9, 35, 31, and 44 short rotation period stars are excluded by plot limits of the 0.5-hour, 2-hour, and 10-hour cadence models, which comprise just ∼1% of the test set. Of the cadences we test, the 24-hour model is the only model that exhibits a markedly different performance behavior across the range of rotation periods. We take note of two significant differences in the performance of this model compared to the models trained on the higher cadence light curves. The first is that similarly to the shorter baseline models, the model trained on the 24-hour cadence light curves is biased at longer rotation periods, with rotation periods longer than 40 days being under-predicted. For this model we also find that the rotation periods of a larger fraction of more quickly rotating stars, with $P_{\rm rot} < 5$ days, are predicted incorrectly. As evident in fractional difference plot for the 24-hour model in Figure 4.9, the rotation periods for a more significant number of fast rotating stars are over-predicted, in some cases by up to ∼40 days. Additionally, 89 short period stars (comprising ∼2% of the test set) are excluded by the plot limits due to their large over-predictions. The more significant degradation in model performance for the stars with the fastest rotation periods makes sense given that one full rotation cycle for these stars is only sampled a few times when the cadence is as sparse as 24 hours.

## 4.8 Discussion

We have systematically explored the recovery of a set of stellar properties from photometric time series data, examining the prediction performance across different baselines and cadences. Our approach, using a 1D CNN model, requires minimal data-processing and no by-hand feature engineering. Using the *Kepler* 97-day, 29.4-minute Q9 data, we first construct a CNN classification model to predict stellar evolutionary state. We then use three catalogues to build training sets to predict the continuous stellar properties of: $\Delta\nu$, $\nu_{\mathrm{max}}$, $\log g$, $T_{\mathrm{eff}}$ and $P_{\mathrm{rot}}$. We implement a CNN regression model, optimizing over a grid of possible hyperparameters, to successfully recover these properties to a high fidelity across the parameter space of the training examples.

Our CNN modeling approach is demonstrative of the information content in the data and how this information is preserved across various baselines and cadences. We expect that our modeling choice is not the primary limitation in our prediction precision. Rather, we expect the information contained in the data, the precision on the input properties and the stellar property range of the training data are the primary drivers of our results. Nevertheless, there are several alternative types of models that can capture the structure of time series data like light curves. Recurrent Neural Networks (RNNs), for example, are a well suited class of neural networks that model temporal data structure, using a recurrence relationship between new outputs of the model and the previous states of the model. Similar to CNNs, RNNs include weight sharing from different parts of the time series throughout the model training process (Lipton et al. 2015).

Indeed one downside of CNN models, compared to RNN models, is that the shape of the input data must be similar across the entire dataset. CNNs do not lend themselves to working with unevenly sampled data, and as discussed in Section 4.3.2.1, to overcome this issue we take the simplest approach and replace missing flux time steps in the time series with zeros values. This results in good model performance, however more well-motivated imputation approaches could be tested, including: simple interpolation of the light curves fluxes between points as done for the ACF processing, modelling-based imputation, providing a missing value mask a second input channel to the CNN, or testing the use of alternative modeling approaches like RNNs, however RNN models are typical more difficult to train than CNN models. We leave exploring these options as a task for future work. This paper establishes a baseline of performance expectation, with the adopted model, hyperparameter optimisation choices and other assumptions like the zero-imputation of missing data.

Another alternative choice is to take a generative, rather than a discriminative, approach to the modeling presented here. Generative models take a probabilistic approach to learning the joint distribution of the data ($X$) and label ($Y$), $P(X,Y)$, which is then used to infer $P(Y|X)$. Generative models also include methods that learn a (typically lower dimensional) latent representation of the data itself from which the original data vector can be generated (e.g. variational autoencoders), with the latent space informing $P(Y|X)$. A generative approach lends itself better to understanding the data generation process, which is arguably more aligned with science goals than a discriminative approach. However, as discussed in Ismail Fawaz et al. (2018), generative models are typically less accurate than discriminative models when it comes to performance on a specific task, and generative models for times series data are not trivial to implement in practice. Future work could certainly include taking a generative approach to the problem. This would perhaps promote understanding of the data generation process, as well as enable the derivation of well-motivated, datapoint-by-datapoint probability distributions for the inferred labels using fast inference methods like variational inference (Blei et al. 2016), delivering effective errors on the stellar properties inferred for each individual star.

Lastly, the discriminative end-to-end nature of many deep learning approaches (including our own) makes these models difficult to interpret, often hindering their ability to be useful for understanding the underlying physical theory. There are numerous definitions of what it means for deep learning to be interpretable, as well as numerous proposed methods for making some of these interpretations (e.g. Simonyan et al. 2013; Yosinski et al. 2015; Binder et al.

2016; Montavon et al. 2017). Specifically, for our case of predicting stellar properties from time series data, it would be insightful to know what features of the light curves contributed most to the prediction of a particular stellar property. For instance, if we could identify the most relevant variation timescales for making a prediction, we could make connections between the quality of the stellar property recovery and our physical understanding of stellar physics. This would allow us to confirm existing theories of how the internal physical processes of stars are imprinted in light curves, as well as have the potential to reveal new connections between stellar properties and stellar variability in the time domain.

## 4.9 Conclusion

We have implemented a 1-dimensional convolutional neural network (CNN) architecture to estimate stellar properties from photometric time series data. Constructing training sets based on the 29.4-minute cadence *Kepler* Q9 data and high-quality stellar property catalogs, we predict evolutionary states, stellar properties ($T_{\text{eff}}$ and $\log g$), asteroseismic parameters ($\Delta\nu$ and $\nu_{\text{max}}$) and rotation periods ($P_{\text{rot}}$) for main sequence and red-giant stars. We compare the quality of predictions based on learning directly from the time series data to learning from transformations of the data, including the ACF and the frequency domain. We also examine how the prediction quality varies with the baseline of observations, training models based on 97, 62, 27, and 14 days of data. For rotation period, which is of particular interest for gyrochronology, we further examine how the prediction quality varies with the cadence, training models based on time series data with an observation every 0.5, 2, 10, and 24 hours. The main results of this work are summarized as follows:

- Training a CNN model to classify stellar evolutionary state, we are able to distinguish red giant stars from main sequence and sub-giant stars to an accuracy of ∼90%.

However, the model is not as successful at distinguishing between main sequence and sub-giant stars, with each of these stellar types having a classification accuracy < 60%. We suspect that this is due to the more subtle physical differences (and how these are manifested in the time domain) between main sequence and sub-giant stars and the limited quality of the training labels, as the border between main sequence and sub-giant stars is sensitive to metallicity.

- Based on one quarter of *Kepler* long-cadence data, our CNN regression model recovers $\nu_{\rm max}$ and $\Delta\nu$ to an rms (fractional rms) precision of $\sim$12 $\mu$Hz (0.2) and $\sim$0.9 $\mu$Hz (0.14), respectively, and log $g$ to an rms precision of $\sim$0.06 dex (0.02), for red giant stars trained with the Yu et al. (2018) catalog. Using the Pande et al. (2018) catalog as a training set, we predict $T_{\rm eff}$ with relatively little bias across evolutionary states (with $T_{\rm eff}$= 4500 - 6500 K), to an rms precision of $\sim$300 K (0.05). We also predict log $g$ across the range log $g$ = 2 - 4.5 dex to an rms precision of $\sim$0.22 dex (0.07). This performance is in part limited by the precision of the training labels, with a mean reported uncertainty of $\sim$0.25 dex (compared to 0.01 dex for the red giant log $g$ estimates).

- For main sequence stars, based on a single quarter of *Kepler* long-cadence data, our CNN regression model predicts rotation periods unbiased from $\approx$5 to 40 days, with an rms precision of $\sim$5.2 days (4.06). Our model becomes biased in parameter spaces with few training examples in the McQuillan et al. (2014) catalog (e.g. $P_{\rm rot} \gtrapprox 40$ days), and over-predicts $\sim$8% of short-period stars in the test set with $P_{\rm rot} < 6.2$ days. Removing these short-period stars, the fractional rms of the predictions is 0.3. We also imprecisely predict stellar mass without bias for $M_* > 0.7$ $M_\odot$, with an rms precision of $\sim$0.16 $M_\odot$ (0.28). The performance is in part limited by the large input uncertainty on mass values.

- For the stellar properties listed above, we compare the performance of the CNN model based on the 97-day time domain data to fully-connected neural network models based on the ACF and frequency domain representations of the same data. We find that the CNN model trained on the original time series data outperforms the models based on the other two data representations. This implies that more information can be gleaned from deep learning models that work closer to the raw data, as transformations and feature engineering of data often results in lost information.

- To inform expectations of what can be delivered from observations made by TESS, LSST, and future missions, we train our CNN model to recover stellar properties from light curves with shorter baselines (62, 27, and 14 days). We find that we can predict stellar properties remarkably well for TESS-like data (27-day baseline), including, for red giant stars, $\log g$ to an rms precision of $\sim$0.07 dex (0.03), $\Delta\nu$ to an rms precision of $\sim$1.1 $\mu$Hz (0.18), and $\nu_{\mathrm{max}}$ to an rms precision of $\sim$17 $\mu$Hz (0.3). Based on the Pande et al. (2018) training set, we predict $\log g$ to an rms precision of $\sim$0.21 dex (0.06) and $T_{\mathrm{eff}}$ to an rms precision of $\sim$300 K (0.05).

- We predict rotation periods for main sequence stars up to $\approx$35 days days based on 27-day and even 14-day data, with an rms precision of $\lessapprox$ 6 days (3.5). Removing the $<$ 10% of short-period stars that are over-predicted, the fractional rms is $<$ 0.38. Investigating light curves with longer cadences (2, 10, and 24 hours), we find that even for observations spaced 1 day apart for 97 days, we can predict stellar rotation to an rms precision of $\sim$6.2 days (4.69), unbiased over the range from $P_{\mathrm{rot}} \approx$ 5 - 35 days. Removing the $<$ 25% of over-predicted short-period stars with $P_{\mathrm{rot}} <$ 5.2 days, the fractional rms is $\approx$0.40.

With the results described above, we have established a baseline of performance for the stellar property information that can be extracted from this light curve data alone. Our modeling approach is generalizable to other time domain surveys, as well as other stellar property catalogs. The method presented here is not proposed to replace asteroseismology measurements from high-quality data. Instead, we predict these properties to demonstrate the capability of our approach, as well as the prospect of transferring the relationships establish with high-quality data to lower-quality data, where these measurements are more difficult to make.

Our ability to predict stellar properties is in part subject to the uncertainty on the input properties, and we expect given more precisely derived properties we could improve our predictions in some cases. To better determine some stellar properties, we could also incorporate other information such as photometry from *Gaia*, 2MASS and WISE, as well as stellar parallaxes from *Gaia*. For example, adding photometric information to the model would improve the precision with which we can recover $T_{\text{eff}}$.

We make our model code publicly available at https://github.com/kblancato/theia-net, which could be used to produce a rotation period catalogue, as well as other stellar property catalogues, for the TESS mission. Some more immediate improvements to our approach could include expanding the extent, as well as the precision quality, of the stellar property training sets, and incorporating *Gaia* photometric and parallax information in the model. More substantial aspects that could be investigated include adapting the model to permit unevenly sampled time series data, taking a generative modeling approach, incorporating the data errors on both the light curves and stellar properties, and interpreting the model in a physically meaningful way.

Looking forward, in the coming years ongoing and future missions will deliver time domain data for millions of stars. Extracting stellar properties from this data will be a rich pursuit, enabling the exciting potential of Galactic archaeology in the time domain.

*Acknowledgements.* The authors would like to thank Kathryn Johnston, David Blei, Gabriella Contardo, Maryum Sayeed, and Adam Wheeler for helpful feedback and discussions. We also thank Travis Berger for providing us his revised *Gaia-Kepler* stellar property catalog. We are grateful to the members of the Flatiron Institute's Scientific Computing Core for their support of the Flatiron's Rusty cluster, which was used to train all of the models for this work.

K.B. is supported by the NSF Graduate Research Fellowship under grant number DGE 16-44869. K.B. thanks the LSSTC Data Science Fellowship Program, her time as a Fellow has benefited this work. M.N. and D.H. are supported by the Alfred P. Sloan Foundation. R.A. acknowledges support from NASA award: 80NSSC20K1006. The research was supported by the Research Corporation for Science Advancement through Scialog award #26080.

This research was partially conducted during the Exostar19 program at the Kavli Institute for Theoretical Physics at UC Santa Barbara, which was supported in part by the National Science Foundation under Grant No. NSF PHY-1748958.

*Software:* `PyTorch` (Paszke et al. 2017), `scikit-learn` (Pedregosa et al. 2011), `Astropy` (Price-Whelan et al. 2018).

## 4.A   Data density

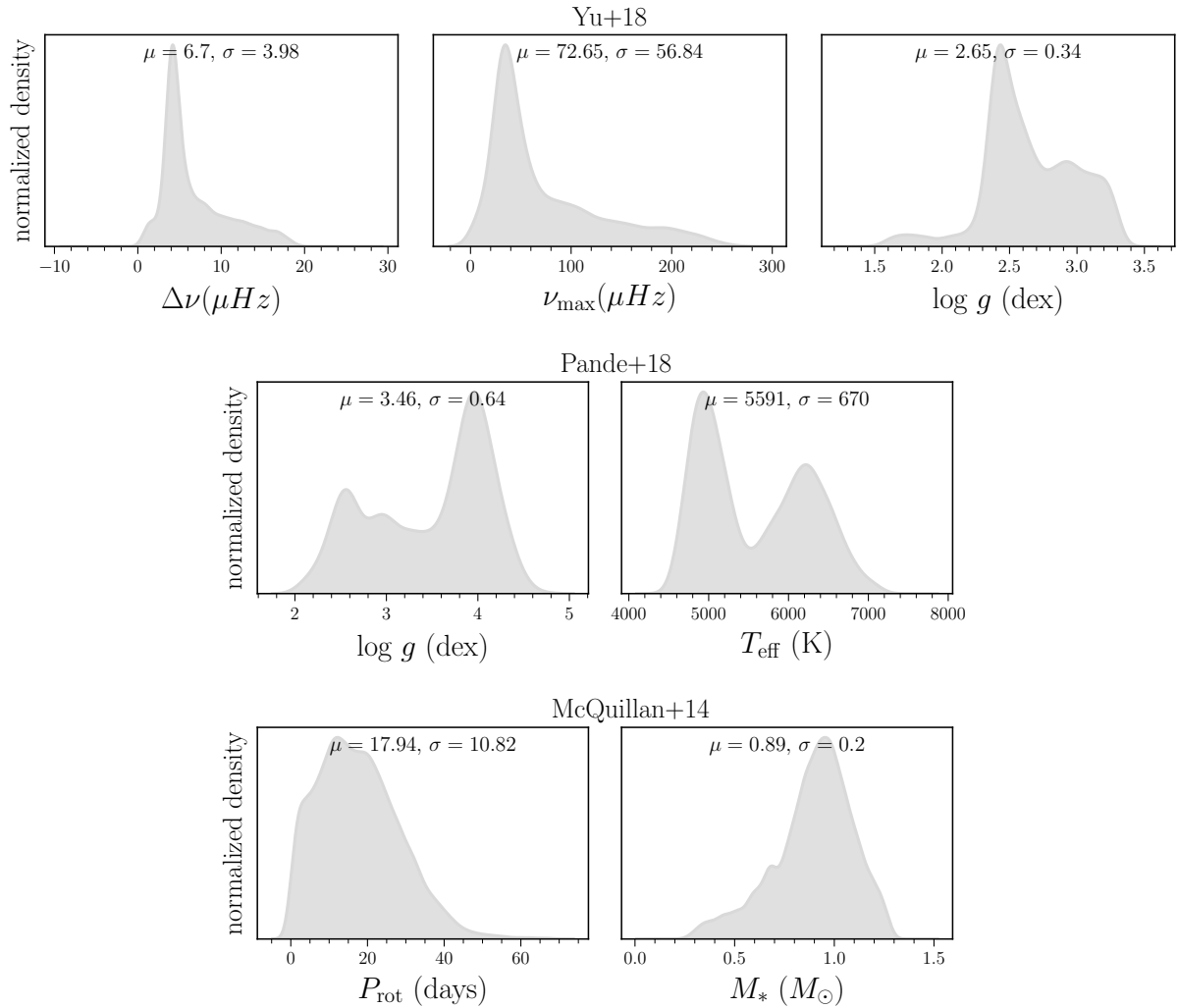

Figure 4.10: The distribution of stellar properties for the Yu et al. (2018) RGB sample (***top row***: $\Delta\nu$, $\nu_{\max}$, and $\log g$), the Pande et al. (2018) RGB and sub-giant sample (***middle row***: $\log g$ and $T_{\text{eff}}$), and the McQuillan et al. (2014) main sequence sample (***bottom row***: $P_{\text{rot}}$ and $M_*$)

|              |              | 97 days |                 |              | 62 days |                 |              |
|--------------|--------------|---------|-----------------|--------------|---------|-----------------|--------------|
|              |              | $r^2$   | $\Delta$        | rms          | $r^2$   | $\Delta$        | rms          |
| Yu+18        | $\Delta\nu$  | 0.95    | -0.17(-0.009)   | 0.89(0.14)   | 0.94    | -0.02(0.036)    | 0.97(0.17)   |
|              | $\nu_{max}$  | 0.96    | -3.51(-0.002)   | 11.85(0.2)   | 0.96    | -1.14(0.022)    | 12.02(0.22)  |
|              | $\log g$     | 0.97    | 0.008(0.004)    | 0.057(0.02)  | 0.97    | 0.006(0.004)    | 0.057(0.02)  |
| Pande+18     | $\log g$     | 0.89    | 0.07(0.026)     | 0.22(0.07)   | 0.92    | 0.01(0.006)     | 0.18(0.06)   |
|              | $T_{eff}$    | 0.79    | -66(-0.008)     | 309(0.05)    | 0.80    | -55(-0.007)     | 297(0.05)    |
| McQuillan+14 | $M_*$        | 0.40    | 0.009(0.057)    | 0.155(0.28)  | 0.38    | 0.012(0.062)    | 0.158(0.29)  |

|              |              | 27 days |                 |              | 14 days |                 |              |
|--------------|--------------|---------|-----------------|--------------|---------|-----------------|--------------|
|              |              | $r^2$   | $\Delta$        | rms          | $r^2$   | $\Delta$        | rms          |
| Yu+18        | $\Delta\nu$  | 0.93    | -0.3(-0.042)    | 1.06(0.18)   | 0.929   | 0.12(0.076)     | 1.09(0.23)   |
|              | $\nu_{max}$  | 0.91    | -3.21(0.01)     | 17.08(0.3)   | 0.94    | 0.22(0.064)     | 14.44(0.26)  |
|              | $\log g$     | 0.96    | 0.006(0.003)    | 0.071(0.03)  | 0.94    | -0.0(0.002)     | 0.086(0.04)  |
| Pande+18     | $\log g$     | 0.89    | -0.03(-0.003)   | 0.21(0.06)   | 0.89    | -0.05(-0.011)   | 0.21(0.06)   |
|              | $T_{eff}$    | 0.79    | -30(-0.002)     | 306(0.05)    | 0.77    | -35(-0.003)     | 320(0.06)    |
| McQuillan+14 | $M_*$        | 0.37    | 0.008(0.06)     | 0.159(0.3)   | 0.36    | 0.006(0.058)    | 0.16(0.3)    |

Figure 4.11: Model performance as a function of observation baseline (as described in Section 4.6.3). The fractional bias and fractional rms is indicated in parenthesis.

# Chapter 5

# Concluding remarks

This Dissertation makes unique contributions in three distinct subfields of astrophysics, namely: stellar astrophysics, Milky Way Galactic archaeology, and the evolution of massive galaxies. This work is united by a common goal: to decode the information contained in starlight. We follow this pursuit across multiple spatial and temporal scales, and along the way, we make use of a diversity of quantitative tools, including empirical, analytical, and numerical approaches. The methodological approaches that are taken in this work, as well as the physical interpretations put forth, are important building blocks in the pursuit to view the Milky Way, and its place in the Universe, more clearly. This work also lays down a foundation for many future endeavors. In what follows we first summarize the main insights of this Dissertation, and then end with a brief perspective on how it sits within the context of past, current, and future work.

## 5.1 Summary of contributions

Starting at the largest scales, in Chapter Two we use the Illustris cosmological simulation to investigate the physical origins of initial mass function (IMF) variability among, and within, massive elliptical galaxies. The IMF, which describes the relative number of stars formed at each stellar mass in a newly formed stellar population, has long been thought a universal feature of star-formation. However, more recently, there has been mounting observational evidence for systematic IMF variations, which suggest that the IMF scales with global galaxy properties, such as velocity dispersion and metallicity, as well as with galactocentric radius within individual galaxies. Understanding the origin of IMF variations is not only relevant to star-formation physics, as the IMF also influences nearly all observations we make of galaxies, such as luminosity and metallicity.

To examine the origin of these suggested IMF variations within a cosmological context, we assign stellar particles in the Illustris simulation with individual IMFs that depend on various physical conditions, such as velocity dispersion, metallicity, or SFR, at the time and place the stars are formed. We then follow the assembly of these populations from $z = 127$ to $z = 0$ and examine how galaxy-wide properties are linked to the properties of the stellar populations that compose a galaxy's stellar content. Our main result is that applying the observed relations between IMF and galactic properties to the conditions at the individual star-formation sites does not result in strong enough IMF variations between galaxies to match the observed trends. The origin of this is related to the hierarchical nature of massive galaxy formation, which has implications for the reliability of the strength of the observed IMF trends, the current ability of cosmological simulations to capture certain physical conditions in galaxies, and theories of star-formation working to explain the physical origin of a variable IMF.

Next, in Chapter Three, we narrow our focus to our own Galaxy and explore how the ratio of two $\alpha$-abundances, namely [Mg/Si], probes the chemical and dynamical evolution of the Milky Way's disk. Despite belonging to the same elemental family, sharing common nucleosynthetic lineages, there are subtle differences in the production mechanisms among the various individual $\alpha$-elements. For instance, hydrostatic $\alpha$-elements, including Mg, are produced in the outermost shells of massive stars during the hydrostatic burning phase and are dispersed into the ISM through CC-SN. In contrast, explosive $\alpha$-elements, including Si, are produced in shells that lie closer to the cores of massive stars during the explosive nucleosynthesis leading up to the CC-SN explosion. Intrigued by these slight distinctions in their posited origins, we explore what the relative enrichment of Mg and Si, that stars inherit from their natal ISM gas at the time of stellar birth, can tell us about how the Milky Way may have arrived at its current state.

We examine the [Mg/Si] abundance across an ensemble of stars, with each star currently residing and having been formed, in various locations in the Galaxy, as informed by their measured astrometric properties from *Gaia*, and with each star also having been formed at various points in time in the Galaxy's evolution, as informed by their spectroscopically inferred stellar ages from APOGEE data. In our exploration, we find slight but significant relationships between [Mg/Si] and $\alpha$-enhancement, stellar age, [Fe/H], location in the Galaxy, and orbital actions. Through the lens of this single abundance ratio, we build up a view of the Milky Way's chemical and dynamical history. To interpret this empirical perspective, we attempt to predict the Mg and Si abundances of stars with analytical models of the Galaxy's gas enrichment history. We find that we are not able to reproduce the abundances for the stars that we fit, highlighting the tensions between the nucleosynthetic yield tables, the chemical evolution models, and the data. This suggests a more data-driven approach to yield tables and chemical evolution modeling as a path forward in the pursuit of deriving insight from large spectroscopic surveys. Additionally, the approach of this work is complementary to that taken in Ness et al. (2019b). In this work, we find that only the [Fe/H] abundance, in addition to stellar age, is necessary to recover a star's orbital properties, and that including the additional set of abundance ratios provided by the APOGEE spectra does not add significantly to this determination. By both isolating and deeply investigating specific features of datasets, as well as determining what features do not add value to a particular phenomenon we are interested in understanding, we can begin to make more sense of highly-dimensional data.

Lastly, at the finest spatial and temporal scales we consider, in Chapter Four we develop a data-driven deep learning method to derive stellar properties from photometric time series data. High-cadence stellar photometry, from missions such as *Kepler*, has revealed how stellar variability, on multiple timescales, contains signatures of stellar oscillations, convection, and rotation, which are linked to fundamental stellar parameters such as mass, radius, and age. The empirical relationships that have been established from this data, as well as the interpretation of them from a theoretical perspective, has advanced our understanding of stellar interiors, and of the Milky Way's stellar populations. Stellar time-domain data that will be delivered by TESS, LSST, and other future missions will strengthen these existing understandings, as well as reveal new connections between time domain variability, stellar physics, and the formation history of the Milky Way.

Given the volume and variety of the forthcoming data, progress in these pursuits will require the development of fast and automated methods for deriving stellar properties from variable quality stellar light curves. To lay groundwork for this, we use *Kepler* light curves and various stellar property catalogs as training sets, and build convolutional neural network models to homogeneously predict a collection of stellar properties (evolutionary state, $\Delta\nu$, $\nu_{\max}$, $\log g$, $T_{\rm eff}$ and $P_{\rm rot}$) directly from time series data, across various baseline and cadences. Subject to the uncertainty on the training labels, as well as various modeling choices and assumptions made, we find that stellar property information is preserved in light curves with baselines as short as 14 days of continuous observations, as well as cadences as sparse as one observation per day.

Data-driven methods for estimating stellar properties from time-domain data will provide stellar parameter estimates for millions of stars throughout the Galaxy. This will facilitate the exciting endeavor of Galactic archaeology in the time domain. Furthermore, by exploring models that can learn from light curve data closer to its original form, as we do in this Chapter, there is the possibility of extracting insight that may be lost when transforming the data into alternate representations, which is often done to make data easier for us to visualize. The approach we begin to explore here, in combination with the development of a meaningful framework for model interpretation that is carefully guided by our specialized domain knowledge, has the potential to enable new perspectives on various pursuits of stellar astrophysics, such as gyrochronology.

## 5.2 Looking forward

Stars carry with them signatures of their intrinsic properties, as well as the environments in which they were formed. By examining these signatures, and how they vary across populations of stars, we are able to gain insight into larger, often unobservable, astrophysical phenomena. However, these signatures are often encoded, across multiple spatial and temporal scales, in the observations we make of stars. While the primary ways in which we measure starlight have remained fundamentally unchanged since the times of early astronomers, innovations in astronomical instrumentation and large observing campaigns have significantly advanced the quality and quality of our stellar observations. With a careful eye, Annie Jump Cannon examined the spectra of over two hundred thousand individual stars – laborious, yet important, work that underpins much of modern stellar astrophysics. Today, we have now observed the properties of millions of stars, and, with the next decade of planned surveys, our collection of stellar observations will grow by several more orders of magnitude.

The growing volume of stellar observations has been occurring simultaneously with and is being facilitated by developments in our computational capabilities. With rising computational power and algorithmic sophistication, we now have many new ways to dissect and find meaning in large and diverse astronomical datasets. Some of these same tools, being employed across various pursuits of society, are confronting us with new and complex challenges, which will require a diversity of perspectives to make sense of. This Dissertation is a testament to the current state of this pursuit in astronomy, and an early inquiry into the possibilities to come.

The investigations and results in this Dissertation suggest that extracting insight from these stellar observations will continue to be a rich undertaking, requiring both playful exploration and careful formulation of specific questions, as well as a strengthened commitment to physical interpretation. This will be a creative and collaborative pursuit, from which we will gain new perspectives on our own Milky Way and its place in the Universe.

Gaia Collaboration, Babusiaux, C., van Leeuwen, F., Barstow, M. A., Jordi, C., Vallenari, A., Bossini, D., Bressan, A., Cantat-Gaudin, T., van Leeuwen, M., Brown, A. G. A., Prusti, T., de Bruijne, J. H. J., Bailer-Jones, C. A. L., Biermann, M., Evans, D. W., Eyer, L., Jansen, F., Klioner, S. A., Lammers, U., Lindegren, L., Luri, X., Mignard, F., Panem, C., Pourbaix, D., Randich, S., Sartoretti, P., Siddiqui, H. I., Soubiran, C., Walton, N. A., Arenou, F., Bastian, U., Cropper, M., Drimmel, R., Katz, D., Lattanzi, M. G., Bakker, J., Cacciari, C., Castañeda, J., Chaoul, L., Cheek, N., De Angeli, F., Fabricius, C., Guerra, R., Holl, B., Masana, E., Messineo, R., Mowlavi, N., Nienartowicz, K., Panuzzo, P., Portell, J., Riello, M., Seabroke, G. M., Tanga, P., Thévenin, F., Gracia-Abril, G., Comoretto, G., Garcia-Reinaldos, M., Teyssier, D., Altmann, M., Andrae, R., Audard, M., Bellas-Velidis, I., Benson, K., Berthier, J., Blomme, R., Burgess, P., Busso, G., Carry, B., Cellino, A., Clementini, G., Clotet, M., Creevey, O., Davidson, M., De Ridder, J., Delchambre, L., Dell'Oro, A., Ducourant, C., Fernández-Hernández, J., Fouesneau, M., Frémat, Y., Galluccio, L., García-Torres, M., González-Núñez, J., González-Vidal, J. J., Gosset, E., Guy, L. P., Halbwachs, J. L., Hambly, N. C., Harrison, D. L., Hernández, J., Hestroffer, D., Hodgkin, S. T., Hutton, A., Jasniewicz, G., Jean-Antoine-Piccolo, A., Jordan, S., Korn, A. J., Krone-Martins, A., Lanzafame, A. C., Lebzelter, T., Löffler, W., Manteiga, M., Marrese, P. M., Martín-Fleitas, J. M., Moitinho, A., Mora, A., Muinonen, K., Osinde, J., Pancino, E., Pauwels, T., Petit, J. M., Recio-Blanco, A., Richards, P. J., Rimoldini, L., Robin, A. C., Sarro, L. M., Siopis, C., Smith, M., Sozzetti, A., Süveges, M., Torra, J., van Reeven, W., Abbas, U., Abreu Aramburu, A., Accart, S., Aerts, C., Altavilla, G., Álvarez, M. A., Alvarez, R., Alves, J., Anderson, R. I., Andrei, A. H., Anglada Varela, E., Antiche, E., Antoja, T., Arcay, B., Astraatmadja, T. L., Bach, N., Baker, S. G., Balaguer-Núñez, L., Balm, P., Barache, C., Barata, C., Barbato, D., Barblan, F., Barklem, P. S., Barrado, D., Barros, M., Bartholomé Muñoz, L., Bassilana, J. L., Becciani, U., Bellazzini, M., Berihuete, A., Bertone, S., Bianchi, L., Bienaymé, O., Blanco-Cuaresma, S., Boch, T., Boeche, C., Bombrun, A., Borrachero, R., Bouquillon, S., Bourda, G., Bragaglia, A., Bramante, L., Breddels, M. A., Brouillet, N., Brüsemeister, T., Brugaletta, E., Bucciarelli, B., Burlacu, A., Busonero, D., Butkevich, A. G., Buzzi, R., Caffau, E., Cancelliere, R., Cannizzaro, G., Carballo, R., Carlucci, T., Carrasco, J. M., Casamiquela, L., Castellani, M., Castro-Ginard, A., Charlot, P., Chemin, L., Chiavassa, A., Cocozza, G., Costigan, G., Cowell, S., Crifo, F., Crosta, M., Crowley, C., Cuypers, J., Dafonte, C., Damerdji, Y., Dapergolas, A., David, P., David, M., de Laverny, P., De Luise, F., De March, R., de Martino, D., de Souza, R., de Torres, A., Debosscher, J., del Pozo, E., Delbo, M., Delgado, A., Delgado, H. E., Diakite, S., Diener, C., Distefano, E., Dolding, C., Drazinos, P., Durán, J., Edvardsson, B., Enke, H., Eriksson, K., Esquej, P., Eynard Bontemps, G., Fabre, C., Fabrizio, M., Faigler, S., Falcão, A. J., Farràs Casas, M., Federici, L., Fedorets, G., Fernique, P., Figueras, F., Filippi, F., Findeisen, K., Fonti, A., Fraile, E., Fraser, M., Frézouls, B., Gai, M., Galleti, S., Garabato, D., García-Sedano, F., Garofalo, A., Garralda, N., Gavel, A., Gavras, P., Gerssen, J., Geyer, R., Giacobbe, P., Gilmore, G., Girona, S., Giuffrida, G., Glass, F., Gomes, M., Granvik, M., Gueguen, A., Guerrier, A., Guiraud, J.,

M., Bellas-Velidis, I., Benson, K., Berthier, J., Blomme, R., Burgess, P., Busso, G., Carry, B., Cellino, A., Clementini, G., Clotet, M., Creevey, O., Davidson, M., De Ridder, J., Delchambre, L., Dell'Oro, A., Ducourant, C., Fernández-Hernández, J., Fouesneau, M., Frémat, Y., Galluccio, L., García-Torres, M., González-Núñez, J., González-Vidal, J. J., Gosset, E., Guy, L. P., Halbwachs, J. L., Hambly, N. C., Harrison, D. L., Hernández, J., Hestroffer, D., Hodgkin, S. T., Hutton, A., Jasniewicz, G., Jean-Antoine-Piccolo, A., Jordan, S., Korn, A. J., Krone-Martins, A., Lanzafame, A. C., Lebzelter, T., Löffler, W., Manteiga, M., Marrese, P. M., Martín-Fleitas, J. M., Moitinho, A., Mora, A., Muinonen, K., Osinde, J., Pancino, E., Pauwels, T., Petit, J. M., Recio-Blanco, A., Richards, P. J., Rimoldini, L., Robin, A. C., Sarro, L. M., Siopis, C., Smith, M., Sozzetti, A., Süveges, M., Torra, J., van Reeven, W., Abbas, U., Abreu Aramburu, A., Accart, S., Aerts, C., Altavilla, G., Álvarez, M. A., Alvarez, R., Alves, J., Anderson, R. I., Andrei, A. H., Anglada Varela, E., Antiche, E., Antoja, T., Arcay, B., Astraatmadja, T. L., Bach, N., Baker, S. G., Balaguer-Núñez, L., Balm, P., Barache, C., Barata, C., Barbato, D., Barblan, F., Barklem, P. S., Barrado, D., Barros, M., Barstow, M. A., Bartholomé Muñoz, S., Bassilana, J. L., Becciani, U., Bellazzini, M., Berihuete, A., Bertone, S., Bianchi, L., Bienaymé, O., Blanco-Cuaresma, S., Boch, T., Boeche, C., Bombrun, A., Borrachero, R., Bossini, D., Bouquillon, S., Bourda, G., Bragaglia, A., Bramante, L., Breddels, M. A., Bressan, A., Brouillet, N., Brüsemeister, T., Brugaletta, E., Bucciarelli, B., Burlacu, A., Busonero, D., Butkevich, A. G., Buzzi, R., Caffau, E., Cancelliere, R., Cannizzaro, G., Cantat-Gaudin, T., Carballo, R., Carlucci, T., Carrasco, J. M., Casamiquela, L., Castellani, M., Castro-Ginard, A., Charlot, P., Chemin, L., Chiavassa, A., Cocozza, G., Costigan, G., Cowell, S., Crifo, F., Crosta, M., Crowley, C., Cuypers, J., Dafonte, C., Damerdji, Y., Dapergolas, A., David, P., David, M., de Laverny, P., De Luise, F., De March, R., de Martino, D., de Souza, R., de Torres, A., Debosscher, J., del Pozo, E., Delbo, M., Delgado, A., Delgado, H. E., Di Matteo, P., Diakite, S., Diener, C., Distefano, E., Dolding, C., Drazinos, P., Durán, J., Edvardsson, B., Enke, H., Eriksson, K., Esquej, P., Eynard Bontemps, G., Fabre, C., Fabrizio, M., Faigler, S., Falcão, A. J., Farràs Casas, M., Federici, L., Fedorets, G., Fernique, P., Figueras, F., Filippi, F., Findeisen, K., Fonti, A., Fraile, E., Fraser, M., Frézouls, B., Gai, M., Galleti, S., Garabato, D., García-Sedano, F., Garofalo, A., Garralda, N., Gavel, A., Gavras, P., Gerssen, J., Geyer, R., Giacobbe, P., Gilmore, G., Girona, S., Giuffrida, G., Glass, F., Gomes, M., Granvik, M., Gueguen, A., Guerrier, A., Guiraud, J., Gutiérrez-Sánchez, R., Haigron, R., Hatzidimitriou, D., Hauser, M., Haywood, M., Heiter, U., Helmi, A., Heu, J., Hilger, T., Hobbs, D., Hofmann, W., Holland, G., Huckle, H. E., Hypki, A., Icardi, V., Janßen, K., Jevardat de Fombelle, G., Jonker, P. G., Juhász, Á. L., Julbe, F., Karampelas, A., Kewley, A., Klar, J., Kochoska, A., Kohley, R., Kolenberg, K., Kontizas, M., Kontizas, E., Koposov, S. E., Kordopatis, G., Kostrzewa-Rutkowska, Z., Koubsky, P., Lambert, S., Lanza, A. F., Lasne, Y., Lavigne, J. B., Le Fustec, Y., Le Poncin-Lafitte, C., Lebreton, Y., Leccia, S., Leclerc, N., Lecoeur-Taibi, I., Lenhardt, H., Leroux, F., Liao, S., Licata, E., Lindstrøm, H. E. P., Lister, T. A., Livanou, E., Lobel, A., López, M., Managau, S., Mann, R. G., Mantelet, G., Marchal, O., Marchant, J. M.,

J. M., Moitinho, A., Mora, A., Muinonen, K., Osinde, J., Pauwels, T., Petit, J. M., Recio-Blanco, A., Richards, P. J., Rimoldini, L., Sarro, L. M., Siopis, C., Smith, M., Sozzetti, A., Süveges, M., Torra, J., van Reeven, W., Abbas, U., Abreu Aramburu, A., Accart, S., Aerts, C., Altavilla, G., Álvarez, M. A., Alvarez, R., Alves, J., Anderson, R. I., Andrei, A. H., Anglada Varela, E., Antiche, E., Arcay, B., Astraatmadja, T. L., Bach, N., Baker, S. G., Balaguer-Núñez, L., Balm, P., Barache, C., Barata, C., Barbato, D., Barblan, F., Barklem, P. S., Barrado, D., Barros, M., Barstow, M. A., Bartholomé Muñoz, S., Bassilana, J. L., Becciani, U., Bellazzini, M., Berihuete, A., Bertone, S., Bianchi, L., Bienaymé, O., Blanco-Cuaresma, S., Boch, T., Boeche, C., Bombrun, A., Borrachero, R., Bossini, D., Bouquillon, S., Bourda, G., Bragaglia, A., Bramante, L., Bressan, A., Brouillet, N., Brüsemeister, T., Brugaletta, E., Bucciarelli, B., Burlacu, A., Busonero, D., Butkevich, A. G., Buzzi, R., Caffau, E., Cancelliere, R., Cannizzaro, G., Cantat-Gaudin, T., Carballo, R., Carlucci, T., Carrasco, J. M., Casamiquela, L., Castellani, M., Castro-Ginard, A., Charlot, P., Chemin, L., Chiavassa, A., Cocozza, G., Costigan, G., Cowell, S., Crifo, F., Crosta, M., Crowley, C., Cuypers, J., Dafonte, C., Damerdji, Y., Dapergolas, A., David, P., David, M., de Laverny, P., De Luise, F., De March, R., de Martino, D., de Souza, R., de Torres, A., Debosscher, J., del Pozo, E., Delbo, M., Delgado, A., Delgado, H. E., Di Matteo, P., Diakite, S., Diener, C., Distefano, E., Dolding, C., Drazinos, P., Durán, J., Edvardsson, B., Enke, H., Eriksson, K., Esquej, P., Eynard Bontemps, G., Fabre, C., Fabrizio, M., Faigler, S., Falcão, A. J., Farràs Casas, M., Federici, L., Fedorets, G., Fernique, P., Figueras, F., Filippi, F., Findeisen, K., Fonti, A., Fraile, E., Fraser, M., Frézouls, B., Gai, M., Galleti, S., Garabato, D., García-Sedano, F., Garofalo, A., Garralda, N., Gavel, A., Gavras, P., Gerssen, J., Geyer, R., Giacobbe, P., Gilmore, G., Girona, S., Giuffrida, G., Glass, F., Gomes, M., Granvik, M., Gueguen, A., Guerrier, A., Guiraud, J., Gutiérrez-Sánchez, R., Hofmann, W., Holland, G., Huckle, H. E., Hypki, A., Icardi, V., Janßen, K., Jevardat de Fombelle, G., Jonker, P. G., Juhász, Á. L., Julbe, F., Karampelas, A., Kewley, A., Klar, J., Kochoska, A., Kohley, R., Kolenberg, K., Kontizas, M., Kontizas, E., Koposov, S. E., Kordopatis, G., Kostrzewa-Rutkowska, Z., Koubsky, P., Lambert, S., Lanza, A. F., Lasne, Y., Lavigne, J. B., Le Fustec, Y., Le Poncin-Lafitte, C., Lebreton, Y., Leccia, S., Leclerc, N., Lecoeur-Taibi, I., Lenhardt, H., Leroux, F., Liao, S., Licata, E., Lindstrøm, H. E. P., Lister, T. A., Livanou, E., Lobel, A., López, M., Managau, S., Mann, R. G., Mantelet, G., Marchal, O., Marchant, J. M., Marconi, M., Marinoni, S., Marschalkó, G., Marshall, D. J., Martino, M., Marton, G., Mary, N., Matijevič, G., Mazeh, T., Messina, S., Michalik, D., Millar, N. R., Molina, D., Molinaro, R., Molnár, L., Montegriffo, P., Mor, R., Morbidelli, R., Morel, T., Morris, D., Mulone, A. F., Muraveva, T., Musella, I., Nelemans, G., Nicastro, L., Noval, L., O'Mullane, W., Ordénovic, C., Ordóñez-Blanco, D., Osborne, P., Pagani, C., Pagano, I., Pailler, F., Palacin, H., Palaversa, L., Panahi, A., Pawlak, M., Piersimoni, A. M., Pineau, F. X., Plachy, E., Plum, G., Poggio, E., Poujoulet, E., Prša, A., Pulone, L., Racero, E., Ragaini, S., Rambaux, N., Ramos-Lerate, M., Regibo, S., Riclet, F., Ripepi, V., Riva, A., Rivard, A., Rixon, G., Roegiers, T., Roelens, M., Romero-Gómez, M., Rowell, N., Royer, F., Ruiz-Dern, L., Sadowski, G., Sagristà

Breddels, M. A., Brouillet, N., Brüsemeister, T., Bucciarelli, B., Budnik, F., Burgess, P., Burgon, R., Burlacu, A., Busonero, D., Buzzi, R., Caffau, E., Cambras, J., Campbell, H., Cancelliere, R., Cantat-Gaudin, T., Carlucci, T., Carrasco, J. M., Castellani, M., Charlot, P., Charnas, J., Charvet, P., Chassat, F., Chiavassa, A., Clotet, M., Cocozza, G., Collins, R. S., Collins, P., Costigan, G., Crifo, F., Cross, N. J. G., Crosta, M., Crowley, C., Dafonte, C., Damerdji, Y., Dapergolas, A., David, P., David, M., De Cat, P., de Felice, F., de Laverny, P., De Luise, F., De March, R., de Martino, D., de Souza, R., Debosscher, J., del Pozo, E., Delbo, M., Delgado, A., Delgado, H. E., di Marco, F., Di Matteo, P., Diakite, S., Distefano, E., Dolding, C., Dos Anjos, S., Drazinos, P., Durán, J., Dzigan, Y., Ecale, E., Edvardsson, B., Enke, H., Erdmann, M., Escolar, D., Espina, M., Evans, N. W., Eynard Bontemps, G., Fabre, C., Fabrizio, M., Faigler, S., Falcão, A. J., Farràs Casas, M., Faye, F., Federici, L., Fedorets, G., Fernández-Hernández, J., Fernique, P., Fienga, A., Figueras, F., Filippi, F., Findeisen, K., Fonti, A., Fouesneau, M., Fraile, E., Fraser, M., Fuchs, J., Furnell, R., Gai, M., Galleti, S., Galluccio, L., Garabato, D., García-Sedano, F., Garé, P., Garofalo, A., Garralda, N., Gavras, P., Gerssen, J., Geyer, R., Gilmore, G., Girona, S., Giuffrida, G., Gomes, M., González-Marcos, A., González-Núñez, J., González-Vidal, J. J., Granvik, M., Guerrier, A., Guillout, P., Guiraud, J., Gúrpide, A., Gutiérrez-Sánchez, R., Guy, L. P., Haigron, R., Hatzidimitriou, D., Haywood, M., Heiter, U., Helmi, A., Hobbs, D., Hofmann, W., Holl, B., Holland , G., Hunt, J. A. S., Hypki, A., Icardi, V., Irwin, M., Jevardat de Fombelle, G., Jofré, P., Jonker, P. G., Jorissen, A., Julbe, F., Karampelas, A., Kochoska, A., Kohley, R., Kolenberg, K., Kontizas, E., Koposov, S. E., Kordopatis, G., Koubsky, P., Kowalczyk, A., Krone-Martins, A., Kudryashova, M., Kull, I., Bachchan, R. K., Lacoste-Seris, F., Lanza, A. F., Lavigne, J. B., Le Poncin-Lafitte, C., Lebreton, Y., Lebzelter, T., Leccia, S., Leclerc, N., Lecoeur-Taibi, I., Lemaitre, V., Lenhardt, H., Leroux, F., Liao, S., Licata, E., Lindstrøm, H. E. P., Lister, T. A., Livanou, E., Lobel, A., Löffler, W., López, M., Lopez-Lozano, A., Lorenz, D., Loureiro, T., MacDonald, I., Magalhães Fernandes, T., Managau, S., Mann, R. G., Mantelet, G., Marchal, O., Marchant, J. M., Marconi, M., Marie, J., Marinoni, S., Marrese, P. M., Marschalkó, G., Marshall, D. J., Martín-Fleitas, J. M., Martino, M., Mary, N., Matijevič, G., Mazeh, T., McMillan, P. J., Messina, S., Mestre, A., Michalik, D., Millar, N. R., Miranda, B. M. H., Molina, D., Molinaro, R., Molinaro, M., Molnár, L., Moniez, M., Montegriffo, P., Monteiro, D., Mor, R., Mora, A., Morbidelli, R., Morel, T., Morgenthaler, S., Morley, T., Morris, D., Mulone, A. F., Muraveva, T., Musella, I., Narbonne, J., Nelemans, G., Nicastro, L., Noval, L., Ordénovic, C., Ordieres-Meré, J., Osborne, P., Pagani, C., Pagano, I., Pailler, F., Palacin, H., Palaversa, L., Parsons, P., Paulsen, T., Pecoraro, M., Pedrosa, R., Pentikäinen, H., Pereira, J., Pichon, B., Piersimoni, A. M., Pineau, F. X., Plachy, E., Plum, G., Poujoulet, E., Prša, A., Pulone, L., Ragaini, S., Rago, S., Rambaux, N., Ramos-Lerate, M., Ranalli, P., Rauw, G., Read, A., Regibo, S., Renk, F., Reylé, C., Ribeiro, R. A., Rimoldini, L., Ripepi, V., Riva, A., Rixon, G., Roelens, M., Romero-Gómez, M., Rowell, N., Royer, F., Rudolph, A., Ruiz-Dern, L., Sadowski, G., Sagristà Sellés, T., Sahlmann, J., Salgado, J., Salguero, E., Sarasso, M., Savietto, H., Schnorhk, A., Schultheis, M., Sciacca, E., Segol,